\documentclass[a4paper,11pt,times,authoryear]{Classes/PHDTHESISPSNPDF}

\input{Preamble/preamble}

\title{Study of Solar Jets and Related Flares} 

\author{Reetika Joshi}

\dept{Department of Physics (CAS)}

\university{Kumaun University, Nainital, India}

\supervisor {\large Prof. Ramesh Chandra}

\renewcommand{\submissiontext}{A Thesis\\
Submitted for the Degree of}
\degreetitle{Doctor of Philosophy\\
in Physics}

\degreedate{31 March, 2021} 

\ifdefineAbstract
\fi

\ifdefineChapter
\fi

\begin{document}

\frontmatter

\maketitle


\begin{dedication} 

\centering
{\Huge \it Dedicated to my loving family and teachers\dots}

\end{dedication}

\begin{declaration}

I declare that this written submission represents my ideas in my own words and where others' ideas or words have been included, I have adequately cited and referenced the original sources. I also declare that I have adhered to all principles of academic honesty and integrity and have not misrepresented or fabricated or falsified any idea/data/fact/source in my submission. I understand that any violation of the above will be cause for disciplinary action by the University and can also evoke penal action from the sources which have thus not been properly cited or from whom proper permission has not been taken when needed.
~~\\
~~\\


\end{declaration}
\begin{acknowledgements}     
{Completion of my doctoral thesis have been done with the assistance, patience, and support of many individuals.}
{First and foremost, I offer my sincere gratitude to my thesis advisor Prof. Ramesh Chandra for his confidence in me and supporting me during these past five years. In his company of keen and scholastic observations, disciplinary adherence to rules, commitment to highest standards, and many more happy moments of negotiations and sharing, I realised that true happiness exists in learning together through interactions. 
I will be indebted to him my entire life for the freedom he provided to pursue various projects without any objection, for his unfailing care, and endless help throughout my Ph.D. journey.
I am also grateful to him for motivating me in difficult situations and allowing me to travel for attending a number of national and international scientific conferences. This research journey with him had been a process of learning along with self inquiry for me and I definitely enjoyed it.  
I am sure he will stand by my side in my future endeavours.}

{I would like to express my gratitude to the Head, Department of Physics, Prof. H.C. Chandola for his support and encouragement. I am thankful to him for providing me with all the necessary facilities for my research work in the department. I will always be grateful to him for taking the initiative to conduct various conferences and seminars in our department, which really helped me to gain more confidence and leadership qualities. Additionally, I extend my gratitude to Dr. Wahab Uddin of ARIES, who motivated me towards the solar observations and encouraged me to pursue a quality research. I also thank to Dr. Hum Chand for sowing the seeds for scientific research during my Master's project with him.}

{A special thanks to the Department of Science and Technology (DST) for the INSPIRE fellowship. I feel honoured to express my gratitude to the Indo-French Centre for the Promotion of Advanced Research (IFCPAR/CEFIPRA) for a Raman Charpak Fellowship, under which I worked at 
Observatorie de Paris, France. 
I am also thankful to the Scientific Committee on Solar Terrestrial Physics (SCOSTEP) for the SCOSTEP Visiting Scholarship.}

{Furthermore I would like to thank my teachers in the department Prof. Sanjay Pant, Prof. Shuchi Bisht, Dr. Alok Durgapal, Dr. Bimal Pande, Dr. Seema Pande, Dr. Grish Chandra, Dr. Rajkumar  for their continuous support and teaching me Physics, and all other faculties who have taught us during our research coursework. I also thank Prof. B.L. Sah, Director HRDC team at Kumaun University Nainital, for conducting an interactive research workshop during our course work to learn the research methodology and inter-disciplinary research approach.
I would like to thank the non-teaching staff of the Physics department for their cooperation and building a smooth working environment.
Many thanks to our administrative and accounts staff of the university for their timely help to complete the administrative works.
No research exists without the library, the centre of learning resources. I express my gratitude to all the library staff in our campus and university for their services. I would like to say a warm thank to solar physics group at ARIES and their computer staff, specially to Navin sir who helped me for the IDL installation and other technical issues during my early Ph.D. days.}

{The thesis would not have come to a successful completion without a good collaborative research work. In this regard, I would like to say a special thank to Drs. Guillaume Aulanier and Brigitte Schmieder for accepting me as a Raman Charpak fellow and providing me all the facilities for conducting a part of my research in France. I will always be grateful to him and other group members at Observatorie de Paris, Dr. Pascal D\'emoulin, Dr. Veroniqu\'e Bommier, Dr. Etienne Pariat for their valuable discussions and to Dr. Fernando Moreno Insertis for the fruitful suggestions which helped me to improve my understanding about solar jets. I can't thank enough to Dr. Brigitte Schmieder for all the important scientific discussions 
and teaching me the observational techniques and spectroscopic analysis. 
I really appreciate her endless efforts to make my stay smooth 
in Paris even during the pandemic COVID19 and caring for me like a family member. She was and remains my best role model for a scientist, mentor, and teacher. Furthermore, I express my heart-felt gratitude to Dr. Guo Yang for inviting me to Nanjing University and Prof. Yuming Wang for hosting me under SCOSTEP visiting program. 
I would also like to extend my appreciation to Prof. Ivan Zhelyazkov and the group in Sofia university for the fruitful discussions.
I thank to all the co-authors of our papers for their valuable inputs and discussions.}

{I also thank my friends and seniors Deependra sir, Geeta, Mona, Deepak, Nupur mam, Nisha mam, Harmeen mam, Shahid sir, Mahesh sir, Divya di, Arti di, Aabha di, and many more for providing support and friendship that I needed. 
I am always delighted with the company of my junior colleague Pooja, who is more like a sister to me, for her selfless affection and being a constant support. 
I am thankful to Dr. Arun K. Awasthi, Dr. Devendra Bisht, and Nisha for helping me to get settled during my visit to China. Many many thanks to my friends in Paris Tomin, Sasi, Sanket, Abhijit, and Suchitra for such a joyful stay.}

{The main two pillars behind my joyful Ph.D. journey are my two siblings: Kumkum (Badi), and Kamal (Bhaiyya). They provided me the safe space where I could comfort, and release my pressure. 
  Not only personal, but they both contribute to my academic learnings by every day discussions and mutual corrections. They made my tough days sail through smoothly and made my happy days happier.} 
 {And at last, I am the most fortunate person to have a family who are pillars of strength:  my parents (Maa-Papa), and my grandmother (Nani). Their encouragement, trust, support, and constant motivation made me what I am today.
My journey and achievements are not only mine but shared with them.}\\
\\

~~~~~~~~~~~~~~~~~~~~~~~~~~~~~~~~~~~~~~~~~~~~~~~~~~~~~~~~~~~~~~~~~~~~~~~~~~~~~~~~~~~~~~~~~~~~~~~~~~~~~~~~~~~~~~~~~~{\bf Reetika Joshi}
\end{acknowledgements}



\chapter*{\centering Preface}
Solar jets are impulsive and collimated plasma ejections that flow with high velocity along open magnetic field lines. Solar flares at the footpoint of the jet are believed to promote the force for pushing the plasma material upward.
These jets associated with solar flares may be a source for transporting a significant mass and energy from the lower solar atmosphere to the upper coronal heights and consequently heating the solar corona and accelerating the solar wind.
In this way, solar jets are key tools to probe the broad dimensions of solar heliospheric problems and the Sun-Earth connections, therefore the thesis is centered on the ``{\it study of solar jets and related flares}”. 





Chapter \ref{c1} deals with the general introduction about the Sun, review of literature concerning the observations, and the existing theoretical models of solar jets. A description about the various space as well as  ground based observatories, and data analysis techniques  
are also presented in this chapter.

A detail analysis of confined and eruptive solar flares and associated jets from solar active region (AR) NOAA 12035 is presented in chapter \ref{c3}. The solar flares show a transition from eruptive to confined behaviour. 
To study the connectivity of the different flux domains and their evolution, we compute a potential magnetic field of the AR. Quasi-separatrix layers (QSL) are obtained from the magnetic field extrapolation. These flares tend to be more-and-more confined when the overlying field gradually become less-and-less anti-parallel, as a direct result of changes in the photospheric flux distribution. The observed solar jets show a slipping motion from one reconnection site to other. The slippage of jets is explained by the complex topology of the AR with the presence of a few low-altitude null points, many quasi-separatrix layers and their interactions.

 A case study of multi-temperature coronal jets for emerging flux MHD models is presented in chapter \ref{c4}. The jets from AR NOAA 12644 at the western solar limb were observed in all the hot filters of Atmospheric Imaging Assembly \citep[AIA,][]{Lemen2012} and in the transition region temperatures by  Interface Region Imaging Spectrograph \citep[IRIS,][]{Pontieu2014}. 
 In the pre-phase of the jets, quasi-periodic intensity oscillations are observed, which are in phase with the small ejections; they have a period between 2 to 6 minutes, and are reminiscent of acoustic or MHD waves.
The jets are initiated at the top of a canopy-like double-chambered structure with cool emission on one
and hot emission on the other side.
The hot jets are collimated in the hot temperature filters, have high velocities (around 250 km s$^{-1}$) and accompanied by the cool surges and ejected kernels. 
The cool surge with kernels show a direct alignment with the cool ejection and plasmoids which have described in theoretical models.
This series of hot jets and cool surges provides a good evidence for the 2D and 3D MHD models (\citealt{Moreno2008}; \citealt{Moreno2013}) that result from magnetic flux emergence.

The role of solar jets for triggering and driving the large scale  
solar eruptions is explained in chapter \ref{c5}. A two step filament eruption from AR NOAA 12297 and a narrow coronal mass ejection (CME) from AR NOAA 11731 are discussed as the two different case studies. The two step filament eruption starts with a push by a small jet from the AR and destabilizes the filament. After this perturbation from the jet, the filament starts to erupt and stops after reaching a high altitude of about 120 Mm and stays in a meta stable stage for 12 hours. An another jet activity from the same location again push the filament and finally it erupts. This eruption is followed by the largest geomagnetic storm of solar cycle 24.  In the second case study, a 
jet starts to erupt from AR NOAA 11731 and deflected in an another direction after reaching an altitude of about 80 Mm and results  as a narrow CME. To explain the relation between
the jet and the CME, the coronal potential field extrapolation is done, which shows that the jet eruption follows exactly the
same path of the open magnetic field lines from the source region which provides a way to the jet  to escape from the solar surface.

Observations of a twisted solar jet from NOAA AR 12736 and comparison with the numerical simulations from Observationally driven High-order scheme Magnetohydrodynamic (OHM) code (\citealt{Aulanier2005OHM, Aulanier2010}) is done in chapter \ref{c6}.
We observed the existence of the long flux rope (FR) near the jet base. 
It is found from the observations and MHD simulations that, there is a twist transfer to the solar jet during the extension of the stable FR to the reconnection site. 
The fast extension of the FR towards the site of reconnection due to photospheric surface motions gives the possibility of the FR arcades to reconnect with magnetic pre existing field lines at the ‘X’-point current sheet without the eruption of the FR. 
We concluded that, the reconnection would start in the low atmosphere in the bald patch reconnection region and extend along the current sheet formed above to an `X'-point. 

The fine structure and dynamics of a GOES B6.7 class solar flare and a jet with IRIS spectroscopic techniques are explained in chapter \ref{c7}. IRIS spectras at the flare and jet base are observed in the spectral ranges of Mg II, C II and Si IV ions and the Doppler velocities from Mg II lines are computed by using a cloud model technique. These high spectral resolution observations of IRIS lines and continuum emissions allow  us to propose a stratification model for the white-light mini flare atmosphere with multiple layers of different temperatures in a reconnection current sheet. It is the first time that we could quantify the fast speed (possibly Alfv\'enic flows) of cool clouds ejected transverse to the jet direction  by using the cloud model technique. We conjecture that the ejected clouds come from  plasma which was trapped between the two emerging magnetic flux regions before  the reconnection or be caused by  chromospheric-temperature (cool) upflow  material like  in a surge, during reconnection. IRIS spectral profiles at the reconnection site show a gradient in the spectra along the jet base indicating the formation of a rotating structure during the magnetic reconnection.


\clearpage
\onehalfspacing
\chapter*{\centering  List of publications}
\begin{enumerate}

\item {\it Multi-thermal atmosphere of a mini-solar flare during magnetic reconnection observed with IRIS}.\\
   \bluebf {Joshi, Reetika}, Schmieder, B., Tei, A., Aulanier, G, Chandra, R., Heinzel, P., {\it Astronomy and Astrophysics}, 2021, 645, A80. 

\item {\it The role of small-scale surface motions in the transfer of twist to a solar jet from a remote stable flux rope}.\\
    \bluebf{Joshi, Reetika}, Schmieder, B., Aulanier, G, Bommier, V., Chandra, R., {\it Astronomy and Astrophysics}, 2020, 642, A169. 
   
    \item {\it Cause and kinematics of a jetlike CME.}\\
     \bluebf{Joshi, Reetika}, Wang, Y., Chandra, R., Zhang, Q., Liu, L., Li, X., {\it Astrophysical journal}, 2020, 901, 94. 

    \item {\it Case-study of multi-temperature coronal jets for emerging flux MHD models.}\\
    \bluebf{Joshi, Reetika}, Chandra, R., Schmieder, B., Moreno-Insertis, F., Aulanier, G., Noberga - Siverio, D., Devi, P., {\it Astronomy and Astrophysics}, 2020, 639, A22. 

    \item{\it Slippage of jets explained by the magnetic topology of NOAA active region 12035}.\\
    \bluebf{Joshi, Reetika}, Schmieder, B., Chandra, R., Aulanier, G, Zuccarello, F. P., {\it Solar Physics}, 2017, 292, 152.

    \item {\it Observational evidences of current sheet formation and loop contraction during a prominence eruption}.\\
   Devi, P., D\'emoulin, P., Chandra, R., \bluebf{Joshi, Reetika}, Schmieder, B., Joshi, B., {\it Astronomy and Astrophysics}, 2021, 647, A85.
   \newpage
   \item{\it Variation of chromospheric features as a function of solar cycles 15–23: implications for meridional flow}.\\
    Devi, P., Singh, J., Chandra, R., Priyal, M., \bluebf{Joshi, Reetika} 2021, Solar Physics, 296, 49.
    
    \item {\it Development of a Confined Circular-cum-parallel ribbon flare and associated pre-flare activity}.\\
    Devi, P., Joshi, B., Chandra, R.,  Mitra, P. K., Veronig, A. M.,  \bluebf{Joshi, Reetika}, 2020, {\it Solar Physics}, 295, 75.
    
  
    
     \item {\it How rotating solar atmospheric jets become Kelvin-Helmholtz unstable?}\\
     Zhelyazkov, I., Chandra, R., and \bluebf{Joshi, Reetika}, {\it Frontiers in Astronomy and Space Sciences}, 2019, 6, 33.
    \item{\it Observations of  two successive EUV waves and their mode conversion}.\\
    Chandra, R., Chen, P. F., \bluebf{Joshi, Reetika}, Joshi, B., Schmieder B., {\it Astrophysical journal}, 2018, 863, 101.
    \item{\it Solar jet on 2014 April 16 modeled by Kelvin-Helmholtz instability}.\\
    Bogdanova, M., Zhelyazkov, I., \bluebf{Joshi, Reetika,} Chandra, R., {\it New Astronomy}, 2018, 63, 75.
    \item {\it Multiple solar jets from NOAA AR 12644}.\\
    \bluebf{Joshi, Reetika} and Chandra, R., {\it IAUS-340} 2018, 177-178.
    \item{\it Two Step Filament Eruption During 14-15 March 2015}.\\
    Chandra, R., Filippov, B., \bluebf{Joshi, Reetika}, Schmieder, B., {\it Solar Physics}, 2017, 292, 81.
    \item{\it The transition from eruptive to confined flares in the same active region}.\\
    Zuccarello, F. P., Chandra, R., B. Schmieder, G. Aulanier, \bluebf{Joshi, Reetika}, {\it Astronomy and Astrophysics}, 2017, 601, A26.

    \end{enumerate}
       
\chapter*{\centering Participation in  conferences}
\begin{enumerate}

\item Oral presentation on ``{\it Twist transfer to a solar jet from a big flux rope detected in the HMI magnetogram}'' in 43$^{rd}$ COSPAR Assembly, during 28 January-4 February 2021.

\item Oral presentation on ``{\it Multiple solar jets from rotating active region}'' in Astronomical Society of India (ASI) at Jaipur, India during 06-10 March 2017.

\item Oral presentation on ``{\it Solar jets: small scale events of the solar atmosphere}'', in Advances in Physics from Small to Large Scales (APSLS), in Nainital during 27-28 March, 2018.

\item Oral presentation on ``{\it Solar jets : small scale mass eruptions from the Sun}'' in Young Astronomer’s Meet (YAM)–2018 at Ahmadabad, India during 24-28 September 2018.

\item Oral presentation on ``{\it How solar jets can trigger the filament eruption}'' in the International conference on Exploring the Universe: Near Earth Space Science to Extra-Galactic Astronomy, in Kolkata India during 14-17 November 2018.

\item Oral presentation on ``{\it Solar jets: an observational overview}'', in Advances in Physics from Small to Large Scales (APSLS2019), in Nainital during 14-16 March, 2019.

\item Oral presentation on ``{\it Solar jets: small scale solar ejections}'' in NASA Heliophysics Summer School (HSS) 2019, at Colorado, Boulder, USA during 23-30 July 2019.

\item Poster presentation on ``{\it Multiple hot and cool jets on 2017 April 04}'' in  International Astronomical Union (IAU) Symposium at Jaipur India during 19-24 February, 2018. 

\item Poster presentation on ``{\it Two step filament eruption triggered by a solar jet}'' at Solar Physics Summer School in Leh, India during 10-16 June 2019.

\item Poster presentation on ``{\it Quasi periodic oscillations in the pre–phases of recurrent jets highlighting plasmoids in current sheet}'', in European Geo-sciences Union (EGU), during 4–8 May 2020.
\end{enumerate}


\tableofcontents

\listoffigures

\listoftables


\printnomenclature

\mainmatter

\chapter{Introduction} \label{c1}

Solar jets are ubiquitous transient collimated mass outflows in the solar atmosphere over a wide range of sizes from small scale nanojets to a few solar radii, embedded in the solar chromosphere to solar corona (\citealt{Shibata2007}; \citealt{Raouafi2016}; \citealt{Joshi2020MHD}; \citealt{Shen2021}). Observed as impulsive and sharp edged collimated  plasma ejections, jets can be originated from active regions (\citealt{Sterling2016}; \citealt{Joshi2020FR}) to quiet regions \citep{Hong2011} and  are frequently accompanied by solar flares. These flares at the jet base provide the force to propagate the plasma material upward  and sometimes accompanied by coronal mass ejections (CMEs). CMEs are the giant clouds of solar plasma propagating outward in the heliosphere and usually associated with large scale solar eruptions \citep{RChandra2017} and occasionally with the solar jets (\citealt{Shen2012}; \citealt{Joshi2020ApJ}). These jets act as a source for transporting a significant mass and energy from the lower solar atmosphere to the upper coronal heights and consequently heating the solar corona and accelerating the solar wind. 

Magnetic reconnection is believed to be the triggering reason behind jets (\citealt{Shibata1995}; \citealt{Jiajia2014}; \citealt{Joshi2020FR}). Different mechanisms are offered for the trigger of solar jets by magnetic reconnection between the  emergence of magnetic flux and  environment, or induced by twisted photospheric motions bringing the system to instability (\citealt{Pariat2009}; \citealt{Moreno2013}). For the first possibility, magnetic reconnection can take place as a result of magnetic flux  emergence from the solar interior. In these emerging flux MHD models, the cool plasma is advected  over the emergence domain without passing near the reconnection site and  flowing  along the reconnected magnetic field lines (\citealt{Nobrega2018}). An another interpretation to drive the jet onset is the injection of helicity through photospheric motions. In this mechanism, the presence of twist motions under a pre-existing null point induces reconnection with the ambient quasi potential flux  and initiates the helical jets (\citealt{Pariat2009}).

Despite the great progress made on both the observational and theoretical fronts, the underlying physics which trigger and drive these events is not completely  clarified, for example: How do these small-scale solar jets evolve into large-scale CMEs and contribute to solar wind acceleration and coronal heating? How the magnetic reconnection occurs for the photospheric jets? The prime physical mechanism (either magnetic flux emergence or cancellation) responsible to trigger the jets still needs more observational evidences (\citealt{Pariat2007}; \citealt{Nistico2009}; \citealt{Moreno2013}; \citealt{Sterling2015};   \citealt{Joshi2020IRIS}).


 This thesis includes active region jet studies with different triggering mechanisms to set off the jet initiation, related flares, and associated large scale eruptions (filament eruptions and CMEs) 
 and mounts a strong observational evidence to validate the numerical experiments for the magnetic flux emergence models.

The present chapter includes the review of literature, existing jet models, observational data sets, reduction techniques, and existing  scientific problems regarding the solar jet eruptions.

\begin{figure}[t!]
\centering
\includegraphics[width=0.7\textwidth]{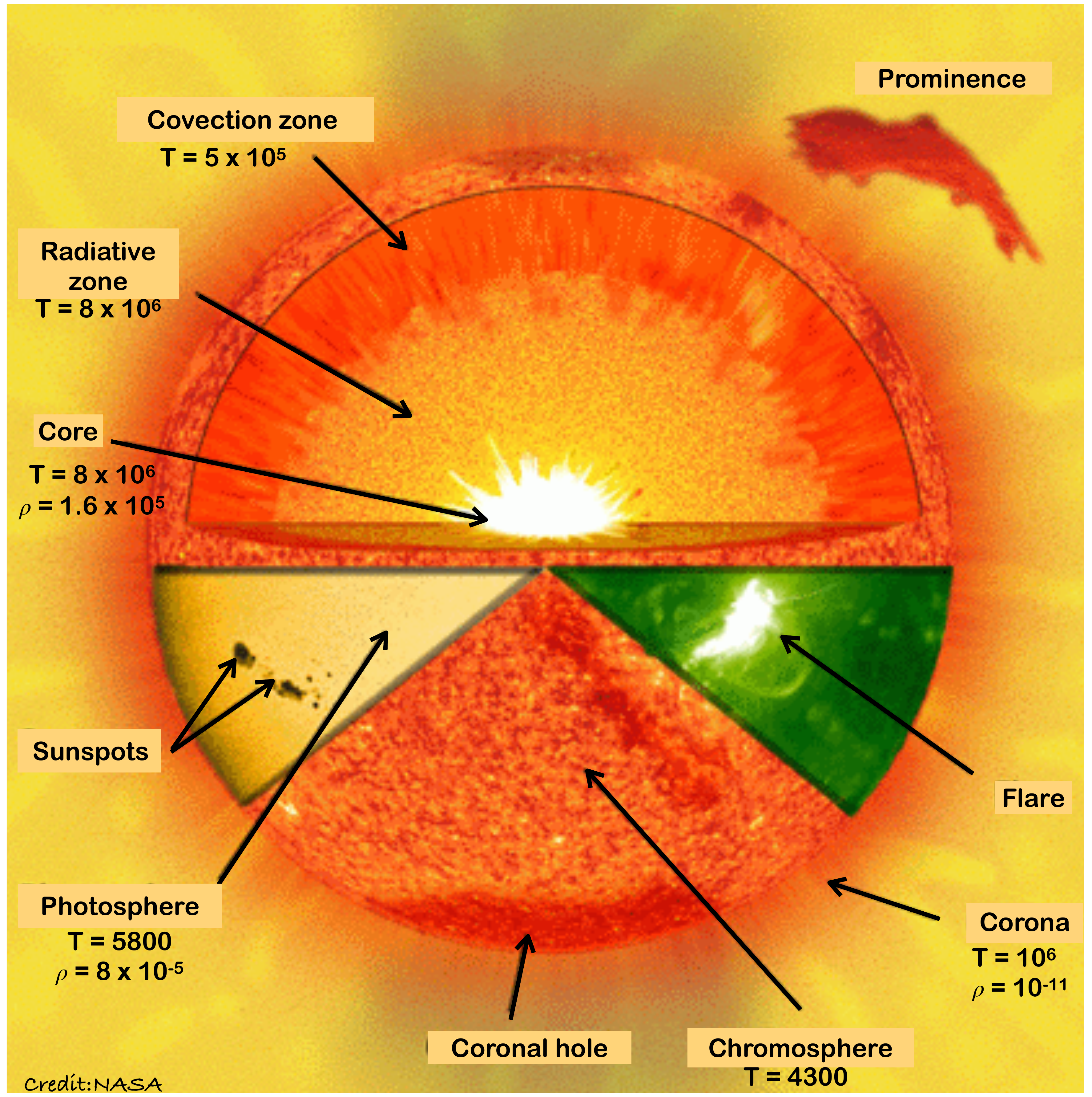}
\caption[Different layers of the Sun from solar interior (core, radiative zone, and convective zone) to exterior (photosphere, chromosphere, and corona).]{Different layers of the Sun from interior (core, radiative zone, convective zone) to exterior (photosphere, chromosphere, corona). T and $\rho$ values are in units of Kelvin and kg m$^{-3}$.}
\label{layers}
\end{figure}
\section{The structure of the solar atmosphere}
\label{atm}
The Sun is a hot plasma ball abundant with hydrogen (71 $\%$), helium (26 $\%$), and also containing some other elements  {\it i.e.} oxygen, carbon, nitrogen, neon, magnesium, and iron. It is a G2 V star in the main sequence where the hydrogen
is smashing into helium in its core. The physical properties of the Sun are presented in Table \ref{tab:ch1} (\citealt{Seeds}; \citealt{Bhatnagar}). The Sun consists a series of co-centeric spherical shells of 
different temperature and density. The overall structure of the Sun 
is presented in Figure \ref{layers}.
\begin{table}[h!]
\centering
  \caption{Physical properties of the Sun.}
\label{tab:ch1}
\setlength{\tabcolsep}{8pt}
\begin{tabular}{ll}
     \specialrule{.1em}{0.1em}{0.1em}
      Age & 4.6 $\times$ 10$^9$ years \\
      Mass & 1.99 $\times$ 10$^{30}$ kg ($\approx$ 330 times of the Earth)\\
      Diameter & 1.38 $\times$ 10$^{8}$ m ($\approx$ 110 times of the Earth) \\
       Surface gravity & 274 m s$^{-2}$ \\
       Surface Temperature & 5800 K\\
Equatorial rotation period & 24.5 days\\
Escape velocity at surface & 618 km s$^{-1}$\\
Luminosity & 3.86 $\times$ 10$^{26}$ W\\
Average density & 1.41 g cm$^{-3}$\\
Spectral type & G2 V\\
Absolute magnitude & + 4.83\\
Apparent magnitude & - 26.74\\
   \specialrule{.1em}{0.1em}{0.1em}
    \end{tabular}
\end{table}

\begin{figure}[ht!]
\centering
\includegraphics[width=0.8\textwidth]{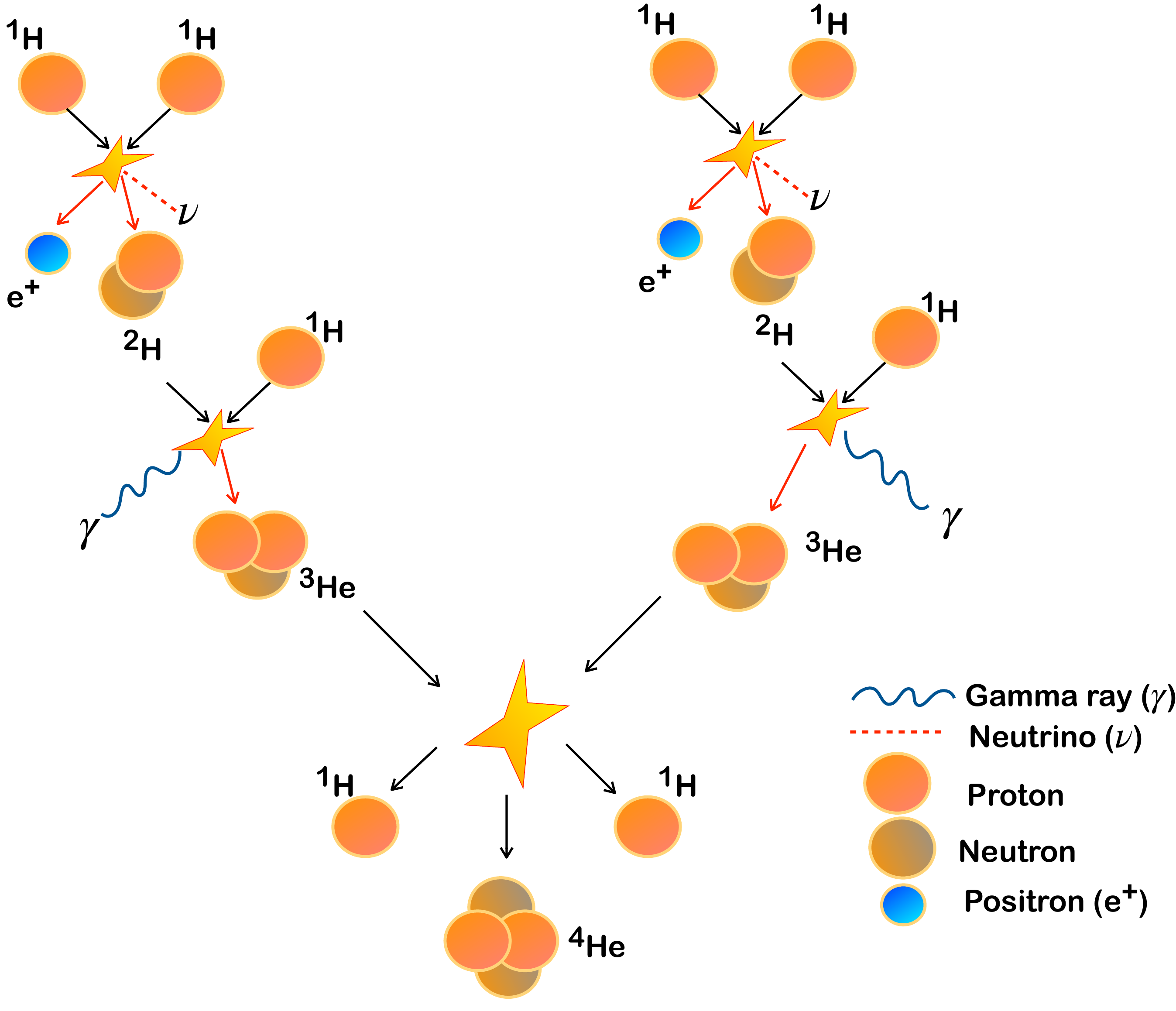}
\caption[Nuclear fusion reaction mechanism inside the Sun's core.]
{The proton--proton chain reaction to form one helium nuclei 
from four protons. 
The neutrinos do not involve in the heating process and escape from the surface \citep{Seeds}.}
\label{core}
\end{figure}

\subsection{The solar interior}
The interior of the Sun is divided into three regions, {\it i.e.} the core, radiative zone, and the convective zone.
The core is a gigantic reactor of radius 150 Mm (\citealt{Priest}), where the nuclear fusion takes place and 
the hydrogen ($^1$H) nuclei converted into helium ($^4$He) nuclei by proton--proton chain reaction and emit two particles: positron (e$^+$) and neutrino ($\nu$). Energy is released in the form of gamma rays (\citealt{Seeds}).
The proton--proton chain reaction is explained in Figure \ref{core} and can be expressed as:
\begin{equation}
    4^{1} H \longrightarrow ^{4}He + 2e^+ + 2\nu + 26.7 ~~MeV
    \label{eq:core}
\end{equation}

The energy generated in the core continuously leaks outwards
by the radiative diffusion process across the radiative zone. In the radiative
zone, the photons are absorbed and emitted many times and take many
years to cross this layer.
The energy flowing outwards as radiation then encounters to the outer
layer of the Sun where the gas is not completely ionized.
In this region, the gas is not very transparent to the radiation, so 
the hot blobs of the gas starts to rise and cool blobs sink.
In this way, the energy is transported by the convection method.

According to the {\it standard model} for solar interior, the 
pressure ($p$), temperature ($T$), and density ($\rho$) are functions of radial distance ($r$) from the Sun's center and the 
co--centric spherical shells are in hydrostatic and thermal equilibrium (\citealt{Stix2002}; \citealt{Arnab}). The {\it standard model} is based on the following basic equations:\\
(1) {\it Perfect Gas Law}, (2) {\it A hydrostatic force balance equation}, and (3) {\it Steady state energy balance equation}, given as follow (\citealt{Seeds}; \citealt{Bhatnagar}):

\begin{equation}
    p~= \frac{k_B}{m}~ \rho~ T
    \label{eq:gas}
\end{equation}
where, $m$ is the mean particle mass, and $k_B$ is the Boltzman's constant.
\begin{equation}
    \frac{dp}{dr} ~= -~\rho(r)~ g(r)
    \label{eq:force}
\end{equation}
 here, gravitational acceleration $g(r)$~=$~M(r)~ G$/$r^2$, and $G$~=~ 6.67$\times$10$^{-11}$ N m$^2$ / kg$^{2}$. 
\begin{equation}
    \frac{dL(r)}{dr}~=4~\pi~r^2 ~ \rho ~ \epsilon
    \label{eq:energy}
\end{equation}
where, $L(r)$ shows the outward flow of thermal energy, {\it i.e.} luminosity, and $\epsilon$
is the rate of energy produced per unit mass. Outside the core this value becomes equal to the solar luminosity ($L_{\odot}$). The solution for these equations are given with the two boundary conditions, {\it i.e}, $M=0$, $\frac{dT}{dr}=0$ at the center ($r=0$), and $M=M_{\odot}=1.989\times10^{30}$ kg, $L=L_{\odot}=3.846\times10^{26}$ Watt, at the solar surface ($r=R_{\odot}$). The solution gives an estimation about the size of the core (0.25 $R_{\odot}$) and the base of the convection zone, which extends to 0.7 $R_{\odot}$. At this height of $\approx$ 0.7 $R_{\odot}$ ($T=10^6$ K) convection instability occurs \citep{Arnab}. This onset of instability is explained in Figure \ref{blob}. Let us consider vertically stratified plasma with pressure ($p(r)$), density ($\rho(r)$), temperature ($T(r)$) is in hydrostatic equilibrium and an elementary plasma blob moves in upward direction. The blob is supposed to be in equilibrium with the surrounding, such that it maintained the horizontal pressure (\citealt{Bhatnagar}; \citealt{Arnab} \citealt{Priest}). It rises up with a buoyancy force under the condition: 
\begin{figure}[t!]
\centering
\includegraphics[width=0.8\textwidth]{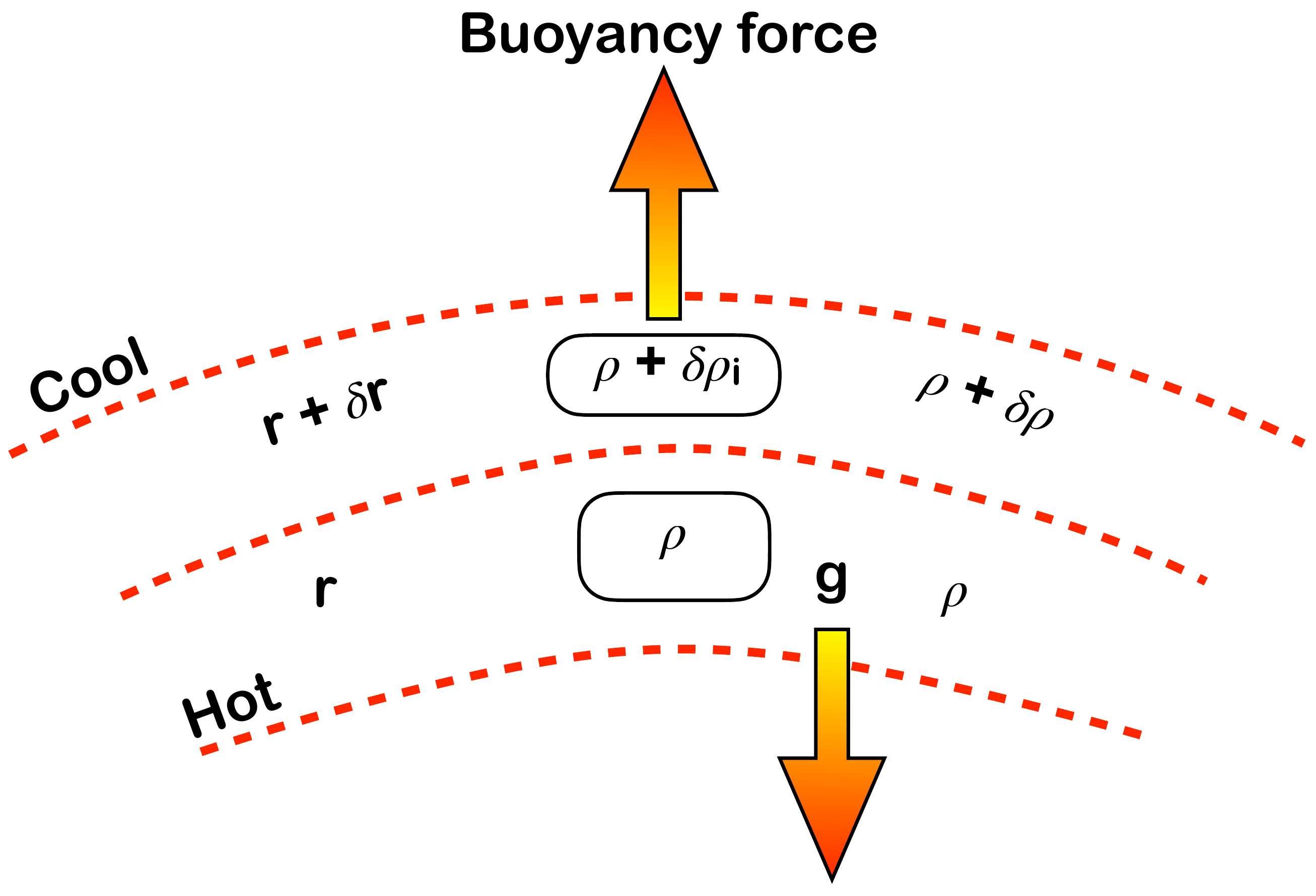}
\caption[The upward motion of the plasma blob from $r$ to 
$r + \delta r$.] 
{The upward motion of the plasma blob from $r$ to 
$r + \delta r$. The upward buoyancy force and downward gravitational force are shown with two arrows up and down respectively.}
\label{blob}
\end{figure}
\begin{equation}
    \delta \rho_i < \delta \rho
    \label{eq:rho}
\end{equation}
where, $\delta \rho_i$ is the change in the density inside the plasma blob and $\delta \rho$ is the change in the density in the ambient medium.\\
Differentiating Equation \ref{eq:gas} for plasma blob and surrounding, we get 
\begin{equation}
    \frac{\delta p_i}{p}=\frac{\delta \rho_i}{\rho} + \frac{\delta T_i}{T}~~,~~
    \frac{\delta p}{p}=\frac{\delta \rho}{\rho} + \frac{\delta T}{T}
\end{equation}

To maintain the condition for horizontal pressure and using Equation \ref{eq:rho},
\begin{equation}
    \delta T > \delta T_i
\end{equation}
Hence,
\begin{equation}
     \left\lvert \frac{dT}{dr} \right\rvert> \left\lvert \frac{dT_i}{dr} \right\rvert
     \label{eq:temp}
    \end{equation}
Therefore, the temperature of ambient medium ($T$) is falling faster with the height then the temperature fall of the plasma blob ($T_i$). By assuming the blob motion so rapid (adiabatic) and from the equations \ref{eq:gas}, \ref{eq:energy} the blob properties can be written as: 
\begin{equation*}
     p_i= \frac{k_B}{m}~ \rho_i~ T_i,
\end{equation*}
\begin{equation*}
    \frac{dp_i}{dr} ~= -~\rho_i~ g,
\end{equation*}
\begin{equation*}
    \frac{p_i}{\rho_i^\gamma}=constant
\end{equation*}
On differentiating and combining these three equations, we get
the {\it adiabatic temperature gradient}
\begin{equation}
     - \frac{dT_i}{dr}>\frac{\gamma - 1}{\gamma}\frac{gm}{k_B}
    \end{equation}
On comparing this with Equation \ref{eq:temp}, the convection instability criteria becomes:
\begin{equation}
    \left\lvert \frac{dT}{dr} \right\rvert< \frac{\gamma - 1}{\gamma}\frac{gm}{k_B}
    \label{convection}
\end{equation}
This condition for convection instability is well known as {\it Schwarzschild criterion}. For a monoatomic gas, the constant factor $\frac{\gamma-1}{\gamma}=\frac{2}{5}$.  
\begin{figure}[t!]
\centering
\includegraphics[width=0.8\textwidth]{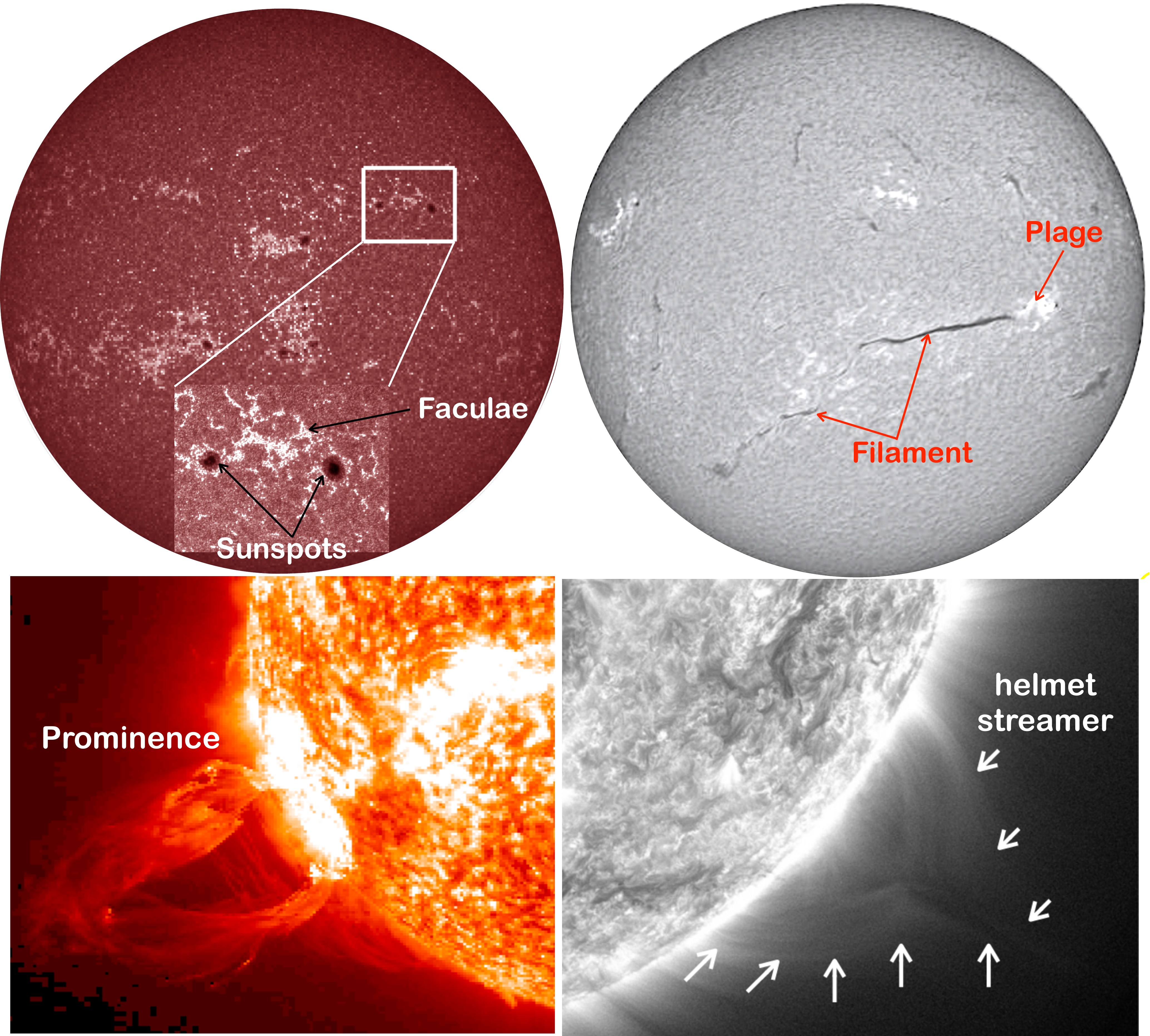}
\caption[Photospheric, chromospheric, and coronal features.]{Photospheric (sunspots, faculae), chromospheric (filament, plage), and coronal (prominence, helmet streamer) features with multiwavelength observations.
}
\label{features}
\end{figure}
\begin{figure}[t!]
\centering
\includegraphics[width=0.8\textwidth]{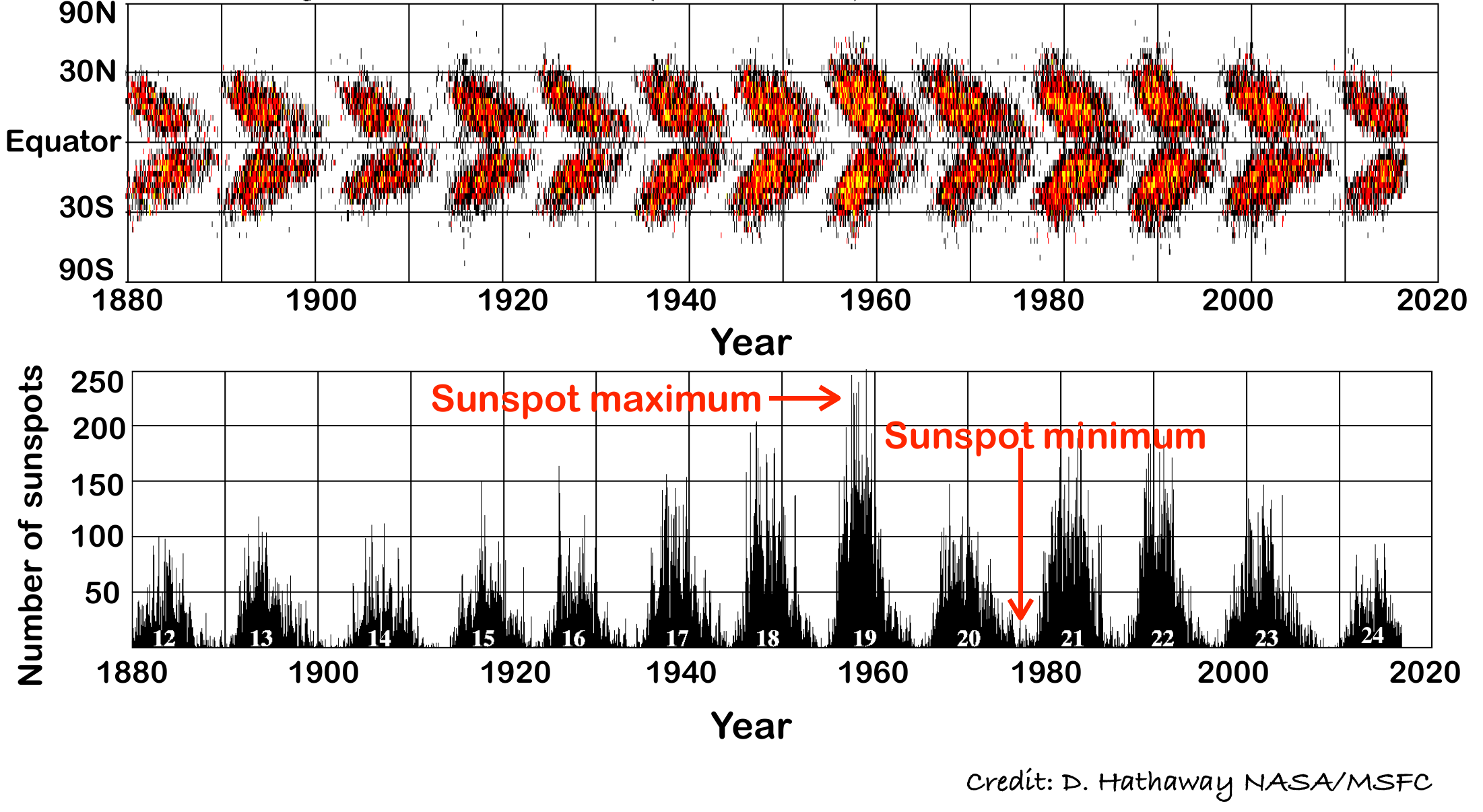}
\caption{Maunder butterfly diagram and variation of sunspots on the solar disk.}

\label{butterfly}
\end{figure}
 
\subsection{Photosphere}
The photosphere (named after a Greek word for light) is the visible surface of
the Sun with thickness of about a few 100 km (\citealt{Priest}). It is the surface of unit optical
depth ($
\tau_\nu$), which is defined as (\citealt{Arnab}):
\begin{equation*}
    d\tau_\nu=\alpha_\nu~~ ds
\end{equation*}
Along the path from s$_o$ to s, it can be stated as:
\begin{equation*}
   \tau_\nu = \int_{s_0}^s \alpha_\nu (s)~ ds
\end{equation*}
An optical thick medium extinguishes the light passing through it, whereas the optically thin medium does not change it. The optically thick ($\tau_\nu>>1$) surface may consider as an opaque medium and the optically thin ($\tau_\nu<<1$) layer acts as a transparent medium.
Photosphere is the region from where most of the Sun's visible light at 
5000 \AA\ is emitted with the surface temperature of about 5800 K (\citealt{Bhatnagar}).
The photosphere is enclosed with several types of convective motions 
{\it namely}, granulations, supergranulations and with bright patches 
near the limb called faculae. Sunspots, which are cool and dense spots,
are also usually appear on the photosphere (Figure \ref{features}) and survive over a time scale
of weeks to months. The sunspot numbers are 
the direct indication of the solar activity and follow an 11--year cycle, by
reaching to maximum and minimum. This 11--year solar cycle with 
changing the location of sunspots traces a beautiful {\it Maunder 
butterfly diagram}, presented in Figure \ref{butterfly}.  
\begin{figure}[t!]
\centering
\includegraphics[width=0.8\textwidth]{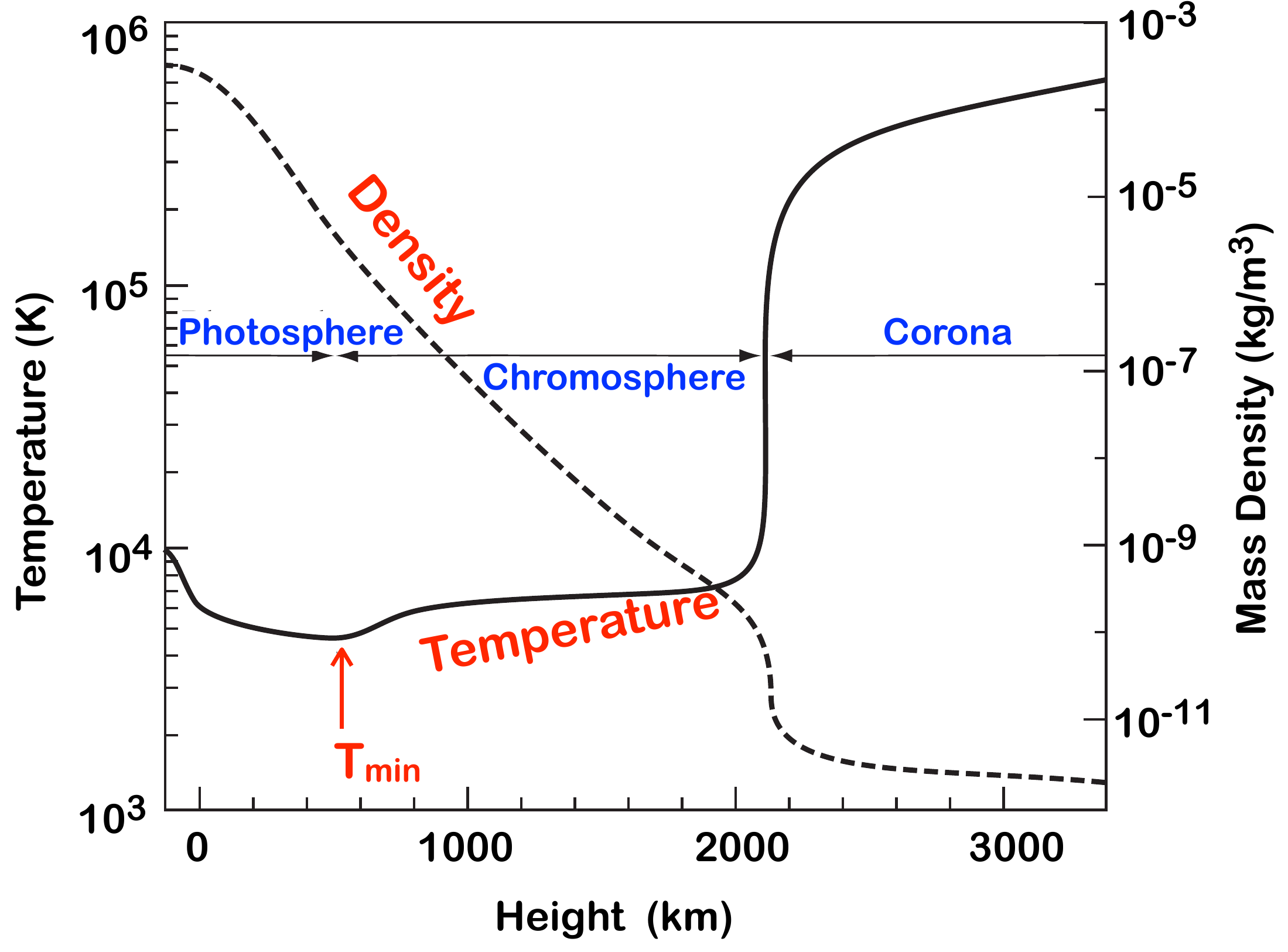}
\caption{The graph shows the mean variation of density and temperature
with solar atmospheric height.
}
\label{temp_density}
\end{figure}

\subsection{Chromosphere}
Chromosphere lies above the photosphere. Its name comes from the 
Greek word {\it chroma}, which means ``color".
The chromosphere is 1000 times fainter
than the photosphere and visible as a pink layer during a total solar 
eclipse. This colorful appearance is from the three bright 
emission lines (red, blue, and violet) of 
hydrogen Balmer H$_\alpha$ emissions.
According to the Vernazza–Avrett–Loeser (VAL, \citealt{Vernazza1981}; \citealt{Avrett2008})
model, 500 km above the photosphere,
the temperature gradually rises from 4300 K to 10,000 K. In the mean time 
the density drops down by a factor of 10$^{6}$ in the same height.
This sudden transition of temperature from lower to higher values in the 
upper chromosphere, and the drop in the density
values are presented in Figure \ref{temp_density}.
One can observe the chromospheric features {\it i.e.}
 spicules (flame like jets), 
filaments/prominences etc. with the appropriate
optical filters.

\subsection{The solar corona}
Solar corona is the outermost layer of the Sun. There is a sharp gradient of 
temperature and density from chromosphere to corona. 
These two regions are connected with a narrow region (a few 100 km thick),
called the transition region. The temperature sharply increases from 
10$^4$ to 10$^6$ K and density drops to 10$^{-11}$ kg m$^{-3}$, presented
in Figure \ref{temp_density}. 
This unusual behaviour of increasing the temperature while
going out from the energy source is still an open question 
in the solar physics, know as the {\it coronal heating problem} (\citealt{Parker1955}; \citealt{Parker1988}).
Solar corona is optically thin region and unlike chromosphere and 
photosphere, radiation is not absorbed over the solar disk 
while passing the corona. 
Coronal features {\it i.e.} plumes, coronal loops, prominences, helmet streamers  are  usually observed with X-ray observations and presented in Figure \ref{features}. 
Solar wind, which is a flow of gases, 
 streams off mainly from the solar corona (coronal holes)
with a speed of more than 500 km s$^{-1}$ (\citealt{Bhatnagar}).
This solar wind perturbs the Earth's magnetic field and 
pumps energy in the radiation belts. To study the Solar atmosphere and the solar magnetic field  is very 
important for the prediction of space--weather and the 
Sun--Earth connections.

\begin{figure}[ht!]
\centering
\includegraphics[width=\textwidth]{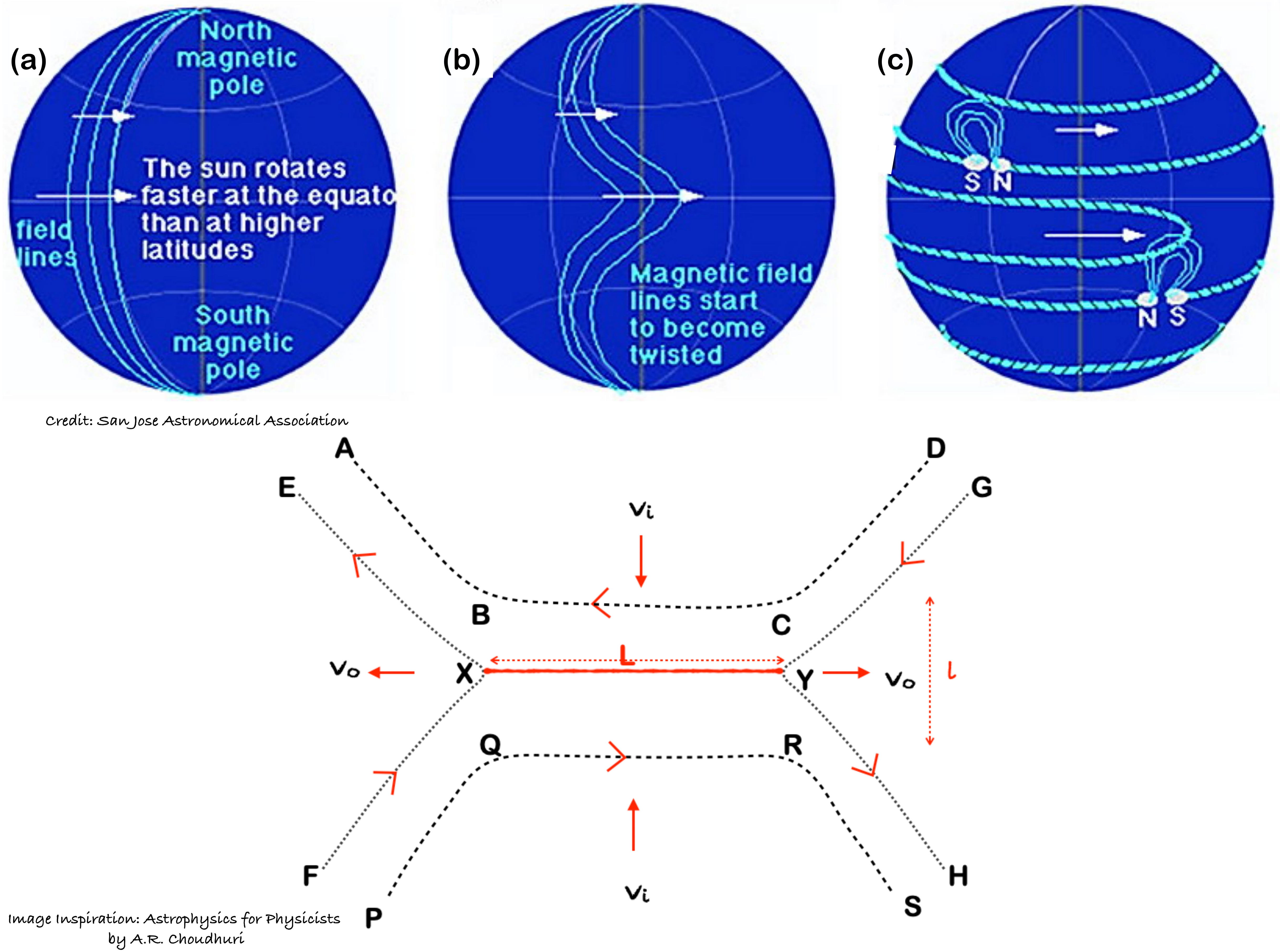}
\caption[Solar magnetic field and Hale's law.]
{Top row: The magnetic field inside the Sun rotates with the highly conductive and rotation material. The polarity of the magnetic loops reverse between the two hemispheres. Bottom row: Current sheet (red solid line XY) formation, during magnetic reconnection process between two oppositely oriented magnetic field lines.}

\label{babcock_reconnection}
\end{figure}

\section{Solar magnetic field}
\label{ch1_mag}
The Sun is driven by a strong magnetic field of strength $\approx$ 10$^{5}$ Gauss generated in the tacholine (thin layer between radiative and convective zone). The presence of magnetic field on the Sun and its interaction with the plasma,
creates a wonderland of fascinating solar activities.
The magnetic field on the Sun is generated by the energy 
flowing outwards with the moving current of highly ionized 
gas. The electrically conducting gas undergoes rotation and stirred by 
convection (\citealt{Seeds}; \citealt{Bhatnagar}). The conversion of this outward flowing energy due 
to rotational and convection motion, into a magnetic field is known as {\it 
dynamo effect} (\citealt{Choudhuri1995}; \citealt{Schrijver2002}; \citealt{Hathaway2003}; \citealt{Baumann2006}). The {\it dynamo effect} is the main 
mechanism for the generation of solar magnetic field deep under the photosphere.
The photosphere rotates faster at the equatorial heights (24.5 days for one rotation)
 and slower at the higher latitudes (at 45$^\circ$ latitude, 27.8 days for one rotation). This different 
speed of rotation at different latitudes is well known as the {\it differential
rotation} and explained in terms of Babcock model (\citealt{Babcock1961}) in Figure \ref{babcock_reconnection} (top row) which gives rise to the Sun's 11 year cycle. 

\begin{figure}[t!]
\centering
\includegraphics[width=0.8\textwidth]{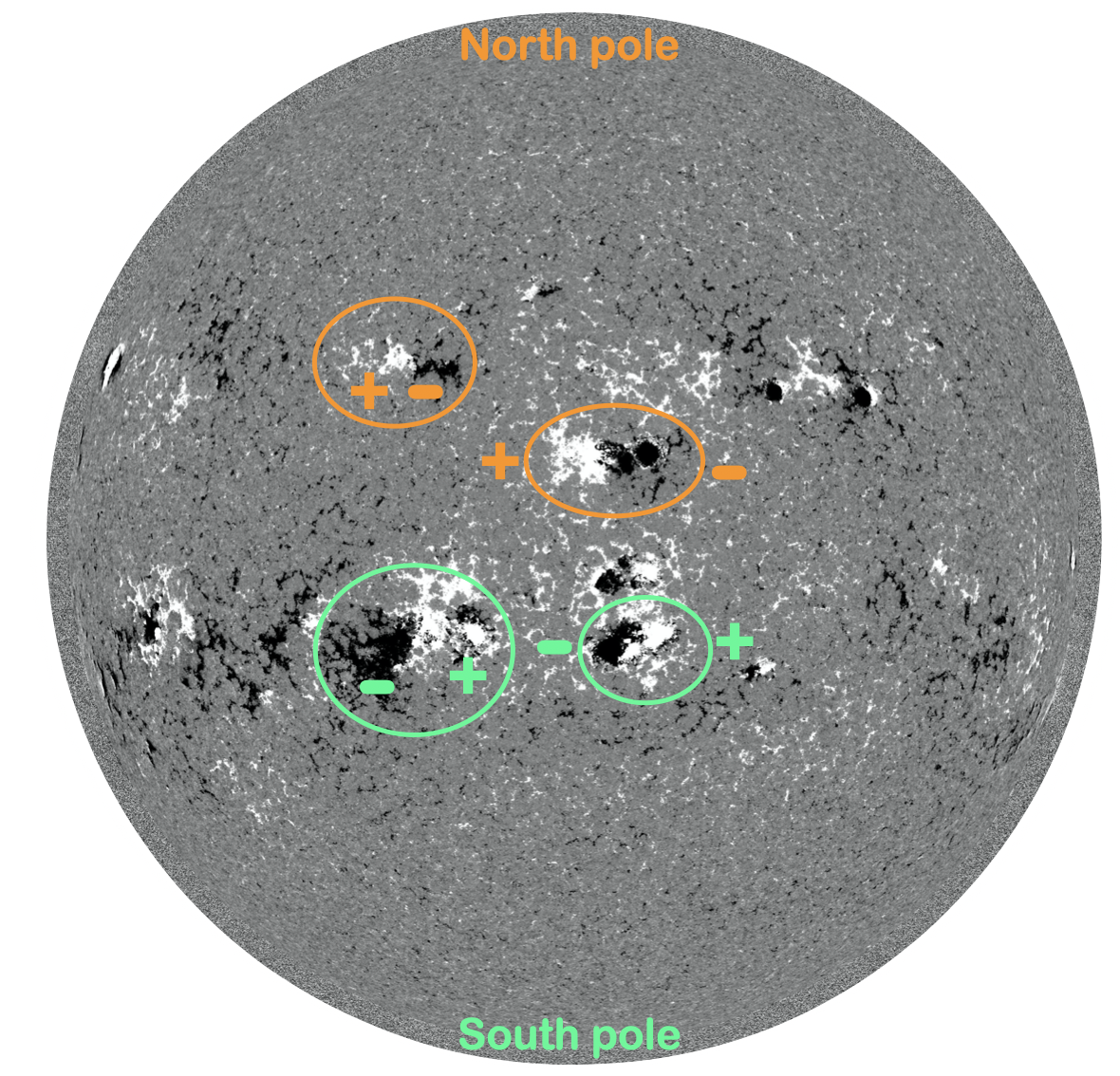}
\caption[Photospheric magnetic field observed with SDO/HMI on April 16, 2014.]
{Photospheric magnetic field observed with HMI on April 16, 2014. Black and white patches show the negative and positive polarities respectively.}
\label{fig:hmi}
\end{figure}

The cyclic variations of different features (sunspots, quite, active, and enhanced networks, plages) on the solar disk are known as solar cycle, discovered by \citealt{Schwabe1843}. Later on, \citealt{Hale1908} showed that the sunspots are strongly magnetized, and the complete magnetic cycle spans two solar cycles {\it i.e.} 22 years, before coming to its original position. This cyclic variation is because of the change of  toroidal to poloidal magnetic field then again from poloidal to toroidal magnetic field.
An observation of photospheric magnetic field on April 16, 2014 (solar cycle 24) with SDO/HMI instrument is shown in Figure\ref{fig:hmi}. The white and black patches are for positive and negative polarities respectively. In the northern hemisphere the negative polarity is leading and followed by the positive polarity. This configuration is similar through out the northern hemisphere with the negative magnetic polarity as in the front and opposite to the southern hemisphere, where positive magnetic polarity is leading polarity. 

From the time of Galileo, it has been observed and believed that the Sun has dark patches (now called as  sunspots), on its surface and contain a central region with concentrated magnetic field of strength upto 0.3 T. The magnetic field generates a Lorentz force, which contains two components {\it namely}: magnetic pressure force and a tension force (\citealt{Priest}). The ratio of plasma pressure to magnetic pressure is a dimensionless quantity, called {\it plasma beta} `$\beta$' (\citealt{Chandrasekhar1952}):
\begin{equation}
    \beta=\frac{2\mu_{o}p_{o}}{B^2}
\end{equation}
 The value of $\beta$ decreases with altitude above the solar surface. In the photosphere ($\beta>1$), plasma pressure dominates over the magnetic force. Below the solar surface it increases with depth and reached upto 10$^5$ near the convection zone. In the solar corona  $\beta<<1$, hence the energetics and dynamics are governed by the magnetic field (\citealt{Stix2002}; \citealt{Solanki2006}).

Magnetic pressure acts from high pressure region to low pressure region. Magnetic tension applies a restoring force with the curved magnetic field and magnetic waves propagate along with the field lines, in a same way as the wave moves in a string.
This force may store energy, and when the magnetic field becomes unstable this stored energy released and results in various violent eruptions. 
\cite{Chandrasekhar1952} laid a foundation of how the convection process changes with the magnetic field, which is known as the {\it magnetoconvection}. The tension force due to magnetic field always opposes the convection motion of the gas. So, if there is any magnetic field present inside the convection region, it try to get swept in the confined regions. In these confined regions, convection process exhibits the suffocation to act naturally due to the magnetic tension, but in the other regions with no magnetic field, it takes place easily. The magnetic field lines are bundled up by the convection inside the sunspot regions. As in the central part magnetic tension will perturb the convection process, so the heat transport process will not take place easily and it lead to a cooler surface at the center. Hence, the sunspot appear dark at the surface related to the outer surrounding.
\cite{Parker1955} explained that a region of high concentrated magnetic field surrounded with a low magnetic field (known as {\it magnetic flux tube}) may become buoyant and the condition is stated as (\citealt{Arnab}):
\begin{equation}
    p_e=p_i+\frac{B^2}{2\mu_o} ~~~~and~~~~ p_i<p_e
    \label{fluxtube}
\end{equation}
where, $p_e$ and $p_i$ are the gas pressure values outside and inside of the flux tube, respectively.
With the temperature `$T$', and the density $\rho_e$ and $\rho_i$ inside and outside of the flux tube. Using the ideal gas equation for pressure and temperature relation:
\begin{equation}
    R\rho_eT = R\rho_iT +\frac{B^2}{2\mu_o}
\end{equation}
And using the relation $p_e = R\rho_eT$, we get
\begin{equation}
    \frac{\rho_e-\rho_i}{\rho_e}= \frac{B^2}{2\mu_o~p_e}
\end{equation}
Thus the plasma present in the interior of the flux tube must be lighter and buoyant. In this way, the flux tube should become buoyant and rise against the gravitational field. 
The solar photosphere is embedded with magnetic field 
by turbulent convection motions, concentrated in the magnetic flux tubes. The buoyant flux tubes rise in the convection zone by obeying the Schwarzschild criteria given in equation \ref{convection} and expelled out in the chromosphere and imposed beautiful magnetic loop structures in the upper chromosphere and in the solar corona (\citealt{Stix2002}). In the emerging process of flux tubes, they form: sunspots in the ARs of magnetic field strength B$\approx$ 10$^3$ Gauss, coronal loops in the photospheric footpoints of strength B$\approx$ 10$^2$ Gauss, and in the large coronal heights with strength B$\approx$ 10 Gauss (\citealt{Solanki2006}).
The flow of magnetic field inside the flux tube may explain by the induction equation in basic magnetohydrodynamics (MHD) (\citealt{Arnab}):
\begin{equation}
    \frac{\partial \vec B}{\partial t}= \vec \nabla \times (\vec v \times \vec B) + \eta \nabla^2 B
    \label{induction}
\end{equation}
here, $\eta=$ $\frac{1}{\mu_o\sigma}$ is magnetic diffusivity, an approximate reversal of electrical conductivity ($\sigma$).
Let us suppose the plasma parameters have the value: `$\vec{B}$' for magnetic field, `$\vec{v}$' for velocity, and `L' for the length scale where the magnetic field vary gradually, then the first term $\vec \nabla \times (\vec v \times \vec B)$ in equation \ref{induction} become $v$B/L, and $\eta \nabla^2 B$ become $\eta$ B/$L^2$. The ratio of these two values is a dimensionless quantity, called {\it magnetic Reynolds number} ($R_e$):
\begin{equation}
    R_e=\frac{vB}{L}/\frac{\eta B}{L^2} ~~~~\approx~~~~ \frac{vL}{\eta}
    \label{Raynold}
    \end{equation}
The value of $R_e$ is usually smaller than unity for laboratory plasma and much larger than unity for the astrophysical plasma system. So, for the astrophysical system, the second term in equation \ref{induction} can be negligible and become: 
\begin{equation}
    \frac{\partial \vec B}{\partial t}= \vec \nabla \times (\vec v \times \vec B)
\end{equation}
This leads to the flux freezing condition (\citealt{Alfven1942}), which gives a clear picture of the behaviour of magnetic field in astrophysical systems. The magnetic flux is believed to be frozen in the plasma and moves with the plasma flow (\citealt{Arnab}).

As the diffusion (second term in equation in \ref{induction}) of magnetic field would be a slow process and can be neglected for the astrophysical systems as explained above but, in some cases it is found that the magnetic energy evolved in an enormous amount very quickly (10$^{26}$ J energy released in case of a flare eruption (\citealt{Benz2008})). Hence even if the magnetic diffusivity is small, the magnetic field gradient may be large enough that the second term in equation \ref{induction} can not be neglected. In such magnetic configurations a concentrated plate/sheet of electric current is supposed to be present in between the two opposite magnetic fields. This 
{\it current sheet} plays an important role for the large and small scale solar eruptions. In the {\it current sheets}, the second term $\eta \nabla^2 B$ in equation \ref{induction} becomes dominant and the magnetic field diffuse/decay at the center. This decay of magnetic field promote the decrease of the magnetic pressure ($B^2/2\mu_o$) in the central region and the plasma from other regions (above and below) will be sucked into the central part. This motion of fresh magnetic field lines towards the central region gives rise to the reorientation of magnetic field lines called, {\it magnetic reconnection} as presented in Figure \ref{babcock_reconnection} (bottom panel). 
In this way, the complete system pushes against the central part due to the plasma situated in the up and down direction, and the plasma in the central region is squeezed out from two sides (with velocity $v_0$ from P and Q in Figure \ref{babcock_reconnection}). In Figure \ref{babcock_reconnection} (bottom panel), the magnetic reconnection process is explained, where the magnetic field lines ABCD and PQRS are moving inward to the center with velocity $v_i$.   The inner part of these lines: BC and QR decay away and the parts AB, PQ and CD, RS move outwards to  form a new set of lines namely: EXF and GYH respectively. These reconnected magnetic field lines EXF and GYH move away from the reconnection region with velocity $v_o$. The current sheet (XY) of length `L' formed in the central part.  This scenario of magnetic reconnection has been theoretically well established by \cite{Parker1957}; \cite{Sweet1958}; \cite{Petschek1964}.
The heat generated due to reconnection can raise the temperature of the plasma, like in case of the solar corona. It is believed that magnetic reconnection takes place in the coronal loops and increase the temperature to million degrees and responsible for violent solar eruptions.

\section{Solar activity}
\label{activity}
Broadly, solar active phenomena are divided into two categories namely: large scale activities (solar filament/prominence eruptions, solar flares, coronal mass ejections (CMEs)) and small scale activities (Ellerman bombs, solar jets). The large scale activities can affect our space weather directly. However, the small scale events are also responsible for the change in the Earth's atmospheric structures   while associated with CMEs. 
\subsection{Large scale solar eruptions}
\begin{figure}
\centering
\includegraphics[width=0.8
\textwidth]{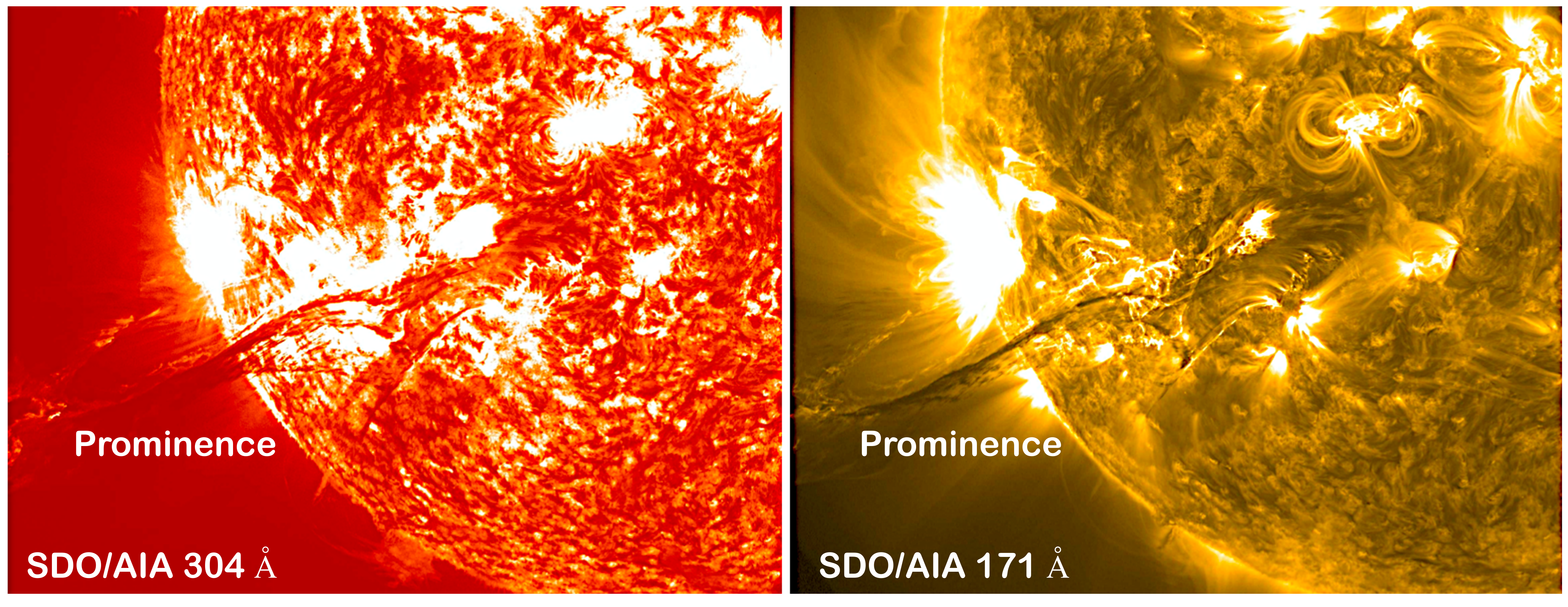}
\caption{AIA observations of a large prominence eruption in AIA 304 \AA, and 171 \AA.
}
\label{filament}
\end{figure}
\begin{enumerate}
    \item 
{\bf Solar filament/prominence eruptions:}
   A Solar filament is a large current system lie above the magnetic polarity inversion line (PIL) (\citealt{Babcock1955,Schmieder2002}). These structures contain dense and cool ($\approx$ 10$^4$ K) plasma against the hot corona and appear as dark thread like structures, when observed on the solar disk. Beyond the solar limb, these structures are termed as prominences and appear as bright cloud like features. So the filaments and prominences are identical structures observed in different locations {\it i.e.} on solar disk or solar limb respectively. They are observed in the chromosphere (H$_\alpha$) or in the lower corona (EUV observations). Filaments formed at the chromospheric height are usually associated with fibrils along the PIL and called spicules when observed over the limb (\citealt{Smith1968, Jean2015}). An EUV observation of solar prominence with SDO/AIA in two wavelengths (304 \AA, and 171 \AA) is presented in Figure \ref{filament}. \citealt{Mackay2010} clarified that the body of the filament consists of three main structural components, {\it namely} (a) spine, (b) barbs, and (c) extreme ends. The long horizontal part of the filament is called the spine, and it forms the main filament structure. From the sides of the spine, there are small branches protrude with an acute angle with respect to the main body. These small branches are called barbs and provide a support to the filament system (\citealt{Schmieder2010}).
   The beginning and end points of the spine are the two extreme ends of the filament. These filament structures are shown in Figure \ref{halpha}, where the huge filament lies on the solar disk observed on March 14, 2015. Filaments are observed in quite as well as ARs and well studied by many authors (\citealt{Ali2007}; \citealt{Gosain2009}; \citealt{Torok2011}; \citealt{Zuccarello2016}; \citealt{Chandra2017}). 
   \begin{figure}[ht!]
\centering
\includegraphics[width=0.8\textwidth]{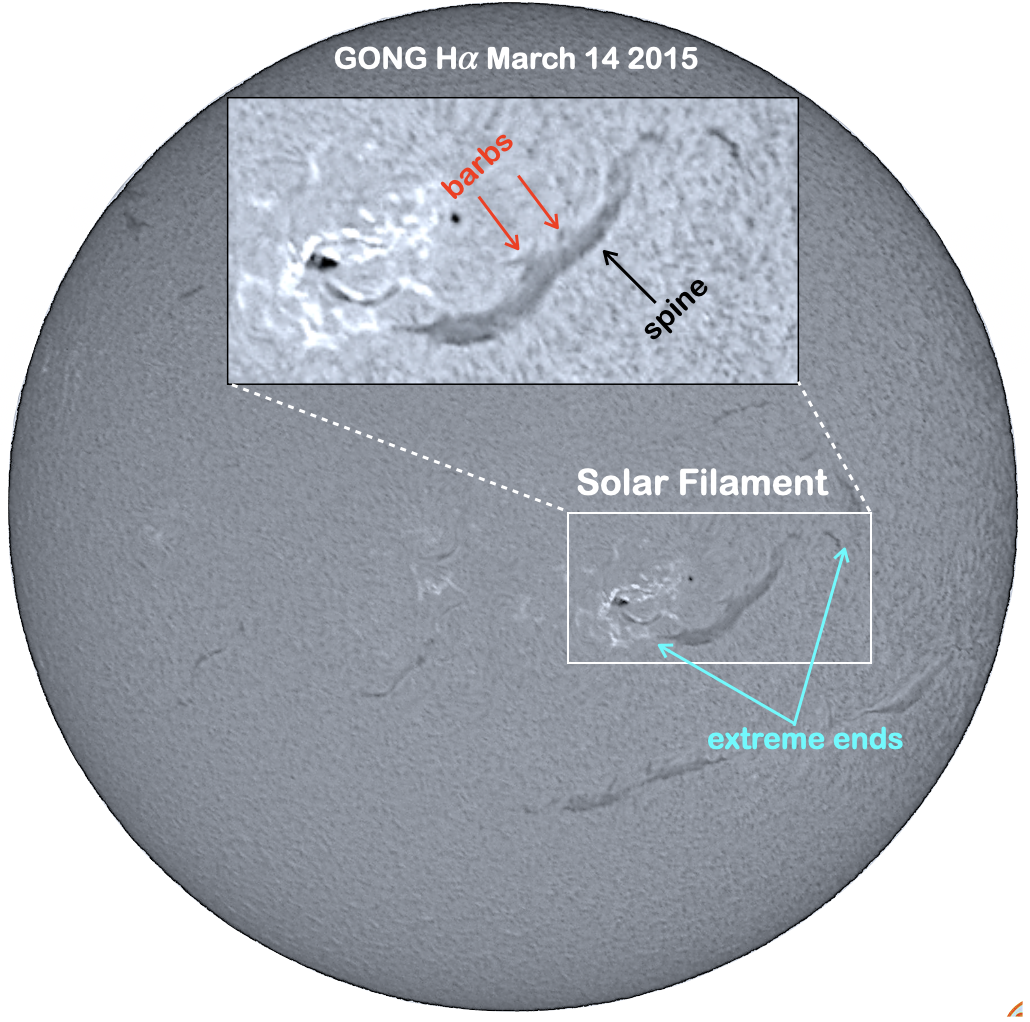}
\caption[A huge filament on the solar disk observed on March 14, 2015 with GONG H$_\alpha$ instrument.]{A huge filament on the solar disk observed on March 14, 2015 with GONG H$_\alpha$ instrument. Three structures of the filament: extreme ends, spines , and barbs are shown with arrows.
}
\label{halpha}
\end{figure}

\begin{figure}[ht!]
\centering
\includegraphics[width=0.7
\textwidth]{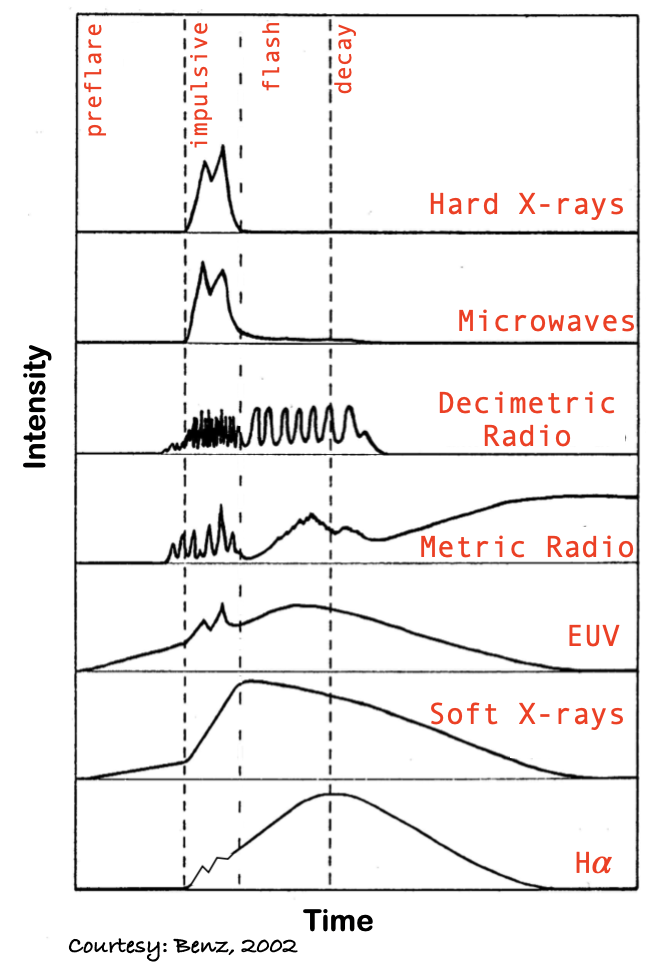}
\caption[Schematic profile for the flare intensity at different wavelengths, {\it i.e.}, in H $\alpha$, soft X-rays, EUV, Metric radio, decimetric radio, microwaves, and hard X-rays.]
{Schematic profile for the flare intensity at different wavelengths (\citealt{Benz2002}).}
\label{flare:intensity}
\end{figure}
\begin{figure}
\centering
\includegraphics[width=0.8
\textwidth]{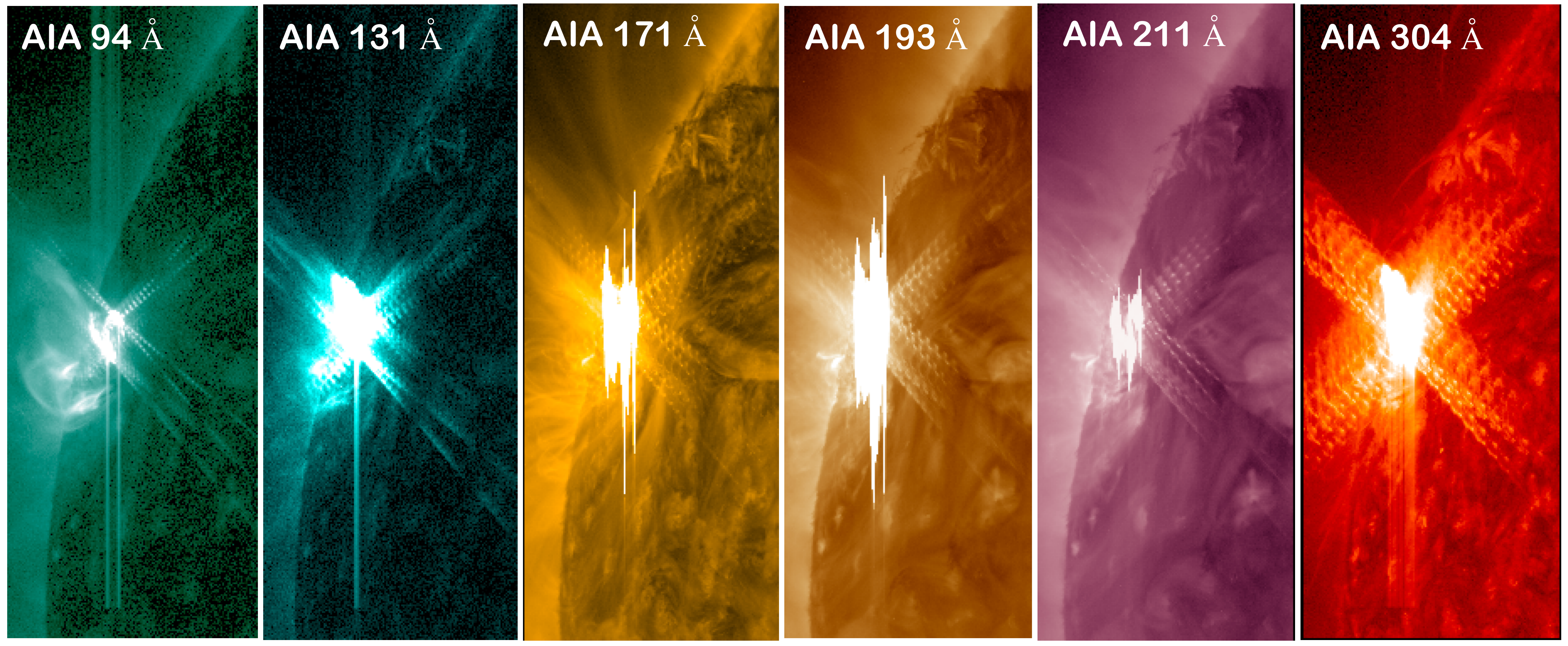}
\caption{Multiwavelength observations of a solar flare observed in six EUV filters of SDO/AIA on the eastern limb on March 9, 2011.}
\label{flare}
\end{figure}
   \item
   {\bf Solar flares:} 
   Solar flares are seen as the bright flashes on the Sun. It is defined as a brightening of any emission observed across the electromagnetic spectra with a lifetime of minutes (\citealt{Benz2008}). The first observation of solar flare was observed in the continuum of white light by R. C. Carrington and R. Hodgson in 1859, afterwards flares are being observed and modelled by many authors (\citealt{Chandra2006}; \citealt{Mandrini2006}; \citealt{Chandra2011}; \citealt{Srivastava2013}; \citealt{Janvier2015}; \citealt{Schmieder2015}; \citealt{Joshi2017}; \citealt{Zuccarello2017}; \citealt{Devi2020}). Usually flares occur in the ARs with a complex geometry of 3D magnetic field (\citealt{Benz2002}; \citealt{Regnier2006}). The different phases of a flare eruption are presented in Figure \ref{flare:intensity}.
   The plasma in the base of the flare region starts to heat up in the preflare phase and appear in the soft X-rays and EUV.
   Electrons and ions are accelerated for the impulsive phase and release an enormous amount of energy. At the chromosphere, hard X--ray footpoints are visible at this phase (\citealt{Hoyng1981}). In the radio band, some of the particles with high energy are kidnapped, and as a result of that intensive emission is being produced. Energy is distributed in decimetre's pulsations in the impulsive phase, where soft X--rays and H$_\alpha$ emissions show a maximum intensity.
   H$_\alpha$ intensity shows a rapid increase in the flash phase.
 According to Geostationary Operational Environmental Satellite (GOES) and National Oceanic and Atmospheric Administration (NOAA) flaring classification to their X--ray brightening between 1--8 \AA\ range, flares are classified as three main categories, {\it namely}: C, M, and X class flares (there are two more classes A and B, for the weaker eruptions). 
In this tenfold classification each class is 10 times more powerful than the previous one. So an X class flare is 10 times stronger than an M class flare and 100 times more powerful than a C class flare. Each category of flare class has further nine subdivisions from C1 to C9, M1 to M9, and X1 to X9 (for the X class the sub classification goes higher and higher upto X28). 
Solar flares are usually accompanied by an emission of high energy particles, with energy release upto 100 MeV. In October 2003, a flare of X28 class was observed and remains as the most strongest flare eruption till now and produced big {\it Halloween storms} (\citealt{Gopalswamy2017}). Recently in September 2017, many X and M class flares were produced and contributed in the large geomagnetic storms, which are responsible for the space-weather phenomenon. An example of solar flare observed in multi-wavelength EUV filters with SDO/AIA on the eastern limb on March 9, 2011 is presented in Figure \ref{flare}. 
\begin{figure}[ht!]
\centering
\includegraphics[width=0.8
\textwidth]{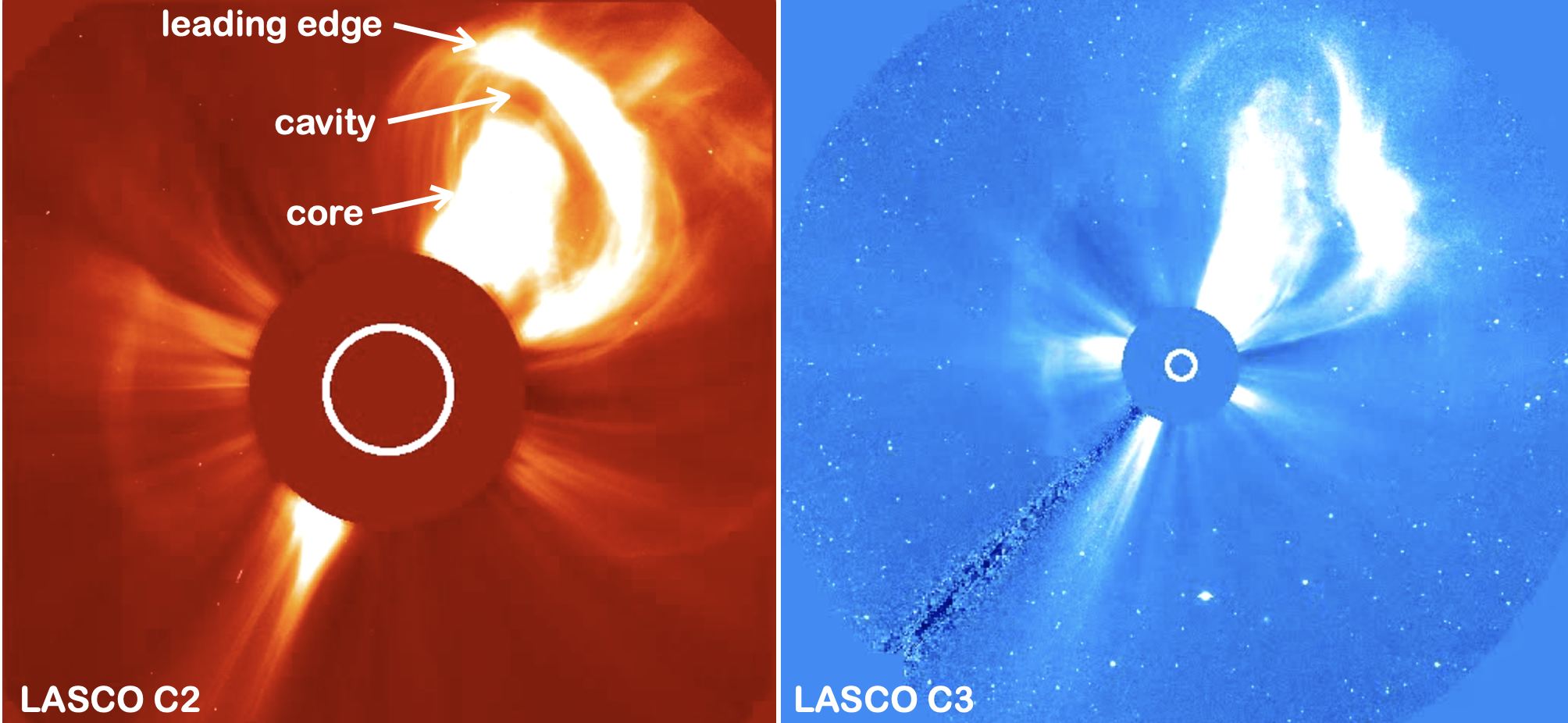}
\caption[A CME observed with SOHO/LASCO instrument in C1 and C2 coronagraph on December 2, 2002.]
{An example of CME observed with LASCO instrument in C2 and C3 coronagraph on December 2, 2002.}
 \label{cme}
\end{figure} 
    \item
    {\bf Coronal Mass Ejections (CMEs):}  
CMEs are the largest scale spectacular eruptions from the solar atmosphere. 
It appears as an outward motion of a big, bright, and detached white light feature in the coronagraph (\citealt{Hundhausen1984}; \citealt{Schwenn1996}; \citealt{Hudson2004}). In the eruption process, a huge mass of plasma ($\approx$ 10$^{11}$-10$^{13}$ kg) is tugged out towards the interplanetary space with speed of 100 km s$^{-1}$ to more than 1000 km s$^{-1}$ (\citealt{Chen2011CME}). In this journey they may interact frequently with the Earth and impacts the terrestrial environment and other high-tech systems in various ways (\citealt{Schwenn2006}; \citealt{Gopalswamy2018}).
Observational studies indicate that CMEs can be observed in many wavebands {\it i.e.}, soft X-rays (\citealt{Rust1983}; \citealt{Gopalswamy1996}), optical and EUV (\citealt{Chen2009}), and in radio (\citealt{Maia1999}). 
During a CME, radio-frequency observations provide first indication of large-scale structural change of the solar corona. They reveal if the energetic electrons are kidnapped in the large coronal structures or moving within the open magnetic field lines.  Observations of type II radio bursts associated with CMEs tells about  the propagation of MHD shock waves. The signatures of type III radio bursts with CME show that open magnetic fields can generate in the ARs as well as in the coronal holes (\citealt{Hudson2004}). 
The CME occurrence rate, with an identification by eye, is available at \url{http://cdaw.gsfc.nasa.gov/CME_list} (\citealt{Gopalswamy2003}). 
 This catalogue shows that the rate of CME occurrence increases from $\approx$  0.5/day near solar minimum to $\approx$ 6/day near solar maximum (\citealt{Yashiro2004}; \citealt{Chen2011CME}).
By the appearance, CMEs can be classified as the narrow CMEs and normal CMEs. Narrow CMEs look like a jet moving along the open magnetic field lines, whereas the normal CMEs contain a three part structure as shown in Figure \ref{cme}. In this structure, a bright front loop is followed by a dark cavity, which again is embedded with a bright core (\citealt{Illing1985}). 
Often, CMEs are accompanied by solar flares and both are considered to be a signature of same magnetic event. Filaments and some times other small scale eruptions, {\it i.e.} solar jets, also follow a CME eruption towards the interplanetary medium (\citealt{Gopalswamy2003}; \citealt{Chandra2017}).

These large scale eruptions are well explained with CSHKP model proposed by \citealt{Carmichael1964}, \citealt{Sturrock1966}, \citealt{Hirayama1974}, \citealt{Kopp1976} and further extended for 3D by \citealt{Aulanier2012}, \citealt{Janvier2013}. CSHKP model explained the flare ribbon formation, their separation, and dynamical flare kernels. To explain the triggering mechanism of solar eruptions, three models were further proposed, {\it namely} {tether cutting} model, {magnetic breakout} model, and {kink instability} model. All these models are based on the instability of the flux rope which results as the loss of equilibrium. For tether cutting mechanism, the magnetic reconnection occurs below the erupting filament while for breakout it occurs high in the corona in null points. Kink instability comes into account where the twist comes in picture and reaches to a critical value.
\end{enumerate}

\subsection{Small scale solar eruptions}
    \begin{enumerate}
    \item {\bf Ellerman Bombs (EBs):}
    EBs are prominent, sudden and short lived bright enhancements in H$_\alpha$ wing images, precisely in the H$_\alpha$ Balmer line ($\lambda=$ 6563 \AA) in ARs (\citealt{Chen2019}; \citealt{Hansteen2019}). This implies that they are formed at photospheric (or a few 100 km above) heights. EBs (also known as moustaches) appear as flame like structures when observed towards the solar limb. They are discovered by Ellerman and described in the discovery paper (\citealt{Ellerman1917}) as: ``a very brilliant, and very narrow band extending four or five \AA\ on either side of the H$_\alpha$ line but not crossing it''. Sometimes with a characteristic of elongated shape, EBs are suspected to be the reminiscent of the ``{\it chromospheric anemone jets}" of \citealt{Shibata2007}. This possibility is well explained by \citealt{Watanabe2011}. They are observed in the emerging flux regions (\citealt{Georgoulis2002}; \citealt{Pariat2004ApJ}; \citealt{Morita2010}; \citealt{Rutten2011}; \citealt{Hansteen2019}). These emerging flux regions are believed to be the host for variety of transient events like EBs. EBs are driven by the interaction of such fields either with photospheric/chromospheric, coronal plasma, or with the pre-existing magnetic field. In the transition region, these compact brightening need high spatial and temporal resolution to be observed hence they are well observed with IRIS in past few years. An example of EBs observed with three IRIS slit jaw images (SJIs) (C II 1330 \AA, Si IV 1400 \AA, and Mg II 2796 \AA) is presented in Figure \ref{ellerman}, highlighted with cyan arrows.
    \begin{figure}
\centering
\includegraphics[width=0.9
\textwidth]{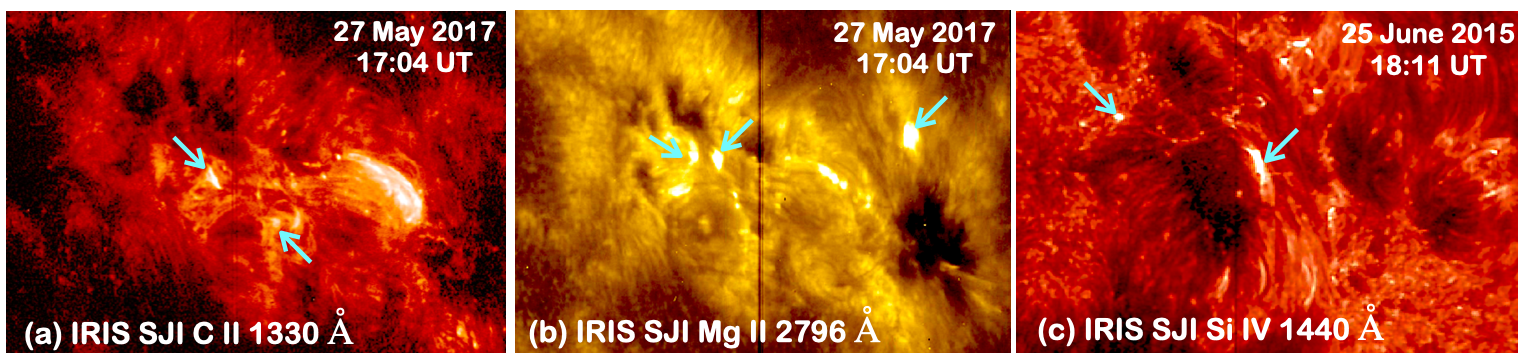}
\caption[Ellerman bombs observed with the slit jaw imager on the IRIS instrument in three transition region wavebands.]
{Ellerman bombs observed with the slit jaw imager on the IRIS instrument in the three (C II, Mg II, and Si IV) transition region wavebands.}
 \label{ellerman}
\end{figure} 
    \item{\bf Spicules, Mottles, and Fibrils:} Small scale activities on the solar chromosphere {\it i.e.} spicules (observed at the limb), mottles (on the quite solar disk), and fibrils (on the ARs on the solar disk) have been observed and modelled for some last decades (\citealt{Beckers1968}; \citealt{Handy1999}; \citealt{Pontieu2007}; \citealt{Rutten2007};  \citealt{Pontieu2012}). It has been observed that the quite sun jets (mottles) and the fibrils share the same triggering mechanisms, though the magnetic field is quite weak at the base of mottles in comparison to fibrils (\citealt{Pontieu2007}). Their presence dominate the highly vigorous chromospheric region, where 90 \% of the non--radiative energy expelled into the outer solar atmosphere and contribute significantly to drive the other solar activities.

    \begin{figure}[t!]
\centering
\includegraphics[width=0.8\textwidth]{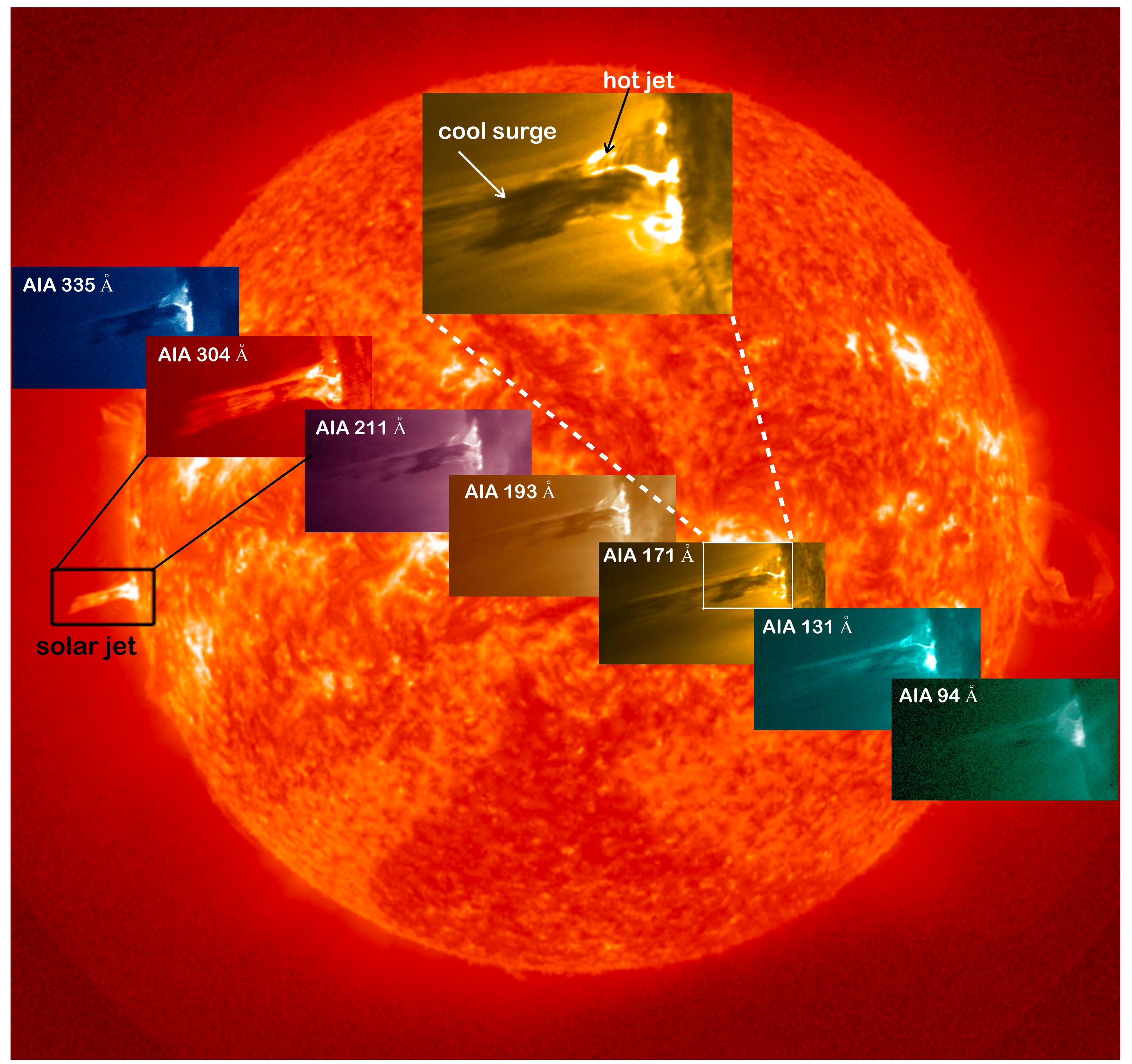}
\caption[Solar jet accompanied by the cool surge on the quite region at the solar limb on January 30, 2015 at 12:59 UT.]
{Solar jet accompanied by the cool surge on the quite region at the solar limb on January 30, 2015 at 12:59 UT observed with multiwavelength SDO/AIA channels.}
\label{fig:jet_QR}
\end{figure}
   \item{\bf Solar Jets and Surges:}
  A Solar jet is defined as a significant amount of plasma ejection from the chromosphere to higher corona. Considered as the most intriguing activity in the solar atmosphere, they have been observed and extensively studied in past few decades (\citealt{Shibata1992}; \citealt{Nistico2009}; \citealt{Moore2010}; \citealt{Shen2012}; \citealt{Liu2014}; \citealt{Sterling2015}; \citealt{Chandra2015}; \citealt{RJoshi2017}). Solar jets are frequently occurring events ($\approx$ 60/day alone in polar coronal holes) (\citealt{Savcheva2007}; \citealt{Srivastava2011}). They have been an attracting research area in solar physics since so long and now again jets are in the limelight as they are considered a possibility to become responsible for the switchbacks (near the Sun magnetic field is abundant with transient, kinked structures) observed with the Parker Solar Probe (PSP) mission (\citealt{Sterling2000}). H$_\alpha$ surges have been analysed since 1973 (\citealt{Roy1973}) and explained as the straight or slightly curved ejections (\citealt{Schmieder1983}; \citealt{Canfield1996}). Surges are considered as the cool counterpart of the solar jets and observational evidences tell us that surges and jets are associated with each other, representing multi--temperature plasma ejections along with the different magnetic field lines (\citealt{Liu2004}; \citealt{Jiang2007}; \citealt{Joshi2020MHD}).
  A multiwavelength observation of solar jet ejection along with the surge is presented in Figure \ref{fig:jet_QR} with SDO/AIA instrument.
  A detail description about the observations and modelling of solar jets is given in Section \ref{ch1_jet_intro}.
  \end{enumerate}

\section{Solar jets: observations and modelling}\label{ch1_jet_intro}


\subsection{Morphological observations}
\begin{figure}[t!]
\centering
\includegraphics[width=0.8
\textwidth]{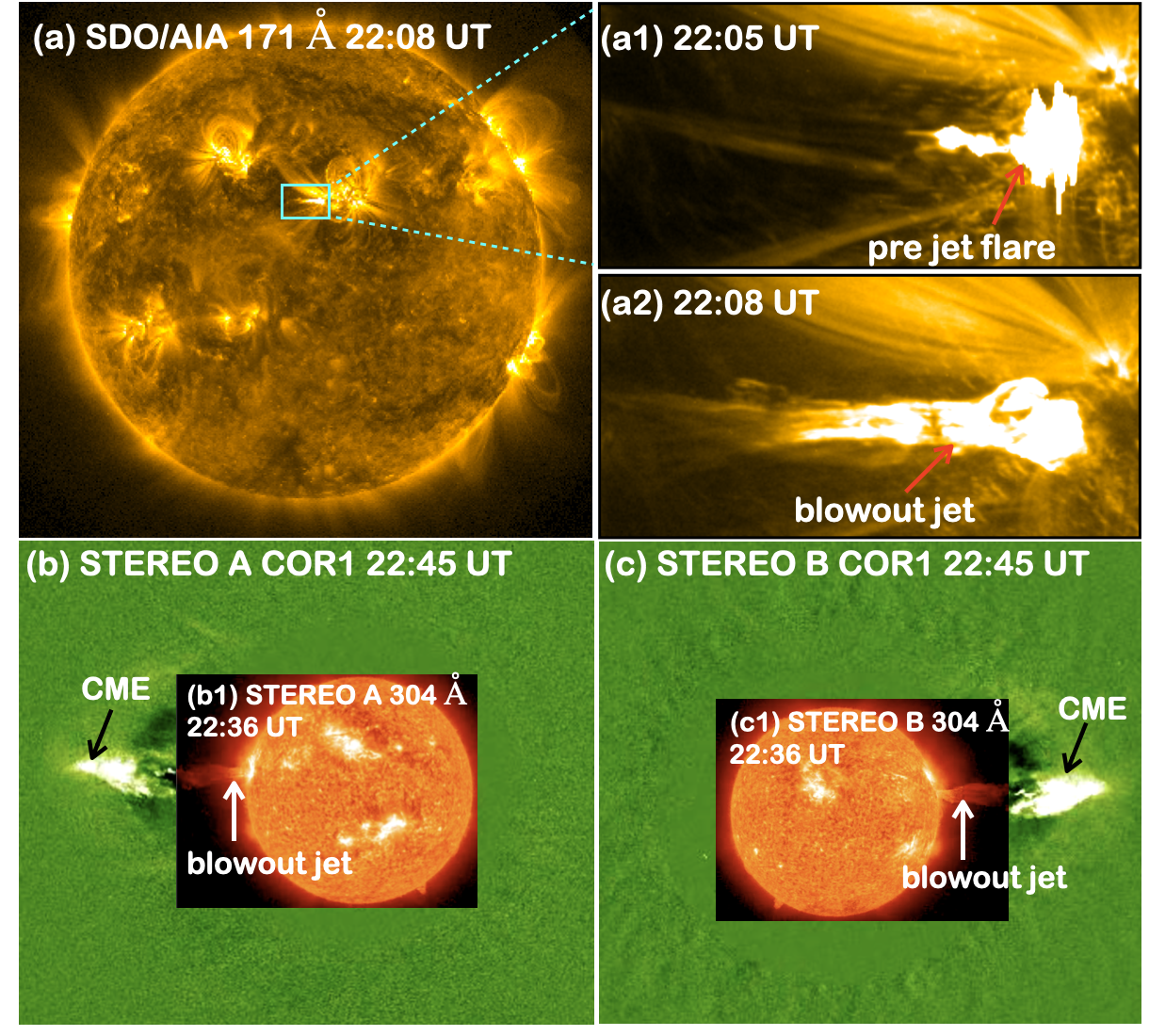}
\caption[Multi instrument observation of a solar jet from AR NOAA 11166 on March 9, 2011.]
{Multi instrument observation of a solar jet and CME from AR NOAA 11166 on March 9, 2011, with SDO, STEREO A and B satellites.}

\label{motivation}
\end{figure}

Solar jet is a common phenomenon of small scale collimated plasma ejection from the solar lower atmosphere towards the solar corona. 
Solar coronal jets are detected throughout the entire solar cycle in a wide wavelength range with different instruments, from X-rays (\citealt{Shibata1992}; \citealt{Chifor2008}) to the extreme ultraviolet (\emph{EUV}) (\citealt{Wang1998}; \citealt{ Alexander1999}; \citealt{innes2011}; \citealt{Sterling2015}; \citealt{Chandra2015}; \citealt{Joshi2020MHD}). Their physical parameters such as height (1-50 $\times$ 10$^{4}$ km), 
lifetime (tens of minutes to one hour),
width (1-10 $\times$ 10$^{4}$ km), and velocity (100-500 km s$^{-1}$) have been studied with these different instruments (\citealt{Shimojo1996}; \citealt{Savcheva2007}; \citealt{Nistico2009}; \citealt{Filippov2009}; \citealt{RJoshi2017}).
A multi instrument observation with AIA EUV 171 \AA, STEREO EUVI 304 \AA\  of an AR jet associated with a GOES C9.4 flare on March 9, 2011 is presented in Figure \ref{motivation}. The associated CME observed with STEREO A and B COR1 is also presented in panels (b-c).
They are observed in active (\citealt{Sterling2016}; \citealt{Chandra2017}; \citealt{Joshi2018}) and quiet regions (\citealt{Hong2011}; \citealt{Panesar2016}). From the previous reported results, it is now well accepted that 68 \% of solar jets are AR jets (\citealt{Shimojo1996}; \citealt{Sterling2017}). \citealt{Raouafi2016} provides a comprehensive review of the coronal jet phenomena, including observations, theory, and numerical simulations.

Solar coronal jets were discovered in the 90’s observed in all the ranges of temperatures from 10$^4$ K to 10$^7$ K in multi-wavelength observations from H$_\alpha$ with ground based instruments (\citealt{Gu1994}; \citealt{Schmieder1995}; \citealt{Canfield1996}) 
 to X-rays (\citealt{Shibata1992}; \citealt{Chifor2008}). Jets often have a helical structure containing both hot and cooler ejected plasma. The co-existence of hot and cool emissions along a jet was first observed with the TRACE observations reported by \citealt{Alexander1999}. 
 \begin{figure}[t!]
\centering
\includegraphics[width=0.7\textwidth]{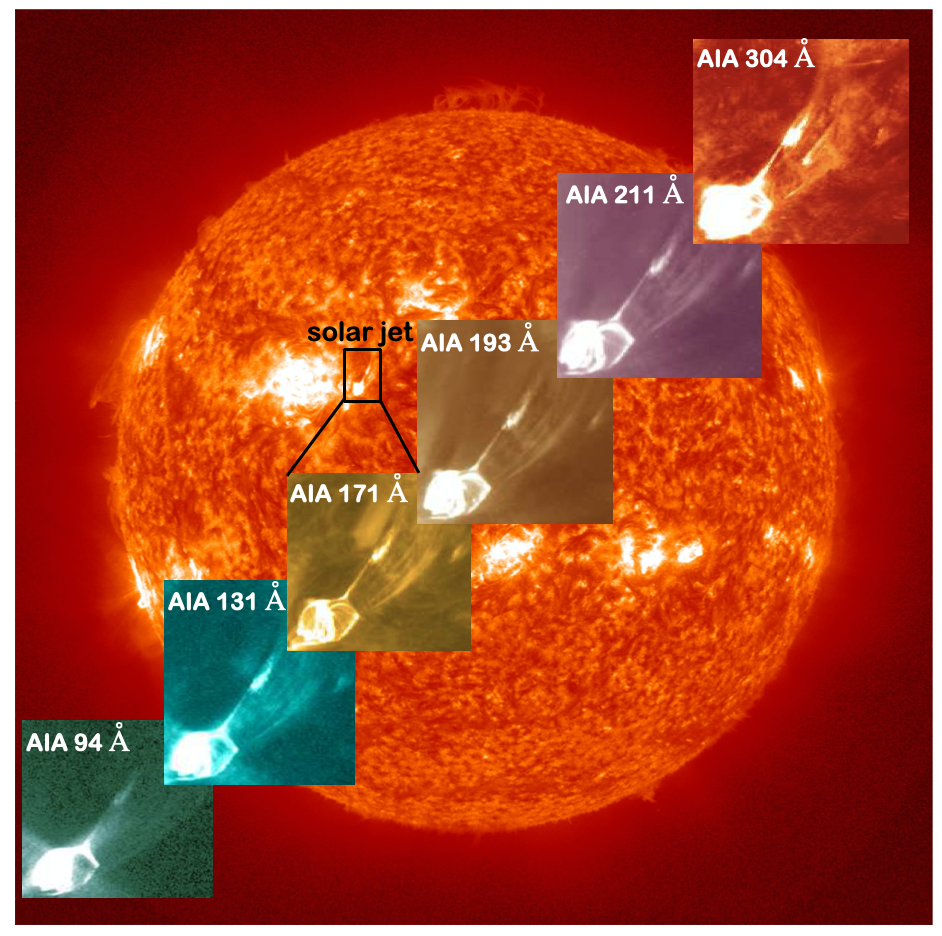}
\caption[Multi-thermal EUV observation of a jet associated with a GOES C class flare on April 28, 2013.]
{Multi-thermal EUV observation of a jet associated with a GOES C class flare on April 28, 2013. The zoom view in  different AIA wavelengths show the jet eruption with a broad bright base.}

\label{fig:jet_AR}
\end{figure}
 Further \citealt{Chae2003} and \citealt{Jiang2007} did the amalgamation  of TRACE observations of coronal hole jets and chromospheric surges with the photospheric magnetic field. 
 These observations also provided evidences that the locations with mini/micro flares observed with RHESSI are associated with the jets. RHESSI observations of jets taking place at the time of standard or mini/micro flares established the strong spatial and temporal relation between jets and flares. Solar flare or the base brightening at the jet footpoint is believed to promote the force for pushing the plasma material upward (\citealt{Joshi2020ApJ}). The hard X-ray emissions at the jet base often associated with the small loop structures energized by the flares. This clearly provides the important role of magnetic reconnection for the triggering of solar jets (\citealt{Krucker2008}; \citealt{Bain2009}; \citealt{Zuccarello2017}).

\begin{figure}[t!]
\centering
\includegraphics[width=0.7\textwidth]{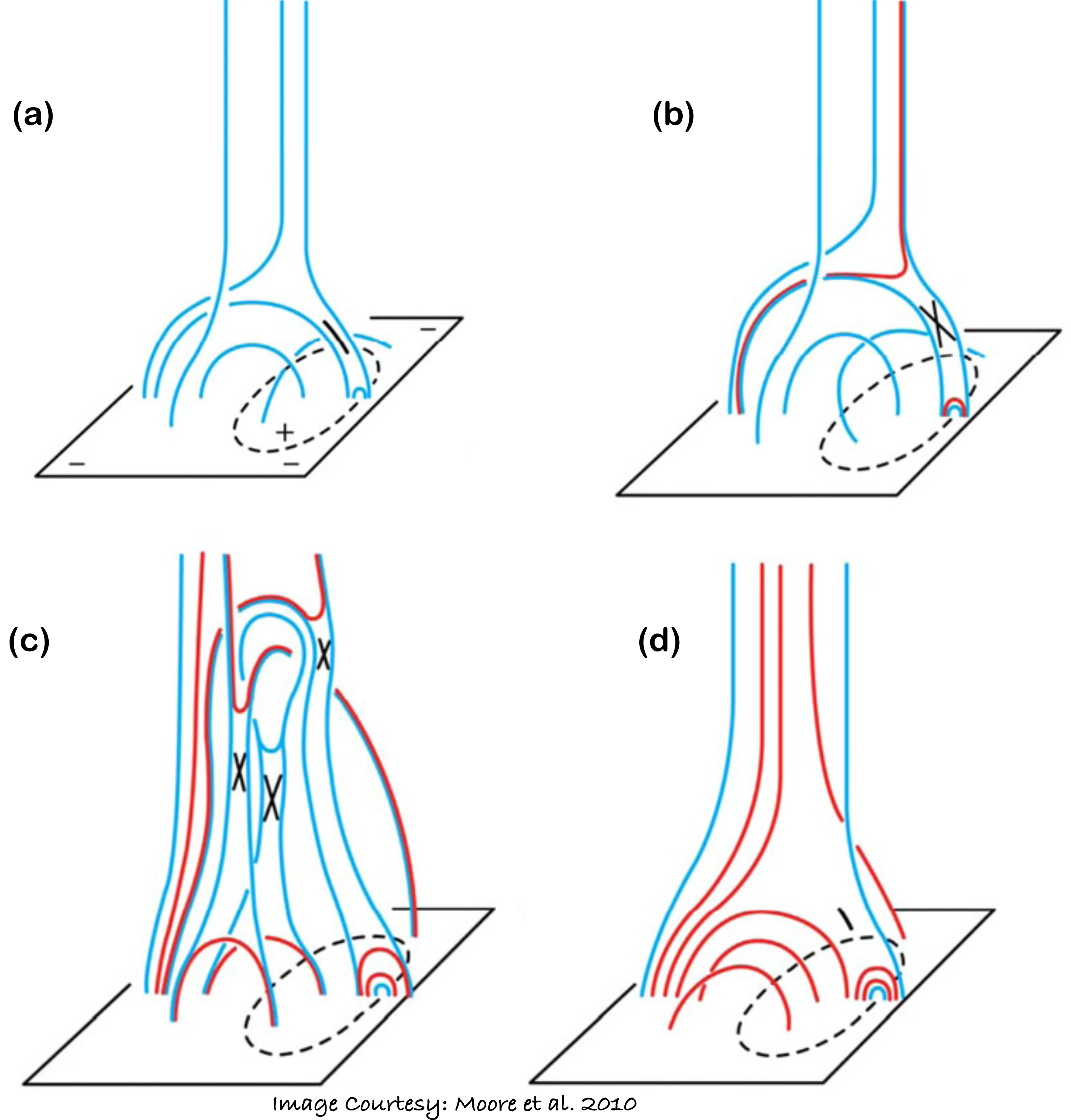}
\caption[Schematic representation of the structure, eruption process, and reconnection of the magnetic field for the classification of jets by \citealt{Moore2010}.]
{Schematic representation of the structure, eruption process, and reconnection of the magnetic field for the classification of jets by \citealt{Moore2010}.}

\label{jetmodel}
\end{figure}

Multi-instrument (AIA, EIS, XRT, and IRIS) observations provided the valuable insights for the classification and morphology of solar  jets.
According to the jet eruption process, (\citealt{Moore2010}) classified solar jets in two sub-classes, {\it i.e.} {\it standard} and {\it blowout} jets by keeping in mind the Hinode/XRT observations. In a standard jet, the core field of the base arch remains close and static whereas in a blowout jet it explodes and results in a breakout eruption. They further clarified that about two-third of the observed X–ray jets fall in the standard picture of jets and one third are of blowout category. For the blowout jets, the base bright points start as a compact feature similar to standard jets but gradually the entire jet base brightens to resemble as bright as the jet base bright point. An example of similar blowout jet is presented in Figure \ref{fig:jet_AR}, where the hot jet erupts with cool counterpart from the solar disk with a circular bright base. The jets with narrow spire fit well in the original jet picture proposed by \citealt{Shibata1992}, hence dubbed as ``standard'' jets. Both the standard and blowout jets start to erupt with the emergence of a magnetic bipole and followed by its reconnection with the pre-existing ambient magnetic field. Specially for the blowout jets, the emerging unstable bipole erupts and blows out the bipole with the surrounding field. This outward motion carries the cool chromospheic material. Hence the in case of the blowout jets, cool material (observed with AIA 304 \AA) is accompanied with the hot jet. The schematic diagram for this classification is presented in Figure \ref{jetmodel}, where red lines are the reconnected field lines and blue lines are those which either have not been reconnected or will never be reconnected.
The dichotomy of coronal jets into two categories is a result of the shear/twist in the base arch of the jet. Blowout jets usually have a high shear/twist in the base to open and erupt (\citealt{Liu2009}; \citealt{Joshi2020ApJ}). Helicity can be transferred from the closed field into the open field due to the magnetic reconnection between them. This ejection of helicity gives birth to the motion of the jet material upwards by nonlinear torsional Alfv\'en waves (\citealt{Pariat2009}; \citealt{Jiajia2019}). The magnetic reconnection between the closed and open field lines is the causatum of magnetic flux emergence and cancellation. The continuous magnetic flux cancellation and emergence destabilize the field at the jet base.

\subsection{Spectroscopic observations of jets and UV bursts}
IRIS spacecraft has revealed several transient small scale  phenomena 
in the solar atmosphere  
such as
UV bursts (\citealt{Young2018}), IRIS bombs or IBs (\citealt{Peter2014}; \citealt{Grubecka2016}; \citealt{Chitta2017}; \citealt{Tian2018}), explosive events (\citealt{Kim2015};   \citealt{Gupta2015};  \citealt{Huang2017};  \citealt{Chen2019}; \citealt{Ruan2019}) blow jets (\citealt{Shen2017})
 and bidirectional outflow jets (\citealt{Ruan2019}). 
UV bursts are very tiny bright points with a bright core less than 2$\arcsec$. Their lifetime is short ($\approx$ 10 s) but with possibly recurrent enhancements during one hour giving the impression of
 flickering (\citealt{Pariat2007}). With IRIS instrument the chromospheric C II and Mg II lines are frequently observed in the UV bursts and mainly  in the quiet chromosphere as well as in solar flares and jets (\citealt{Leenaarts2013a};  \citealt{Rathore2015}). They  are optically-thick lines and need a radiative transfer approach to determine the physical quantities of  plasma.
  The Mg II h and k resonance lines in the quiet Sun are formed over a wide range of chromospheric heights. They usually appear as doubly peaked profiles with a central reversal.
  IRIS  spectral data allow to make many  progresses on the plasma diagnostics in flares.
\citealt{Kerr2015} and \citealt{Liu2015}
 recently discussed the emission of chromospheric  lines as observed in solar flares. They said about these lines that: ``They appeared as redshifted, single-peaked profiles, however some pixels present 
 a net blue asymmetry''. The blue asymmetry can be explained by down-flowing plasma absorbing the red peak emission and not by strong blueshift emission
 (\citealt{Berlicki2005}). IRIS spectroscopic and imaging  observations of jets reveal bidirectional outflows (extended wings in chromospheric and transition line profiles) in transition region lines at the jet base, implying  explosive  magnetic  reconnection processes  (\citealt{Li2018};  \citealt{Ruan2019};  \citealt{Joshi2020IRIS}).
  
\subsection{Transverse motion, prejet oscillations, and rotation in jets}
The sideways motion of coronal jets have been studied in several observations (\citealt{Shibata1992};  \citealt{Canfield1996};  \citealt{Savcheva2007};  \citealt{Chandrashekhar2014};  \citealt{RJoshi2017}). Polar coronal hole jets were studied by \citealt{Savcheva2007}, where more than half jets move in the transverse direction with a speed of 35 km s$^{-1}$. Similarly \citealt{Shibata1992}, found the side-way motion of an X-ray jet with 20-30 km s$^{-1}$. This transverse speed decreases with an increase of height and sometimes the shifting of jet footpoint also show a whip like motion, followed with an expansion of closed magnetic field lines and explained in the mentioned studies.
The transverse motion of jets may explain with two mechanisms {\it namely} expanding ({\it curtain like spires}) motions and oscillatory motions. For the first possibility, of {\it expanding motions}, theoretically it has been found that with an Alfv\'en speed of 1000 km s$^{-1}$, the reconnected flux moves with $\approx$ 100-1000 km s$^{-1}$. This experimental value is much more larger than the observed speed {\it i.e.} 20-50 km s$^{-1}$. Hence it may give a hint about the expansion of the reconnection region rather than the motion.
The second flavour of transverse motion, {\it oscillations}, is used to determine the temperature and magnetic field in the solar corona (\citealt{Cirtain2007}). For the temperature estimation, \citealt{Morton2012} studied a dark jet oscillations with a time period of 1 minute and inferred a temperature of less than 3$\times$10$^4$ K from kink mode oscillations (\citealt{Raouafi2016}). For the magnetic field approximation \citealt{Chandrashekhar2014b} explained a coronal hole boundary jet oscillating with a time period of 3.6 minutes. They inverted this time period value to 1.2 Gauss magnetic field strength.   

For the acceleration mechanisms of the solar eruptive events and the energetics of the solar jets, axial speed is the key parameter to evaluate. From the UV/EUV observations, it has been found that the observed speed of the ondisk solar jets is 10-20$\%$ smaller than the actual speeds. Hence the apparent speeds calculated with the imaging instruments are the lower limits of the jet speed. The usual observed axial speed for solar jets (200 km s$^{-1}$) is comparable with the coronal sound speed (\citealt{Shimojo1996};  \citealt{Savcheva2007};  \citealt{Raouafi2016}).
Hence it is believed that these bright and hot jets are a result of chromospheric evaporation and is responsible for the acceleration of solar jets. On the basis of measured apparent speeds and temperatures, \citealt{Sako2013} classified a number of jets into thermal or magnetic dominated category. They reported that most of the AR jets are thermally driven jets unlike the quite region and coronal hole jets. 

A common property for solar jets is to display a  twisting motion  or rotation (\citealt{Raouafi2016};  \citealt{Joshi2020MHD}). Twist of jets can be due to the  helical motions (\citealt{Nistico2009};  \citealt{Patsourakos2008}). 
Helical structures and motions are believed to play an important role for storing free magnetic energy and results as the torsional waves or instability.
This complete process transfers the magnetic energy into the upper heliosphere and convert the energy to thermal/kinetic energies. Magnetic reconnection between the twisted closed FRs and open field lines gives rise to the untwisting process, which is believed to be a signature for the rotational motion of solar jets.
Twisting motions have been found in a large velocity range of jets and surges (\citealt{Chen2012};  \citealt{Hong2013};  \citealt{Zhang2014}). In the study done by \citealt{Schmieder2013},  a jet revealed a striped pattern
of dark and bright strands propagating along the jet with apparent damped oscillations across the jet. They concluded that this is suggestive of an (un)twisting motion in the jet, possibly Alfv\'en wave. 
The physical mechanism behind the untwisting process may explain as: 
A pre existing or newly emerged closed flux system which contain the twist inside of it, reconnects with the ambient open magnetic field lines.
During this interaction of field lines, the twist from closed system could be transferred to open field lines. The jet motion follows the same path as with the open magnetic field lines and hence twist could pass to the jet from the open field (\citealt{Jiajia2019};  \citealt{ Joshi2020FR}). 
With the jet rotation speed `$v_{rot}$', and jet width `d', the rotational period can be estimated as:- 
\begin{equation*}
    T_{rot}\approx \frac{\pi d}{v_{rot}}
\end{equation*}
If the life time for rotational motion is `T', then the total number of turns a jet may propagate is $\frac{T}{T_{rot}}$, and the twist will be $2\pi \frac{T}{T_{rot}}$. In some recent studies done by \citealt{Moore2013,Jiajia2019}, it has been found that the rotating jet shows $\approx$ 1.3 turns (twist=2.6$\pi$). This value is in agreement with the theoretical values provided by \citealt{Hood1981}, and in the numerical simulations of \citealt{Pariat2009}.
Spectroscopic data also provide signatures for detecting the twist for the on disk solar jets with different Dopplershift measurement at different heights. It is difficult to study the rotational motion for the on disk effect sue to a line of sight (LOS) effect. In the same way, blue and red shifts observed along the axis of a jet in H$_\alpha$ as well as in Mg II lines were interpreted by the existence of twist along the jet (\citealt{Ruan2019}). A further spectroscopic analysis is done by \citealt{Joshi2020FR} using IRIS Mg II, C II, and Si IV spectras to analyse the transfer of twist from a stable FR to the jet.

The observational analysis
from the previous studies
 has revealed that the jet evolution might be 
preceded by some wave-like or oscillatory disturbances (\citealt{Pucci2012};  \citealt{Li2015};  \citealt{Bagashvili2018};  \citealt{Joshi2020MHD}). \cite{Pucci2012} studied the X-ray jets observed with {\it Hinode} and
found that most of the jets are associated with oscillations of the coronal emission in bright points at the base of the jets and a detail analysis is done by \citealt{Madjarska2019}. 
They concluded that the pre-jet oscillations are the result of a change  in jet base area or 
temperature of the pre jet activity region. Recently, a statistical analysis of pre jet oscillations of coronal hole jets has been carried out by \citealt{Bagashvili2018}, and \citealt{Joshi2020MHD}. They reported that most of the selected jets in their study  were preceded by 
pre jet intensity oscillations of some 12-15 minutes before  the onset of the jet.  
They tentatively suggested that these quasi-periodic intensity
oscillations  may be  the result of MHD wave generation through rapid temperature variations and shear flows associated with  local reconnection events (\citealt{shergelashvili2006}). Quasi-oscillatory variations of intensity can be a signature of MHD wave excitation processes, which are generated by very rapid dynamical changes in velocity, temperature, and other parameters which manifest the apparent non-equilibrium state of the medium in which the oscillations are sustained (\citealt{Zaqarashvili2002};  \citealt{Shergelashvili2005};  \citealt{Joshi2020MHD}).

\subsection{Theoretical models of jet formation}
Magnetic reconnection is believed to trigger the activation of the jet eruption established with different theoretical models  (\citealt{Yokoyama1995};  \citealt{Archontis2004}; \citealt{Pariat2015}). 
Reconnection is the process of restructuring the magnetic field lines and 
can occur in 2D 
(\citealt{Filippov1999};  \citealt{Pontin2005}) or in 3D configurations
(\citealt{Demoulin1993};  \citealt{Filippov1999};  \citealt{Longcope2003};  \citealt{Priest2009};  \citealt{Masson2009}). Coronal null is an area which is suitable for building up a thin and strong current sheets where magnetic reconnection can occur in an explosive manner.
 In a 2D magnetic null-point configuration, magnetic field lines contained in a plane and with opposite orientations approach each other across an `X'-point 
 and instantaneously change connectivity. The results are hybrid field lines that expelled away from the `X'-point, typically with velocities of about the Alfv\'en speed.
 In 3D there is a whole variety of possible patterns ({\it e.g.} spine-fan, torsional, or separator reconnection pattern). In many cases the underlying structure is known as a fan-spine configuration around a central null-point. The field lines from inside the fan surface are joined to open field lines from just outside with ensuing connectivity change. Changes in the remote connectivity of magnetic field lines may also take place in regions with strong spatial gradients of the field components that are called quasi-separatrix layers ({\it QSLs}) (\citealt{Mandrini2002};  \citealt{Zuccarello2017}). 
Two main principal scenarios have been investigated for the reconnection of magnetic field lines, which is responsible to drive the coronal jets: {\it flux emergence scenario} and {\it instability onset scenario}. 

In the flux emergence scenario, the newly emerging magnetic field from below photosphere reconnects with the ambient coronal field. In this principle the driver of the jet is the vertical or horizontal flow of plasma in the neighbourhood of the reconnection site. In this way there is an increase of magnetic flux in or near to the jet region. The simulations have been done by \citealt{Yokoyama1995,Yokoyama1996} in 2D and by \citealt{Moreno2008}; \citealt{Torok2009}; \citealt{Moreno2013}; \citealt{Archontis2013}.
The second scenario for driving the jet is onset of instability or the loss of equilibrium, where the non-potential and stressed flux below the null point reconnects with the quasi potential and ambient flux exterior to the fan surface, after a threshold. Unlike the flux emergence models, there is no increase in the vertical magnetic flux inside or in the neighbourhood of the jet because the reconnection happens due to the rearrangement of coronal fields not due to the plasma flows. This scenario is modelled by \citealt{Pariat2009, Pariat2010}; \citealt{Dalmasse2012}.

Magnetic reconnection can take place as a result of a process of magnetic flux  emergence from the low solar atmosphere or interior. In typical magnetic flux emergence processes, 
the emerging magnetised plasma interacts with the pre-existing ambient coronal magnetic field, thus providing a favourable condition for magnetic reconnection, and therefore for the occurrence of solar jets.
The observations indicate that the expansion of the region in which the magnetic flux emerges leads to reconnection with the ambient quasi-potential field and magnetic cancellation 
(\citealt{Gu1994};  \citealt{Schmieder1996};  \citealt{Liu2011};  \citealt{Guo2013}).
A number of numerical models have simulated this process
(\citealt{Yokoyama1996};  \citealt{Moreno2008};  \citealt{Torok2009};  \citealt{Archontis2013}; \citealt{Moreno2013}; \citealt{Nobrega2016};  \citealt{Ni2017}).
 
 \begin{figure}[ht!]
  \centering
  \begin{subfigure}[b]{0.8\textwidth}
    \includegraphics[width=\textwidth]{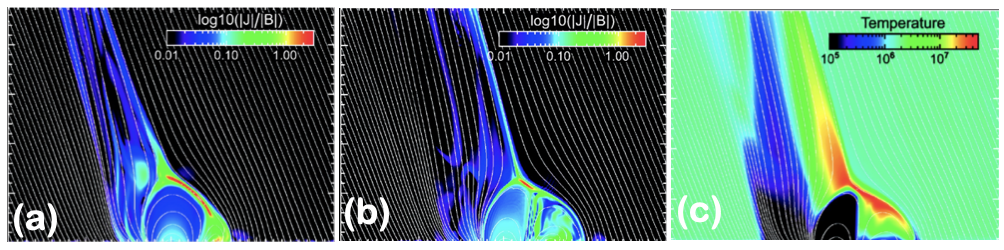}
    \caption{Current sheet formation and an inverted-Y (Eiffel tower) type jet eruption process. The jet structure appears mainly, with T $\approx$ 3 $\times$ 10$^7$ K (at the reconnection site) and $\approx$ 10$^7$ K (upward pointing jet) at two different locations.}
    \label{vault}   
  \end{subfigure}             
  \begin{subfigure}[b]{0.6\textwidth}
    \includegraphics[width=\textwidth]{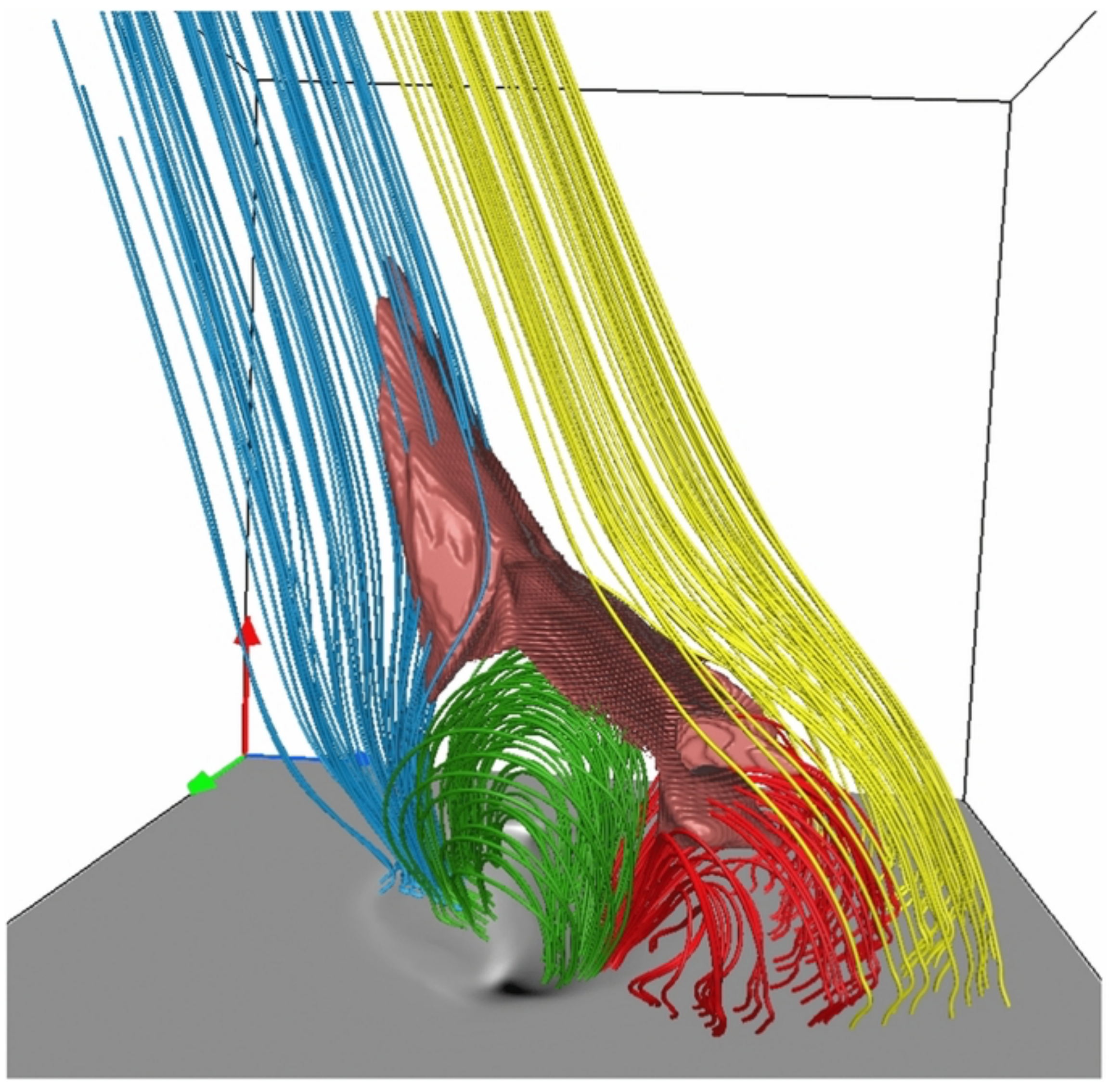}
    \caption{Blue, green, red, and yellow, field lines in the selected jet region are presenting the basic connectivity domains in the experiment done by \citealt{Moreno2013}. 
    }
    \label{3D_emergence}
  \end{subfigure}             
  \caption[3D magnetic flux emergence MHD models]{3D magnetic flux emergence models given by \citealt{Moreno2008}.}
  \label{moreno}
\end{figure}

The 2D magnetic simulations for flux emergence scenario was done by \citealt{Yokoyama1995, Yokoyama1996}, with two type of magnetic configurations {\it e.g.} anemone type and two sided type during the reconnection. In their simulations, they successfully reproduced hot and cool jets simultaneously, which is similar to the observations where a hot EUV jet is followed by a cool H$_\alpha$ surge. They explained that the hot jet is not a direct outcome of the reconnection. Its a two step process: (a) the reconnection outflow reacts with the coronal magnetic fields. This collision produces a fast mode MHD shock. (b) The outflow material is deflected/ switched back by this shock and become the hot jet moving parallel to the large scale coronal field. This 2D model also predicts about the plasmoids (magnetic islands) ejection from the current sheet during the jet propagation and the rate of reconnection increases during the plasmoid ejection. In the  2D model it is hard to observe  the plasmoid though it is responsible for the fast reconnection. Recently in favour of this 2D model, \citealt{Joshi2020MHD} observed the motion of plasmoid blobs in AIA and IRIS observations.

A 3D model for the flux emergence scenario was first proposed by \citealt{Moreno2008} with an inserted magnetic tube near the bottom of the domain and leads to form the buoyant $\Omega$ loops. The reconnection between the rising loops and the coronal plasma material gives rise to a concentrated and curved current sheet which leads to the outflow of the jet. In this model a split-vault structure is clearly shown to form below the jet, and it contains two chambers: the chamber containing previously emerged loops with a decreasing volume, and the chamber containing reconnected loops with a increase in volume as a result of reconnection. This structure is also confirmed in radiation MHD simulations by \citealt{Nobrega2016} and observations by \citealt{Joshi2020MHD}. A three dimensional view of the current sheet formation and two vault structures from the model of \citealt{Moreno2008} is presented in Figure \ref{vault}.
 The observations that motivated those models were either X-ray jets observed by Hinode (\citealt{Moreno2008}), 
 or cool surges observed in chromospheric lines and bright bursts in transition region lines (\citealt{Nobrega2017}). The jets in this model  show a horizontal motion due to the gradual decrease in the connectivity of the emerged loops turning to reconnected loops with a speed of 10 km s$^{-1}$.
 Recently these models  have been compared with simultaneously observed hot jets and cool surges by \citealt{Joshi2020MHD} with SDO and IRIS observations.

 The 3D magnetic flux emergence model is further analysed in detail by \citealt{Moreno2013} and presented in Figure \ref{3D_emergence}. In this 3D model, the jet flow has a shape like hollow semi cylinder with fast (300-400 km s$^{-1}$) and slow streams. Many null points and plasmoids were identified in this 3D jet model. The speciality of the jet is a complete modification in 3D of the 2D model provide by \citealt{Yokoyama1995, Yokoyama1996} and explained above. In this model a wall of dense and cool plasma material is also observed along with the hot EUV jet. This dense, cool and slower part of the jet in the model has a speed of $\approx$  50 km s$^{-1}$ and meets the expectations with the cool jets introduced by \citealt{Yokoyama1995}. \citealt{Torok2009} proposed a $\beta=0$  MHD simulation for the flux emergence, with a two step reconnection process. They proposed the drift of null points from location to another. This leads to a displacement of reconnected field lines foot-point and connection with the pre-existing background field, which is observationally confirmed by \citealt{RJoshi2017}. 

The another instability onset scenario is based on the kink instability occurrence which reveals that the generation of solar jet is a multi phase process of energy storage and dissipation. In the energy storage phase, a highly localized and thin current sheet formation takes place at the null point. The inclination angle of the background field with the vertical direction plays an important role to decide whether the jet would be straight or helical.  
Large inclination angles introduce asymmetries into the growing and strengthening of current sheet and sufficiently inclined field lines start to reconnect quasi steadily. This process is followed by a slow energy release and generates an outflow of straight jet of tension driven outflows. This straight jet is due to the retraction of newly reconnected magnetic field lines  as in  the ``{\it slingshot effects}'' (\citealt{Dalmasse2012};  \citealt{Pariat2015}).
For small inclination angle of background field, the energy storage rate exceeds over the slow energy release process and hence magnetic field continues to accumulate and no straight jet is generated. 
In this case the explosive energy release process is driven by the ideal kink-like instability and a broad, highly dynamic current sheet produces an impulsive release of enormous energy and helicity. 
A broad helical jet driven by the large amplitude, torsional Alfv\'en waves moves along with the newly reconnected open field lines.
The stored free energy and helicity is carried away by these Alfv\'en waves. The helical jet generation depends strongly on the precursor of the straight jet {\it i.e.}, a strong straight jet reduces the amount of energy released during the helical jet and delays the triggering of the helical jet. This 3D modelling and simulations based on the instability scenario was further carried out by \citealt{Karpen2017}. The Alfv\'en waves, responsible for the motion of helical jets can propagate out into the inner heliosphere and the signatures may be detected by the Parker solar probe and solar orbiter spacecrafts in the corona, as earlier done with STEREO. This straight/helical jet classification may represent a similar morphology but different nomenclature for the standard/blowout scheme given by \citealt{Moore2010}. As the straight jets are well collimated narrow plasma ejections with an inverted ``Y'' shape structure and less (or no) rotational features, exactly similar to the standard jets. On the other hand the helical jets, very similar to blowout category are broad and strongly rotating jets containing more mass and energy than the straight jets.

\section{Data sets and reduction techniques }
To study the solar jets and related flares, we have collected data with different space borne satellites and ground based observatories and analysed it. A brief description about the used instruments, and the data analysis techniques are given in this section.
\begin{table}[h!]
\centering
  \caption[Classification of three different instruments onboard SDO.]{Classification of three different instruments onboard SDO (\citealt{Pesnell2012}).}
\label{tab:SDO}
\setlength{\tabcolsep}{12pt}
\begin{tabular}{ll}
  \specialrule{.1em}{0.1em}{0.1em}
      Instruments & Description \\
      \specialrule{.1em}{0.1em}{0.1em}
      AIA&  Rapid cadence full-disk UV-EUV solar images\\
      HMI&  Full-disk Dopplergrams\\
      & Full-disk LOS magnetograms\\
      & Full-disk vector magnetogram\\
      EVE& Rapid cadence EUV spectral irradiance\\
     \specialrule{.1em}{0.1em}{0.1em}
    \end{tabular}
\end{table}
\subsection{Solar Dynamics Observatory (SDO)}
\label{sec:SDO}
\begin{figure}[t!] 
\centering    
\includegraphics[width=0.8\textwidth]{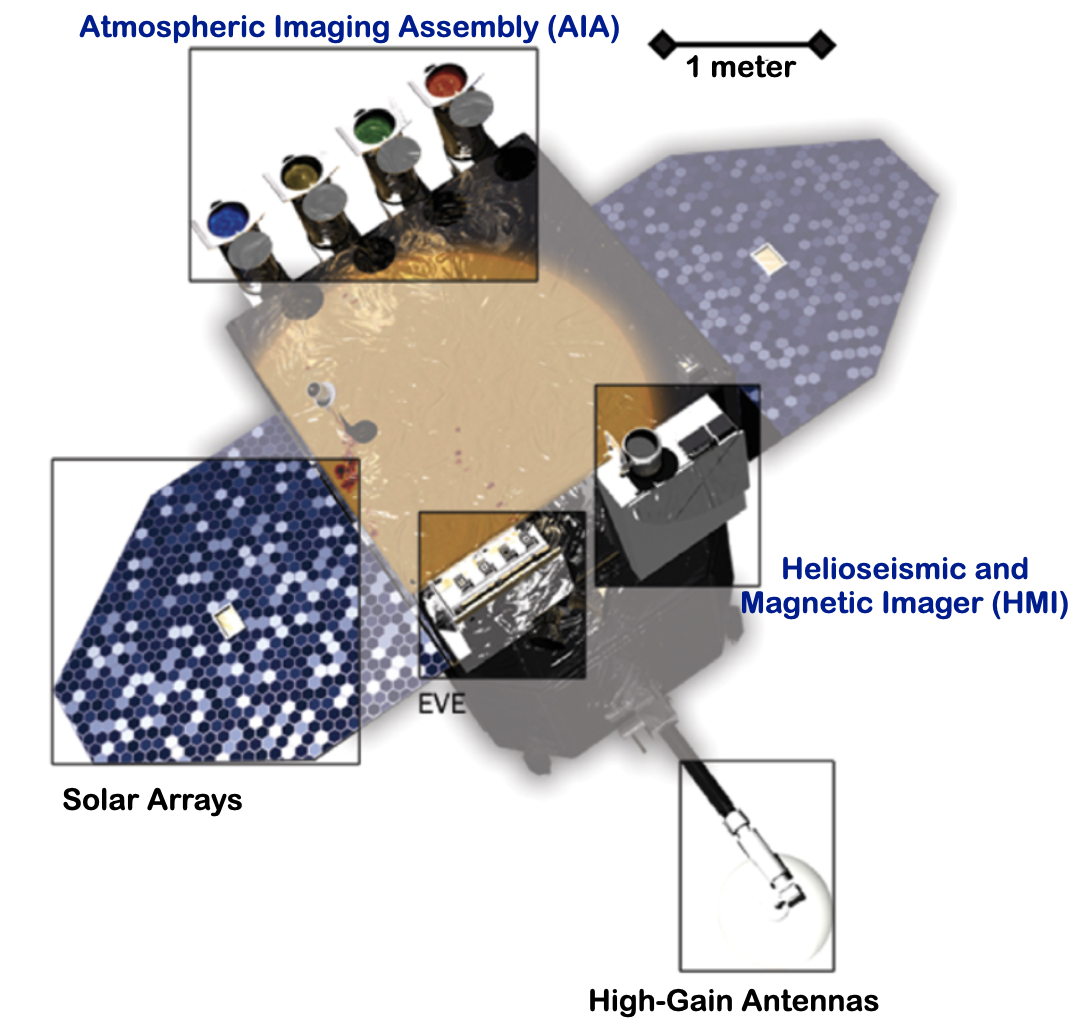}
\caption[Three instruments (AIA, EVE, and HMI) onboard SDO spacecraft.]{Three instruments (AIA, EVE, and HMI) onboard SDO spacecraft (Image credit: NASA).}
\label{im_SDO}
\end{figure}

Solar Dynamics Observatory (\citealp[SDO,][]{Pesnell2012}) is the a space mission under NASA's Living With a Star (LWS) Program. It was launched on February 11, 2010 from Kennedy space center in Florida. The main objective of SDO is to understand the solar variations and their effects on the Earth's climate and other technological arrangements. SDO is also working to determine the generation of magnetic field on the solar surface and its evolution. This mission ensembles three scientific instruments: the Atmospheric Imaging Assembly \citep[AIA,][]{Lemen2012}, Extreme Ultraviolet Variability Experiment (EVE), and Helioseismic and Magnetic Imager \citep[HMI,][]{Schou2012}. A brief classification of three instruments onboard SDO is given in table \ref{tab:SDO} Along with these three instruments, SDO contains a spacecraft bus and a ground station for monitoring. SDO is transmitting 150,000 fulldisk solar images with 9000 spectras. The observational studies presented in this thesis are done with AIA and HMI instruments onboard SDO. An image of the SDO spacecraft with AIA and HMI is presented in Figure \ref{im_SDO}. Full resolution science data for AIA and HMI are freely available at the Joint-SOC (JSOC) Stanford website\footnote{http://jsoc.stanford.edu}.

\begin{enumerate}
    \item  {\bf Atmospheric Imaging Assembly (AIA)}: AIA provides multiple high resolution full-disk image of the solar corona and transition region upto 0.5 R$_\odot$  of the solar limb with spatial resolution of pixel size 0\farcs6 and temporal resolution of 12 second. It consists of four generalised Cassegrain telescopes with a 20 cm primary mirror and an active secondary mirror on each telescope. Each of the four telescope has a field of view (FOV) of $\approx$ 41 $\arcmin$ circular diameter and observes the full-disk over a 4k $\times$ 4k CCD. 
\begin{figure}[ht!] 
\centering    
\includegraphics[width=0.8\textwidth]{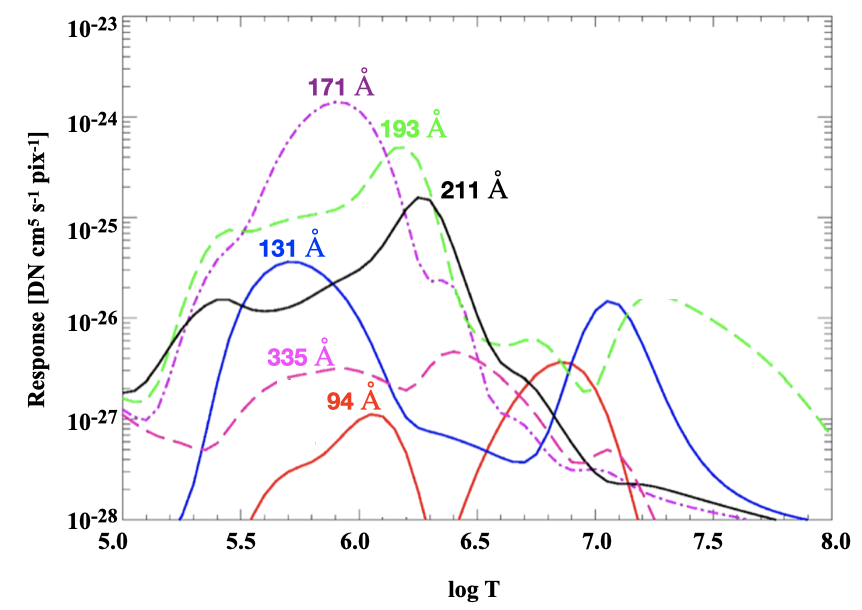}
\caption[AIA response function curve for six EUV channels.]{Response function curve dominated by Fe emission lines 
(\citealt{Lemen2012}).}
\label{AIA_response}
\end{figure}
\begin{table}[ht!]
\centering
  \caption[EUV and UV channels in AIA instrument with the primary ions.]{EUV and UV-visible channels in AIA instrument with the primary ions (\citealt{Lemen2012}).}
\label{tab:AIA}
\setlength{\tabcolsep}{6pt}
\begin{tabular}{llll}
      \hline
       
      Channel & Primary ion(s) & Region of atmosphere& log T \\
      \hline
      4500 \AA& continuum& photosphere& 3.7\\
      1700 \AA& continuum& temperature minimum, photosphere& 3.7\\
       304 \AA& He II& chromosphere, transition region& 4.7\\
        1600 \AA& C IV, continuum& transition region, upper photosphere& 5.0\\
         171 \AA& FeIX& quiet corona, upper transition region& 5.8\\
          193 \AA& Fe XII, XXIV& corona and hot flare plasma& 6.2, 7.3\\
           211 \AA& Fe XIV& active-region corona, photosphere& 6.3\\
            335 \AA& Fe XVI& active-region corona, photosphere& 6.4\\
             94 \AA& Fe XVIII& flaring corona photosphere& 6.8\\
             131 \AA& Fe VIII, XXI& transition region, flaring corona& 5.6, 7.0\\
      \hline
    \end{tabular}
\end{table}

\begin{figure}[ht!] 
\centering    
\includegraphics[width=0.7\textwidth]{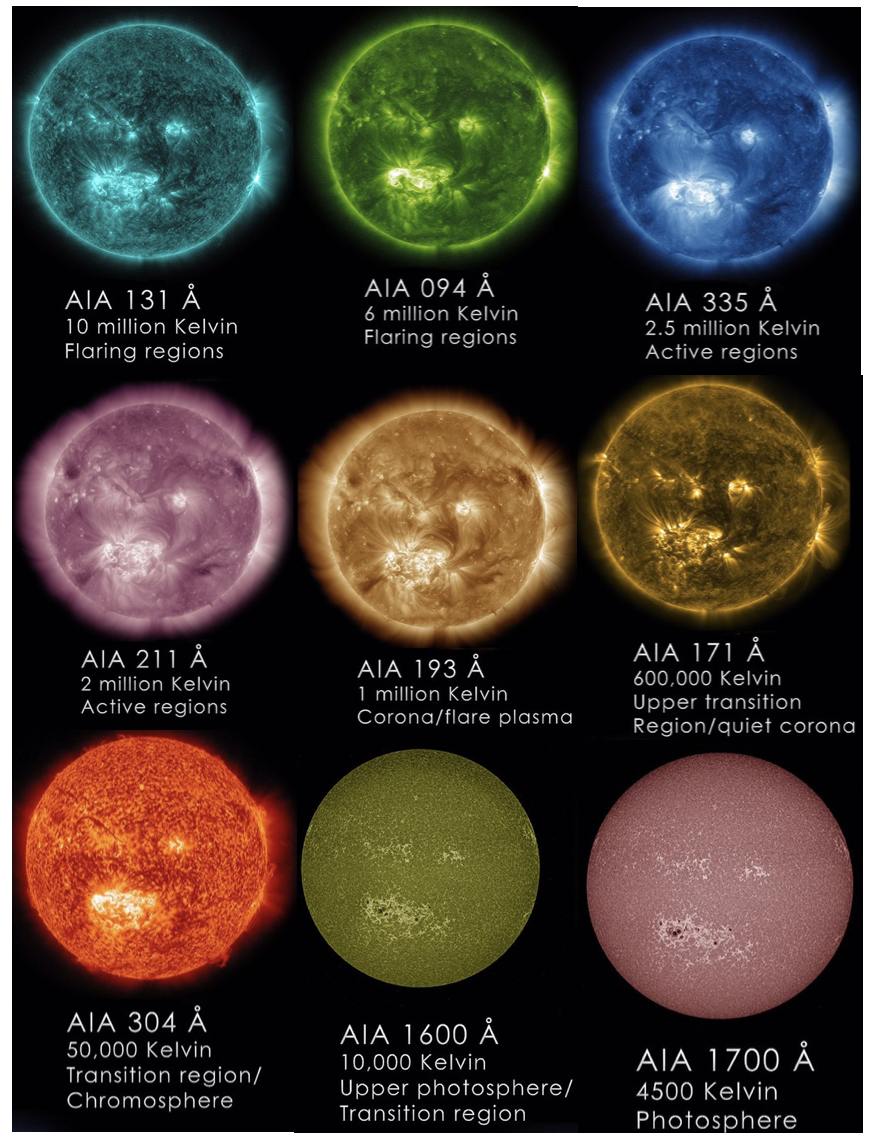}
\caption[Solar images from AIA observations of seven EUV and two UV wavebands.]{Solar images from AIA observations of EUV/UV wavebands (\citealt{Lemen2012}).}
\label{im_AIA}
\end{figure}
A guiding telescope is mounted on the side of the telescope tube with an achromatic refractor, a band-pass entrance filter focused at 5700 \AA, and a Barlow-lens assembly.  To provide the low thermal coefficient of thermal expansion, the telescope mirrors are fabricated on zerodur substrates. 
AIA provides seven the information of the solar surface in seven EUV and three UV channels. The seven EUV bands are centered at different lines emitting at different temperatures from 335 \AA\ (Fe XVI),  304 \AA\
(He II, T$\sim$0.05 MK), 171 \,\AA\ (Fe IX, T $\sim$0.6 MK), and 193 \,\AA\ (Fe XII, T $\sim$0.6 MK) to hotter temperatures of  94 \,\AA{} (Fe XVIII, T$\sim$6.3 MK), 131 \AA\ (T$_1$= 10 MK and T$_2$= 0.64 MK), and 211 \,\AA\ (T$_1$ $\sim$ 20 MK and T$_2$ $\sim$1.6 MK). 
Two UV wavebands are centered at 1600 \AA\ (C IV), 1700  \AA, and 4500 \AA\ (continuum). 
Responses to different solar emissions in the AIA instrument can compute with with the CHIANTI solar spectral model (\citealt{Dere2009}). The response function curve for the six EUV wavebands are dominated by iron emission lines and are presented in Figure \ref{AIA_response}. 
A comparative and detail description of these UV and EUV channels is presented in table \ref{tab:AIA}. The temperature for different wavebands varies from 6$\times$10$^4$ K to 2$\times$10$^7$ K. 
The multi-wavelength observations from the AIA instrument in UV and EUV channels are presented in Figure \ref{im_AIA}. 

\item{\bf Helioseismic and Magnetic Imager (HMI):}
HMI is focused to map the magnetic fields and velocity fields at the solar surface and use to probe the oscillations at the photosphere with 6173 \AA\ Fe I absorption line. It measures the Dopplershifts to determine the surface velocity at the photosphere and creates a fulldisk Dopplergram with a temporal resolution of 45 second and pixel size of 0\farcs5. These Dopplergrams (maps of solar surface velocity) are important to study the interior of the Sun. 
HMI contains an optics package, an electronics box, and a harness to connect both of them. Optical package ensembles a front-window filter, a telescope, waveplates for polarimetry , one blocking filter, an image stabilization system, five stage Lyot filter, two wide field tunable Michelson interferrometers, and a pair of 4k $\times$ 4k cameras. The Lyot filter is based on the same basic design as of  Michelson Doppler Imager \citep[MDI,][]{Scherrer1995}.
The fulldisk longitudinal magnetic field are measured using the Zeeman effect with the splitting of the spectral lines. 
\begin{table}[ht!]
\centering
  \caption[Data sources for high resolution HMI SHARP data.]{Data sources for high resolution HMI SHARP data (\citealt{Bobra2014}).}
\label{tab:SHARP}
\setlength{\tabcolsep}{3pt}
\begin{tabular}{lll}
      \hline
     No.& Description & Source \\
      \hline
     1& Updated plots of near & \url{jsoc.stanford.edu/data/hmi/sharp/dataviewer}\\
      &real time SHARP data&\\
     2& Detail of SHARP product & \url{http://jsoc.stanford.edu/doc/data/hmi/sharp/sharp.htm}\\
     3& Overview of JSOC series& \url{http://jsoc.stanford.edu/jsocwiki/DataSeries}\\
     4& Guidelines for HMI data &
      \url{http://jsoc.stanford.edu/jsocwiki/PipelineCode}\\
      & processing&\\
     5& Technical note on SHARP & \url{http://jsoc.stanford.edu/jsocwiki/sharp_coord}\\
      &coordinate system, mapping,&\\
     & and transforming vectors& \\
     6& Detail of HARP data series&  \url{http://jsoc.stanford.edu/jsocwiki/HARPDataSeries}\\
     7& Comprehensive guide to SDO & \url{http://www.lmsal.com/sdouserguide.html}\\
    & data analysis and methods &\\
      
      \hline
    \end{tabular}
\end{table}

For the magnetic field extrapolations and more precise magnetic field observations, the Space-weather HMI Active Region Patches (SHARPs) data is used. The detail analysis of HMI SHARP data is given by \citealt{Bobra2014}. This SHARP data series is centered to characterize the distribution of magnetic field and focused on a small spatial scale. Thus the SHARP data provides a deviation from the global/potential magnetic field configuration. SHARP follow and automatically detect each single AR patch of solar magnetic field during its complete life cycle from appearance on the solar disk to disappearance (\citealt{Turmon2014}). This SHARP data series provide indices, photospheric vector magnetic field data maps of Doppler velocity, maps of continuum intensity, LOS magnetic field and other variables. This SHARP data is available for the individual AR on the solar disk with 12 minute cadence and  the real time parameter source are given in table \ref{tab:SHARP}. 
\end{enumerate}

\begin{figure}[ht!] 
\centering    
\includegraphics[width=0.7\textwidth]{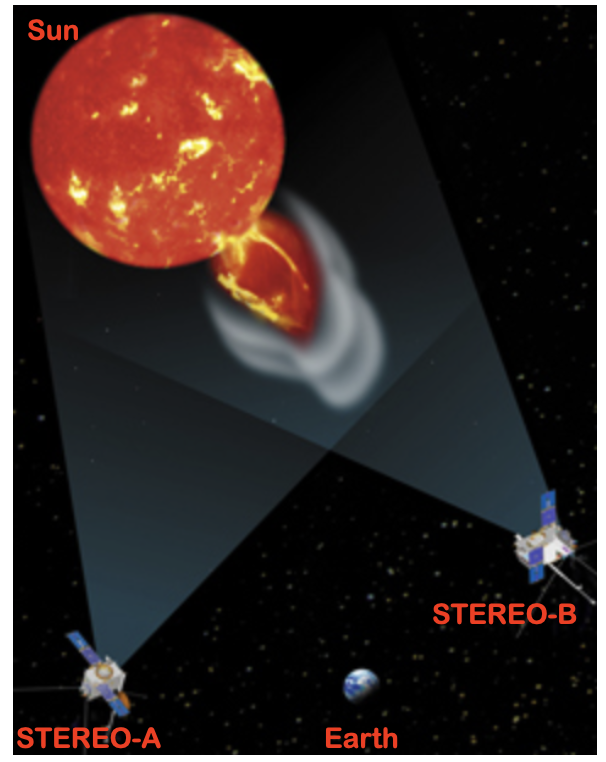}
\caption{A schematic view of the Sun observed with STEREO twin spacecrafts and from the Earth (image credit: NASA).}
\label{im_STEREO}
\end{figure}
\subsection{Solar Terrestrial Relations Observatory (STEREO)}
Solar Terrestrial Relations Observatory mission \citep[STEREO,][]{Howard2008} is the first ever space mission onboard the Sun Earth Connection Coronal and Heliospheric Investigation (SECCHI)  to observe the inner heliosphere from photosphere to the complete vicinity of our Earth.  
The STEREO mission includes two identical refrigerator sized spacecrafts: STEREO-A (ahead) and STEREO-B (behind). Both of these spacecrafts observe the Sun from two different angles in different orbits. STEREO-A spacecraft orbits moderately inside the Earth's orbit and takes 347 days to complete one revolution around the Sun. STEREO-B orbits slightly outwards to Earth's orbit and takes 387 days to complete one revolution of the Sun. In each year, both twin spacecrafts (STEREO-A and STEREO-B) move apart from each other by $\approx$ 44$^\circ$. with these two different orientations of spacecrafts, STEREO provides a multi-dimensional view of the Sun from the Earth (SDO). A schematic view of the Sun observed with STEREO twin spacecrafts and from the Earth is presented in Figure \ref{im_STEREO}.  

The SECCHI optical telescope set consists of five telescopes maintaining a broad range of FOV from the solar surface till interplanetary space to know the Sun-Earth connection. These five telescopes are of three types: 
     (a) EUV Imager (EUVI) - EUVI is used to observe the images of the Sun from solar chromosphere to low corona out to 1.7 R$_\odot$ in four different emission lines in 171 (Fe IX), 
195 (Fe XII), 284 (Fe XV), and in 304 (He II) \AA.
    (b) COR1 and COR2 coronagraphs -  COR1 and COR2 are the visible light Lyot coronagraphs act as the imager from inner to outer corona from 1.4 to 15 R$_\odot$. These are arranged in two telescopes, as there is a large change of coronal brightness with changing height.
    (c) Heliospheric Imagers (HI1 and HI2) - HI1 and HI2 are the third type of telescopes in the SECCHI instrument  setup and observe the coronal imaging from 15 R$_\odot$ out to the Earth's radius at 215 R$_\odot$. 

\subsection{Interface Region Imaging Spectrograph (IRIS)}
IRIS is NASA's Interface Region Imaging Spectrograph and was built by Lockheed Martin Solar and Astrophysics Laboratory (LMSAL) in CA, USA. It was launched on June 27, 2013 into a Sun-synchronous orbit and contains a 19 cm UV telescope along with a slit based dual-bandpass imaging spectrograph and described in \citealt{Pontieu2014}. 
IRIS observes the interface region between the relatively cool (6,000 K; photosphere) solar surface and the hot ($\approx$ millions of degrees; corona) outer atmosphere. It takes images in four different wavelengths in the ultraviolet range. These passbands are each sensitive to  plasmas of different temperatures: Mg II wing (2830 \AA, 6,000 K), Mg II k (2796 \AA, 10,000 K), C II (1330 \AA, 25,000 K), Si IV (1400 \AA, 80,000 K). 
The IRIS mission includes a detail analysis of this highly structured and complex interface region with strong radiative-MHD codes. The importance of this interface region relies on the fact that the mechanical energy, which is a driver of solar activity  and solar atmospheric heating, is converted into radiation and heat in this region. A small amount of energy leaks from this region through the coronal heating and accelerating solar wind. The transition between low to high plasma $\beta$ occurs in between of the photosphere and corona. Therefore in the interface region, there is always a competition between the plasma and the magnetic field for dominance. This race between plasma and magnetic field yields various impacts {\it e.g.} wave generation,  mode coupling, reflection, and refraction. The evidence for the supersonic and super Alfv\'enic motions is provided by the plasma flow from partially ionized chromosphere to fully ionized corona. Non local thermodynamic equilibrium (non-LTE) effects dominate the radiative transfer in the partial opaque chromosphere. In this way to explain the radiation process and to determine the energy balance state needs advanced computer modelling. Hence the high cadence observational data is required. 

\begin{table}
\centering
  \caption{IRIS data sources and URLs.}
\label{tab:IRISsource}
\setlength{\tabcolsep}{1pt}
\begin{tabular}{lll}
      \hline
    No.& Detail& Source location\\
    \hline
    1& Website& \url{http://iris.lmsal.com}\\
     2& Operations& \url{http://iris.lmsal.com/operations.html}\\
      3& Data search& \url{http://iris.lmsal.com/search/}\\
       4& Recent view& \url{http://www.lmsal.com/hek/hcr?cmd=view-recent-events&instrument=iris}\\
        5& IRIS today& \url{http://iris.lmsal.com/iristoday}\\
      \hline

 \hline
    \end{tabular}
\end{table}


\begin{table}[h!]
\centering
  \caption{Multi-thermal coverage of IRIS.}
\label{tab:IRISparameters}
\setlength{\tabcolsep}{8pt}
\begin{tabular}{cccccc}
      \hline
     Ion& Wavelength& FOV& Pixel& log T& Pass\\
     & (\AA)& ($\arcsec \times \arcsec$)& $(\arcsec)$& (log K)& band\\
      \hline
     Mg II wing& 2820& 175$^2$& 0.1679& 3.7-3.9& NUV\\
     O I& 1355.6& 175$^2$& & 3.8& FUV 1\\
     Mg II h/ Mg II k& 2803.5/ 2796.4& 175$^2$& 0.1679& 4.0& NUV\\
     C II/ C II& 1334.5/ 1335.7& 175$^2$& 0.1656& 4.3& FUV 1\\
     Si IV/ Si IV& 1402.8/ 1393.8& 175$^2$& 0.1656& 4.8& FUV 2\\
     O IV/ O IV& 1399.8/ 1401.2& 175$^2$& & 5.2& FUV 2\\
     Fe XII& 1349.4& 175$^2$& & 6.2& FUV 1\\
     Fe XXI& 1354.1& 175$^2$& & 7.0& FUV 1\\
      \hline
    \end{tabular}
\end{table}


IRIS provides a new and advanced platform for high spatial/temporal resolution observations, efficient MHD simulation codes, and massively parallel supercomputers (\citealt{Pontieu2014}). The IRIS data is freely available tio the scientific community and the data sources are presented in table \ref{tab:IRISsource}. 
The multi-thermal coverage of the IRIS instrument is presented in table \ref{tab:IRISparameters} and it's main features are as follows: 
\begin{figure}[t!] 
\centering    
\includegraphics[width=0.8\textwidth]{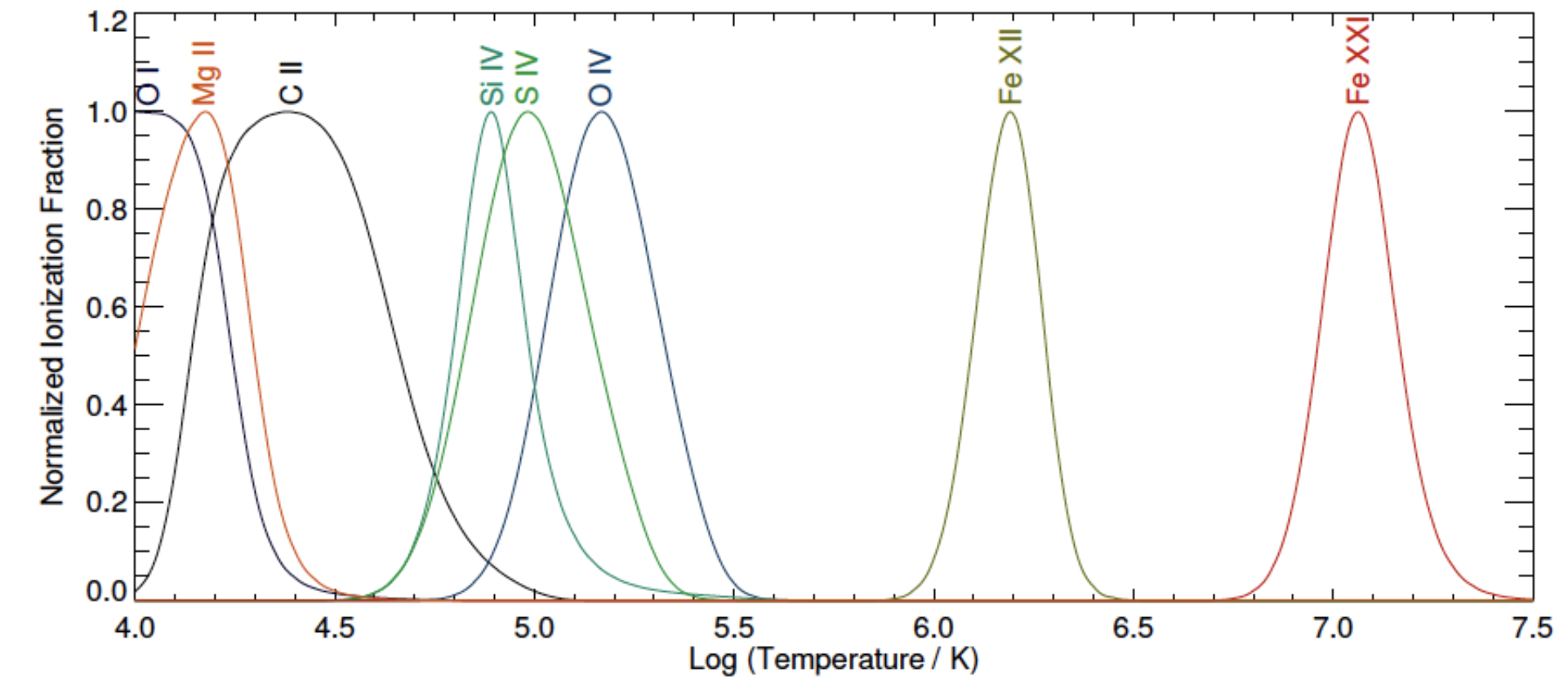}
\caption[Ionization fraction curve for different ions]{Ionization fraction curve for different ions, {\it e.g.} O I, Mg II, C II, Si IV, S IV, O IV, Fe XII, Fe XXI, along with temperature observed with IRIS.
It is taken from the ionization equilibrium file of chianti\_v7.ioneq (\citealt{Dere1997}; \citealt{Landi2013})
}
\label{iristemp}
\end{figure}

\begin{enumerate}
    \item It contains a 19 cm Cassegrain telescope with a dual range UV spectrograph and slit-jaw imager with 0\farcs16 pixels and four 2061 $\times$ 1056 CCDs. Slit-jaw imager includes four passbands: two in transition region lines (C II 1335 \AA\ and Si IV 1400 \AA), one in chromospheric region (Mg II k 2796 \AA), and one in photospheric region line (2830 \AA). These passbands cover a FOV of 175$\arcsec \times$ 175$\arcsec$.
    \item A spectrograph with 0\farcs33 wide and 175$\arcsec$  slit. It covers two FUV passbands from 1332-1358 \AA\ and 1389-1407 \AA\ along with an NUV passband from 2783-2835 \AA, from photospheric temperature (5000 K) to coronal temperatures (1-10 MK).
    \item Rastering of slit across the solar surface upto 21$\arcmin$ from the solar disk center with many slit-jaw choices.

    IRIS spectroscopic observations are available in two far ultraviolet channels (FUV) and one near ultraviolet band (NUV).  The spectrum is obtained by the Czerny-Turner spectrograph and recorded with three CCD cameras (one for each channel).  The FUV spectra (C II, and Si IV) have a spectral sampling of 12.8 m\AA\ and a spatial sampling of 0.167\arcsec per pixel along the 0.33\arcsec per pixel wide slit.  The FUVS channel, at wavelengths of 1332-1358 \AA, contain spectral lines of formation temperature at log T = 3.7 - 7.0, including two C II (1335 Å) lines. The FUVL channel, at 1390-1406 Å shows lines which correspond to the formation temperature log T = 3.7 - 5.2 along with emission from Si IV (1395 \AA) and O IV (1403 \AA).  These  multi temperature IRIS lines are explained with wavelength and temperature in Figure \ref{iristemp}. 
    The NUV observation at 2785 - 2835 \AA, are provided with 26 m\AA\ spectral resolution and with the same spatial scale as FUV. This channel provides a study of lines of formation temperature at log T = 3.7 - 4.2, including the Mg II k (2796 \AA) and Mg II h (2803 \AA) lines along with the Mg line wings. 
\end{enumerate}

\subsection{Large Angle Spectroscopic Coronagraph (LASCO)}
Large Angle Spectroscopic Coronagraph \citep[LASCO,][]{Brueckner1995} is a three coronagraph package, jointly launched with \emph{SOlar  and Heliospheric Observatory} \citep [SOHO,] [] {Domingo1995}.  
 \begin{table}[ht!]
\centering
  \caption{A brief description of LASCO C1, C2, and C3 coronagraphs.}
\label{tab:LASCO}
\setlength{\tabcolsep}{12pt}
\begin{tabular}{ccccc}
      \hline
     Coronagraph& FOV (R$_\odot$)& Occulter type& Objective element& Pixel size\\
     \hline
    C1& 1.1 - 3.0& internal& Mirror& 5.6$\arcsec$\\  
    C2& 1.5 - 6.0& external& Lens& 11.4$\arcsec$\\  
    C3& 3.7 - 30.0& external& Lens& 56.0$\arcsec$\\  
      \hline
    \end{tabular}
\end{table}

The LASCO spacecraft is providing the dynamics of solar coronal structures, and their  geomagnetic effects. C2 and C3 coronagraphs are presently operating since 
1996 and the FOV  of these two coronagraphs are presented in table \ref{tab:LASCO}. The C1 coronagraph was ceased operating during 1998, when the control of SOHO spacecraft was lost. The data from LASCO satellite is available at \url{https://cdaw.gsfc.nasa.gov/CME_list/}.

\subsection{H$_\alpha$ observations}
The Global Oscillations Network Group \citep[GONG,][]{Harvey1996} project contains a network of six instruments for acquiring the low noise and long-term stable data. It is a community based mission for the vast study of the internal structure and dynamics of the Sun using helioseismology.  GONG is operating and contributing to clarify the conditions in the solar interior with the knowledge of acoustic wave propagation through the Sun. The six different locations are chosen such that the long term observations from different sites will not disturbed by the daily setting of the Sun. These six observing networks are: {\it Big Bear Solar Observatory (BBSO)} in California USA, {\it Mauna Loa Observatory} in Hawai, USA, {\it Learmonth Solar Observatory} in Australia, {\it Udaipur Solar Observatory (USO)}, India, {\it Observatorio del Teide} in Tenerife, Spain, and {\it Cerro Tololo Interamerican Observatory}, in Chile. The GONG project is very important and less costly as all the six networks are the ground based observatories. Space borne satellites are of course free from the terrestrial effects but expansive to implement and difficult to maintain. On the other hand ground based observatories can be repaired easily but the data contains terrestrial atmospheric disturbances.

\begin{figure}[t!]
\centering    
\includegraphics[width=0.8\textwidth]{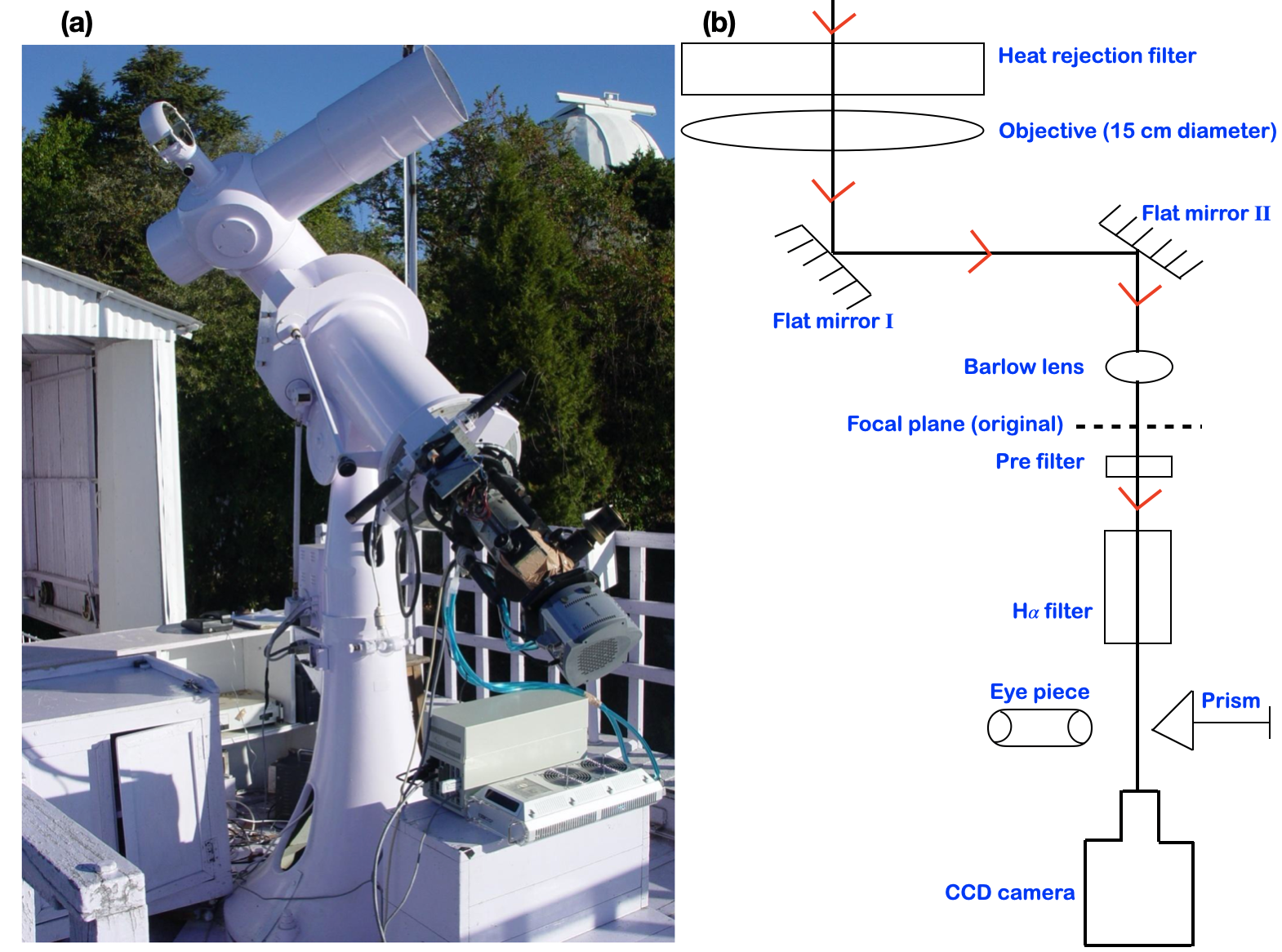}
\caption[15 cm Coud\'e refractor solar tower telescope at ARIES Nainital.]{15 cm Coud\'e refractor solar tower telescope with an optical ray diagram.}
\label{im_ARIES}
\end{figure}

The ground based H$_\alpha$ observations for solar jets and flares are also observed with 15 cm, f/15 Carl Zeiss Coud\'e refractor solar tower telescope located at Aryabhatta Research Institute of Observational Sciences (ARIES) Nainital (latitude = 29.4$^\circ$ North, longitude = 79.7$^\circ$ East, altitude $\sim$ 2000 m) India. An image of the 15 cm Coud\'e telescope installed on 35 feet height of the ARIES solar tower  is presented in Figure \ref{im_ARIES} (a). It is a modest size telescope with the optical filter centered at H$_\alpha$ and an objective lens of 15 cm clear aperture  (focal length = 225 cm). The image sizes are enlarged by a factor of 2 using the Barlow lens. H$_\alpha$ images were recorded with a 16-bit (385$\times$576 pixels) CCD camera of Wright instrument and having the image resolution of 1$\arcsec$ per pixel. The optical ray diagram of the observing system of the telescope and the instruments are presented in Figure \ref{im_ARIES} (b).
Two highly reflected aluminium coated, and coal protected plane mirrors (flat mirror I and II in Figure \ref{im_ARIES} b) are located at the intersection of the declination axis and polar axis of the Coud\'e configuration. This configuration results in a stationary (prime) focus at the lower end of the polar axis.  At the prime focus, an image of the Sun of diameter $\sim$ 22 mm is produced by the refractor and after getting the image enhancement with a factor of 2, the final image enlarged to a size of 44 mm diameter. The heat rejection filter (Figure \ref{im_ARIES} b) above the objective lens reflects the heat radiation from the intense incoming radiation and keeps the telescope tube cool and maintains the better image qualities. 

\subsection{Data analysis and reduction techniques}
\label{data_analysis}
All the used observational data is analysed with SolarSoftWare  IDL \citep[SSWIDL,][]{Freeland1998}. It is  a  set  of  software  libraries  for  IDL  that  has  been  developed  by  the  solar
community over the last few years aimed for providing the tools to analyse data from solar physics missions
and observatories to read and analyse data
from different instruments. The SSWIDL libraries can be obtained from
\url{http://www.lmsal.com/solarsoft/sswdoc/index_menu.html}. The data is provided in the Flexible Image Transport System (FITS) format from the websites given in Section \ref{sec:SDO}, analysed with SSWIDL by the read\_sdo routine can be seen and stored as, {\it e.g.}:\\
{\it IDL>~~ read\_sdo,`aia\_file.fits', index, data}\\
The AIA data supplied by the website is usually at level 1. For the bad pixel correction and some additional calibrations (translation, rotaion, and scaling) of this level 1 data can be improved to leven 1.5 by using aia\_prep.pro routine in IDL platform, {\it e.g.}:\\
{\it IDL>~~ aia\_prep,`aia\_file.fits',-1, index, data}

The raw images with basic keywords from the IRIS instruments are converted to level 0 image files and these files are reoriented in the increasing order of wavelengths to constitute level 1 data. However, we use the level 2 data which is ``{\it science ready}'' and can be downloaded from the website provided in table \ref{tab:IRISsource}. This has already been corrected for the dark current flat fielding, cosmic ray spikes, and wavelength calibration. There are two types of files available: slit-jaw and spectrograph. These IRIS level 2 files are processed and save in memory with read\_iris\_l2.pro routine in SSWIDL, {\it e.g.}:\\
{\it IDL>~~ read\_iris\_l2,`iris\_SJI.fits',index,data}\\
The available three slit jaw files for C II, Si IV, Mg II, and one raster files are capable to produce the image sequence and spectral profiles by iris\_xfiles and iris\_xcontrol routines in SSWIDL. Intensities in DN s$^{-1}$ pixel$^{-1}$ are calculated by dividing the calibrated data by exposure time. 
For the calibration of intensities we use the routines {\it IRIS\_GET\_RESPONSE} and {IRIS\_GET\_CALIB} available in SSW.  We convert the intensity count rates to the intensity values in {\it erg s$^{-1}$ sr$^{-1}$ cm$^{-2}$ \AA$^{-1}$} with the IRIS instrument response version 005. 
The routine in IDL for the IRIS intensity calibration ({\it e.g.} for Mg II) is:\\
{\it IDL>~~ 
Calibrated\_intensity=iris\_getwindata(datafile, 2796.2, /calib, /perang)}


Observations of solar disk with two different viewing angles (SDO and STEREO), provide the same structures at different locations. We used the tie-pointing method in SCC\_MEASURE routine of 
SECHHI available in SSWIDL to locate the positions of these same structures in two different images (\citealt{Thompson2006};  \citealt{Gosain2009}). In this routine we reconstruct the three dimensional picture of the ejecting feature by clicking the same 
feature on both STEREO and SDO images. A corresponding LOS crosses the identified feature in both images and the point at which this LOS meets is the tie-point. 
Both SDO and STEREO images use the World Coordinate System keywords and are useful to determine the position of the specific solar feature and appears as a `+' sign. 

For analysing the localized variation of power within a time range, we used the wavelet analysis technique. 
The periodogram provides the power spectral distribution as a function of frequency and the maximum peak in the power spectral distribution is considered as  the frequency of the signal. 
The first step to get the wavelet analysis is to find the Fourier transform of the given time series. This is followed with the selection and normalization of the function, which must have zero mean and localized in time and frequency space. Finally wavelet transform is calculated with the determination of the cone of influence and Fourier wavelength. The 95\% confidence contours are choosen by assuming a background Fourier power spectrum (\citealt{Torrence1998}).  

The distribution of plasma in different temperatures is derived using the differential emission measure (DEM) technique (\citealt{Labrosse2010}):
\begin{equation}
    DEM(T)=n_e^2~~ \frac{dh}{dt}~~~~ [cm^{-5} K^{-1}]
    \label{eq:dem}
\end{equation}
Integrating the line intensity in an emitting volume and cross sectional area along the line of sight, it is found that the line intensity is proportional to the square of electron density ($n_e$).
To derive the DEM from fix AIA EUV filters (94, 131, 171, 193, 211, 335 \AA), we have used the regularized inversion method developed by \citealt{Hannah2012}. 
To find the DEM, we need to invert equation \ref{eq:dem} using the observed line intensities (\citealt{Veronig2019}). From the AIA images, we construct the DEM maps and calculate the total emission measure (EM) and the mean plasma density. Using the filling factor unity, the mean plasma density is calculated as follows:
\begin{equation}
    n = \sqrt{\frac{EM}{h}}
\end{equation}
If the height `h' of the eruption does not vary much during the eruptive event, the relative density `n'with respect to an earlier eruption time ($t_0$) can be given as (\citealt{Veronig2019}):
\begin{equation}
    \frac{n(t)}{n(t_0)} \propto \sqrt{\frac{EM (t)}{EM (t_0)}}
\end{equation}

In addition to these explained methods, we have developed several routines in SSWIDL to interpret the SDO, STEREO, IRIS, LASCO, and GONG H$_\alpha$ observational data.\\

 While the jets have been observed and modelled in the past few decades, there are several questions which needs to be addressed, {\it e.g.,} 
What is the relationship between solar jets and flares? 
Are blowout jets a perceptible subclass of X-ray/EUV jets, or do most quiescent jets towards the end of their lives develop an intense phase with flux rope eruptions?
What is the role of solar jets {to increase the population} of solar energetic particles? 
Are the hot and cool jets have same or different triggering  mechanisms? Does the magnetic environment modify the trigger process? 
How the rotation is created in the solar jets
{Does the size and temporal spectrum of the so far observed solar jets can extend to even smaller events? The answer could have a direct implication on our 
understanding of the unsolved mystery of coronal heating. 
Spectra of different chromospheric and transition region lines at the jet reconnection site could explain the physical mechanism with the precise information of plasma parameters (temperature, density, velocity) and the multi layer heating process for the solar eruptions. The formation of plasmoids in the current sheet at the jet source region with high accuracy using the spectral capabilities of IRIS instrument  can serve as the validation for the various theoretical models and experiments.


To address the above questions, the objectives, work done, and the outline of the thesis are presented in the next section.
}


\section{Outline of the thesis}\label{outline}
Over the last decade, solar jets become the key interest area of 
research with the space borne satellites (SDO, STEREO, and IRIS) as well as with the ground based
observatories 
and revealed the dynamical activities of jets in 
chromosphere and solar corona.
X--ray jets from Yohkoh Soft X--ray telescope (SXT) are believed to be the most energetic solar jets (\citealt{Shibata1992}), and share their common properties  {\it i.e.} their impulsive nature, magnetic field configuration, and CME association  with  standard solar flares. 
Considering the important contributions of solar jets and related solar flares in the Sun-Earth connection, this thesis is focused to explain the pre-existing interrogations with high spatial and temporal observations and theoretical MHD models, {\it i.e.}:
\begin{itemize}
    \item
To understand the trigger mechanism of solar jets and its relation with solar flares.
\item
To explain the dynamics and kinematics of the solar jets.
\item
Do all jets triggered at high altitude magnetic reconnection or it can occur at any height of the solar atmosphere. This can be understand by the magnetic topology. Therefore we aim to study the magnetic topology at the location of solar jets and flares.
\end{itemize}


To probe the jet initiation and acceleration process from observational and theoretical point of view, this thesis presents studies for AR jets. The physical parameters of the solar jets and clear association with other large scale events {\it i.e.} solar filament eruptions, flares and CMEs are established with different case studies.
For the magnetic reconnection at the jet base the present work put forth a good observational evidence for the magnetic flux emergence MHD models. The organization of the thesis is as follows: 

A case-study regarding the evolution of key topological structures of solar flares along with an investigation of eleven recurring solar jets is presented in chapter \ref{c3} (\citealt{Zuccarello2017}; \citealt{RJoshi2017}).  The transition from eruptive to confined flares was observed between 2014 April 15-16. During the two days of observation a filament evolved from two separated filaments on April 15 to a single S-shaped filament on April 16. Contemporaneously with this evolution we observed the presence of significant shear motions that were the results of clockwise/counterclockwise motions of the two magnetic polarities where the arcade that supports the filament was anchored. To study the topology of the active region we performed two potential field extrapolations, one on April 15 and one on April 16, and computed the QSLs. We found that a closed fan-like QSL exists around the location of the filament on both days. The presence of circular, closed fan-QSLs indicates the presence of a (quasi-) separator in the corona. 
The discerning element between full and failed eruption behavior being determined by the mutual inclination of the flux systems involved in the process, namely the erupting flux and the overlying field. Along with theses flare observations,  this chapter also gives an investigation of eleven recurring solar jets originated from two different sites (site 1 and site 2) close to each other ($\approx$ 11 Mm) in the same AR.  The jets of both sites have parallel trajectories and slipped towards the south direction with a speed between 100 and 360 km s$^{-1}$.  The evolution of the jets indicates that different jets have not only
different speeds but their speed also varies with different wavelengths. We interpret it as the multi-temperature and multi-velocity structures in the solar jets. Our calculated values of the speeds, widths
and lifetimes are consistent with earlier reported values
in the literature.
We observed that the average lifetime is longer in 304 \AA\ than in shorter wavelength observations, which suggests that the cooler component of jets have a longer lifetime in comparison to the hotter component.
To study the connectivity of the different flux domains and their evolution, a potential magnetic field model of the AR, at the jet base have been computed. QSLs are retrieved from the magnetic field extrapolation and we explained the slippage of jets due to the interaction of many QSLs and the presence of multi null points at the jet location. We have observed the magnetic flux emergence followed by cancellation at site 1 on 15 April 2014. Moreover, on 16 April 2014, flux emergence and cancellation are recurrent in both jet sites. The observation of cool and hot material in our study supports the hypothesis of small filament eruption and a universal mechanism for eruptions.

Chapter \ref{c4} depicts the detail analysis of the AR NOAA 12644 on April 4, 2017 from where six recurrent jets were observed in all the hot filters of AIA as well as cool surges in IRIS slit-jaw high spatial and temporal resolution images (\citealt{Joshi2020MHD}). The temperature and the emission measure of the jets using the filter ratio method is done with the different AIA filters. The fluctuations for the pre-jet phases by analysing the intensity oscillations at the base of the jets with the wavelet technique is also well explained. This series of jets was initiated at the top of a canopy-like double-chambered structure with cool emission on one and hot emission on the other side. The hot jets were collimated in the hot temperature filters, have high velocities ($\approx$ 250 km s$^{-1}$) and were accompanied by the cool surges and ejected kernels that both move with $\approx$ 45 km s$^{-1}$. In the pre-phase of the jets, quasi-periodic intensity oscillations at their base in phase with small ejections with a time period of 2 to 6 minutes were observed and explained as the reminiscent of acoustic or MHD waves. We conclude that our observations of the EUV jets and surges  constitute a clear case-study for comparison with the experiments  developed  to study flux emergence events such as the MHD  models  of \citealt{Moreno2008}; \citealt{Moreno2013}; \citealt{Nobrega2016}. Many observed structures were identified in their models: the reconnection site with two vaults,   hot jets accompanied by surges, ejections of plasmoids in parallel with the development of the cool surges;  the velocity of the hot jets  and of the cool surge, in particular, fit quite well with the predicted velocity in the models.
 The similarities between the observations and the numerical models based on magnetic flux emergence are no proof, of course, that the observed jets are directly caused by episodes of magnetic flux emergence through the photosphere into the solar corona. Because of the limb location of the current observations, there is no possibility of ascertaining whether magnetic bipoles are really emerging at the photosphere and causing the jet activity.
  The cool surge with kernels is comparable with the cool  ejection and plasmoids that naturally appears in the models. Hence this study serves a clear observation with high spatial AIA and IRIS instrument to validate the numerical experiments for flux emergence MHD models of the theoretical scientists.

The role of solar jets for triggering and driving the large scale solar eruptions is extended in chapter \ref{c5}, with two different case studies of solar cycle 24, one on March 14-15, 2015 from AR NOAA 12297 and the other on April 28, 2013 from AR NOAA 11731 (\citealt{Chandra2017}; \citealt{Joshi2020ApJ}). An interesting two step filament eruption during March 14-15, 2015 and associated halo CME are studied with the observations from AIA, HMI instruments onboard SDO, and SOHO/LASCO satellites. 
The filament shows first step eruption on March 14, 2015 when it gets hit by a small solar jet from the same location and it stops to rise after attaining $\sim$ 125 Mm projected height. It remains  at this height for 12 hrs. Finally it was again pushed by an another jet and the meta-stable filament completely ejected on March 15, 2015. In this way the jet activities in the AR during both days helped  the filament for its de-stabilization and eruption. 
The filament location was in favorable for observations between the central meridian and the limb. This allowed us to have reliable photospheric magnetic-field data and to observe the ascending 
trajectory of the eruptive filament. The disadvantage was the absence of STEREO observations from another point of view, but we use the rather reliable method of measuring of the filament spine height on the disk. 
 Our conclusions for this are based on the detailed distribution of the decay index over the AR. The distribution of the decay index provides an opportunity for the two-step eruption.
The estimated heights of the filaments show that the western filament is within the zone of stability during all times of observation. That is why it kept its position despite the strong disturbance and intensive internal motions. The southern filament was close to the threshold of stability during all time before the abrupt acceleration in the second step. Very likely the eastern part of the initial filament becomes slightly unstable at the beginning of the first step. It moves rather slowly in the direction of the zone of stability and is able to find a new position for stable equilibrium there. However, it was also close to the threshold of stability and the next disturbance (jet) from the AR causes the full eruption.  In this way, a small jet disturbed and triggered the filament to form a complete eruption. Afterwards the filament follows a CME and becomes the largest geomagnetic storm of solar cycle 24. In the second case study, a multi--viewpoint and multi--wavelength analysis of an atypical solar jet based on the data from SDO, STEREO, and SOHO/LASCO coronagraphs is done. It is usually believed that the CMEs are developed from the large scale solar eruptions  in the lower atmosphere. However, the kinematical and spatial evolution of the jet on 2013 April 28 guide us that a jet was clearly associated with a narrow CME having a width of about 25$^\circ$, with an approximate speed of 450 km s$^{-1}$. Even the jet speed was lower than the escape speed at the solar surface, we observed the clear CME associated with the jet by all the space-borne coronagraphs. The possible mechanism for the jet continuously accelerating to reach the escape speed and form the narrow CME was that the falling back material made the upward material of the jet moving faster to keep the momentum of the whole jet conserved. We concluded that the observed speed of the CME was containing the speed of the CME center and the expansion speed, and was much larger than the jet speed, because the different parts of the erupting structures were being measured. The speed of CME center (the trajectory followed by the jet) was 220 km s$^{-1}$ and was equivalent to the speed of the jet (200 km s$^{-1}$). This provides a clear view of the jet-CME association. To better understand the link between the jet and the CME, the coronal potential field extrapolation was done from  the  line  of sight magnetogram  of the AR. The extrapolations present that the jet eruption follows exactly the same path of the open magnetic field lines from the source region which provides the route for the jet material to escape from  the solar surface towards the outer corona. These studies of large scale eruptions caused by the solar jets gives worth evidence that some jets can escape from the solar disk towards the solar corona to form a CME, which may contribute to the solar wind acceleration.

The solar jet study in chapter \ref{c6} concerns with the  multi wavelength observations of a jet and mini flare occurring in the AR 12736 around 02:04 UT on March 22, 2019 (\citealt{Joshi2020FR}). Usually jets contain the hot and cool ejected plasma that is denser than the solar corona. This chapter explains, how the twist was injected into a jet, with the AIA, HMI, IRIS observations and comparison with the MHD simulation of \citealt{Aulanier2005OHM,Aulanier2005}. The IRIS slit positions were located exactly at the reconnection location and the nature of the jet reconnection site is characterised using them. The magnetic history of the AR is followed based on the analysis of the HMI vector magnetic field computed with the UNNOFIT code.  We found that, this AR is the result of the collapse of two emerging magnetic flux regions overlaid by arch filament systems. In the magnetic field maps, the evidence suggests a pattern of the long sigmoidal flux rope along the polarity inversion line between the two  emerging magnetic flux regions, which is the  site of  the reconnection. Before the jet, an extension of the FR was present and a part of it was detached and formed a small bipole with a bald patch region, which dynamically became an `X'-current sheet over the dome of one emerging flux region where the reconnection took place.  A comparison with numerical MHD simulations confirms the existence of the long FR. The cool material follows different paths than the hot and acts as a wall in front of the hot jet. It resembles the  surges that accompany  jets in the MHD simulations. 
Mini flares in this AR were  observed frequently changing the connectivity at  this  interface region from a bald patch region to a current sheet region and vice-versa. The main bipole was the result of collision between two emerging fluxes.  The negative polarity was sliding, extending towards  the South creating a small bipole  and cancelled with the positive polarity.  We did not make a non linear force free field  magnetic extrapolation, but  preferred to  use 
 directly  the horizontal vector magnetic field  (vec B) observations to relate the small bipole  and the BP to the origin of the jet (both positioned at the same place). We compared the observed  magnetic field  pattern  and the values of the electric current density J$_{z}$  with synthetic J$_{z}$ and vec B data from the  MHD model of FR \citep{Aulanier2010,Zuccarello2015}  to infer the location of  FR. Our detailed observation analyses suggest that the jet reconnection  occurred in  a  bald patch current sheet and in the rapidly formed  above null point current sheet, driven by the  moving magnetic polarity that carried twist from the remote flux rope and injected it into the jet. The initial flux rope remains stable during the reconnection process. We concluded that the reconnection  site  is heated at all the temperatures and the  hot jet is expelled towards the West  side in twisted field lines.

Chapter \ref{c7} put on display the spectroscopic analysis of a miniflare and the associated jet with the IRIS instrument with its high spatial and temporal resolution, which brings exceptional plasma diagnostics of
solar chromospheric and coronal activity during magnetic reconnection (\citealt{Joshi2020IRIS}).
A spatio-temporal analysis of IRIS spectra observed in the spectral ranges of Mg II, C II and Si IV ions are proceed and the Doppler velocities from Mg II lines are computed by using a cloud model technique. Strong asymmetric Mg II and C II line profiles with extended blue wings observed at the reconnection site are interpreted by the presence of two chromospheric temperature clouds, one explosive cloud with  blueshifts at 290 km s$^{-1}$ and one cloud with smaller Dopplershift (around 36 km s$^{-1}$). We used the cloud model  technique as a simple diagnostics tool for velocity calculations. Such models are used to derive  true velocities which usually differ from those obtained from Doppler shifts, and this is the basic idea behind the cloud
model. 
Simultaneously at the same location (jet base), strong  emission of several transition region lines (O IV and Si IV) and the emission of the Mg II triplet lines of the Balmer-continuum are observed. 
Emission at the minimum temperature was detected with AIA 1600 \AA\ and 1700 \AA\ filters and confirmed the low level heating.
Mg II  and  C II  lines are good diagnostics for detecting plasma at chromospheric temperature (T $<$ 20,000 K). We identified the bilateral outflows in the bald patches, and the bald patch current sheet was transformed to an `X'-point current sheet during the reconnection. 
The absorption  of  identified chromospheric lines in Si IV broad profiles have observed and analysed. In fact 
Si IV  profiles are  striped by  absorption lines formed  at  photospheric  temperatures and clearly show the presence of cool plasma  over transition region temperature material in the  reconnection site. This cool plasma is certainly due to the cool clouds, either the ejected fast cloud  feed  by trapped material in bald patch or by surge plasma.
All this event is  finally embedded  in the corona.
This  demonstrated  the possibility of having successive layers in the atmosphere with different velocities and temperatures in the current sheet region.  With such observations of  IRIS line and continuum emission, we proposed a stratification model for the white-light mini flare atmosphere with multiple layers of different temperatures along the line of sight, in a reconnection current sheet. It was an important result and for the first time that one could quantify the fast speed (possibly Alfv\'enic flows) of cool clouds  ejected transverse to the jet direction  by using the cloud model technique. It is concluded that the plasma of ejected clouds could come from the trapped region, where the cool material was stucked before reconnection between the two emerging flux regions or be caused by a chromospheric-temperature (cool) upflow during the magnetic reconnection.

\chapter
{Solar flares and recurrent solar jets from AR NOAA 12035}
\label{c3}
\ifpdf
    \graphicspath{{Chapter3/Figure/}{Chapter3/Figure/}}
\fi
\section{Introduction}
Solar flares are sudden and violent releases of magnetic energy in the solar atmosphere and can be divided as eruptive flares, when plasma is ejected from the solar atmosphere, resulting in a CME , and as confined flares when no CME is associated with the flare (\citealt{Schmieder2015};  \citealt{Janvier2015}). 
In addition to full and failed eruptions, there are cases where only a part of the filament is erupted, such events are defined as partial erupting events. Partial eruption may or may not be associated with a CME (\citealt{Gibson2006};  \citealt{Liu2008};  \citealt{Tripathi2013};  \citealt{Kliem2014};  \citealt{Zhu2014}). The most energetic flares are commonly eruptive (\citealt{Yashiro2005}), even though confined, non-eruptive X-class flares have been reported (\citealt{Thalmann2015};  \citealt{Sun2015};  \citealt{Harra2016}) as well as CMEs  with associated only C-class flares (\citealt{Romano2014};  \citealt{Chandra2016}).

The CSHKP model (\citealt{Carmichael1964};  \citealt{Sturrock1966};  \citealt{Hirayama1974};  \citealt{Kopp1976}) and its extension in three-dimensions (\citealt{Aulanier2012};  \citealt{Janvier2013};  \citealt{Janvier2015}) can explain several observational signatures of the fully (or failed) eruptive flares, such as the presence of X-ray sigmoids,  flare ribbons, and  brightening motions along the ribbons themselves. In particular, \cite{Savcheva2015,Savcheva2016} have shown that the flare ribbons often coincide with the photospheric signature of quasi-separatrix layers \citep[QSLs,][]{Demoulin1996}, i.e., thin layers characterized by a sharp gradient in the connectivity of the magnetic field. The brightening motions along the ribbons have been interpreted as the signatures of the slipping reconnection of the magnetic field lines through the {\it QSL} (\citealt{Aulanier2006};  \citealt{Janvier2013}; \citealt{Dudik2016}).  High resolution observations and interpretation of the slipping motion of plasma material in the AR are required to explain the magnetic behaviour of the region (\citealt{Zuccarello2017}; \citealt{RJoshi2017}). 

The morphology and evolution of  flare ribbons can also give information on the overall topology of the system. \cite{Masson2009} have shown that circular flare ribbons are associated with the presence of a null-point topology in the corona, while parallel ribbons moving away from  each other have been interpreted as  an indication of quasi-separator reconnection occurring higher and higher in the corona (\citealt{Aulanier2012}). Different triggering mechanisms have been proposed (\citealt{Forbes1991};  \citealt{Chen2011};  \citealt{Aulanier2014};  \citealt{Filippov2015};  \citealt{Schmieder2015}), but essentially the equilibrium of a magnetic FR embedded in an overlying magnetic field is determined by two competing effects: the outward-directed magnetic pressure between the FR and photosphere, and the inward-directed magnetic tension of the overlying field. 
In the torus instability or catastrophic loss of equilibrium model (\citealt{Forbes1991};  \citealt{Kliem2006};  \citealt{Demoulin2010};  \citealt{Kliem2014}) it  is the onset of an ideal MHD instability that leads to the disruption of this equilibrium, while in the breakout model (\citealt{Antiochos1999};  \citealt{Lynch2008};  \citealt{Zuccarello2008};  \citealt{Zuccarello2009};  \citealt{Karpen2012}) it is the onset of a resistive instability.   
Assuming an overlying external field $B_{ex}$ that scales with the height $z$ from the photosphere as $B_{ex} \propto z^{-n}$, in the torus instability model the system becomes unstable when the apex of the axis of the magnetic FR reaches a critical height $z_{cr}$ where the decay index $n$ of the external overlying field $B_{ex}$ becomes larger than a critical value $n_{cr}$. The results of several MHD simulations place $n_{cr}$ in the range $[1.3-1.75]$ (\citealt{Torok2005}; \citealt{Fan2007}  \citealt{Torok2007};  \citealt{Isenberg2007};  \citealt{Aulanier2010};  \citealt{ Kliem2013};  \citealt{Amari2014};  \citealt{Inoue2015};  \citealt{Zuccarello2016}).  Attempt to estimate the decay index at the onset of solar eruptions have also been made both using limb observations as well as stereoscopic observations (\citealt{Filippov2001};  \citealt{Guo2010};  \citealt{Filippov2013};  \citealt{Zuccarello2014};  \citealt{McCauley2015}). 
Contrary to the torus instability model that does not require any particular magnetic field topology, the breakout model requires a multi-flux distribution. The eruption begins when a resistive instability sets in at the breakout current sheet that exists between the arcade that confines the FR and the overlying field (\citealt{Karpen2012}). This reconnection removes the confining flux by transferring it to the neighboring flux domains. As a result, the magnetic tension of the confining field decreases resulting in an eruption. For the breakout model to work two conditions must be satisfied: the presence of a null-point or quasi-separator in the corona, 
and the flux of the confining arcade must be larger than the flux of the overlying field. 
Due to the nature of the problem, i.e., evidence of reconnection occurring higher up in the corona, observational studies that clearly support the breakout model are quite rare (\citealt{Aulanier2000};  \citealt{Mandrini2006};  \citealt{Chandra2009};  \citealt{Chen2016}). Both models address the triggering of the eruption, but what determines if the eruption results in a CME or in a failed eruption? Many questions have to be answered, {\it i.e.}:
How does the trigger mechanism affect the eruptive/failed behavior of the flare? How important is the magnetic environment of the AR?

In some of the studies circular ribbons have been 
observed at the base of solar jets (\citealt{Wang2012}). This topology
supports the presence of magnetic null-points above  jet locations. However, using  photospheric magnetic field extrapolations, 
\citealt{Mandrini1996} and  \citealt{Guo2013} explained the 
jets by the presence of bald patch (BP) regions along 
 BP separatrices  are regions 
where magnetic field lines  are not anchored on the
 photosphere but are tangent  between two  
different magnetic regions. Flux cancellation at the jet locations is 
frequently proposed as the driver of jets
(\citealt{Innes2010};  \citealt{Liu2011};  \citealt{Innes2016};  \citealt{Adams2014};  \citealt{Young2014};  \citealt{Cheung2015};  \citealt{Chen2015}). \citealt{Zhang2000}) proposed such flux cancellation
 between oppositely directed magnetic 
field  to explain macrospicules and microscopic jets.
\citealt{Adams2014} also found  magnetic flux convergence and cancellation along the PIL, where the jets were initiated. 
In \citealt{Guo2013} the cancelling flux occurred at the edge of EMF 
during its expansion. Therefore, it is still 
discussed if the jets are due to the process of flux emerging or cancelling or both.

In this chapter we present a detail study of the AR NOAA 12035 on April 15-16, 2014 with AIA, HMI space borne instruments and chromospheric observations from ARIES solar telescope in H$_\alpha$ (\citealt{Zuccarello2017}; \citealt{RJoshi2017}). This study shows the evolution of key topological structures, such as spines and fans which determined the eruptive versus non-eruptive behavior of the series of eruptive flares, followed by confined flares. An interesting slippage of jets from one location to another is also observed and explained with complex topology of the region with the presence of a few low-altitude null points and many {\it QSLs}, which could intersect with one another.

\section{Flare and CME observations}\label{ch3_obs1}

The AR NOAA 12035 appeared at the East limb on  April 11, 2014 with a $\beta$ magnetic configuration and crossed the West limb on April 23, 2014. During its disk passage it produced many small--to--medium class solar flares. 
The AR turned into a $\beta\gamma$ magnetic configuration on April 13, 2014.
During the disk passage on April 15-16, 2014 the AR (located S15, E20 to E08) produced  eruptive and compact flares, respectively.
The description of these confined and eruptive flares is given in table \ref{recurrent}. 

\begin{table*}[t!]
\caption[Details of compact and eruptive flares on April 15-16, 2014.]{Details of compact and eruptive flares. The `--' indicates small flares that are not reported by GOES, `$X$' indicate CMEs visible in LASCO, but not associated with the filament activity/eruption.}
\label{recurrent}
\begin{center}
\resizebox{\textwidth}{!}{
\begin{tabular}{cccccccccccccc}
\specialrule{.1em}{0.1em}{0.1em}
 \multicolumn{14}{c}{\bf April 15, 2014}\\

\specialrule{.1em}{.1em}{.1em}
Flare No.&1&2&3&4&5&6&7&8&9&10&11&12&13\\
\specialrule{.1em}{1em}{1em}
Flare Onset (UT)& 05:56&06:15&06:59&09:15&12:37&14:37&16:56&17:53&19:22&20:55&21:39&22:48&23:40\\

Flare Class&--&--&--&C8.6&C3.6&--&--&C7.3&--&--&--&--&--\\
Related CME&No&No&No&10:24&14:00&No&14:00&18:48&No&No&No&No&No\\
\specialrule{.2em}{1em}{1em}
 \multicolumn{14}{c}{\bf April 16, 2014}\\
 \specialrule{.1em}{1em}{1em}
Flare No.&14&15&16&17&18&19&20&21&22&23&24&25&26\\
\specialrule{.1em}{.1em}{.1em}
Flare Onset (UT)& 01:10&02:42&03:20&03:48&05:02&06:37&07:14&08:36&09:20&10:42&12:42&17:30&19:54\\
Flare Class&C1.9&--&--&--&--&C1.8&--&C5.2&--&--&C7.5&C2.0&M1.0\\
Related CME&No&No&No&No&No&{\it X}&No&No&No&No&No&No&{\it X}\\
\specialrule{.1em}{.1em}{.1em}
\end{tabular}
}
  \end{center}
\end{table*}
\begin{figure*}
\centering
\includegraphics[width=1.0
\textwidth]{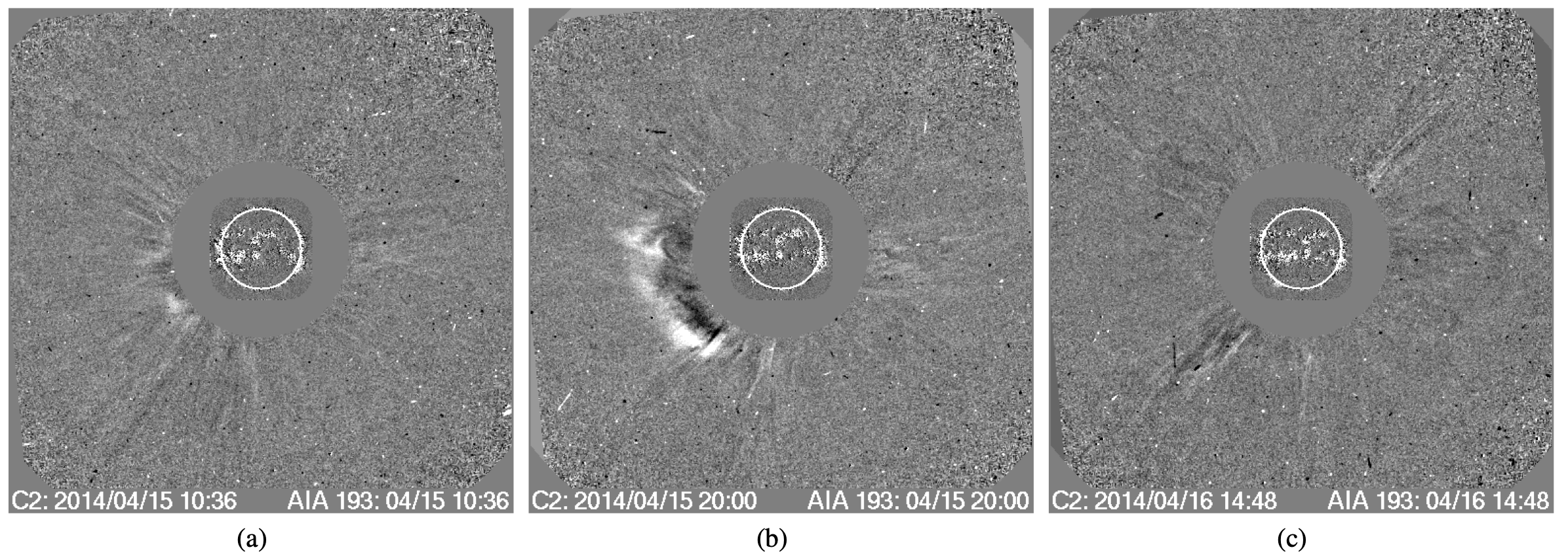}
\caption[SOHO/LASCO running difference images showing the associated CMEs for the eruptive flares.]{SOHO/LASCO C2 running difference images showing the associated CMEs for the eruptive flares. The last panel shows that no CME could be detected for the failed eruption on April 16, 2014.
}
\label{Fig:OBS-CME}
\end{figure*}
\subsection{LASCO/CME observations}

All the eruptive flares occurred on April 15. The CMEs associated with two of these eruptive flares are presented in Figures~\ref{Fig:OBS-CME}. The CME associated with the C8.6 X-ray flare at 09:15~UT  is first seen in LASCO C2 coronagraph at 10:36 UT (Figure~\ref{Fig:OBS-CME} (a)), and is characterized  by a narrow  angular width of 27$^{\circ}$ and an average speed of 274 km s$^{-1}$.  The CME associated with GOES C7.3  X-ray flare that occurred at 17:53 UT is visible in LASCO C2 {field of view (FOV)} at 20:00~UT (Figure~\ref{Fig:OBS-CME} (b)), and  has an angular width of 179$^{\circ}$ and an average speed of 360 km s$^{-1}$. On April 16 no CMEs associated with  the flares in the AR are observed. One example of the corona observed two hours after the flare that occurred at 12:42~UT is presented  in  Figure~\ref{Fig:OBS-CME} (c) to show that no CME is detectable. However, we note that two CMEs  are recorded on April 16  (Table \ref{recurrent}). After a detailed inspection of the LASCO observations we identified  a poor CME around 06:30~UT that it is  too early to correspond to the flare at 06:38~UT, and a narrow CME directly towards  the south at 20:00~UT that is  again  too early to correspond to the M1.6 flare. These CMEs could correspond to jet activity that characterize the eastern part of the AR.

\begin{figure}[ht!]
    \centering
    \includegraphics[width=1.0
\textwidth]{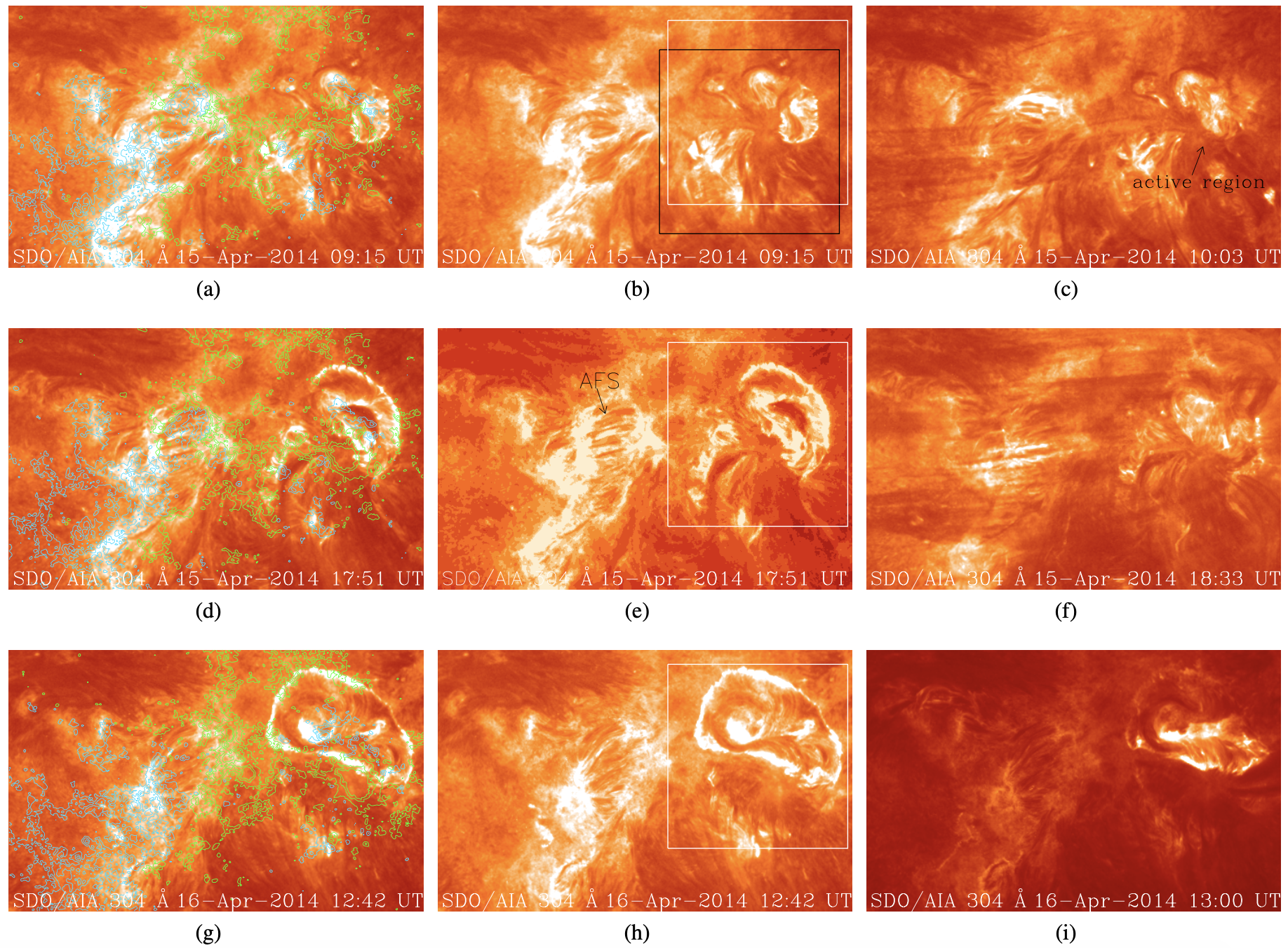}
    \caption[Two examples of the C-class eruptive flares in the AR 12035 on April 15, 2014.]
    {Two examples of GOES C-class eruptive flares in the AR 12035 on April 15, 2014  in AIA 304 \AA\ (two top rows) and one example of confined flare on April 16, 2014 (bottom row).}
    \label{Fig:OBS-15-04}
\end{figure}
\subsection{SDO/AIA observations}

All the flares  of the AR 12035 considered in this study and listed in Table \ref{recurrent} were well observed by  the AIA instrument on board SDO. Apart from the last one (flare~26 in Table~\ref{recurrent}), they are all low energy events, and correspond,  for the strongest ones, to C--class flares. During April 15-16, 2014 the AR produced six eruptive and twenty confined flares. Two of the eruptive flares, productive of CMEs,  that occurred on April 15 are shown in the first two rows of Figure~\ref{Fig:OBS-15-04} (in AIA 304~\AA\ wavelength).

\begin{figure}[ht!]
    \centering
    \includegraphics[width=\textwidth]{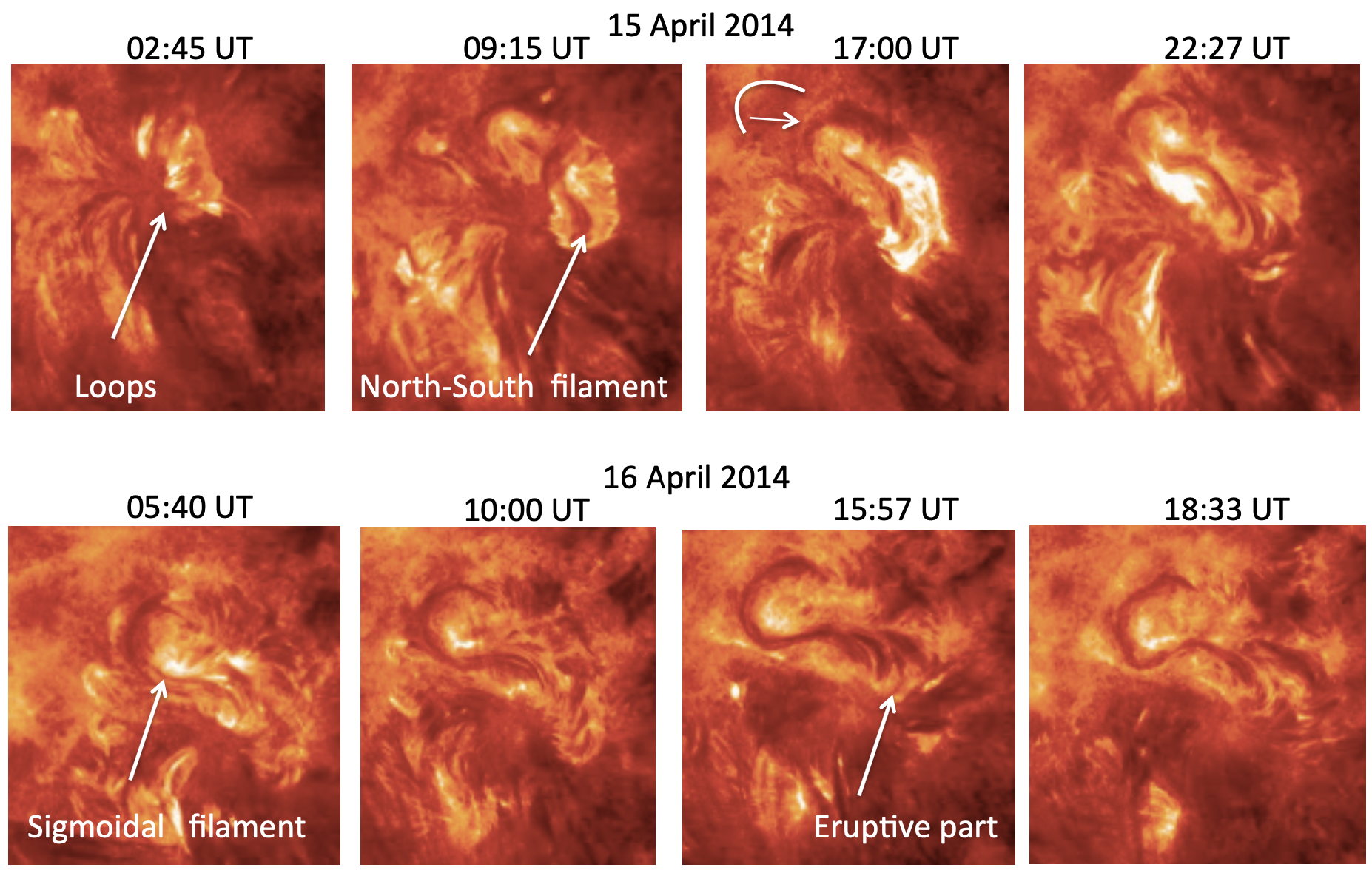}
    \caption[Evolution of the filaments in the north part of the AR 12035 between April 15-16 in AIA 304 \AA.]
    {Evolution of the filaments in the north part of the AR 12035 between April 15-16 in AIA 304 \AA. The FOV corresponds to the black box in Figure \ref{Fig:OBS-15-04}.}
\label{Fig:filament}
\end{figure}

Figure~\ref{Fig:OBS-15-04} presents the environment of the region around the time of the flare, and the black box in panel (b) is focused on the AR 12035.   A zoom of its evolution is presented in Figure \ref{Fig:filament}. As an example, the image at 09:15~UT shows a dark north-south oriented filament that has been activated a few minutes before and, consequently bright arcades are observed around it. The flare emission reaches its maximum at 09:23~UT while at 09:36~UT a round shape brightening can also be observed (Figure~\ref{Fig:OBS-15-04}). Finally, between  09:39~UT and 10:03~UT  dark strips  are seen to cross the AR from West to East.
The second flare that we consider here occurred at 17:53~UT, when we see the activation of the filament that started to be more east-west aligned and with a second half-circle shaped filament at the north of it (Figure~\ref{Fig:OBS-15-04} (e)).  Observations show that, after a first failed eruption of the southern threads of the filament, the main body of the filament starts to erupt at 17:51~UT, when  circular  bright arcades on the west of the filament are also visible (Figure~\ref{Fig:OBS-15-04} (e)). During the eruption the  filament  interacted with the environment and, similarly to the other eruptive flares, resulted in the ejection of plasma, visible as dark stripes around 18:32 UT (Figure~\ref{Fig:OBS-15-04} (e)).  These dark, filamentary eruptive structures that can be clearly seen in Figures~\ref{Fig:OBS-15-04} (c) and \ref{Fig:OBS-15-04} (f) had a duration of about 45 minutes, and eventually produced the CMEs. During the two days of observation the filament(s) evolved from being constituted by two separated filaments on April 15 ---one relative-straight and north-south at 09:15 UT oriented  and an upside down U  at the north of the first one well visible at 17:00 UT --- to a complete east-west oriented sigmoidal  filament  on April 16  at 05:40 UT  (Figure \ref{Fig:filament}). All failed eruptions observed on April 16 were initiated by an  asymmetric failed eruption of the southwestern part of the sigmoidal filament.

One example of a compact flare  that occurred on April 16 is shown in the bottom row of Figure~\ref{Fig:OBS-15-04}.  Around the time of the onset of the flare, i.e., at 12:42~UT, we observe an oval shape of brightening  around the AR 12035 with inside the dark sigmoid and many filamentary structures in its  southwestern  end (Figure~\ref{Fig:OBS-15-04} (h) and also Figure \ref{Fig:filament}). Until this time the dynamics is similar to what was observed the day before. However,  at 13:00~UT a bright overlying arcade is seen over the AR 12035 and the dark material inside stops to rise (Figure~\ref{Fig:OBS-15-04} (h)).
The eruption concerned only the southern part of the  sigmoid and did not succeed to drive all the sigmoid to erupt. The two other failed eruptions,  at 10:51~UT and at 20:00~UT, followed  the same scenario. These three events lasted 15 minutes each. The eruption  of 10:51 UT is well observed in H$_\alpha$ and is discussed in detail in the next section. 

\subsection{H$_\alpha$ observations}

\begin{figure}[ht!]
    \centering
    \includegraphics[width=0.7\textwidth]{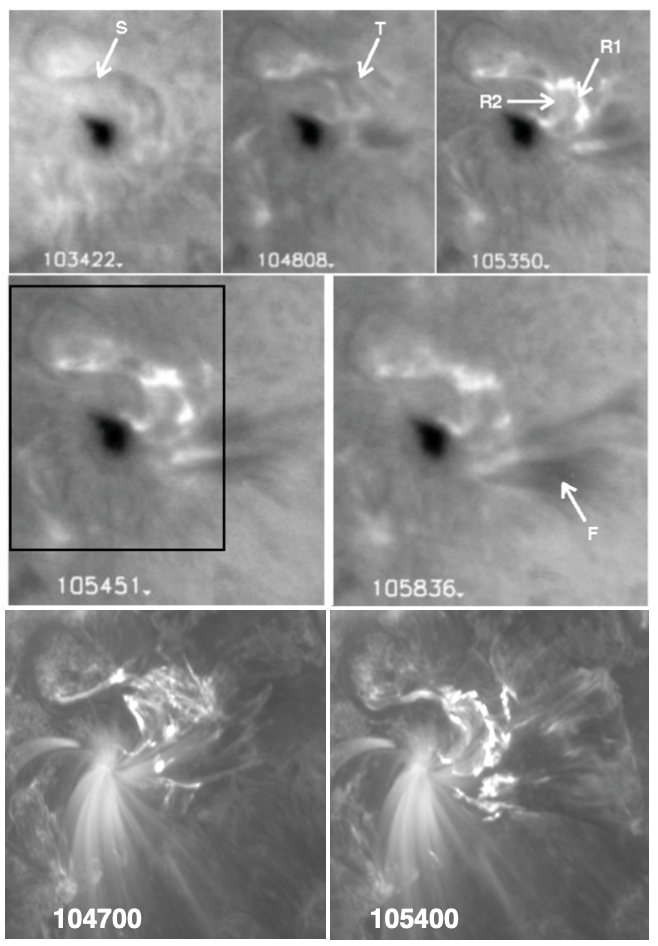}
      \caption[H$_\alpha$ images of the failed filament eruption and of the flare on April 16, 2014 from 10:34 UT to 10:58 UT from ARIES, Nainital telescope.]
      {H$_\alpha$ images of the failed filament eruption and of the flare on April 16, 2014 (from ARIES, Nainital telescope, top two rows). The box in the left middle image represents the FOV of the images in the top row. In the bottom row,  the corresponding images in AIA 171 \AA.}

\label{Fig:Halpha}
\end{figure}

In this subsection we discuss the failed eruption that occurred on April 16 at 10:30~UT and that is well observed in H$_\alpha$ from ARIES, Nainital. The H$_\alpha$ image taken at 10:34~UT on April 16 (Figure~\ref{Fig:Halpha}), shows the S-shaped filament (S) in the north of the AR, which was formed between April 15-16 (Figure~\ref{Fig:filament}).  
Around 10:38 UT, the filament started to  be activated, and at around 10:46~UT it broke in its center. The northern part of the filament remained in its original condition, while the broken part of  it consisted of many threads (T)  that are visible in H$_\alpha$ at 10:48~UT  (Figure~\ref{Fig:Halpha}), when the filament started to erupt in the west direction. 
However, the broken filament's southern foot point remained fixed. Eventually, the erupted part of the filament fell back on the solar surface, resulting  in a failed eruption. Together with the filament eruption close to the breaking location of the filament, we observe the maximum flare brightening  at 10:51~UT corresponding to  a  C2.0 class flare. Later on, we observe two flare ribbons  (R1, R2) at 10:53~UT. Finally, we note that the dark H$_\alpha$ structure with a fan-like shape  (F) did not expand after 10:58~UT (Figure~\ref{Fig:Halpha}).
The AIA 171~\AA\ observations confirm the failed eruption (Figure~\ref{Fig:Halpha}, bottom row).  The filament is visible  in absorption  with a S-shaped  at 10:47~UT and with a side-view of the arcade overlying the western part of the filament during the eruption at 10:54~UT. The ribbons appear as  bright structures along the foot points of the arcades.

\begin{figure}[ht!]
    \centering
    \includegraphics[width=\textwidth]{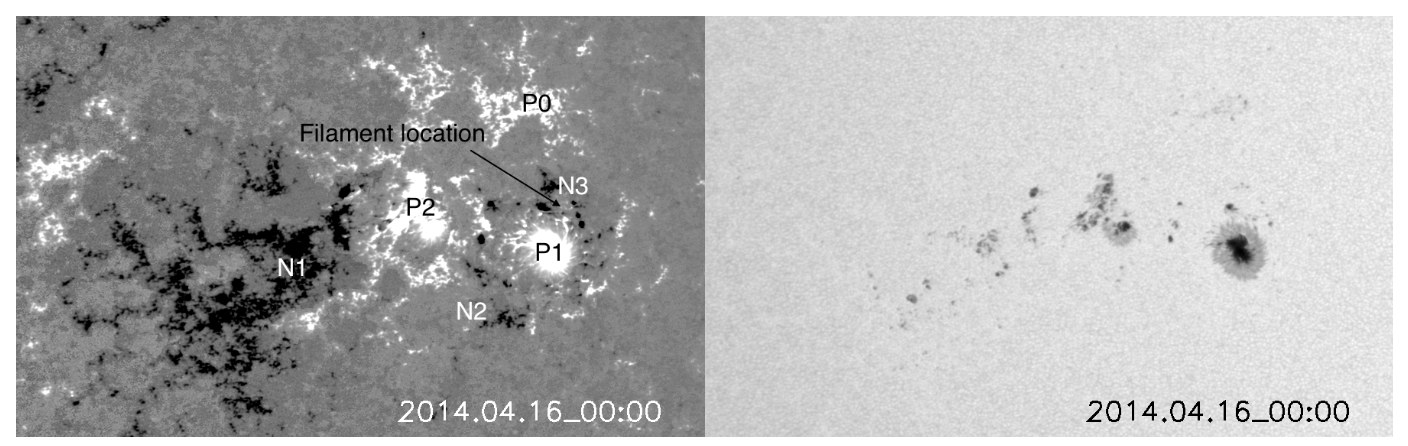}
    \caption[Distribution of the radial component of the magnetic field and of the HMI continuum.]
    {Distribution of the radial component of the magnetic field (left) and of the HMI continuum (right). The color scale for the magnetic field is saturated at $\pm$500 Gauss and black/white indicate negative/positive magnetic field.}

\label{Fig:Br-continuum}
\end{figure}

\subsection{Evolution of the magnetic field at flare site}
\label{Sect:MF}

On April 15, the AR had an overall bipolar structure characterized by a positive leading polarity P1 and a following negative one N1 (Figure~\ref{Fig:Br-continuum}. The leading polarity appeared to be constituted by a preceding compact flux distribution P1, coinciding with the umbra of the leading sunspot (Figure~\ref{Fig:Br-continuum}, right panel), followed by a more disperse polarity P2.  The positive polarity P1 is surrounded by a moat region with frequent bipole flux emergence and cancellation. Consequently, the positive polarity P1 is surrounded by two negative flux distributions, indicated as N2 and N3 in Figure~\ref{Fig:Br-continuum}, left panel.  
The continuum intensity  shows that the sunspot P1 is rotating in the clockwise direction by an angle of about 35$^{\circ}$ during the two days of observations (and about 80$^{\circ}$ between April 14-18). 

\begin{figure}[t!]
    \centering
    \includegraphics[width=0.8\textwidth]{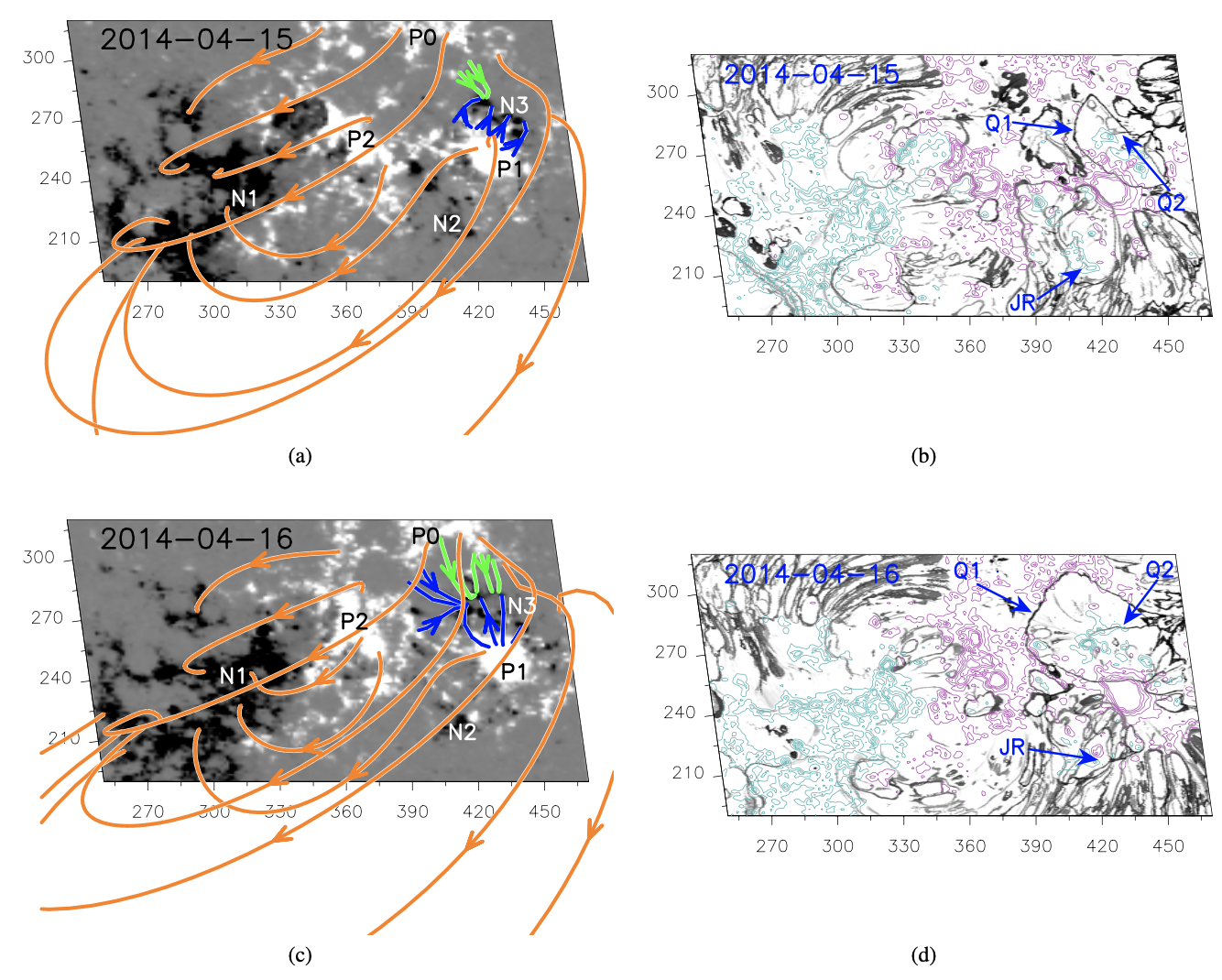}
    \caption{Magnetic field distribution of AR NOAA 12035 together with  some representative potential field lines and {\it QSL} maps for April 15-16.}
\label{Fig:T-Overview}
\end{figure}

Between April 15-16 we observe the emergence of new magnetic flux between the dispersed positive flux P2  and the following negative polarity N1. This region corresponds to the area of an arch filament system (AFS) visible in Figure~\ref{Fig:OBS-15-04} (e). As a result of this process, part of the positive dispersed flux that constitutes the leading polarity is annihilated  and the separation between the negative N1 and positive P2 flux distribution increases. Furthermore, the leading polarity is now characterized  by two compact distributions of positive flux that are well separated 
(P1 and P2 in Figure~\ref{Fig:T-Overview} (c)). 

Starting from about 19:00~UT on April 15,  a succession of bipoles with a larger negative polarity and a weaker positive one is seen to emerge in the north of P1 leading to an accumulation of flux in N3. Subsequently, we observe a northeast migration  with a counter clockwise rotation of the newly emerged flux N3. Therefore, there is a strong shear between the clockwise rotating polarity P1 and the counterclockwise rotation of N3.

Contemporaneously to this migration, small concentrations of magnetic flux are seen to spread from the compact leading polarity P1 in all directions. As a result part of the flux of P1 is canceled with the negative fluxes N3 and N2. The recurrent jets probably originate from the cancellation of N2 and P1 that may lead to magnetic reconnection producing the observed jets around the location of N2. By 10:24~UT on April 16  (Figure~\ref{Fig:T-Overview} (c)) the positive polarity of the AR is constituted of three separate (more or less compact)  distributions of positive flux (P1, P2, P3 in Figure~\ref{Fig:T-Overview} (c)) with a negative intrusion N3 at the north of the leading compact one. The filament that is the subject of this study is located along the PIL between the compact positive polarity P1 and the negative flux distribution N3 (arrow in Figure~\ref{Fig:Br-continuum}, left panel and blue arcades in Figure~\ref{Fig:T-Overview}). 
\section{Magnetic field configuration}
\label{Sect:MFT}
Here we describe the key topological magnetic structures of the AR between April 15-16, 2014 {\it i.e.}, between the time period when the nature of the flares changed from eruptive to confined.  
\subsection{Potential magnetic field extrapolations}
\label{Sect:MFExtr}
To study the connectivity of the different flux domains and their evolution we computed  a potential magnetic field model of the AR (Figure~\ref{Fig:T-Overview}, left columns). Potential configurations give robust information on the topological structures of the coronal field such as separatrices and quasi-separatrices \citep[section~\ref{Sect:QSL} and][]{Demoulin1996}

Since  we are mainly focused on the connectivity of the AR,  we perform the potential extrapolation using a larger FOV provided by the HMI LOS-magnetograms that includes the neighboring ARs rather then the much smaller FOV provided by the HMI SHARP data product. To this purpose the HMI LOS-magnetograms of AR 12035 (and its neighboring ARs) taken at 10:24~UT on April 15-16 have been re-mapped to the disk-center using the mapping software available  through \textit{solarsoft}. As a result of this process the AR 12035 is rotated so that its center is located along the central meridian. 
During this process we also decreased the resolution of the images from the $\approx 0.5\arcsec$ of HMI to $\approx 2\arcsec$. The sub region of the derotated magnetogram (containing both AR 12035 and the neighboring ARs) is then  inserted at the center of a 8 times larger grid padded with zeros. The potential field extrapolation is  performed by applying  the fast Fourier transform method of \citealt{Alissandrakis1981} on this larger grid. 

As Figure~\ref{Fig:T-Overview} (left panels) shows the large scale magnetic field is indeed bipolar as discussed in section~\ref{Sect:MF}, but the part of the AR that displays an increased level of activity is characterized by a more complex connectivity. Essentially, four flux domains are observed: the first connecting the north-most part of the positive polarity P0 to the negative flux N3 at its south (green field lines in Figure~\ref{Fig:T-Overview} (a)),  the second connecting the negative polarity N3 with the leading compact positive polarity P1 (blue field lines), the third connecting this latter with the negative flux N2 at the southeast of it (connecting field lines not shown), and the last one is the large scale field that connects the positive polarities P0, P1 and P2 to the following negative one (N1, orange field lines). The anemone-like structure (blue-green field lines) is embedded in a bipolar field resulting in a \textit{breakout}-like magnetic field configuration, and  evolves from an northwest-southeast elongated structure on April 15 to a more circular one on April 16 (Figure~\ref{Fig:T-Overview}, left panels). 

\subsection{Quasi-separatrix layers}
\label{Sect:QSL}

{\it Quasi-separatrix layers} \citep[QSLs,][]{Demoulin1996} are thin layers characterized by a finite, but sharp, gradient in the connectivity of the magnetic field, and are  defined as regions where the squashing degree $Q$  is \textit{large}  (\citealt{Titov2002}). {\it QSLs} are also locations where current layers easily develop, where (slip-running) magnetic reconnection can occur (\citealt{Aulanier2006};  \citealt{Janvier2013};  \citealt{Dudik2014}), and they often coincide with the position of the flare ribbons (\citealt{Savcheva2012};  \citealt{Savcheva2015};  \citealt{Savcheva2016};  \citealt{Zhao2016}).

According to \citealt{Demoulin1996}, suppose we integrate in both directions along a distance `s', the field line crosses a point P$(x,y,z)$ of the solar corona. Two end points $(x$\arcmin$,y$\arcmin$,z$\arcmin$)$ and $(x$\arcsec$,y$\arcsec$,z$\arcsec$)$
 constitute a vector {\bf D}$(x,y,z)$=\{$X_1,X_2,X_3$\} = \{$x\arcsec-x\arcmin$, y\arcsec-y\arcmin, y\arcsec-y\arcmin\}. As the $QSLs$ are locations of rapid change in field-line linkage, hence for a slight change in point P$(x,y,z)$, the point 
{\bf D}$(x,y,z)$ will shift greatly. To limit this change into a given value of distance `$s$', the region with drastic change can be located with a function defined as: 
\begin{equation}
    {\bf N}(x,y,z,s)=\sqrt{{\sum\limits_{i=1,2,3} \Bigg[
    {\left(\frac{\partial X_i}{\partial x}\right)^2}} + {\left(\frac{\partial X_i}{\partial y}\right)^2} + {\left(\frac{\partial X_i}{\partial z}\right)^2} \Bigg]}
\end{equation}

We put a limit on the `s' either with a physical boundary or by the distance covered by a wave during the magnetic reconnection process, and this establishes as $z=0$ at the photospheric level from low to high plasma beta, hence at photospheric level:

\begin{equation}
   {\bf N}(x,y)=\sqrt{{\sum\limits_{i=1,2}\Bigg[{\left(\frac{\partial X_i}{\partial x}\right)^2}} + {\left(\frac{\partial X_i}{\partial y}\right)^2}\Bigg]}
\end{equation}

This is the norm evaluated at the boundary. The region with high value of {\bf N}(x,y) are the field lines taking part in the {\it QSLs} formation. Therefore, by following these lines, the coronal parts of {\it QSLs} can be traced.
It has been reported that the {\it QSLs} exist when the field components in the z-direction is weak compared to the maximum value in x and y directions (\citealt{Demoulin1996}; \citealt{Demoulin1997}). 
At the {\it QSLs} the breakdown of ideal MHD occurs, which results in reconnection. A change of connectivity of plasma elements happens even when a smooth boundary motion is imposed. Because the imposed boundary velocity values get amplified and the field line velocities exceeds the possible plasma velocities or Alfv\'enic speeds and an electric field component is produced in the layer along the magnetic field with slippinng magnetic field lines. In this way, concentrated currents are naturally formed at {\it QSLs}.

\begin{figure}[t!]
    \centering
    \includegraphics[width=\textwidth]{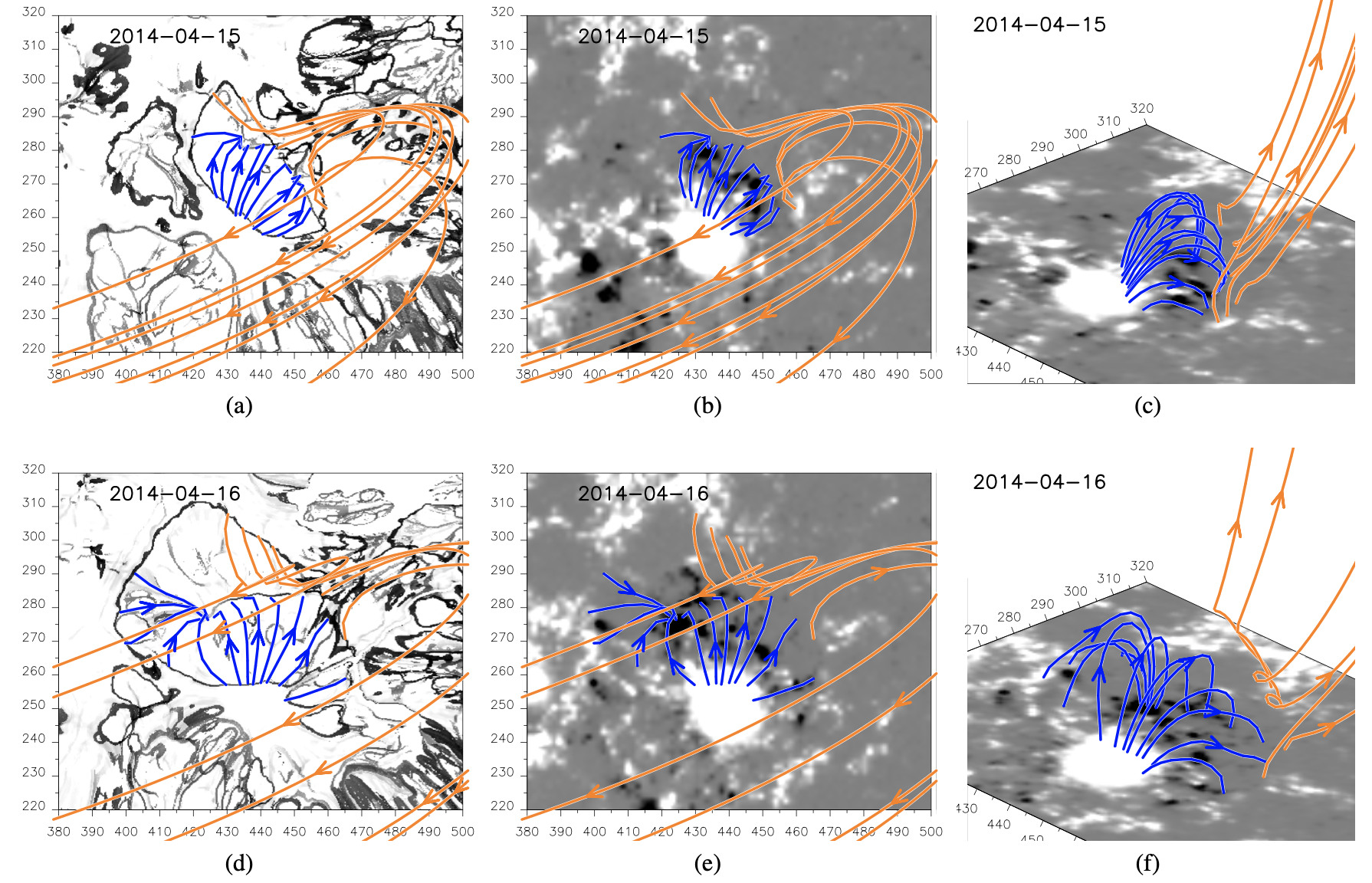}
    \caption{Distribution of the {\it QSL} maps at $z = 0.4$ Mm and of the photospheric magnetic field. The color scale for the magnetic field is saturated at ±300 Gauss and black/white indicates a negative/positive magnetic field.}
    \label{Fig:T-QSL}
\end{figure}

In this work we compute the {\it Q}-factor using the latest version of the topology tracing code (\citealt{Demoulin1996}), where the formula of \citealt{Pariat2012} is implemented. To this purpose we define the plane at $z=0.4~$Mm as the seed plane from which the field lines  are traced. On April 15 an elongated  fan-type {\it QSL} (arrow~Q1, Figure~\ref{Fig:T-Overview} (b)) surrounds the negative magnetic field distribution N3 at the northwest of the compact leading positive polarity P1 and embeds the portion of the PIL where the filament is located. This {\it QSL}  essentially encloses and separates the anemone-like structure (blue-green field lines) from the global/large-scale field (orange field lines) of the AR.  A spine-like {\it QSL} that starts from the northwest part of the fan-{\it QSL} and intrudes towards the central part of it is also observed (arrow~Q2, Figure~\ref{Fig:T-Overview} (b)). Field lines that originate at the north of the spine-{\it QSL} connect to the north-most positive polarity P0  (green lines), while the ones that originate at the south of it  connect to the compact leading positive polarity P1 (blue lines, Figures~\ref{Fig:T-QSL} (a) and \ref{Fig:T-QSL} (b)). These latter are the ones that enclose the filament that is the object of this study. 

On April 16 the fan-{\it QSL} (arrow~Q1, Figure~\ref{Fig:T-Overview} (d)) displays a more circular and less elongated shape, while  the spine-{\it QSL} (arrow~Q2, Figure~\ref{Fig:T-Overview} (d)) now originates from the center of the fan-{\it QSL} circle and extends westward. This magnetic field configuration indicates the presence of an elongated, locally two-dimensional, hyperbolic flux tube (HFT) that separates the two lobes of the anemone-like magnetic field configuration from each other and from the overlying field. This is confirmed by the vertical distribution of $Q$ along the plane passing through the spine-{\it QSL} (Figure~\ref{Fig:T-QSL_vert}, left panels). The 2D cuts show that a local `null-point like' configuration is achieved  in both cases, but the two lobes are more symmetric on  April 16 than they are on April 15. 

A second, less-pronounced, more-complex {\it QSLs} system is present around the region (arrow~JR, Figure~\ref{Fig:T-QSL}) where the recurrent jets are observed. However, this latter does not intersect the {\it QSL} labeled as Q1 suggesting that the jet-producing region and the flaring region are not directly connected to each other (although propagating Alfv\'{e}n waves may still induce a causality connection between the two parts of the AR).

While {\it QSLs} are robust topological features essentially determined by the connectivity of the magnetic field their exact morphology depends on the actual magnetic field model used to compute them (\citealt{Sun2013}). To compute the $Q$-factor we used the simplest magnetic field compatible with the given boundary, i.e., the current-free magnetic field. Despite this very simple assumption we note (1) that the computed fan-{\it QSL} of the flaring region matches  well the circular flare ribbons,  (2) that similarly to the computed fan-{\it QSL} the flare ribbon actually crosses the compact leading polarity, (3) that a brightening is observed approximately at the location of the spine-{\it QSL}, and (4) that the jet-associated brightening do not cross the flare-associated ribbons.This evidence suggests  that the  magnetic field model used is sufficient to capture the key features of the event. 

We note that the discrepancy between the computed fan-{\it QSL} (Figure~\ref{Fig:T-Overview}) and the circular brightening ribbon (Figure~\ref{Fig:OBS-15-04}) is probably also due to the simplistic magnetic field model used. This can be seen from Figure~10 of \cite{Sun2013} where the {\it QSLs} computed using both a potential field  and a non-linear force-free field (NLFFF) are compared. The fan {\it QSL} is relatively round in the potential field model, but displays a more sigmoidal shape in the NLFFF model that actually accounts for the shear present in the configuration. This is compatible with our configuration where the counterclockwise motion of the polarity N3 and the clockwise rotation of the sunspot P1 definitely introduced a degree of shear that the potential field model does not capture.  

\subsection{Decay index estimation}
\label{ch3_dis1}
A parameter that allows the estimation of the stability of a given magnetic field configuration is the decay index. Briefly, a magnetic FR embedded in an external magnetic field ($B_{\text{ex}}$) is unstable to perturbations if the axis of the FR has an height $z$ above the photosphere where the decay index of the external magnetic field (\citealt{Filippov2009}):
\begin{equation}
n=-\frac{d \ln B_{\text{ex}}}{d \ln z},
\label{Eq1}
\end{equation}
is larger than a critical value $n_{\text{cr}}$, that depends on the morphology of the FR (\citealt{Demoulin2010};  \citealt{Zuccarello2015}), and is in the range $n_{\text{cr}} \simeq 1.1-1.75$.    

\begin{figure}[ht!]
    \centering
    \includegraphics[width=\textwidth]{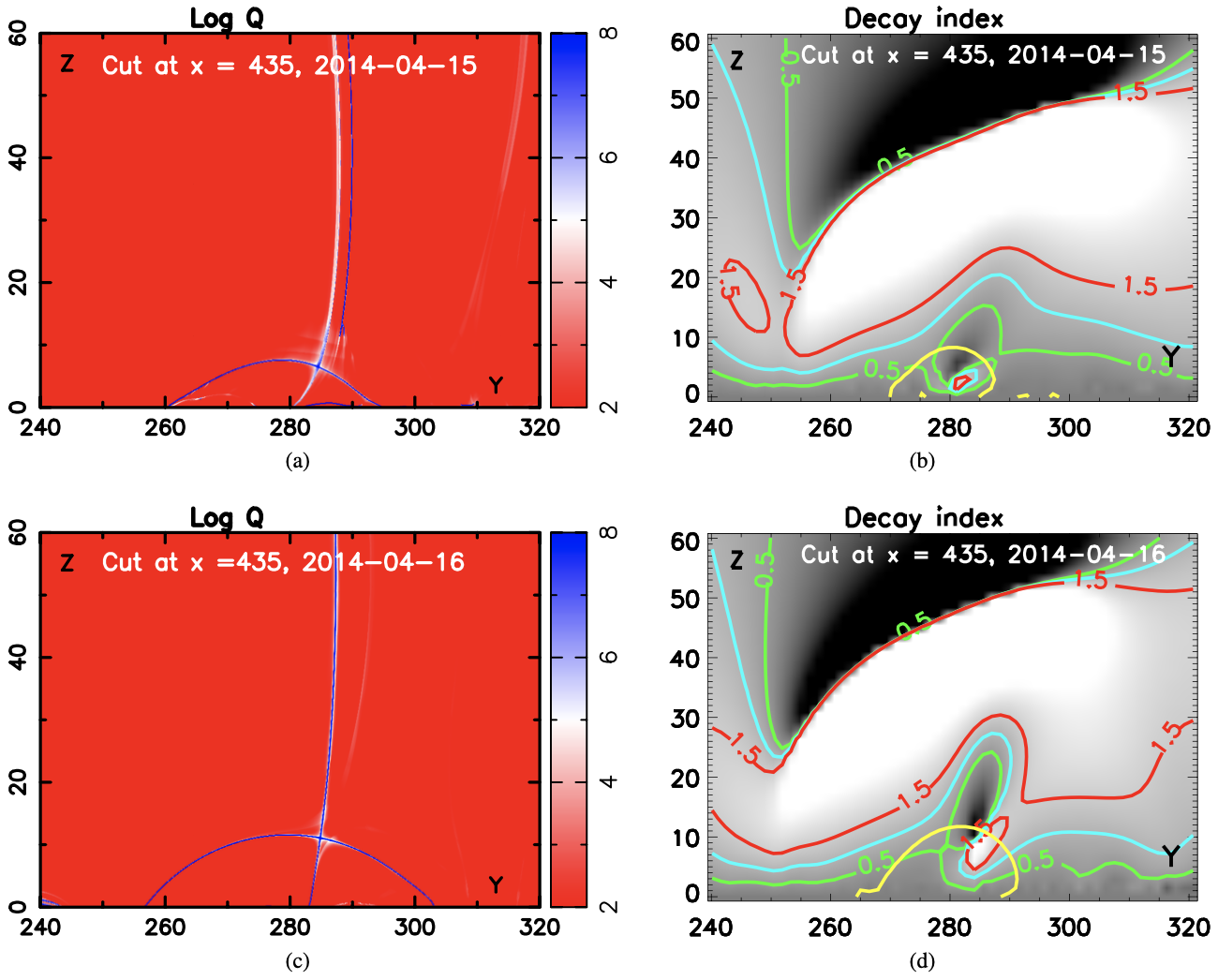}
    \caption[Projected 2D view of the squashing degree $Q$ and of the decay index.]
    {Projected 2D view of the squashing degree $Q$ (left panels) and of the decay index (gray color scale, right panels). The green, cyan, and red contours on the right panels indicate isocontours of decay index n = 0.5, 1, and 1.5, respectively. The yellow contour indicates the polarity inversion line.}
    \label{Fig:T-QSL_vert}
\end{figure}

To evaluate the stability of the magnetic field configuration we computed the decay index (using only the tangential component of the computed potential magnetic field) in all the volume above the flaring region. A vertical cut of decay index  along a plane passing through the approximate position of the HFT is shown in Figure~\ref{Fig:T-QSL_vert} (right panels).  The first conclusion that can be drawn from the figure is that in the proximity of the HFT the decay index changes sign becoming negative  (and reaching very large, negative values) as already shown by \cite{Torok2007} and as expected from its definition (Equation~\ref{Eq1}). 
A second conclusion is that the large scale stability of the magnetic field (away from the HFT) does not change significantly between April 15-16. This can be deduced by comparing the height at which the decay index is larger than the ``nominal'' $n=1.5$ critical value. The decay index for both days shows an initial increase with altitude (i.e., the system is more prone to erupt), followed by a decrease at even larger altitudes (i.e., the system is torus stable). 

As previously discussed, both the circular-shaped photospheric-{\it QSL} (Q1, Figure~\ref{Fig:T-Overview}) and the fan-spine-like distribution of $Q$ (2D vertical cuts of Figure~\ref{Fig:T-QSL_vert}) indicate the presence of a null-point topology in the corona. At the null-point the decay index (Equation~\ref{Eq1}) has a singularity and its validity is limited in this region. The distinction between torus instability or breakout-type reconnection as trigger mechanism for the eruption in this configuration is not at all straightforward  (\citealt{Kliem2014}).  Furthermore, in configurations with a vertical magnetic field (Figure~\ref{Fig:T-QSL}, right columns), the verticality of the field lines itself prevents any tension-related confinement even in a uniform field where the decay index is zero. Therefore, for this complex magnetic field topology the analysis of the decay index does not provide a useful criterion for eruptivity.  

For configuration that displays a coronal null point, the breakout scenario is a valuable mechanism to trigger the eruption. In this scenario the eruption is triggered by the onset of magnetic reconnection, and the efficiency of it also depends on the mutual orientation of the reconnecting fields. \cite{Galsgaard2007} performed a series of MHD simulations of a dynamical flux emergence experiment  aimed to study the role of the mutual orientation between the emerging FR and the overlying field. The authors have shown that when the two system are (nearly) anti-parallel substantial reconnection is observed, while this is not the case when the flux systems are (nearly) parallel. More recent simulations of dynamical emergence have shown that interaction of (nearly) anti-parallel flux systems leads to FR-like eruptions, while this is not the case for (nearly) parallel systems (\citealt{Archontis2012};  \citealt{Leake2014}). 

\section{Recurrent solar jets}
     \label{obs_c3}
The AR NOAA 12035 produced many solar jets during 
April 15-16, 2014 towards the south direction. We have selected
eleven clearly visible jets for our investigation. 
Their description is given in Table \ref{tab:II}. 
These jets were observed by the AIA onboard SDO in different 
EUV and UV wavebands. 
On April 16, 2014  two jets, that we will name jet J${_5}$ 
and J\ensuremath{'}${_5}$ in the next section, 
 were observed in H$_\alpha$ by the 15 cm Coud\'e telescope
operating at ARIES, Nainital, India (Figure~\ref{fig:hal}). 
The pixel size and cadence are  0.58$\arcsec$ and 1 minute respectively. 
For consistency, we have alligned all the images at April 16, 2014 10:30 UT. For identifying the onset and peak time of the jets,
we look into the temporal evolution of intensity at the jet foot-point.
We create a box containing the bright jet base and calculate 
the total intensity inside it. Then this total intensity is 
normalized by the intensity of the quiet region.

\begin{table}[ht!]
\small
\caption[Different physical parameters derived from SDO/AIA data for 
jets on April 15-16, 2014.]{Different physical parameters derived from SDO/AIA data for 
jets on April 15-16, 2014.}
\label{tab:II}
\centering
\begin{tabular}{ccccccc}
\hline
Jet&~~Start/& Speed at different& Height& Width& Lifetime \\
number&end  & wavelengths ($\lambda$) in km s$^{-1}$ & (Mm) & (Mm) & (minutes) \\
          &(UT) &304\AA~~~211\AA~~~193\AA~~~171\AA~~~131\AA~~~~94\AA &  &\\
\hline
J${_1}$ &~~14:55/ & 205~~~~ 295~~~~~ 257~~~~~ 249~~~~~ 268~~~~~ 206& 145 &4.4 &15 \\
        &15:10 &~~~~~~~~~~~~~(252)~~~~~~~~~~~~~~~~~~~~~~~~~&    &    &  \\
\hline
J${_2}$ &~~18:01/ & 234~~~~ 177~~~~~ 199~~~~~ 232~~~~~ 221~~~~~ 235& 124 &5.2 &08 \\
        &18:09 &~~~~~~~~~~~~~(196)~~~~~~~~~~~~~~~~~~~~~~~~~&    &    &  \\
\hline
J${_3}$ &~~06:33/ & 137~~~~ 140~~~~~ 132~~~~~ 153~~~~~ 148~~~~~ 163& 202 &4.3 &23 \\
        &06:56 &~~~~~~~~~~~~~(126)~~~~~~~~~~~~~~~~~~~~~~~~~&    &    &  \\
\hline
J\ensuremath{'}${_3}$ &~~06:10/ & 113~~~~ 121~~~~~ 109~~~~~ 110~~~~~ 105~~~~~ 100& 108 &3.5 &15 \\
        &06:34 &~~~~~~~~~~~~~(100)~~~~~~~~~~~~~~~~~~~~~~~~~&    &    &  \\
\hline
J${_4}$ &~~07:13/ &136~~~~ 139~~~~~ 147~~~~~ 133~~~~~ 164~~~~~ 138& 95 &2.6 &11 \\
        &07:23 &~~~~~~~~~~~~~(147)~~~~~~~~~~~~~~~~~~~~~~~~~&    &    &  \\
\hline
J\ensuremath{'}${_4}$ &~~07:12/ & 305~~~~ 295~~~~~ 325~~~~~ 300~~~~~ 296~~~~~ 303& 116 &4.5 &06\\
        &07:18 &~~~~~~~~~~~~~(322)~~~~~~~~~~~~~~~~~~~~~~~~~&    &    &  \\
\hline
J${_5}$  &~~10:36/ & 183~~~~ 202~~~~~ 174~~~~~ 185~~~~~ 184~~~~~ 182& 217 &6.0 &24\\
        &10:50 &~~~~~~~~~~~~~(174)~~~~~~~~~~~~~~~~~~~~~~~~~&    &    &  \\
\hline
J\ensuremath{'}${_5}$ &~~10:33/ & 275~~~~ 343~~~~~ 364~~~~~ 291~~~~~ 316~~~~~ 241&87  &3.6 &10 \\
        &10:43 &~~~~~~~~~~~~~(357)~~~~~~~~~~~~~~~~~~~~~~~~~&    &    &  \\
\hline
J${_6}$ &~~14:41/ & 197~~~~ 177~~~~~ 192~~~~~ 156~~~~~ 170~~~~~ 171&94  &3.0 &15 \\
        &14:55 &~~~~~~~~~~~~~(187)~~~~~~~~~~~~~~~~~~~~~~~~~&    &    &  \\
\hline
J\ensuremath{'}${_6}$ &~~14:47/ & 343~~~~ 326~~~~~ 340~~~~~ 323~~~~~ 307~~~~~ 304& 152 &4.0 &16 \\
        &14:59 &~~~~~~~~~~~~~(335)~~~~~~~~~~~~~~~~~~~~~~~~~&    &    &  \\
\hline
J${_7}$ &~~16:59/ & 183~~~~ 174~~~~~ 154~~~~~ 187~~~~~ 184~~~~~ 217& 145 & 5.1 &14\\
        &17:13 &~~~~~~~~~~~~~(154)~~~~~~~~~~~~~~~~~~~~~~~~~&    &    &  \\
\hline
\end{tabular}
\end{table}
\begin{figure}[ht!]
\includegraphics[width=\textwidth, clip=]{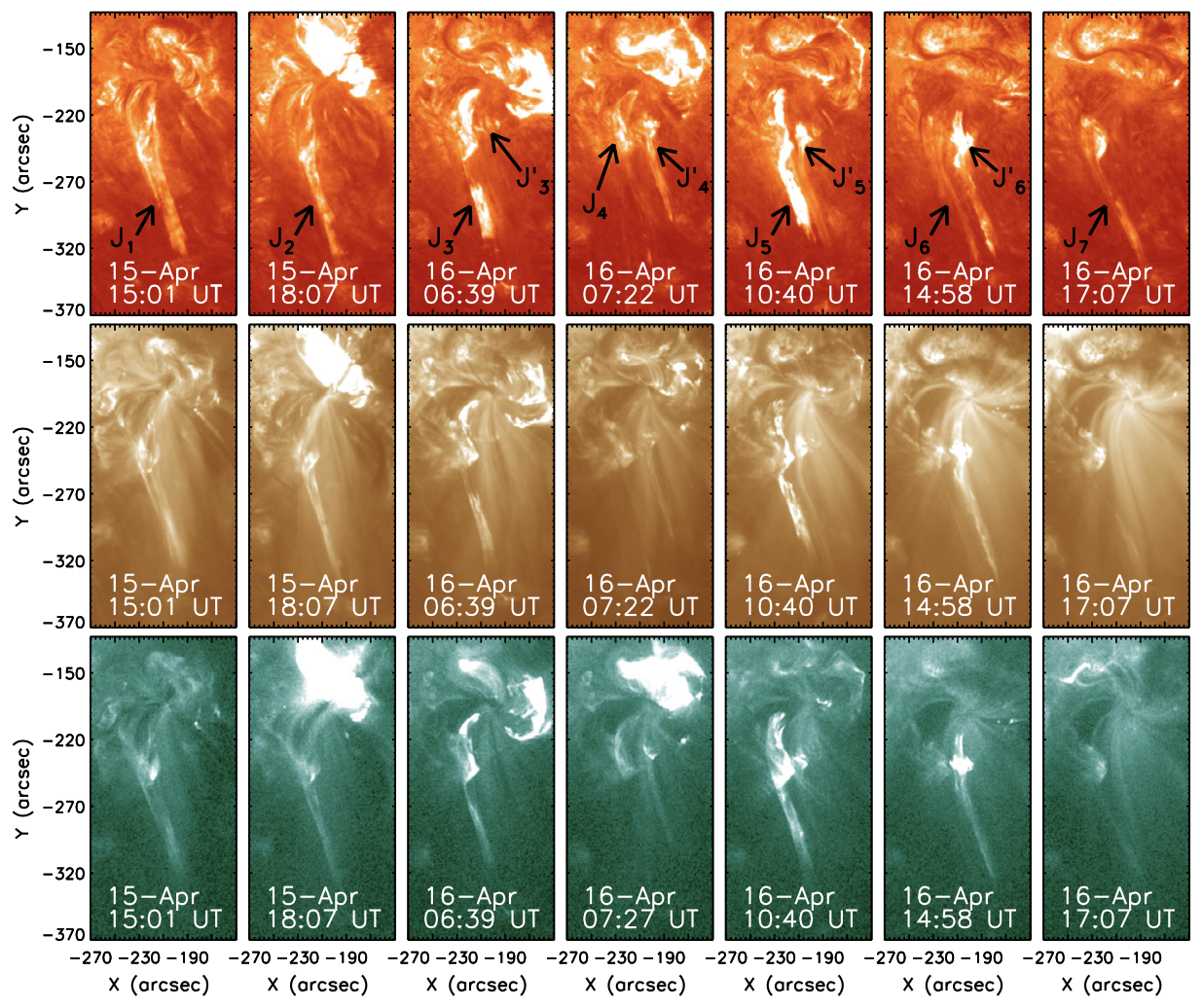}
\caption{Recurrent jets in AR 12035 on 15 and 16 April 2014. The top, middle, and bottom panels show images of AIA filters at 304 \AA\, 193 \AA\, and 94 \AA\, respectively. The arrows indicate the bright emission of the jets in the two different locations.}
\label{fig:jet/morpho}
\end{figure}

\begin{figure}[ht!]
\includegraphics[width=\textwidth, clip=]{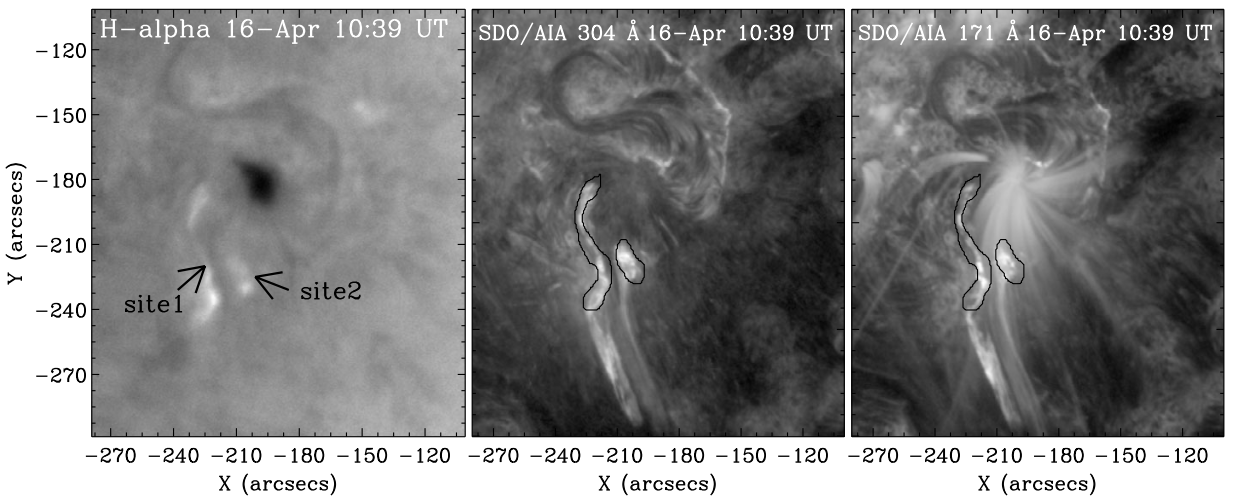}
\caption{Jets J${_5}$  and J\ensuremath{'}${_5}$  from site 1 and site 2, from left to right in H$_\alpha$, observed in ARIES Nainital,  in 304 \AA\ and in 171 \AA\ observed  with SDO/AIA.}
\label{fig:hal}
\end{figure}

\begin{figure}[ht!]
\includegraphics[width=1.0\textwidth, clip=]{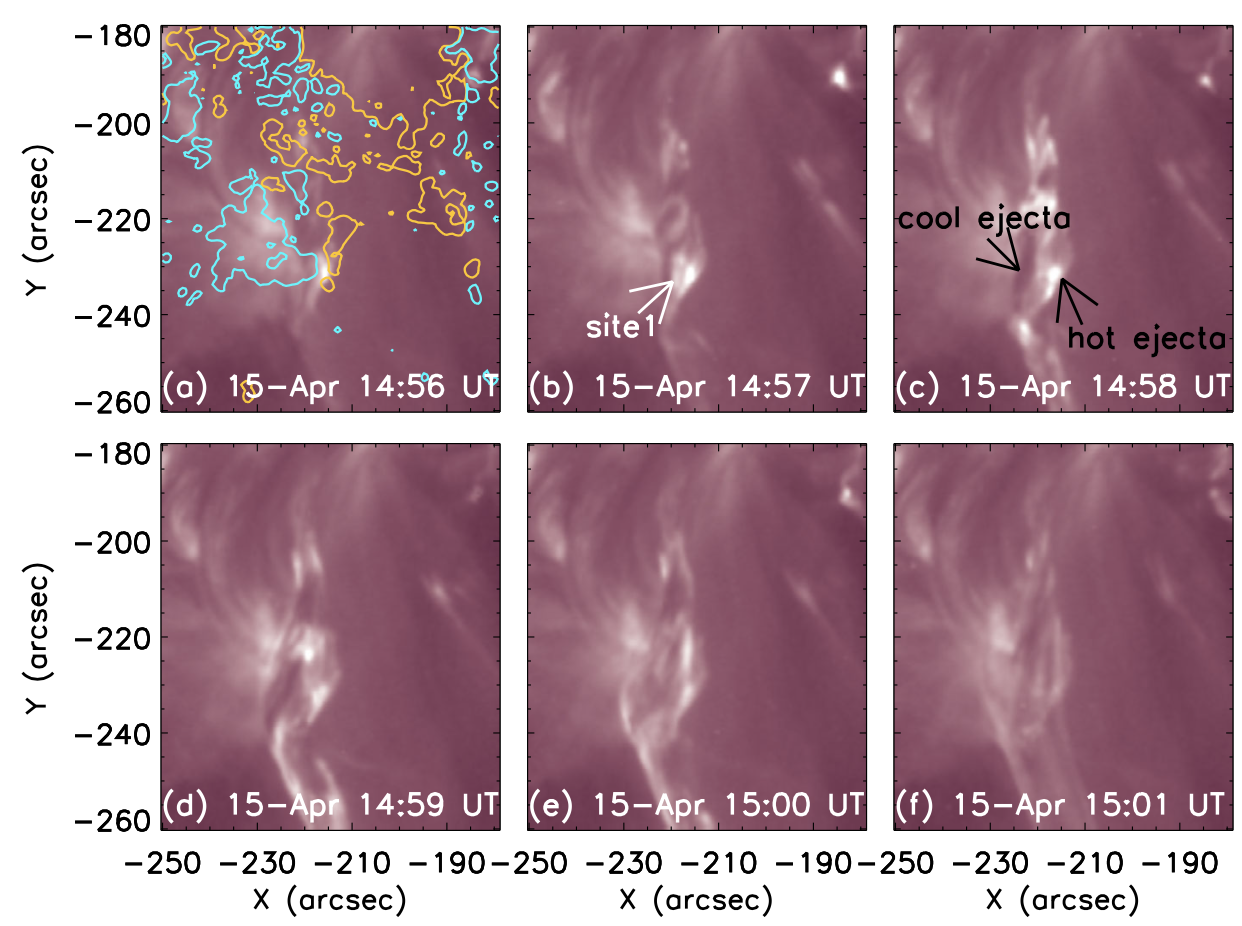}
\caption[Evolution of jet J${_1}$ initiated in site 1 
observed by AIA 211 \AA.]
{Evolution of jet J${_1}$  initiated in site 1 
observed by AIA 211 \AA. The contours on the first image are HMI 
longitudinal magnetic field with
 yellow and cyan colors for positive and negative polarities.}
\label{fig:jet1}
\end{figure}
The eleven selected jets during April 15-16, 2014
are named as J${_1}$--J${_7}$ and  J\ensuremath{'}${_3}$-- J\ensuremath{'}${_6}$ 
 respectively.
Out of eleven jets, the first two jets J${_1}$ and J${_2}$ occur on April 15 and the remaining nine jets 
are on April 16, 2014.
These jets originated from two locations in the south part of the AR NOAA 12035.
One location is at position [X,Y] = [-220$^{''}$, -215$^{''}$] and the other is at 
 [-205$^{''}$, -215$^{''}$] (Figure~\ref{fig:hal} left panel). 
Here, we will refer them as 
site 1 jets (J) and  site 2 jets (J\ensuremath{'}) respectively.  The two sites are at a distance 
of 15$\arcsec$ from each other ($\approx$  11 Mm).
Figure~\ref{fig:jet/morpho} displays images of the eleven jets observed  
with the AIA filters at 304 {\AA} (top), 193 {\AA} (middle) 
and 94 {\AA} (bottom) during their peak phase.
 Almost all the jets are visible in all  EUV channels, which 
indicates the multi-thermal nature of the  jets. 
The onset and end time of each jet are summarized in Table \ref{tab:II}.

\begin{figure}[ht!]
\includegraphics[width=1.0\textwidth, clip=]{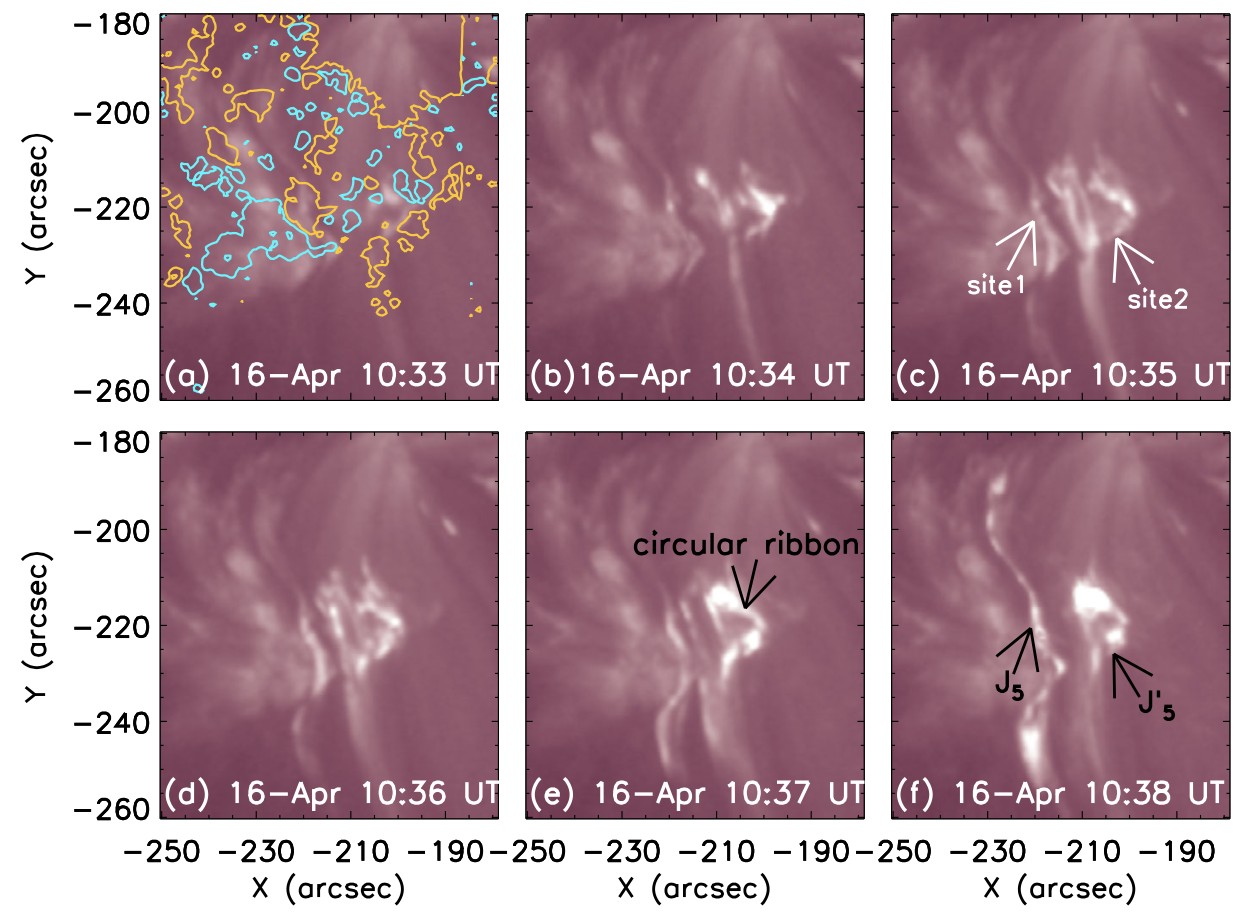}
\caption[Evolution of jets J${_5}$ and J\ensuremath{'}${_5}$  
from site 1 and site 2
observed by AIA 211 \AA.]
{Evolution of jets J${_5}$ and J\ensuremath{'}${_5}$
from site 1 and site 2 (indicated by white arrows) 
observed by AIA 211 \AA. Yellow and cyan colors are for positive and negative polarities respectively.}
\label{fig:jet5}
\end{figure}
Jet J${_1}$ started on April 15, 2014 with a bright base at site 1. 
The zoomed view of evolution of jet 
J${_1}$ in AIA 211 \AA\ is shown in Figure~\ref{fig:jet1}. 
Together with the ejection of bright material, we
 have also observed the ejection of cool and 
dark material in 211 {\AA} (Figure~\ref{fig:jet1}c). 
 The dark jet is due to the presence of cool material
   absorbing the coronal emission. The bright
and cool jet material were rotating clockwise.
After the onset of $\approx$ 2 minutes it started to rotate anti-clockwise. 
This indicates the untwisting of the system to relax to a lower energy state by propagating 
twist outwards. 
The jet J${_2}$ started also with a bright base similar to
 J${_1}$ from the same location.
This jet was also showing untwisting like as in the  case of 
J${_1}$. However, its base was less bright and less 
broader than J${_1}$. On April 16, 2014 J${_3}$ and J\ensuremath{'}${_3}$
 started from  site 1 and site 2 respectively. 
The base of J\ensuremath{'}${_3}$ was like a circular ribbon and broader than J${_3}$.
Jet J\ensuremath{'}${_4}$  is bigger than J${_4}$. 
Jet J${_4}$ peaks almost simultaneously in all EUV wavelengths around 07:17 UT. 
The peak time for J\ensuremath{'}${_4}$ is five minutes later than  J${_4}$, around 
07:21 UT in all wavebands. One difference between 
these jets and  J${_1}$--J\ensuremath{'}${_3}$ was that they have no rotation.
The evolution of jets J${_5}$ and J\ensuremath{'}${_5}$ is presented in 
Figure~\ref{fig:jet5} in AIA 211 \AA. 
In contrast to the other jets, the peak time for jet J${_5}$ and J\ensuremath{'}${_5}$
 is different for different wavebands. The peak
 time for J${_5}$ at 304 \AA\ is 10:38:30 UT, 
and at 94 \AA\ is 10:41 UT. For the case of J\ensuremath{'}${_5}$,
the peak time in 304 \AA\ is 10:37:30 and at 94 \AA\ is 10:38 UT. 
\begin{figure}[ht!]
\includegraphics[width=1.0\textwidth, clip=]{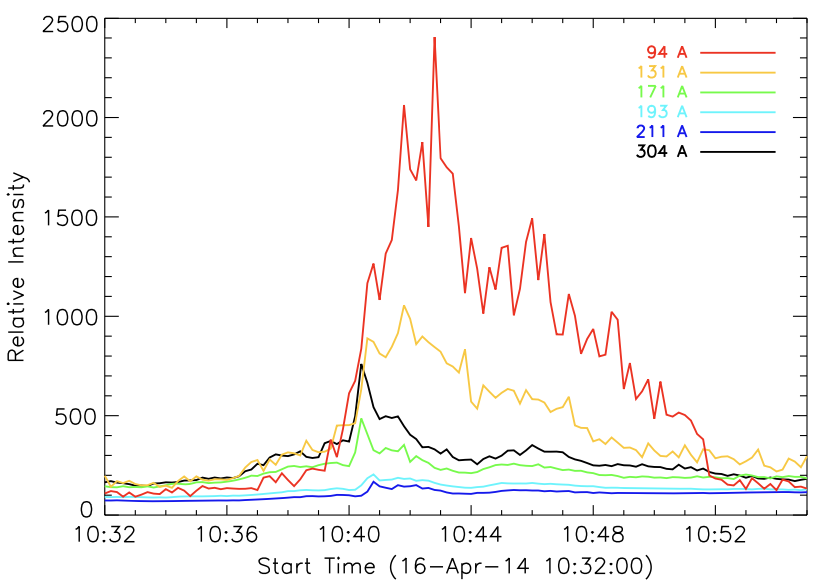}
\caption[Intensity profile of jet J${_5}$ at base location. Different colors
represent different SDO/AIA wavelengths.]{Intensity profile of jet J${_5}$ at base location. Different colors
represent different SDO/AIA wavelengths. The peak of the cooler component is 
earlier than the peak of the hotter component.}
\label{fig:inten}
\end{figure}

\begin{figure}[h!]
\centering
\includegraphics[width=0.92\textwidth,clip=]{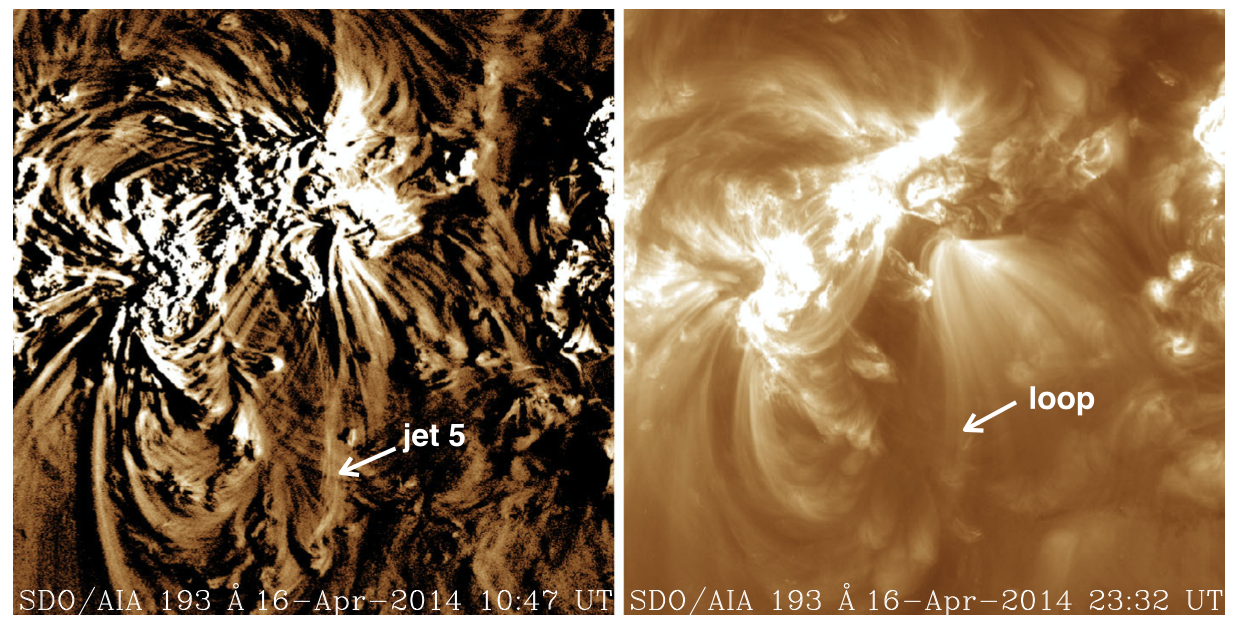}
\caption[Sigmoidal orientation of the  loops in  the AR 
in the AIA  193 \AA.]{Sigmoidal orientation of the  loops in  AIA  193 \AA\, filter and in the difference image.} 
 \label{fig:loop}
\end{figure}
For these jets, the peak for
cool plasma (longer wavelength 304 \AA) appears earlier than the hot plasma 
material (shorter wavelength 94 \AA).
The temporal evolution of flux at the base of jet J$_5$ is shown in  
Figure~\ref{fig:inten}.
This behavior is contrary to  other jets reported in the literature
 where hotter plasma appears before cooler plasma,
 suggesting some mechanism of cooling versus time (\citealt{Alexander1999}).
These are the largest jets among the ones discussed in this study. As the event
progress, interestingly part of jet J\ensuremath{'}${_5}$  was detached from it and 
moved towards  site 1 and finally merged with jet
J${_5}$. 
We have estimated the speed of
 J\ensuremath{'}${_5}$ towards J${_5}$ as $\approx$ 45 km s$^{-1}$.
Merging of the broken part of J\ensuremath{'}${_5}$ 
with J${_5}$ made it bigger as it was ejected in the south 
as well as in the north direction at the same time. 
We have also noticed the circular ribbon at the base of 
J\ensuremath{'}${_5}$, shown in Figure~\ref{fig:jet5} (e).
Looking at the evolution of the J\ensuremath{'}${_5}$ jet, we found
that the jet follows a sigmoidal-shape loop path visible in AIA 193 \AA\ 
(Figure~\ref{fig:loop}).
 These loops are originating from the sunspot and going towards  the south  with a 
sigmoid-shape. The sigmoidal-shape of these loops could be due the the clockwise rotation of big positive 
polarity spot. 
We have observed the clockwise rotation in jet  
J${_5}$ and the calculated rotation speed in four wavelengths
171 \AA\ , 193\AA\ , 211 \AA\ and in 304 \AA. 
The speed varies from 90 km s$^{-1}$ to 130 km s$^{-1}$.
Jet J${_6}$ was small and of weak intensity whereas J\ensuremath{'}${_6}$ had
 a strong intensity and was very bright, wide and had a circular base. 
J\ensuremath{'}${_6}$ started to move towards site 1, but like in the case of 
J\ensuremath{'}${_5}$, it could not reach up to J${_6}$ location. We did not find any rotation in this jet. 
Jet J${_7}$ started from site 1 location and  also showed clockwise rotation.
 In a nutshell, we have found that all jets from  site 1 have similar morphology.
Jets  from site 2 also have similar morphology, but this
is different from the morphology of jets ejected from site 1.
One common feature in all jets of site 2 on 16 April
was that after their trigger they all have a tendency to 
move towards site 1 before or during the ejections. 
 It seems that for the case of J${_5}$ and  J\ensuremath{'}${_5}$, 
there is a connection between these two.
 Another common feature of jets 
originated from site 2 is 
that they all follow the sigmoidal  
loops visible in different AIA wavebands.  
The jets from site 2 occur before or almost simultaneously 
(in the case of J\ensuremath{'}${_4}$) with the jets from site 1.
This is true apart from the jet J$_6$. The main difference is 
that J$_6$ is quite weak among all of the coupled jets.
We have noticed that jets observed in 193 \AA\ 
are thinner than in 304 \AA. All jets started
with a bright base, similar to  common X-ray jet observations.

\subsection{H$_\alpha$ observations}
     \label{obs2_c3}
Two jets-- jet J${_5}$ and J\ensuremath{'}${_5}$ were also observed 
in the H$_\alpha$ line center (6563 \AA) by the solar tower telescope at ARIES, as a bright ejecta.
H$_\alpha$ jets from both jets sites started at  $\approx$ 10:35 UT,
 one minute later than  in the  EUV wavebands and they faded
away  at $\approx$ 10:50 UT. We also observed the dark 
material ejection surge between the bright jets from the two 
sites site 1 and site 2 respectively. 
To compare the spatial location of H$_\alpha$ with EUV images,
 we have over-plotted the
contours of H$_\alpha$ in AIA 304 \AA\ and 171 \AA\ . 
We found  that H$_\alpha$ jets are coaligned with the EUV jets.
 However, the position of the H$_\alpha$ jet is 
only at the origin of the EUV jets. It may be because the H$_\alpha$ jets 
are in  the chromosphere and have 
less height than the EUV jets. The bright
 ejections in H$_\alpha$ are followed by dark material ejections. 
It could be the cooler part of the jet. 
This cooler part is spatially shifted towards the west side from the bright jets.
\subsection{Height-time analysis of the jets with two different methods}
     \label{obs3_c3}

\begin{figure}[ht!]
\includegraphics[width=\textwidth, clip=]{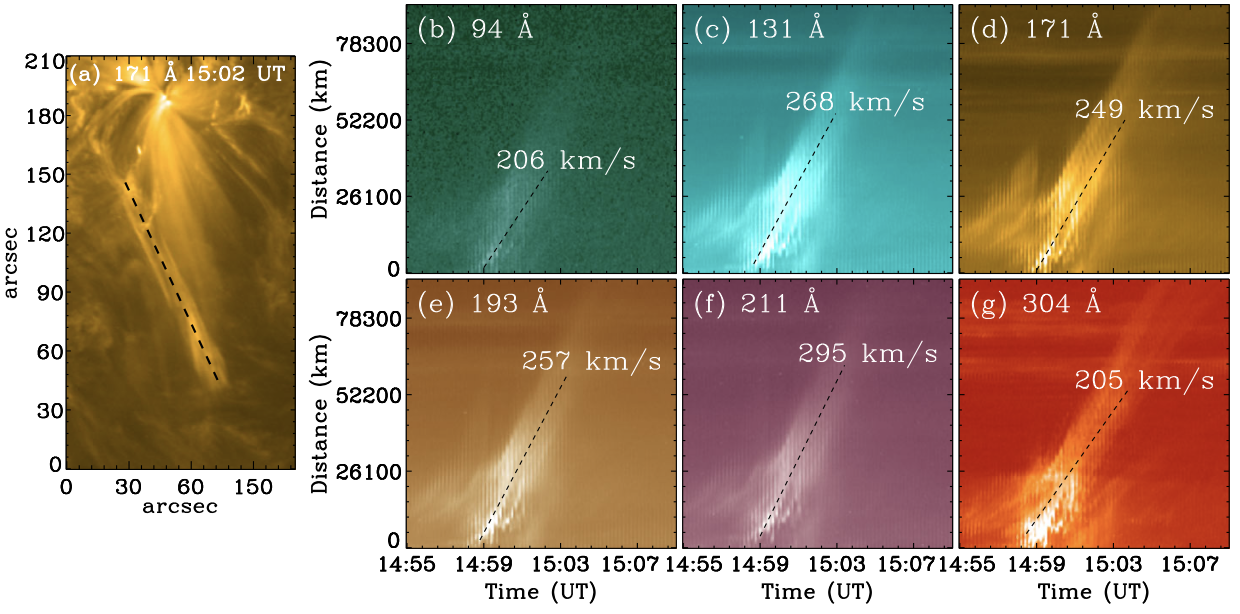}
\vspace{-0.6cm}
\caption[Time-slice analysis of jet J${_1}$ on April 15, 2014]{Time-slice analysis of jet J${_1}$ on April 15, 2014. Left: position of slit along the jet, 
right: velocity comparison at different wavelengths.}
\label{fig:slice1}
\end{figure}

\begin{figure}[ht!]
\includegraphics[width=1.00\textwidth,clip=]{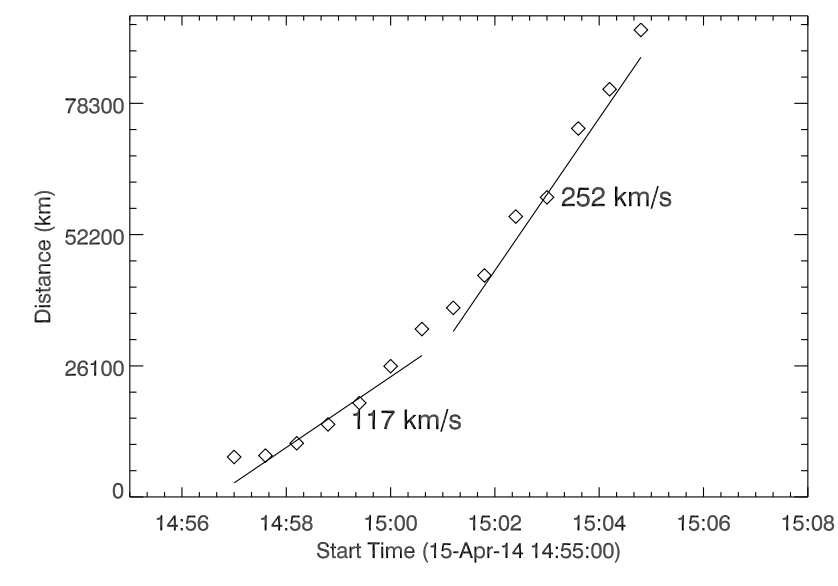}
\caption{An example of a   height--time plot for Jet J${_1}$ derived using the
leading edge procedure.}
\label{fig:leading_edge}
\end{figure}
To understand the kinematics and dynamics of the observed jets, we have made a time-distance analysis in 
different AIA/EUV channels {\it i.e.} 94, 131, 171, 193, 211, and 304 \AA. 
To perform this analysis, we adopted two methods, one is the time--slice technique and 
the other is tracking of the leading edge of the jets. For the time--slice technique, we 
fixed a slit at the center of the jets and observed the motion of plasma along the slit. 
An example of this time--slice analysis of jet J${_1}$ on April 15, 2014  is presented in
 Figure~\ref{fig:slice1}.
In the figure, the left image shows the location of  the slit (drawn as a black dashed line) and the right 
denotes the height-distance maps at different EUV wavelengths. 
Using this time--slice technique, we have computed the heights and projected
 speeds
 for the different structures of all jets at different wavelengths,
shown in the right panel of the figures.
For the same jet of April 15, 2014 at 193 \AA, the height--time plot using the leading 
edge procedure is presented in  Figure~\ref{fig:leading_edge}. 
The curve  has an exponential behavior with  two  acceleration phases.
 We decided to fit  linearly  the beginning of the expansion and then 
the later phase in order to compared the values with those of the other 
methods. Two speed values are derived. The first is 117 km s$^{-1}$ which is 
nearly half of the speed derived  after the acceleration phase 
(252  km s$^{-1}$). The  error is around 10  km s$^{-1}$ according to the points chosen 
for the fits.

The time--slice technique indicates that every jet has
 multi-speed structures 
and the speeds of the jets are different in different EUV wavelengths.
The average speed by the time--slice technique varies in different wavelengths
 for J${_1}$: 205 -- 295 km s$^{-1}$, J${_2}$: 
177--235 km s$^{-1}$, J${_3}$: 132--163 km s$^{-1}$, J\ensuremath{'}${_3}$: 100 --121 km s$^{-1}$, 
J${_4}$: 133 --164 km s$^{-1}$, J\ensuremath{'}${_4}$: 295 --325 km s$^{-1}$, J${_5}$: 174 --202 km s$^{-1}$,
J\ensuremath{'}${_5}$: 275 --364 km s$^{-1}$, J${_6}$: 156 --197 km s$^{-1}$  
J\ensuremath{'}${_6}$: 304 --343 km s$^{-1}$, and for J${_7}$: 154 --217 km s$^{-1}$ respectively.
The dispersion of the values for one jet according to the  different considered wavelengths is 
between 10$\%$ to 50$\%$.
 It is difficult to understand
if it is just the range of the uncertainties of the measurements   or  if it really
 corresponds to the existence of multi-components in the jet  with multi-temperatures and speeds 
or if the slit of the time distance diagram is crossing different components of the jets.
We have compared these results with the speed derived by the leading edge procedure.
 We found  that the fast speed fits with the time-distance derived 
velocity. The time-distance technique with a straight slit ignores the 
first phase of the jets.  The values of speed derived by both methods 
are presented in Table \ref{tab:II}.

We have also computed the lifetimes, widths, and the maximum heights 
of each jet.
The lifetimes vary from 6 to 24 minutes.
 J${_5}$ has the maximum lifetime (24 minutes), whereas J\ensuremath{'}${_4}$ 
has the minimum (6 minutes) lifetime.
In general, we found the narrow and long life--time jets achieved 
more height than the wide and short lifetime jets.
We have also noticed that the lifetimes of jets are longer in 304 \AA\ and in 171 \AA\ than in 94 \AA. 
The width ranges from 2.6--6 Mm. 
The maximum height was attained by J${_5}$ and it was 217 Mm. 

\begin{figure}[t!]
\includegraphics[width=1.00\textwidth,angle=0,clip=]{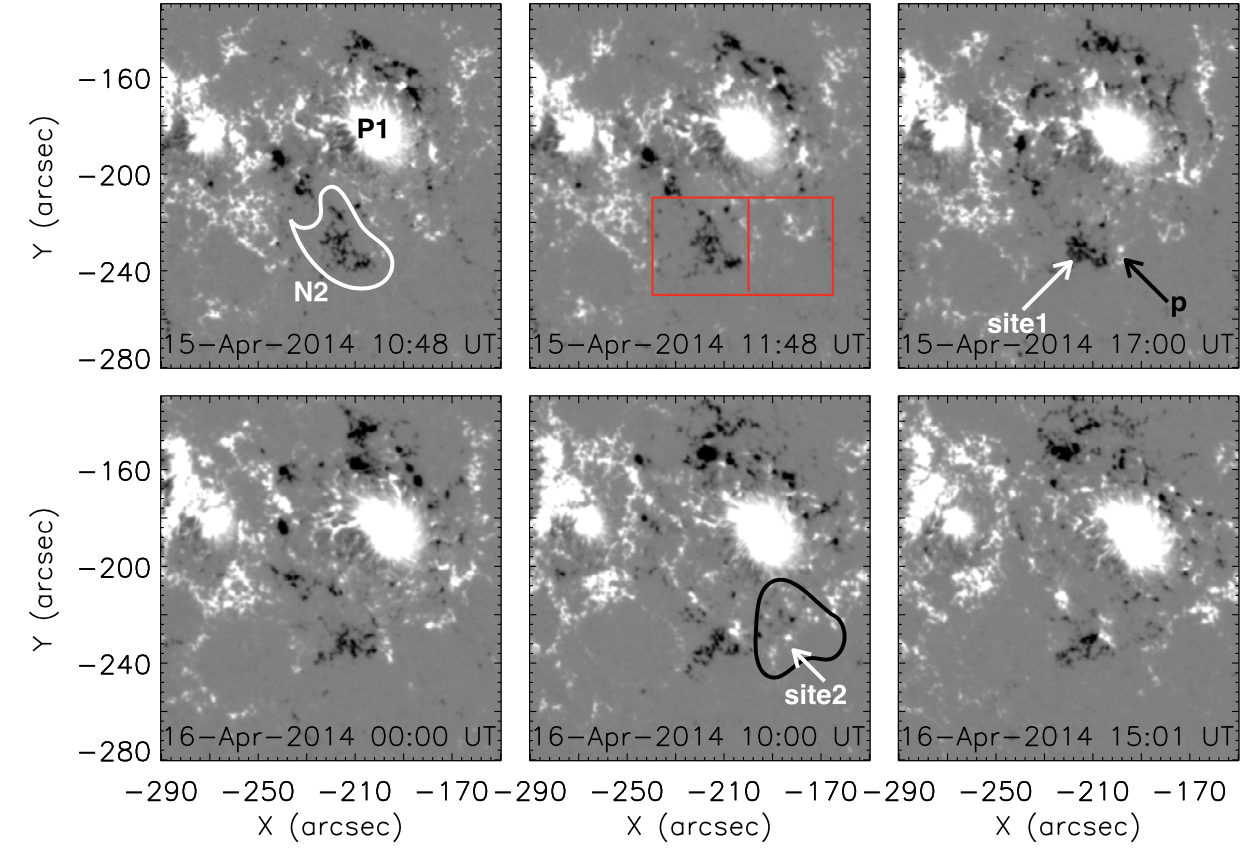}
\caption[Magnetic field evolution of AR NOAA 12035.]{Magnetic field evolution of AR NOAA 12035. The two jet
locations  (site 1 between N2 and p, site 2 at the edge of the supergranule 
indicated by the black contour) are indicated by white arrows.} 
\label{fig:mag}
\end{figure}

\begin{figure}[h!]
\includegraphics[width=\textwidth,clip=]{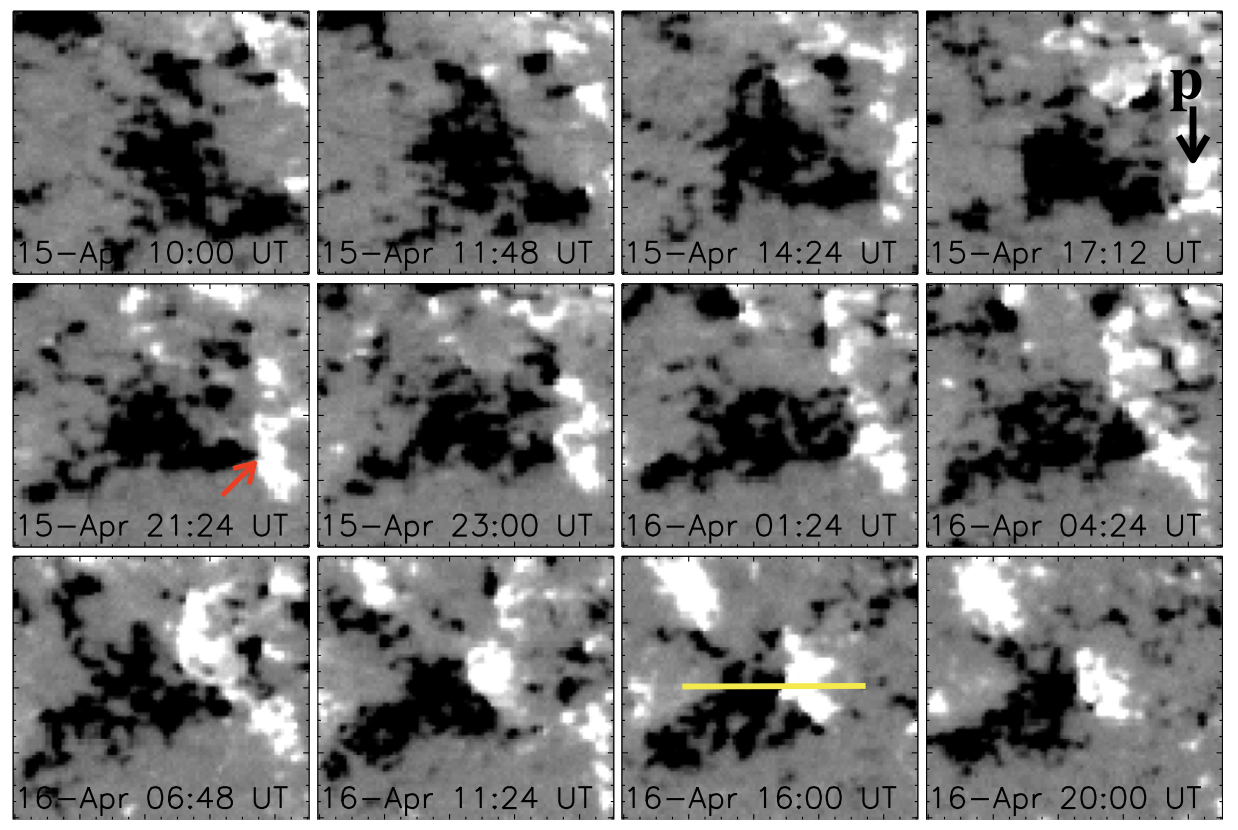}
\caption[Evolution of magnetic field at the site 1 location.]
{Evolution of magnetic field at the site 1 location. 
The flux cancellation at site 1 between the negative polarity and the positive polarity (p) is shown by a red arrow.}
\label{fig:mag1}
\end{figure}

\begin{figure}[h!]
\includegraphics[width=1.0\textwidth,angle=0,clip=]{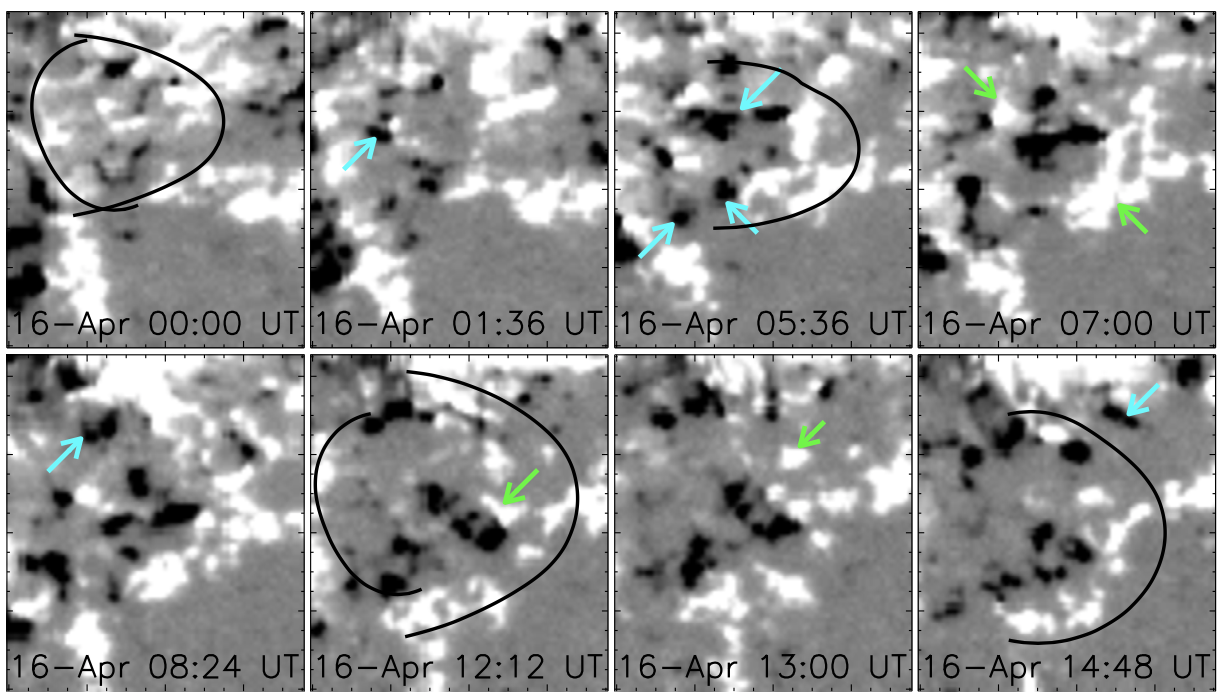}
\caption[Evolution of magnetic field at the site 2 location.]{Evolution of magnetic field at the site 2 location. The positive (negative) flux emergence is shown by green (cyan)
arrows respectively. The roundish black curve indicate supergranule cell.}
\label{fig:mag2}
\end{figure}

\begin{figure}[ht!]
\includegraphics[width=1.00\textwidth,clip=]{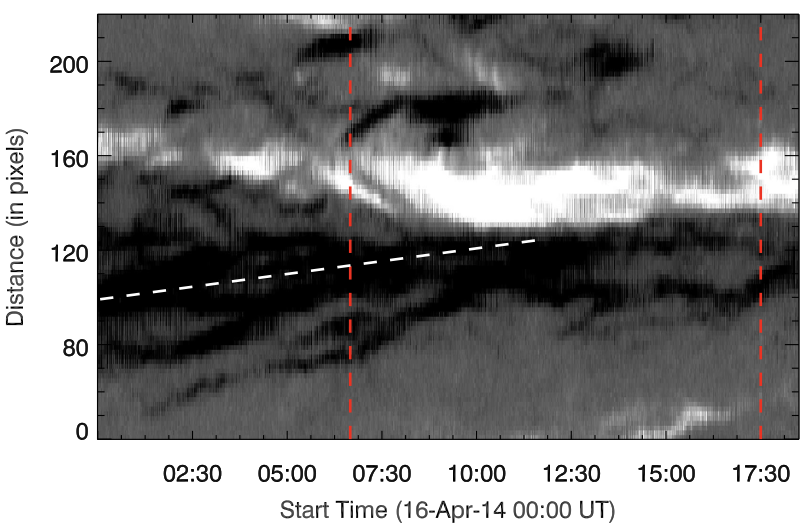}
\caption{Magnetic field evolution along the slice shown in Figure \ref{fig:mag} at site 1. Two vertical lines in the image 
indicate the time of jet activity.}
\label{fig:mag_slice}
\end{figure}

\subsection{Photospheric magnetic field at the jet base}
     \label{obs3}
 Figure~\ref{fig:mag} presents the magnetic field evolution of 
the AR near the jet source region observed by SDO/HMI during April 15-16, 2014.
The AR  consists of two large positive polarity sunspots P1 and P2 followed by the negative 
polarity spot N1.
The positive polarity P1 behaves like a decaying sunspot with a ``moat region''
 around it. When the 
sunspot is close to its decay stage, it loses its polarity by dispersing
 in all directions. This dispersed positive polarity
 cancels a part of the negative magnetic polarity 
left at the site of the jet location.
The origin of jet activity lies between the
  two leading positive polarities and the whole AR shows 
clockwise rotation.
Together with the rotation of the whole AR, the 
western big spot P1 also shows a rotation
 in the same direction as the AR. The sunspot rotation makes the AR
 sheared and the loops connecting the preceding
 positive polarity and the
following negative polarity become sigmoidal 
 (Figure~\ref{fig:loop}). The jets are ejected in the 
south-east direction instead of south direction probably because of the  upper part of the sigmoidal loops.

\begin{figure}[t!]
\centering
\includegraphics[width=1.00\textwidth,clip=]{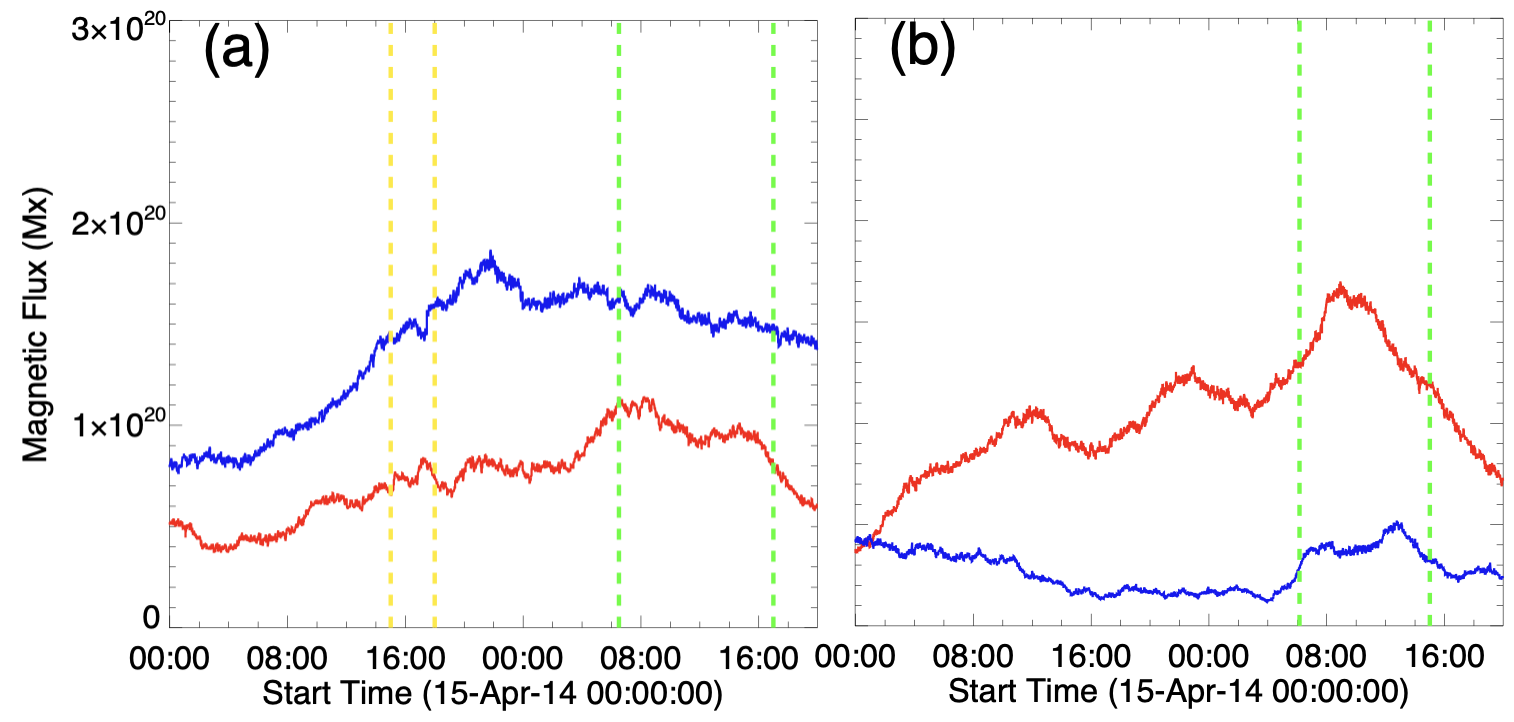}
\caption[Magnetic flux as a function of time calculated over the red box of
 Figure \ref{fig:mag}.]{Magnetic flux as a function of time calculated over the red box of Figure \ref{fig:mag} for the site 1 (a) and site 2 (b) location. Red and blue curves are for positive and negative magnetic flux.}
\label{fig:mag_evo}
\end{figure}

To investigate the magnetic field evolution at the jet's origin, we have 
carefully analyzed the development of different polarities. 
On April 15, we observed the positive polarity P1 and in its south a bunch of 
negative polarity N2 and  a positive polarity p (Figure~\ref{fig:mag}).  
Site 1 of the jet's origin is located between N2 and p. 
Site 2 is located to the west of site 1.
The zoomed view of magnetic flux evolution at site 1 is shown in Figure
\ref{fig:mag1}. In the
figure, we have noticed several patches of emerging flux of positive and negative polarities.
In addition to this emerging flux, we have interestingly found that the 
negative polarity N2 and the positive polarity p came closer and cancelled
each other. The jet's cancelling location is represented by the red arrow.
 To examine the flux cancellation at the jet
site 1, we made  a  time-slice diagram along the slit
shown in Figure \ref{fig:mag1} as the yellow line.
The result  is presented in Figure \ref{fig:mag_slice}.
The positive and negative flux approached each other and cancelled afterwards.
We have drawn the start and end time of the jets from site 1 and this is
shown in the figure by vertical red lines. We noticed that the jet
activity from  site 1 was between this flux cancellation site.
Further, we did quantitative measurements in the box at jet site 1 drawn in Figure \ref{fig:mag} (top, middle panel).
The positive, and negative flux variation as a function of time
 calculated over the box is shown in Figure \ref{fig:mag_evo}(a).
On April 15 the flux is constantly emerging even with some cancellation around 16:00 UT, 
on April 16 the flux decreases due to cancellation.

For site 2, the enlarged version of magnetic field evolution is shown in Figure \ref{fig:mag2}. The emergence of  different  positive
and negative patches is shown by green and cyan arrows respectively.
 Around the site 2 location, we have observed  positive 
polarities surrounded by a kind of supergranule cell.
 Inside this positive
polarity supergranule, small bipoles with mainly negative polarities are 
continuously emerging. 
For the quantitative evolution of the positive, and negative magnetic flux at 
the site 2 location, we have also calculated the magnetic flux as a function of time inside the
red box (top, middle) of Figure \ref{fig:mag}. The variation of magnetic flux with time
 is shown in Figure \ref{fig:mag_evo}(b). We have noticed that the emergence of magnetic field
is continuously followed by the cancellation. At site 2, all the four jets are present on April 16 and we have drawn the onset and end time of these jets. The jet duration over site 2 is
denoted by the green dashed lines in  Figure \ref{fig:mag_evo}(b). We found that the jets from site 2
were lying in between the emergence and the cancellation site.


\begin{figure}[t!]
\centering
\includegraphics[width=0.80\textwidth,clip=]{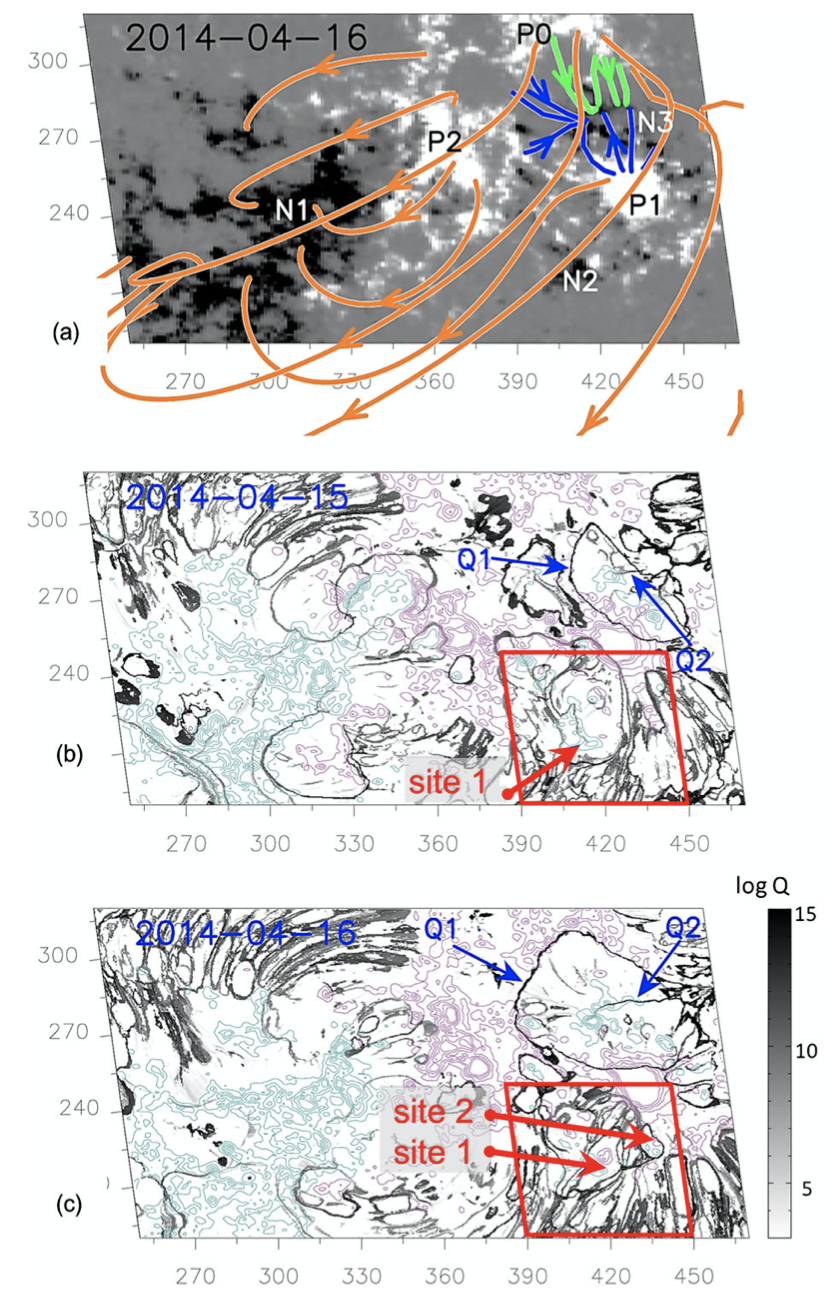}
\caption[Magnetic field  (LOS component) of AR 12035 for April 16, 2014.]
{Magnetic field configuration of AR 12035. Panel (a): two leading positive and following negative polarity. Panels (b-c):  contours of the magnetic field  overlaid with the footprints of the {\it QSL}.}
\label{fig:top}
\end{figure}
\section{Magnetic null points formation at the jet base}
 \label{mag_top}

The  longitudinal magnetic field  maps observed by HMI show a strong 
complexity of the polarity pattern and a fast evolution 
(Figures \ref{fig:mag}, \ref{fig:mag1}, \ref{fig:mag2}) in the two
 sites where the two series of jets are initiated. FRom the AIA observations, we have detected a shift of the jets  from site 2 to site 1 and never from site 1 to site 2.  
It is important to understand why there is a 
slippage of the magnetic field lines. Slippage reconnection has been 
observed in many flares (\citealt{Priest1992};  \citealt{Berlicki2004};  \citealt{Aulanier2005};  \citealt{Dudik2012}).
Commonly it is due to the slippage of magnetic field lines  anchored along 
{\it QSL} structures. The slippage occurs when and 
where the squashing degree is high enough along the {\it QSL} to force the 
reconnection. \citealt{Demoulin1996} and \citealt{Demoulin1998} has shown, 
theoretically and 
observationally, how it can be produced. \citealt{Dalmasse2015}  demonstrated that the \textit{QSLs}
 are robust structures and can be computed in potential configurations. 
Qualitatively the results are very good with this approach
and there is a relatively good fit between the location of \textit{QSL} footprints
 with the observed flare ribbons (\citealt{Aulanier2005}). However when the
 magnetic configuration is  too complex, the \textit{QSL} footprints do not fit 
perfectly and a one to one comparison is  increasingly difficult with smaller and smaller scale polarities
(\citealt{Dalmasse2015}).   
The benefit of using linear force--free field (LFFF)  extrapolation can be weak since {\it QSLs} are
 robust to parameter changes.  In the present case, LFFF extrapolation would perhaps help 
to follow the path of the jets which shows some curvature at their  bases  
(mentioned as a sigmoidal shape) but at the expense of the geometry of loops at large heights.
However, the magnetic field strength  is really fragmented in the jet regions. 
Pre-processing the data would smear  electric currents in their moving weak  polarities,  so it   will
  not help to derive better {\it QSLs}.
Therefore to investigate the magnetic topology of the jet producing regions, 
we use the same
 potential magnetic field extrapolation of AR 12035 calculated in \citealt{Zuccarello2017}. 
This method is based on the fast Fourier transform method of \citealt{Alissandrakis1981} and
 the extrapolation is performed by using a large field-of-view that includes at its center the AR 12035. 
This allows us to identify the key topological structures of the AR (Figure~\ref{fig:top}).
Maps of the squashing degree {\em Q} at the photospheric plane were calculated using the topology tracing code 
topotr (\citealt{Priest1995}, \citealt{Demoulin1996}, \citealt{Pariat2012} for more details).
The locations of the largest values of {\em Q} define the footprints of the {\it QSLs} 
(\citealt{Demoulin1996};  \citealt{Aulanier2005}) and they correspond to regions where electric current layers can easily develop. We have found two {\it QSLs}: the first one (Q1 in Figure~\ref{fig:top} (b) and (c)) 
encircling the positive polarity P1 and separating the magnetic flux system from the external field and a 
second one (Q2 in Figure~\ref{fig:top} (b) and \ref{fig:top} (c)) highlighting
the spine of a high-altitude coronal null-point similarly to what is seen in \citealt{Masson2009}. 
The flares occurred mainly at the north-west edge of the large {\it QSL} (Q1). Since the fan-like {\it QSL} (Q1) encircling the flare region was separated from the complex {\it QSL} system 
around the jet producing region (at the south of P1), we have argued that the jet activity and the 
flares were not really linked to each other, even if their timings seem to be related (\citealt{Zuccarello2017}).

\begin{figure}[t!]
\includegraphics[width=1.00\textwidth,clip=]{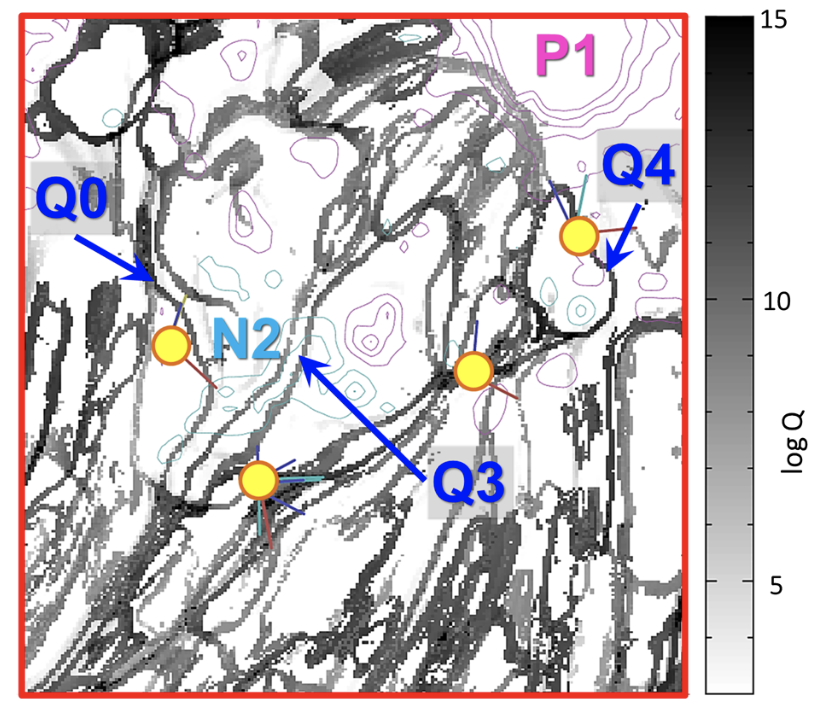}
\caption[Zoom of the jet region 
 inside  the red box  of April 16.]
 {Zoom of the jet region inside  the red box  of April 16 (panel (c) in Figure \ref{fig:top}). Q1 and  Q2 indicate the  {\it QSLs} related to the flares, Q3 and Q4 to the site 1 and  site 2.}
\label{fig:null}
\end{figure}

On April 15 the jets are initiated in site 1 and we find  a well defined {\it QSL} surrounding the region of the jet.
 On April 16 the configuration is much more complicated with 
many {\it QSLs}  which are in the site region  of both jets.  The zoom of the {\em Q} map of April 16
 around the jet producing region (red box of Figure~\ref{fig:top} (c)) is shown in Figure~\ref{fig:null}. 
We find that the two sites of the jets, site 1 and site 2, are respectively inside the {\it QSLs} Q3 and Q4, 
and that both were embedded in a larger {\it QSL}, labeled as Q0 in the figure. 
We also identified several quasi photospheric null points, that are indicated by yellow circles. 
The {\it QSL} map is very complex, and difficult to analyze and compare in detail with the observations due to 
the small scale of the events and fast motion of the small polarities, both the moving p polarity in 
site 1 and the emergence of small bipoles in site 2.
Since it looks quite possible that the 2 {\it QSLs} (Q3 and Q4) intersect or touch each other, 
we conjecture that field line foot points could move from site 2 and site 1 by a sequence of 
reconnections across {\it QSLs} as in \citealt{Dalmasse2015}. 
This could produce the transfer/or tendency of movement of jets from site 2 to site 1, as we have
 observed for jets J$_5$ and J\ensuremath{'}${_5}$ for example (Figure~\ref{fig:jet5}). In the case of the
 other jets from site 2, they also show a tendency of slippage towards site 1.

\section{Results and conclusion}
     \label{result_c3}

The study of the transition from eruptive to confined flares and the recurrent solar jets in active region NOAA 12035 is done in this chapter. The transition occurred between 2014 April 15 and April 16. On April 15, four of the 13 flares observed resulted in a CME, while none of the 13 flares recorded on April 16 resulted in a measurable CME. The slippage of the jets from one reconnection site to other is explained by the complex topology of the region and the intersection of many QSLs. The main results of this study are as follows:

Eruptive to confined behavior of the flares can be attributed to the change of orientation of the magnetic field below the fan with respect to the orientation of the overlaying spine, rather than an overall change in the stability of the large scale field. We found that a closed fan-like QSL exists around the location of the filament on both days. The presence of circular, closed fan QSLs indicate the presence of a (quasi-)separator in the corona. The presence of a null-point topology in the corona, the presence of shear motions that reduced the mutual inclination between the two flux systems achieving a configuration less favorable for reconnection, as well as the non significant change in the theoretical stability (with respect to the torus instability scenario), leads us to the conclusion that the breakout scenario seems the more probable scenario to describe the observed behavior. The discerning element between fully and failed eruption behavior being determined by the mutual inclination of the flux systems involved in the process, namely the erupting flux and the overlying field.

We found two sites for the different jet's activity. 
On April 15 the jets originated from site 1 and we measure  a large increase of emerging flux and small cancellation. On April 16 site 1 
and site 2 are  associated with continuous  emerging magnetic flux followed by  cancellation at the jet time. The kinematics of jets at different  EUV wavebands revealed that the speeds, 
widths, heights and lifetimes of jets
are slightly different at different wavelengths. This can be interpreted as the multi-temperature and multi-velocity structure of
solar jets. In addition to this, most of the jets showed clockwise rotation, which indicates untwisting. As a result of this 
untwisting, the twist/helicity was injected 
in the upper solar atmosphere (\citealt{Pariat2015}).
The injected helicity in the jets may
be  part of the global emergence of  twisted magnetic fields.
During the rotation like in the case of jet J${_1}$, we observed 
the rotating  jet material contains bright as
 well as dark material.
This result is consistent with simulations done by \citealt{Fang2014}.
In their simulation, they found the simulated jet consists 
of untwisted field lines, with a mixture of cold and hot plasma.

We observed the slippage of jets at site 2 namely
 J\ensuremath{'}${_3}$--J\ensuremath{'}${_6}$ towards the eastern site (site 1) and never the reverse movement.
Along with the movement of jets towards site 1, we found, in the case of 
jet J\ensuremath{'}${_5}$ that a part
 detached from it and moved towards the site 1 location and
finally merge into jet J${_5}$. On April 16 both  jet sites are associated with 
the {\it QSLs}. The possible intersection of the two
 {\it QSLs} encircling each site could explain the slip reconnection occurring along the
 {\it QSLs}  which favor the translation of jets from
 site 2 to  site 1. The coronal jets may be due to the eruption of mini filaments (\citealt{Sterling2015}; \citealt{Panesar2018}). According to this hypothesis,  the spire of the jets moves away from the jet base bright point. Our observation of the motion of the broken part of J\ensuremath{'}${_5}$ is away from the jet base. This supports the findings of \citealt{Savcheva2009} and the interpretation proposed by \citealt{Sterling2015}. We have observed the flux emergence followed by flux cancellation at site 1 on 15 April 2014. Moreover, on 16 April 2014, flux emergence and cancellation are recurrent in both jet sites. The observation of cool and hot material in our study supports the hypothesis of small filament eruption and a universal mechanism for eruptions at different scales (\citealt{Sterling2015}; \citealt{Wyper2017}). The observations with high spatial resolution instruments, {\it e.g.} IRIS, will be very useful to explain the jet mechanism with the another scenario of magnetic flux emergence MHD models.
\chapter{Multi temperature coronal jets for emerging flux MHD models}
\label{c4}
\ifpdf
    \graphicspath{{Chapter4/Figs_chapter4/}{Chapter4/Figs/PDF/}}
\fi

\section{Introduction}
\begin{figure*}[ht!]
\centering
\includegraphics[width=\textwidth]{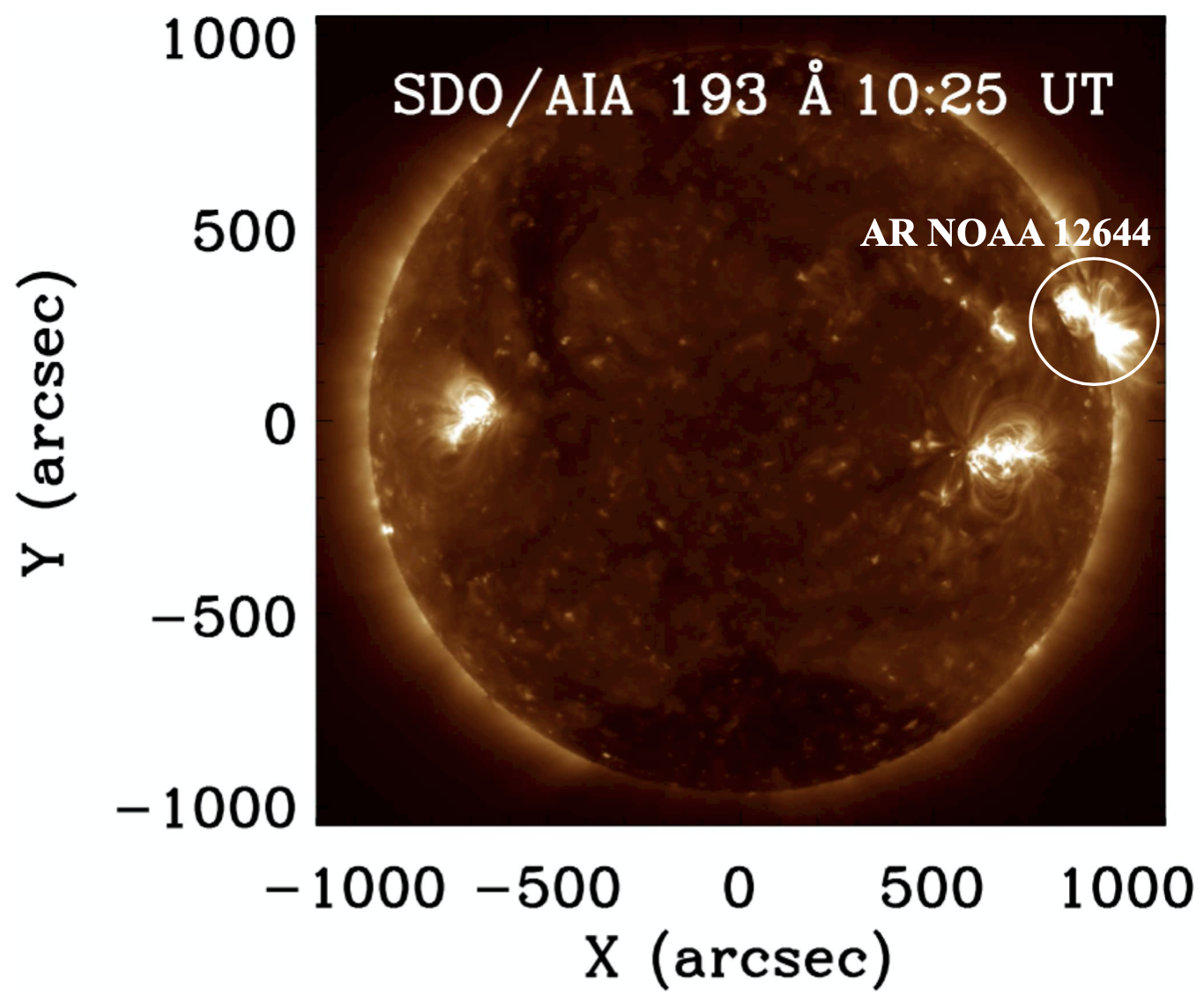}
\caption[Full disk image of the Sun on April 4, 2017.]{Full disk image of the Sun on April 4, 2017.
The solar jets were ejected
from the AR NOAA 12644 shown by the white circle at the west limb.}
\label{full}
\end{figure*}

Hot coronal jets are a basic observed feature of the solar atmosphere whose physical origin is still actively debated.
Here, we found  a series of jets observed 
 in the hot EUV channels of SDO/AIA as well as 
 in cool temperatures with IRIS slit--jaw images. The
jets were ejected from the 
AR NOAA 12644 on April 4, 2017; on that date, the region was
located at the west limb (N13W91) (Figure \ref{full}).
When passing through the central meridian, this region had shown high jet activity alongside episodes of emerging magnetic flux (\citealt{Ruan2019}). 
 The AIA filters allow us to study the temperature and the emission measure of the jets using the filter ratio method. We studied the pre-jet phases by analysing the  intensity oscillations at the base of the jets with the wavelet technique. A fine co-alignment of the AIA and IRIS data shows that the jets are initiated at the top of a canopy-like double-chambered structure with cool emission on one and hot emission on the other side. The hot jets are collimated in the hot temperature filters, have high velocities (around 250 km s$^{-1}$) and are accompanied by the cool surges and ejected kernels that both move at about 45 km s$^{-1}$. In the pre-phase of the jets,  we find quasi-periodic intensity oscillations at their base that are in phase with small ejections; they have a period of between 2 and 6 minutes, and are reminiscent of acoustic or MHD waves.
 This series of jets and surges provides a good case study for test the 2D and 3D MHD models that result from magnetic flux emergence. The double--chambered structure found in the observations corresponds to the cold and hot loop regions found in the models beneath the current sheet that contains the reconnection site. The cool surge with kernels is comparable with the cool  ejection and plasmoids that naturally appears in the models. The location of the AR at the limb in the present observations allows us to visualize the structure of the brightenings from the side and thus facilitates the comparison with the MHD jet models, which motivates for this present study.

We present the multi wavelength EUV observations and the pre-jet oscillations and conclude that this series of jet and surge observations obtained with a high spatial and temporal resolution match important aspects of the expected behaviour predicted by the
 MHD models of emerging flux. We could identify a candidate location for the current sheet and reconnection site and follow the evolution of the cool surge and hot jets with individual blob ejections. This is a clear case-study for the emerging flux MHD jet models (\citealt{Joshi2020MHD}). 
 
\section{Multi-temperature coronal jets}\label{observation}
\subsection{Multi-instrument observations}
In this study, we select six jet eruptions occurring  in the AR NOAA 12644
at the western  solar limb on April 4, 2017.
 We use data from  the AIA and IRIS instruments.
The IRIS target was pointed towards the 
AR NOAA 12644 at the western limb with a field of view
of 126$\arcsec$ x 119$\arcsec$ between  11:05:38 UT and  17:58:35 UT. 
For our current study, we  use the SJIs in the C II 
and 
Mg II k bandpasses obtained  with a cadence of 16s.
 The SJIs picture the chromospheric plasma around 10$^4$ K.

\begin{figure}[ht!]
\centering
\includegraphics[width=0.9\textwidth]{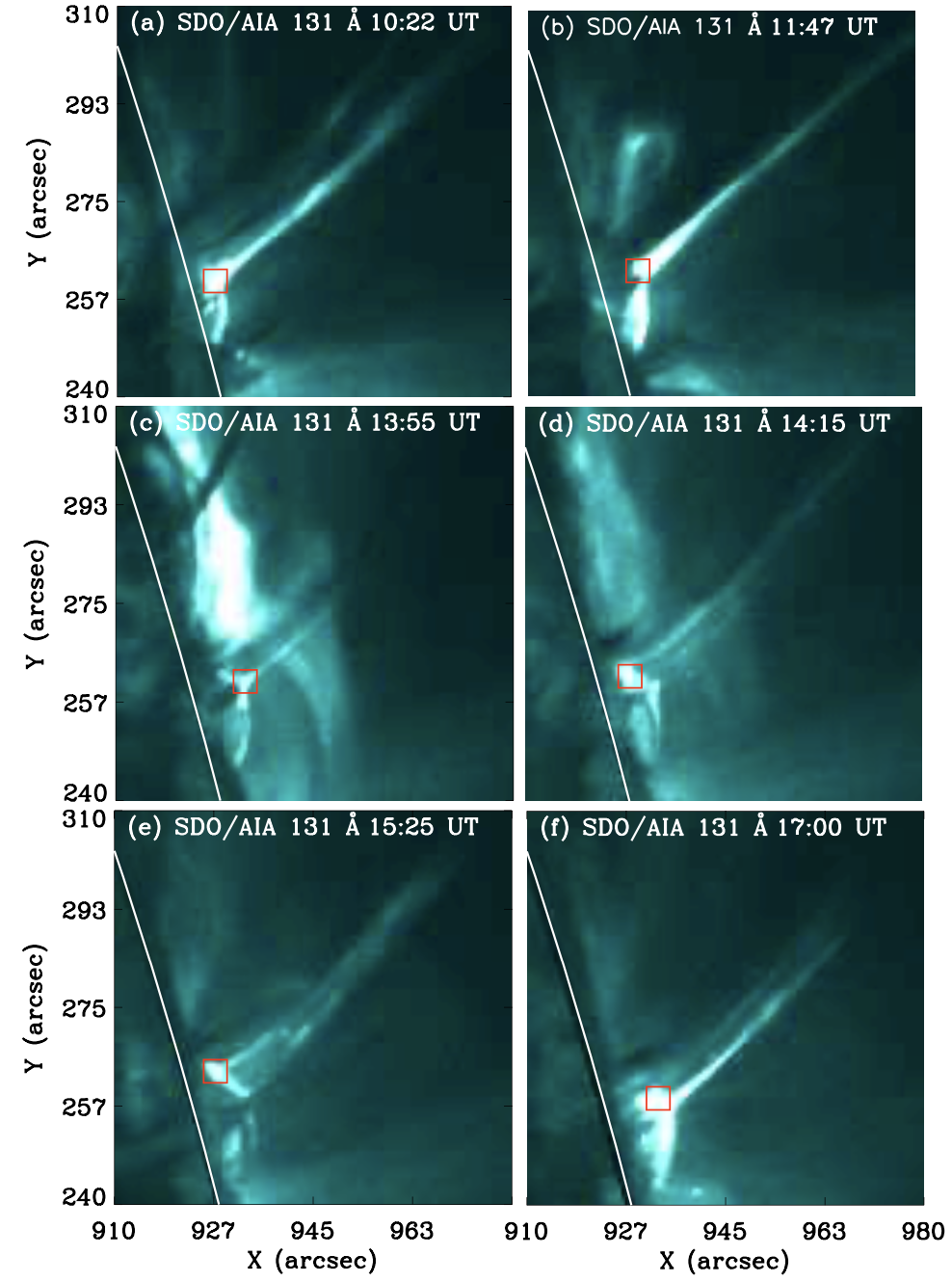}
\caption[Six solar jets ({\it Jet1}--{\it Jet6}) in AIA 131 \AA\ 
filter.]{Six solar jets ({\it Jet1}--{\it Jet6}) in AIA 131 \AA\ 
filter. The red square in each panel
shows the position at which the pre--jet oscillations are measured.}
\label{evolution}
\end{figure}

\subsection{Characteristics of the coronal jets}\label{sec:characteristics}
\label{morpho}
On  April 4, 2017, AR jets were 
observed at the limb between 02:30--17:10 UT with AIA. The observations in different wavelengths of AIA  (131 \AA, 171 \AA, and 304 \AA) reveal that there are two sites of plasma ejections (jets) along the  limb. First, there is a northern site (921$\arcsec$, 264$\arcsec$), where the jets are straight and have their base located behind the limb. Second, there is a site in the south of the field of view (931$\arcsec$, 255$\arcsec$) in which the jets have their base over the limb.
Therefore we study in the present chapter the six main jets originated in the southern site occurring after 10:00 UT. Five of them were also observed by IRIS, whereas the first of them occurred before the IRIS observations. 
These jets reach an altitude
between 30 and 70 Mm; their recurrence period is around 80 minutes, with the exception of two jets which were separated by only 15 minutes. We also noticed many small jets reaching less than 10 Mm height both before and in between the main jets. 
The observed jets in AIA 131 \AA\ are shown in 
Figure \ref{evolution} (a--f). 
 The first main jet, {\it Jet1}, reaches its peak at $\approx$ 10:22  UT 
 with  an average speed of 210 km s$^{-1}$ (panel a). {\it Jet2} (panel b) starts at 11:45 UT and reaches its maximum extent at $\approx$ 11:47 UT. 
 We note  a large 
 filament eruption located in the northern site of the jets which erupts $\approx$ 13:30 UT and falls back after reaching its height. 
Moreover, we could see that the jet and the filament  are not associated with each other. 

{\it Jet3} and {\it Jet4} (panel b and c, respectively) reach  their maximum altitude 
at 13:55 UT and 14:15 UT respectively.
{\it Jet5} (panel e) erupts 
with a broader base and 
 reaches its maximum height at $\approx$ 15:25 UT.
We see a fast  lateral extension
 of the jet base along a bright loop.
{\it Jet6} (panel f) is  ejected at $\approx$ 16:57 UT.  A second instance of filament eruption is observed during the peak phase of {\it Jet6} starting again at the same 
 location of the first one.
 In this case the erupted filament material seems to merge  later
 with the jet material and 
is ejected in the same direction.
However, here the jet is not launched by the filament eruption, because it is not at the jet footpoint. 
\begin{figure*}[t!]
\centering
\includegraphics[width=1.0\textwidth]{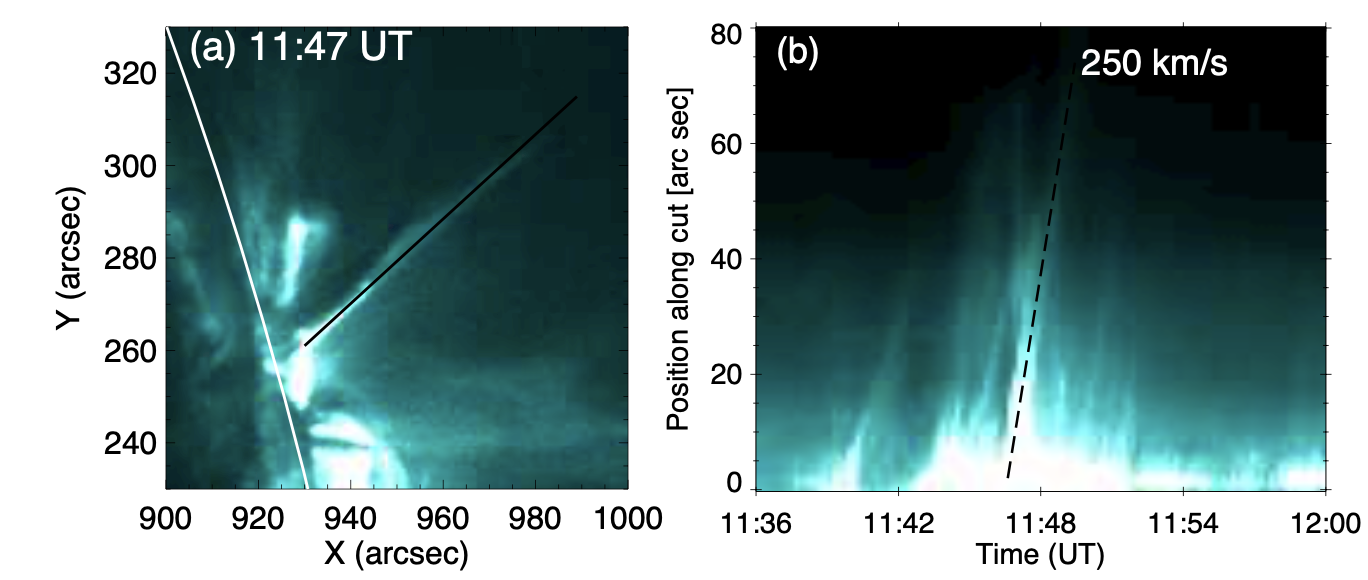}
\caption[An example of timeslice analysis of the jet 1, used
for velocity and height calculations in AIA 131 \AA. ]{An example of timeslice analysis of the jet 1, used
for velocity and height calculations in AIA 131 \AA. In panel (a) the solid black line is the slit location, which we use to make the height--time.}
\label{timeslice}
\end{figure*}

\begin{figure*}[t!]
\centering
\includegraphics[width=1.00\textwidth]{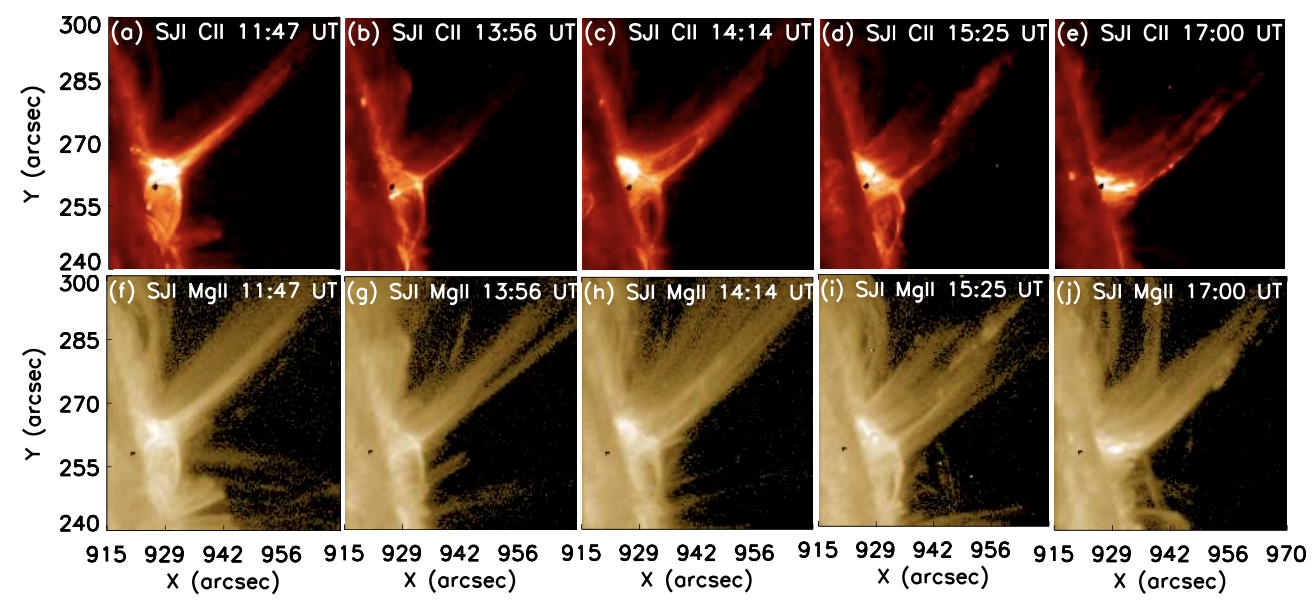}
\caption[IRIS observations of the AR from 11:05 UT to 17:58 UT.]{IRIS observation of the the AR from 11:05 UT to 17:58 UT.
 It covers five jets in our present analysis in CII (top) and MgII k (bottom)
 lines.}
\label{iris}
\end{figure*}

We have computed various physical parameters, namely, height, width, lifetime, speed
 of these jets using the AIA~131~\AA\ data. 
 For the velocity calculation, 
we calibrated
height--time 
of each jet in AIA~131~\AA\ fixing a slit in the middle of the jet plasma flow and calculating the average speed in the flow direction. 
An example of height--time calculation is shown in Figure \ref{timeslice} for {\it Jet2}. 
All computed physical parameters  are listed in Table \ref{table1}.
The maximum height, average speed, width, and lifetime of the observed 
jets vary in the ranges $\approx$ 30--80 Mm, 200--270 km s$^{-1}$, 1--7 Mm, and 
2--10 minutes respectively.

{\it Jet2}--{\it Jet6} were also observed by IRIS in two wavelength, namely, CII (top row of Figure \ref{iris}) and MgII k (bottom row).
The high spatial resolution of IRIS allowed us to make a clear identification of what looks like a null--point structure at a 
height of $\approx$ 6 Mm. In the CII filter we see bright loops above a bright half dome in the northern site of the jet footpoints. 
In the Mg II filter, the northern part of the dome is also bright.
 We  find jet strands all over the northern side of the dome, like a collection of sheets. We will discuss about these jet strands, which are infact cool jets/surges with a lower velocity in section \ref{theory}.
In AIA 131 \AA\ we see clearly, for all the jets, a  bright area which could correspond to a current sheet (CS), possibly containing a null point, with underlying bright loops shaping a dome (Figure \ref{evolution}). However we notice that the bright dome and loops are located on the southern side of the tentative current sheet, whereas the bright loops  in IRIS C II are rather on its northern side. In the following, for simplicity, when referring to observations of this candidate current sheet and possible null point we will sometimes call them `the null point' even though there is clearly no way in which one could detect a zero of the magnetic field (nor the intensity of the electric current) in those temperatures with present observational means.
\begin{figure*}[t!]
\centering
\includegraphics[width=1.0\textwidth]{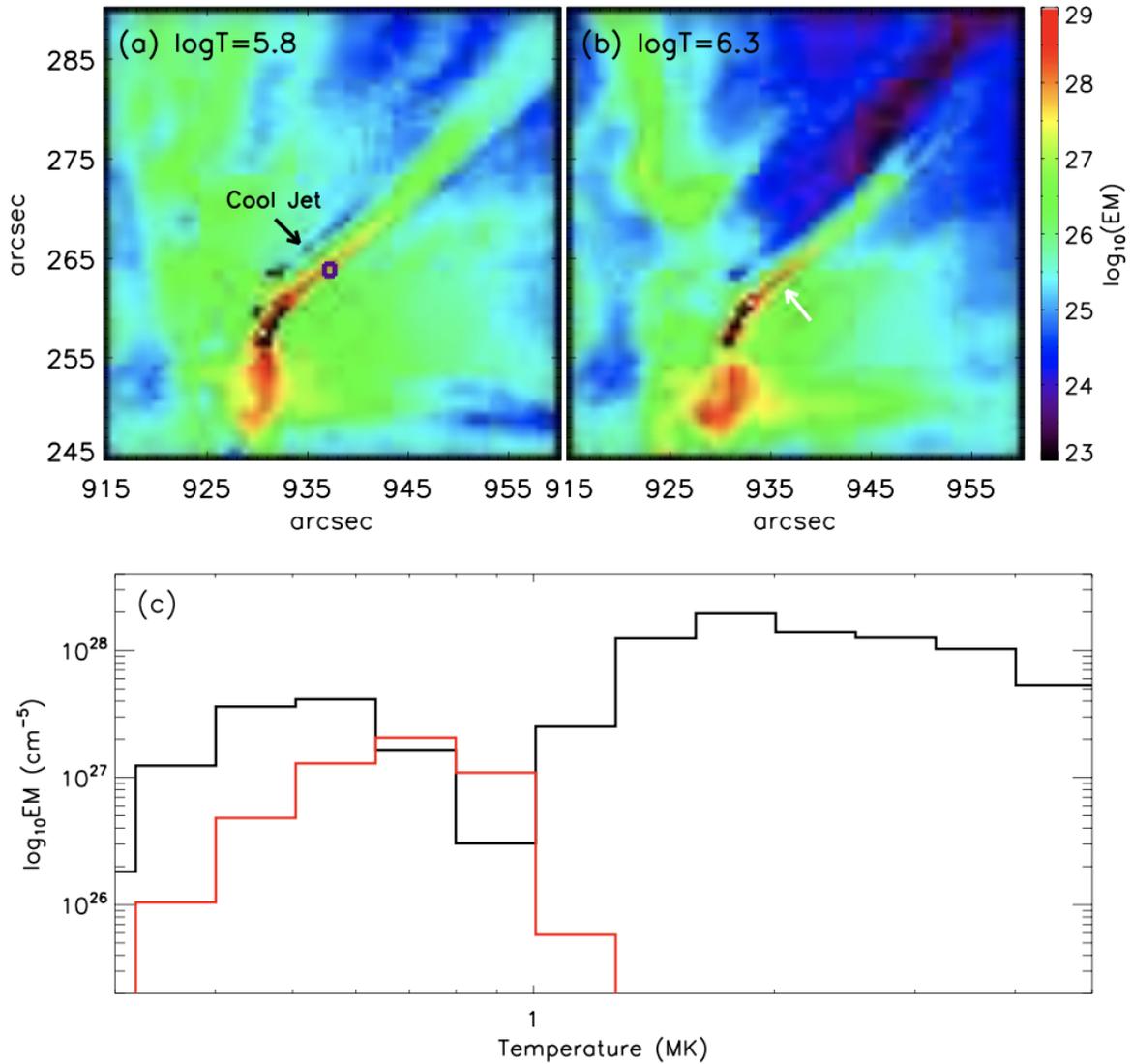}
\caption[Temperature and emission measure of {\it Jet2}.]{Top: Two maps of the AR {\it Jet2}, at two different temperatures (log T= 5.8 and 6.3). Bottom: In panel (c) the red line shows the temperature at
the location of the  blue box in panel (a) before the first jet ejection on April 4, 2017 and the black line shows the temperature of solar AR jet at
11:45 UT.}
\label{emission}
\end{figure*}
Moreover  in  all the hot channels of AIA (131 \AA\, 193 \AA, 171 \AA, 211  \AA) 
and IRIS  C II and Mg II SJIs 
the jets have an anemone (``Eiffel--Tower'' or ``inverted--Y'') structure, with a loop at the base and elongated jet arms (Figures \ref{evolution} and \ref{iris})
as reported in previous events (\citealt{Nistico2009};  \citealt{Schmieder2013};  \citealt{Liu2016}).

In AIA 131 \AA\ we could also see that between the first and the last jet eruption, the tentative current sheet and the jet spine move towards the south--west direction (Figure \ref{evolution}). More precisely, by following the motion of the point with maximum intensity, we determined a drift of 5 $\arcsec$ in less than 6 hours.

\subsection{Temperature and emission measure analysis}\label{DEM}
We have investigated the distribution of the temperature and emission measure (EM)  at the jet spire for all jet events. 
We performed the differential emission measure (DEM)
analysis with the regularized inversion method
introduced by \cite{Hannah2012} using six
AIA channels (94 \AA, 131 \AA,
171 \AA, 193 \AA, 211 \AA, and 335 \AA).
After this process we find the regularized DEM maps as a function of temperature.
We use a temperature range from 
log T(K) = 5.5 to 7 with 15 different bins of width $\Delta$ log T = 0.1. 
 We calculated the EM and lower limit of electron density in the jet spire 
using n$_e$ = $\sqrt{EM/h}$, with $h$ the jet width, assuming that the filling factor equals unity. 
These EM values were obtained by integrating the DEM values over the temperature range
log T(K) = 5.8 to 6.7.
We chose a square box to measure the EM and density 
at the jet spire and at the same location before the jet activity
for each jet.
The example for DEM analysis of 
{\it Jet2} is presented in Figure \ref{emission}, which represents the DEM maps 
at two different temperatures, namely log T (K) = 5.8 (panel a)
and 6.3 (panel b), at 11:45 UT.
We investigate the temperature variation at the jet spire 
during the jet and pre--jet phase. 
 During the  pre--jet phase for {\it Jet2} 
the log EM and the electron density values were 27.3 and 2 x 10$^9$ cm$^{-3}$, whereas for the jet phase the values were 28.1 and 8.6 x 10$^9$ cm$^{-3}$ respectively.
Thus, during the jet evolution the EM value increased by over one order of magnitude and the electron density increased by a factor three at the jet spire. 
We find that the EM and density values increased during the jet phase
in all six jets. The values for all jets are listed in Table \ref{table1}.

\begin{figure*}[t!]
\centering
\includegraphics[width=1.0\textwidth]{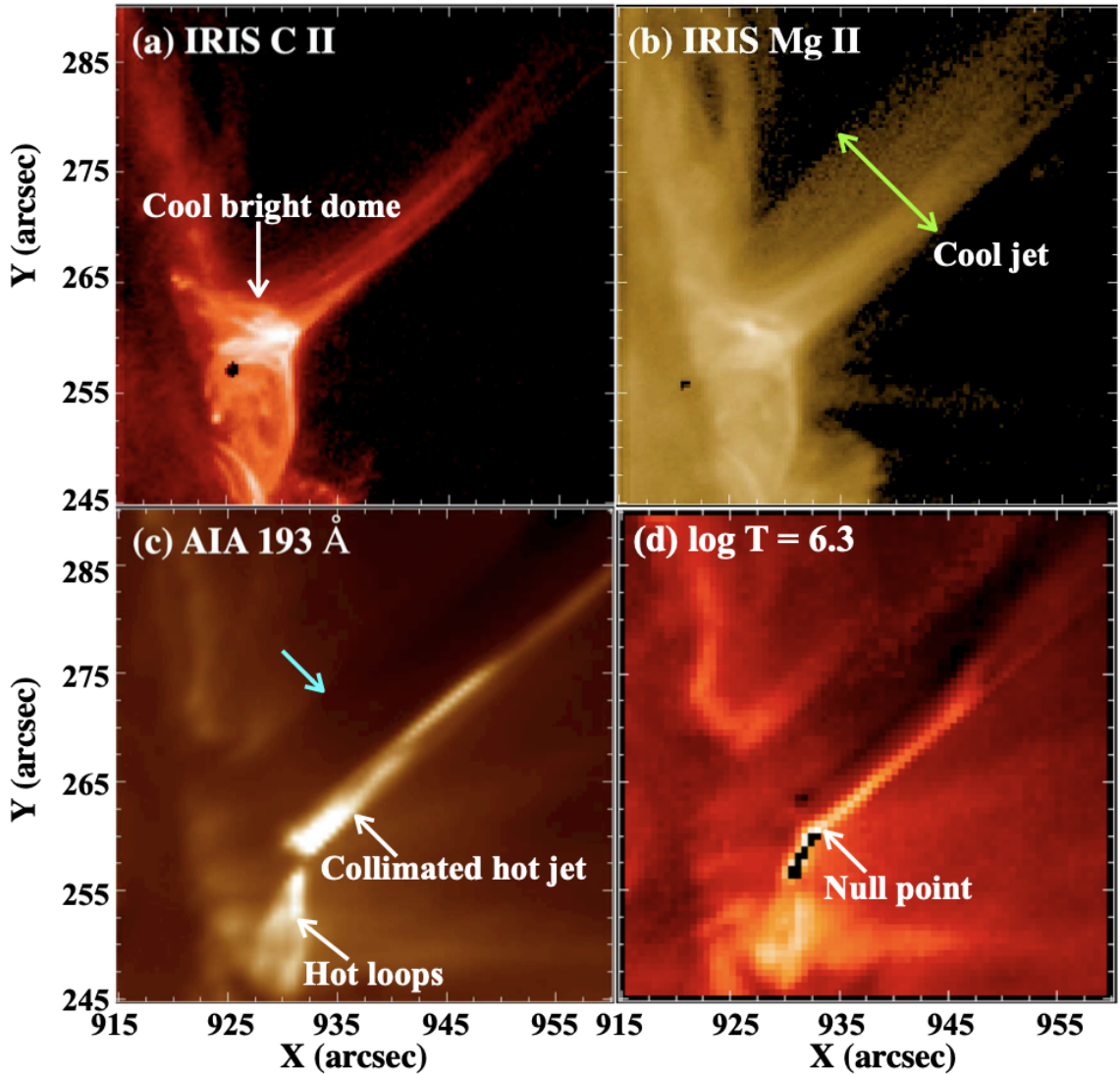}
\caption[Example of {\it Jet2} at 11:45 UT 
observed with IRIS in panel a,b and AIA 193 \AA\ in panel c.]
{Example of {\it Jet2} at 11:45 UT observed with IRIS in panel a,b and AIA 193 \AA\ in panel c. Cool bright dome in the northern side of the null point is shown with a white arrow in panel a.}
\label{loop}
\end{figure*}
\begin{table}[h!]
  \caption{Physical parameters of six studied hot jets from AR NOAA 12644.}
\bigskip
\label{table1}
\setlength{\tabcolsep}{9.5pt}
\begin{tabular}{cccccccc}
      \hline
      \text{Jet} & \text{Jet start}& \text{Jet peak} & \text{Max} 
& \text{Average} &\text{T} &\text{EM}&
  \text{Oscillation}\\
      \text{no.} & \text{time} & \text{time} & \text{height} 
& \text{speed} &\text{ } &\text{(10$^{28}$}&
  \text{period}\\
  \text{ } & \text{(UT)} & \text{(UT)} & \text{(Mm)} 
& \text{(km s$^{-1}$)} &\text{(MK)} &\text{ cm$^{-5})$}&
  \text{(min)}\\
      \hline
       
       1 & 10:15 & 10:22 & 80 & 210 & 1.4 & 1.4 & 6.0\\
       2 & 11:46 & 11:47 & 50 & 245 & 1.8 & 1.9 & 1.5\\
       3 & 13:54 & 13:55 & 40 & 265 & 1.4 & 1.5 & 2.5\\
       4 & 14:12 & 14:15 & 50 & 250 & 1.8 & 1.1 & 2.0\\
       5 & 15:23 & 15:25 & 55 & 235 & 1.8 & 1.3 & 4.0\\
       6 & 16:57 & 17:00 & 70 & 220 & 1.8 & 2.0 & 2.5\\
      \hline
    \end{tabular}
\end{table}
\subsection{ Identification of observed structural elements}\label{theory}
In Section \ref{morpho} we have discussed the morphology of the jets observed with AIA and IRIS. The region below the jet, as seen in different wavelengths, has a remarkably clear structure, resembling those discussed in theoretical models of the past years. For identification with previous theoretical work, in Figure \ref{loop} several structural elements are indicated for the case of the {\it Jet2} observations.  
 In IRIS C II (Figure \ref{loop}, panel a) the brightenings below the jet delineate a double--chambered vault structure, with the main brightening being located in  the northern part 
of the base of the jet.  Only narrow
loops are seen above the southern part of the vault in this wavelength. In the other chromospheric line, IRIS Mg II, we see (panel b) roughly the same scenario, although the general picture is rather fuzzier. The jet, in particular, is no longer narrow but formed by parallel strands issuing from the edge of the northern part of the vault, similar to a comb (Figure \ref{loop} panel b). 

\begin{figure*}[ht!]
\centering
\includegraphics[width=1.0\textwidth]{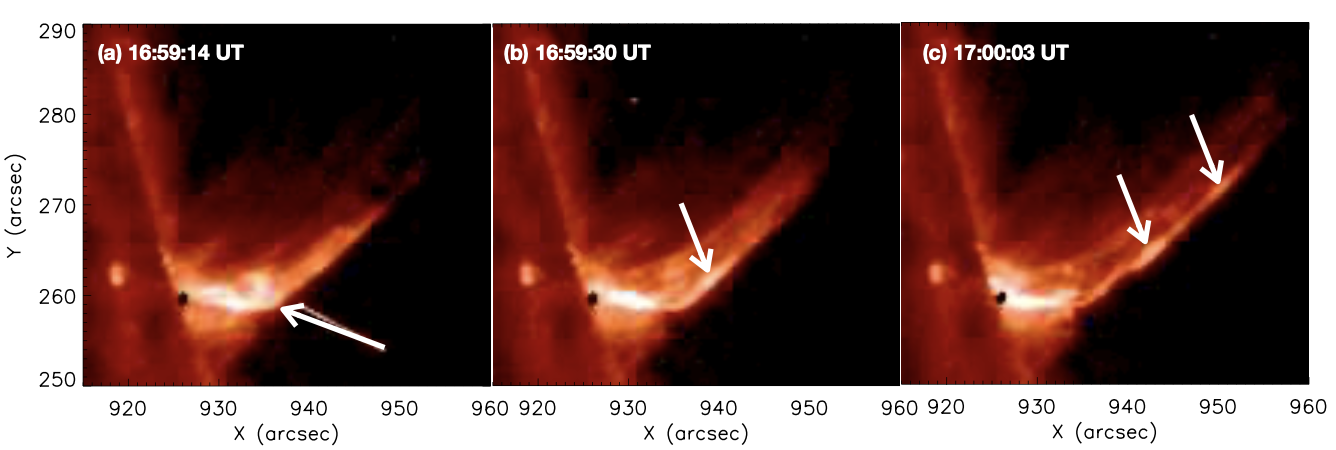}
\caption[Kernels of brightening moving along the {\it Jet6} observed in the IRIS SJI in CII wavelength range.]{Kernels of brightening moving along the {\it Jet6} observed in the IRIS SJI in CII wavelength range (white arrows). The kernels could correspond to untwisted plasmoids.}
\label{kernel}
\end{figure*}

 The assumption of a double-vault structure below the jet is reinforced when checking both the hot-plasma observations (AIA~193~\AA, panel c) and the temperature map obtained through the DEM analysis explained in the previous section (panel d). In those two panels, the southern loops are shown to be bright and hot structures, and the same applies to the point right at the base of the jet, where the temperature reaches $10^6$ K. Additionally, we observe bright kernels moving from time to time along the jets and more clearly visible in {\it Jet4}, {\it Jet5}, and {\it Jet6}. An example of kernels of brightening moving along the {\it Jet6} in IRIS CII is presented in Figure \ref{kernel}. We have computed the velocities of the kernels and find that they are comparable to the mean velocities of the cool jet. The time between  the ejection of two kernels is less than 2 minutes.

  The foregoing structural elements seem to correspond to various prominent features in the numerical 3D models of \citealt{Moreno2008};
  \citealt{Moreno2013}, or in the more recent 2D models of \citealt{Nobrega2016, Nobrega2018}, all of which study in detail the consequences in the atmosphere of the emergence of magnetized plasma from below the photosphere. One can identify the bright and hot plasma apparent in the observations at the base of the jet with the null point and CS structures resulting in those simulations (the scheme in Figure~\ref{null}, right panel): the collision of the emerging magnetized plasma with the preexisting coronal magnetic system leads, when the mutual orientation of the magnetic field is sufficiently different, to the formation of an elongated CS harboring a null point and to reconnection.
As a next step in the pattern identification, the hot plasma loops apparent in the southern vault in the AIA~193~\AA~image and the temperature panels of Figure~\ref{loop} should correspond to the hot post-reconnection loop system in the numerical models (as apparent in Figures~3 and 4 of the paper by \citealt{Moreno2008}, or along the paper by \citealt{Moreno2013}. 
On the other hand, the northern vault appears dark in AIA~193~\AA, and has lower temperatures in the DEM analysis. This region could then correspond to the emerged plasma vault underlying the CS in the numerical models: the magnetized plasma in that region is gradually brought toward the CS where the magnetic field is reconnected with the coronal field. Additional features in the observation that fit in the foregoing identification are:

\begin{figure*}[ht!]
\centering
\includegraphics[width=1.0\textwidth]{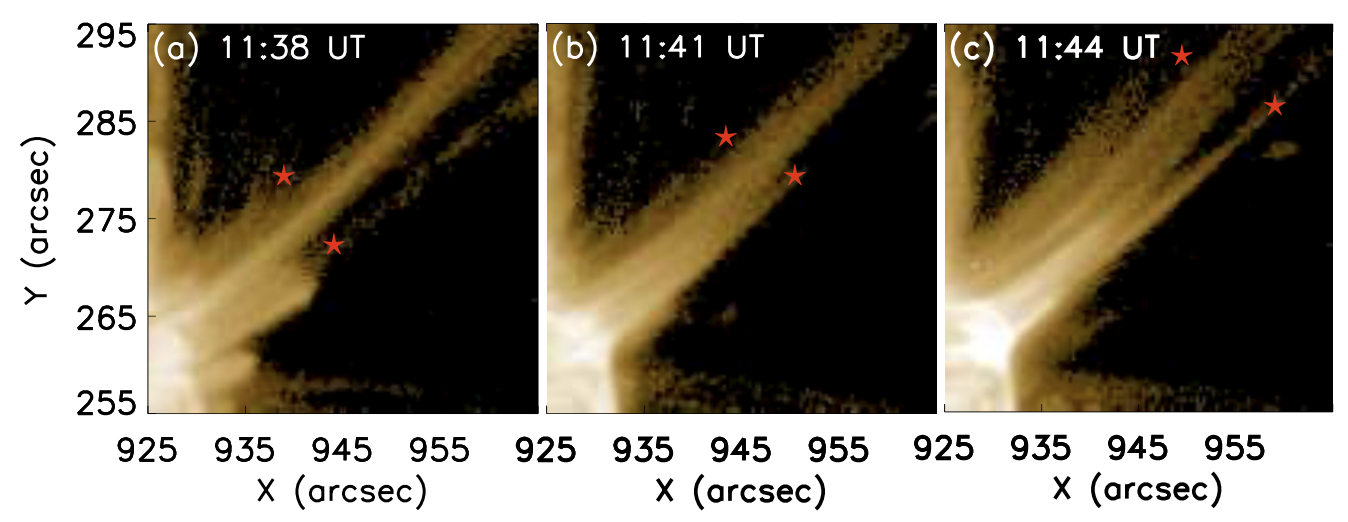}
\caption[The evolution of cool plasma material along both sides of the hot jet.]{
The evolution of cool plasma material along both sides of the hot jet ({\it Jet2}) in IRIS MgII wavelength. The red star shows the leading edge of the cool material ejecting with an average speed of 45 km s$^{-1}$.}
\label{cool}
\end{figure*}

\begin{figure*}[ht!]
\centering
\includegraphics[width=1.0\textwidth]{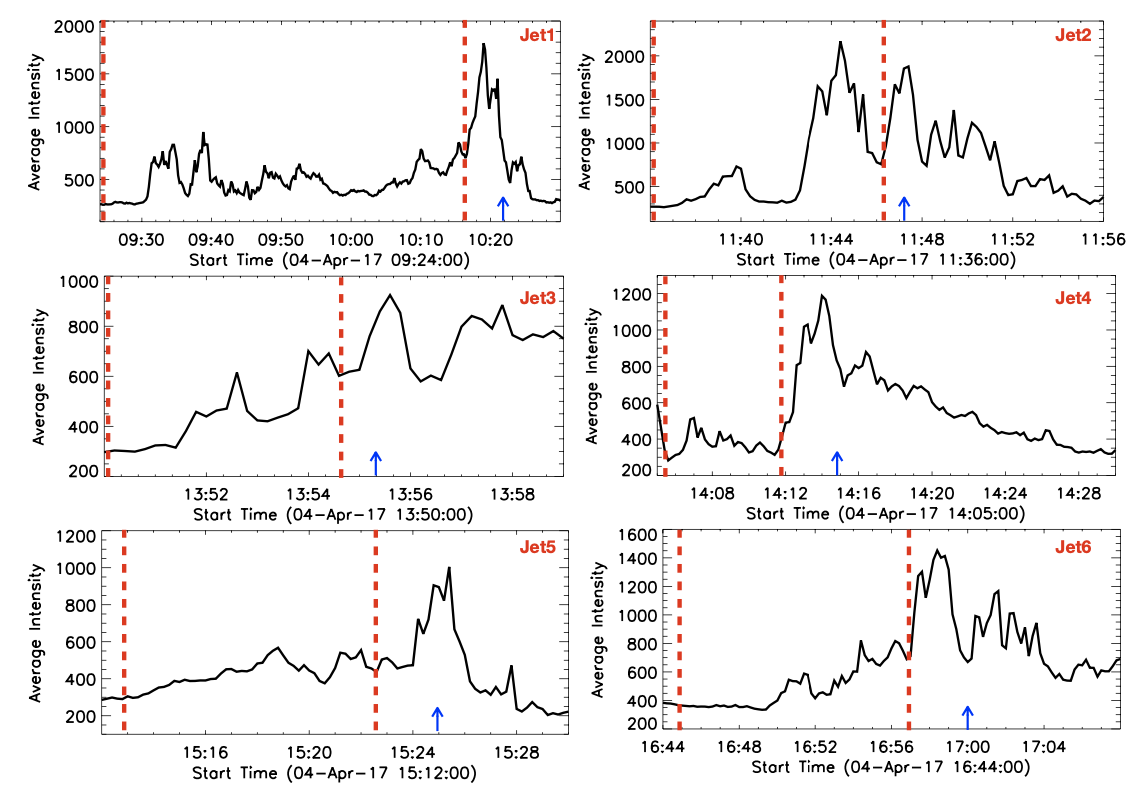}
\caption[Intensity distribution during pre--phase of recurrent jets at the base of each jet in AIA 131 \AA.]{Intensity distribution during pre--phase of recurrent jets at the base of each jet in AIA 131 \AA.
The location of the base of the each jet 
is displayed in Figure \ref{evolution} (red squares).
}
\label{intensity}
\end{figure*}

\begin{enumerate}
    
    \item 
As time proceeds the northern chamber decreases in size while the southern chamber grows. In our observations in the beginning phase of the jets (for instance; {\it jet2} at 11:30 UT) the area of the northern and southern vaults is 1.4 x 10$^{18}$ and  1.16 x 10$^{18}$ cm$^{2}$, respectively, and during the jet phase (11:47 UT), they become
 $1.05 \times 10^{18}$ and $2.2 \times 10^{18}$ cm$^{2}$, respectively.
 This suggests that while the reconnection is occurring, the emerging volume is decreasing whereas the reconnected loop domain grows in size, as in the emerging flux models (\citealt{Moreno2008}; \citealt{Moreno2013}; \citealt{Nobrega2016}).

\item
A major item for the identification of the observation with the flux emergence models is the possibility 
that we also observe a wide, cool and dense plasma surge ejected in the neighborhood of the vault and jet complex.
This  wide laminar jet is observedin the Mg~II  IRIS filter  as an absorption sheet parallel to the hot jet in AIA 193 \AA. The evolution of the cool material along both sides of the hot jet in the IRIS Mg II channel is presented in Figure~\ref{cool} and the leading edge of the cool part is indicated by red stars. 
The cool ejection is generally less collimated than the hot jet and is seen to first rise and then fall, similarly to classical H$_\alpha$ surges.
The velocities measured along the cool sheet of plasma in Mg II are $\approx$ 45 km s$^{-1}$.  
 The ejection of cool material next to the hot jets is a robust feature in different flux emergence models (\citealt{Yokoyama1996}; \citealt{Moreno2008}; \citealt{Nishizuka2008}; \citealt{Moreno2013}; \citealt{MacTaggart2015}; \citealt{Nobrega2017, Nobrega2018}). The cool plasma in the models is constituted by matter that has gone over from the emerged plasma domain to the system of reconnected open coronal field lines without passing near the reconnection site, that is, just by flowing, because of flux freezing, alongside the magnetic lines that are being reconnected at a higher level in the corona. All those models report velocities which match very well the observed value quoted above.
 
\item 
The observed kernels in Figure \ref{kernel} could be plasmoids  created in the CS during the reconnection process. In some of the flux emergence models just discussed, plasmoids are created in the CS domain (see, for example \citealt{Moreno2013}), and they are hurled out of the sheet probably via the melon-seed instability (\citealt{Nobrega2016}), even though they are not seen to reach the jet region. 
Observational evidences of the formation of plasmoids in this kind of scenario
have been found by \citealt{Rouppe2017}. On the other hand,  in the 2D jet model by \citealt{Ni2017}, plasmoids are created in the reconnection site that maintain their identity when rising along the jet spire, possibly because of the higher resolution afforded by the Advanced Mesh Refinement used in the model; this is in agreement with the behavior noted in the present observations as well as in the previous observations of \citealt{Zhang_Ji_2014} and \citealt{Zhang2016} mentioned in the introduction. Plasmoids are also generated in the model by \citealt{Wyper2016}, which is 
a result of footpoint driving of the coronal field rather than flux emergence from the interior.
On the other hand, the formation of the kernels could follow the development of the Kelvin–Helmholtz instability (KHI). The KHI can be produced when two neighboring fluids  flow in same direction with different speed (\citealt{Chandrasekhar1961}). This instability may develop following the shear between the jet and its surroundings (\citealt{Zhelyazkov2019}).

\item
The main brightening at the top of the two vaults seems to be changing position systematically in the observations.
There is a shift in the south--west direction as time advances, and the same displacement is apparent in AIA 131 \AA\ (Figures \ref{evolution} and \ref{iris}), possibly marking the motion of the reconnection site. Such type of observations are also reported in the study of \citealt{Filippov2009}. This shift may be used to compare with the drift of the null point position detected in the MHD models. 

\item
We also notice a significant rise of the brighter point (null point)
between different jet events. The rise of the reconnection site as the jet evolution advances has been found in the MHD emerging flux models of \citealt{Yokoyama1995}; \citealt{Torok2009}.
In the present case, it may be because during each jet event the reconnection process causes a displacement of the null point and jet spine. In this way the next jet event occurs in a displaced location as compared with the previous jet. This could indicate that the magnetic field configuration has some reminiscences of the earlier reconnection and behaving in the same manner afterwards. Another possible reason for this shifting could be as a result of the interaction between different {\it QSLs} as suggested by \citealt{Joshi2017}. However, in the present case because of the limb location of the AR, we could not compute the {\it QSL} locations.
 \end{enumerate}
\section{Prejet intensity oscillations}\label{ocs}
In Section \ref{sec:characteristics} we mentioned that before and in between the six main jets we also observed many small jet-like ejections, with length less than $10$ Mm. Also, in the AIA  131~\AA\ observations we clearly see many episodic brightenings related to the small jets. In the present section we would like to investigate different properties, like the periodicity, of these features. To that end,  
we select a square of size $4 \times 4$ arcsec at the base of the jets where the intensity is maximum, in  the AIA  131 \AA\ data, as shown in  Figure \ref{evolution} and calculate the mean intensity inside the square in the AIA 131 \AA\ channel. 
We compute the relative intensity variation in the base, after normalization by the quiet region intensity. 
 We find that the oscillations start at the jet base some 5--40  minutes before the  main jet activity.

Figure \ref{intensity} shows the intensity distribution at the jet base for all the jets before and during the jet eruption and the pre--jet phase is shown in between two vertical red dashed lines. The right red dashed lines indicate the starting time of the main jets. The blue arrows indicate the time of the maximum elongation of the main jets. We note that the maximum of the brightening at  the jet base  does not  always  coincide exactly  with the start of the jet neither with the maximum extension time.  
In most of the cases the maximum brightening  occurs  before the 
peak time of the jets by a few minutes.
For the smaller jets it is nearly impossible to compute the delay between  brightenings and jets. They appear to be in phase with the accuracy of the measurements.

To calculate the time period of these pre--jet oscillations, we apply a  wavelet analysis technique. For the significance of time periods in the wavelet spectra, we take a significance test 
into 
account and the levels higher than or equal to 
95\% are labeled as real. The significance test and the 
wavelet analysis technique is well described by \citealt{Torrence1998}.
The cone of influence (COI) regions
make 
an important background for the edge effect at the 
start and end point of  the time range (\citealt{Tian2008}; \citealt{Luna2017}).

\begin{figure*}[ht!]
\centering
\includegraphics[width=1.0\textwidth]{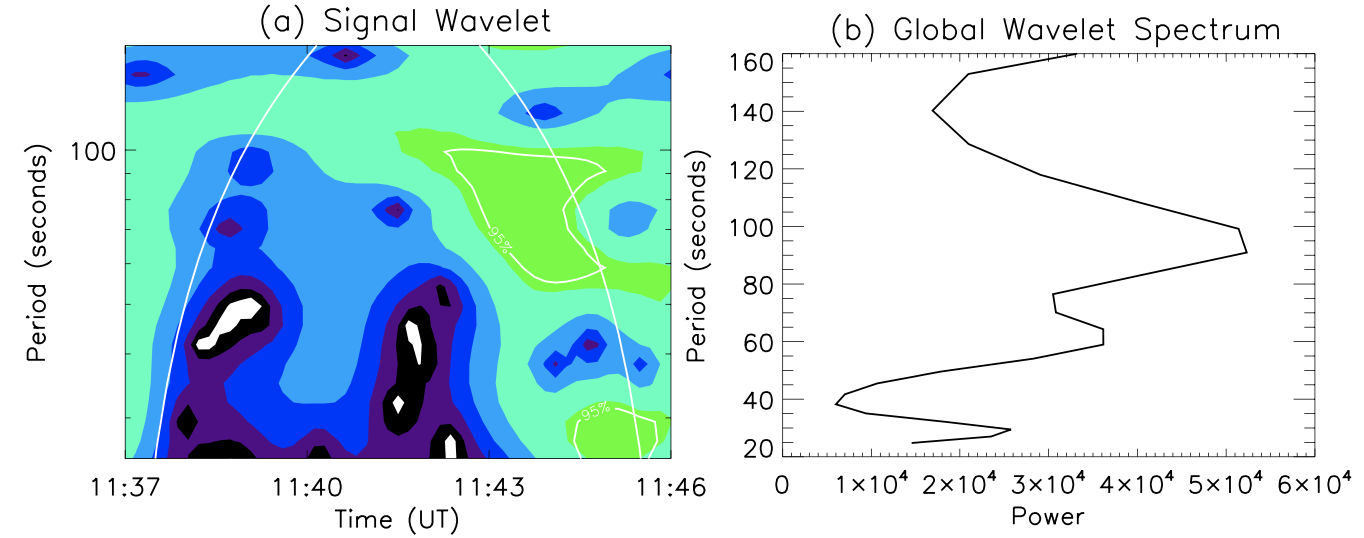}
\caption[An example of wavelet spectrum for the pre–jet intensity oscillations
for {\it Jet2}.]
{An example of wavelet spectrum for the pre–jet intensity oscillations for {\it Jet2}. The solid thick white contours are the regions with the value of wavelet function larger than the 95\% of its maximum value.}
\label{wavelet}
\end{figure*}

\begin{figure*}[ht!]
\centering
\includegraphics[width=\textwidth]{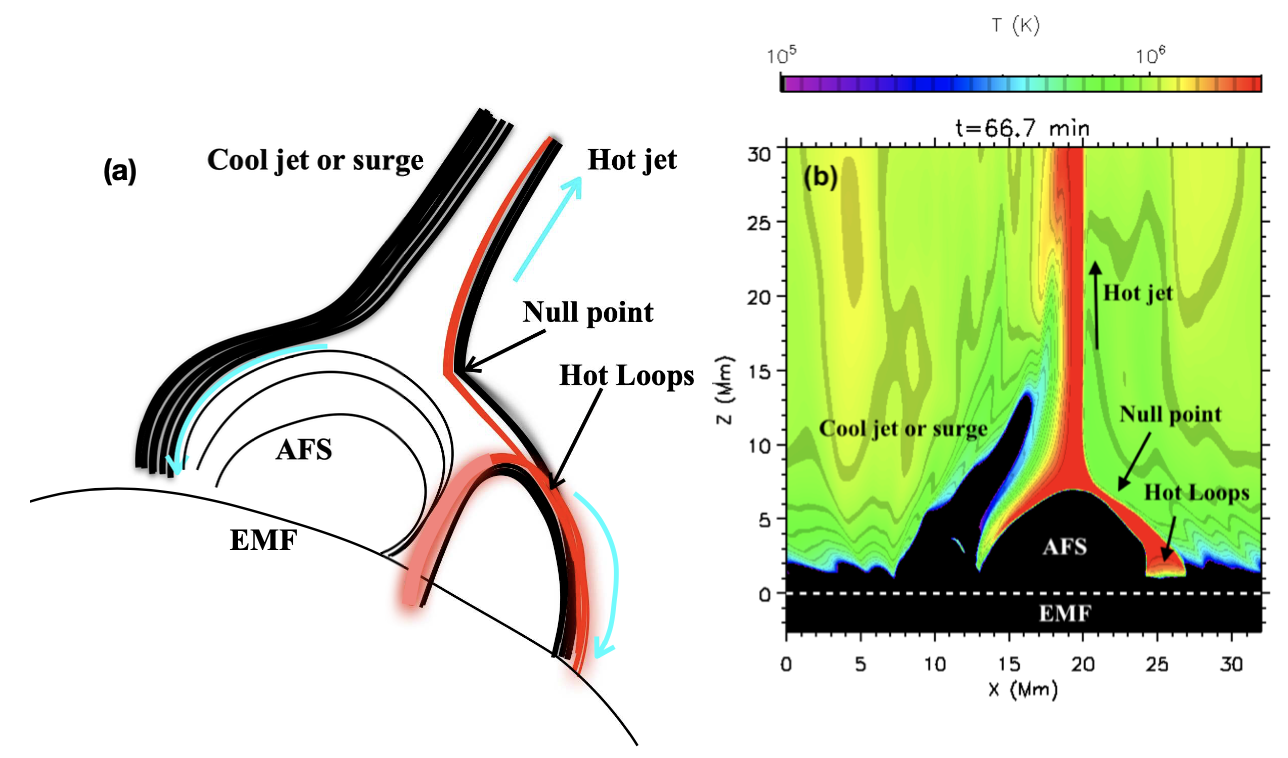}
\caption[Schematic view of the 3D jet 
showing the location of the null point, the cool surge and the hot loops
next to the AFS.]
{Schematic view of the 3D jet 
showing the location of the null point, the cool surge and the hot loops. Panel (a): Schematic view of the 3D jet derived from \citealt{Moreno2008}. Panel (b): 
Temperature map from the numerical experiments by \citealt{Nobrega2017, Nobrega2018}.}
\label{null}
\end{figure*}

The wavelet analysis of the intensity fluctuation at the jet base 
shows that the oscillation period for these pre--jet intensity varies between 1.5 minutes and 6 min; the current values obtained are presented in the last column of Table \ref{table1}. 
An example of wavelet spectrum for the pre--jet activity for {\it Jet2} is presented in Figure \ref{wavelet} (a). The COI region is the outer 
area of the white parabolic curve.
The global wavelet spectrum in panel (b) shows the distribution of power spectra over time.  
 \citealt{Bagashvili2018} investigated the intensity at the base of several jets issued in a coronal hole and  obtained similar results concerning the  periodicity and  duration of the oscillations.

\section{Results and conclusion}\label{res}
This chapter presents observations concerning the structure,
kinematics, and pre-jet intensity oscillations of
 six major jets that occurred on April 4, 2017 in active
   region NOAA 12644. The discussion is based on 
the observational data from AIA and IRIS.
 The conclusion of our main results is as follows:

  A first significant finding of this study is the observation of pre--jet activity,
in particular in the form of oscillatory behavior. Earlier authors had studied the pre--jet activity of quiet region jets observed in the hot AIA filters (\citealt{Bagashvili2018}).  The jets studied by those authors had their origin in coronal bright points 
and the bright points showed oscillatory behavior before the onset of jet activity.  They reported periods for the pre--jet oscillations of around 3 minutes.  Our study deals with AR jets, instead,  also observed in the hot filters of AIA and we find  an oscillatory behavior of the intensity  in a time interval of 5--40 minutes prior to the
  onset of the jet. The period of the  intensity oscillation is
  in the range
  1.5--6 minutes. These values are consistent with
the results reported by \citealt{Bagashvili2018}. They are also
close to  typical periods of acoustic waves in the magnetized solar
atmosphere. This indicates that 
acoustic waves may be responsible
for these observed periods in the occurrence of jets
(\citealt{Nakarikov2005}). 
Quasi-oscillatory variations of intensity can also be the signature of MHD wave excitation processes  which are generated by very rapid dynamical changes of velocity, temperature and other parameters manifesting the apparent non-equilibrium state of the medium where the oscillations are sustained (\citealt{Shergelashvili2005}; \citealt{Shergelashvili2007}; \citealt{Zaqarashvili2002}).
 In 3D reconnection regions like the 
 quasi-separatrix layers, a sharp velocity gradient is likely to be present. The impulsiveness of the jets could lead to such MHD wave excitation. The observed brightness fluctuations could also be due to the oscillatory character of the reconnection processes that lead to the launching of the small jets. Oscillatory reconnection has been found in  theoretical contexts in two dimensions (\citealt{Craig1991}; \citealt{McLaughlin2009}; \citealt{Murray2009}). The latter authors, in particular, studied the emergence of a magnetic FR into the solar atmosphere endowed with a vertical magnetic field. As the process advances, reconnection occurs in the form of bursts with reversals of the sense of reconnection, whereby the inflow and outflow magnetic fields of one burst become the outflow and inflow fields, respectively, in the following one. The period of the oscillation covers a large range, 1.5 to 32 minutes. However, this model is two-dimensional and it is not clear if the oscillatory nature of the reconnection can also be found in general 3D environments.
  
A second significant point in our study is the comparison of 
the observations of the structures and time evolution of the jet complex with numerical experiments of the launching of jets following flux emergence episodes from the solar interior.
Structures like the  double-vault dome with a  bright point at the top where the jets  are initiated as seen in the hot AIA channels and also in the high-resolution IRIS images mimic the structures found in the numerical simulations of   \citealt{Moreno2008} and \citealt{Moreno2013}, who solved the MHD equations in
three dimensions to study the launching of coronal jets following the emergence of magnetic flux  from the solar interior into the atmosphere; they also have similarities with the more recent experiments, in two dimensions, of \citealt{Nobrega2016}, obtained with the radiation-MHD Bifrost
code (\citealt{Gudiksen2011}). In the 3D models, the jet is launched along open coronal field lines
that result from the reconnection of the emerged field with
the preexisting ambient coronal field. Underneath the jet, two vault
structures are formed, one containing the emerging cool plasma and the other a set of hot, closed coronal loops resulting from the reconnection. Overlying the two vaults one finds a flattened CS of Syrovatskii type, which
contains hot plasma and where the reconnection is occurring. The field in the
sheet has a complex structure with a variety of null points; in fact, in its
interior, plasmoids, with the shape of tightly wound solenoids, are seen to
be formed. The reconnection is of the 3D type, in broad terms of the kind
described in the paper by \citealt{Archontis2005}. A vertical cut of the 3D
structure, as in Figure 4 of the paper by \citealt{Moreno2008}, clearly shows
the two vaults with the overlying CS containing the reconnection
site and with the jet issuing upwards from it. The figures in that paper
contained values for the variables as obtained solving the physical
equations; a scheme of the general structure is provided here as well
(Figure~\ref{null}, left panel). As the reconnection process advances, the
hot-loop vault grows in size whereas the emerged-plasma region decreases,
very much as observed in the present study.

An interesting feature in the observations is the tentative detection
  of a surge-like episode next to the jet apparent in the IRIS
  Mg-II time series  in a region that appears dark, 
   in absorption, in the AIA~193~\AA\ observations.
This ejection of dense and cool plasma next to the hot 
  jet, with the cool matter rising and falling, like in an H$_\alpha$ surge,
  also occurs naturally both in the 3D and 2D numerical models cited above
  (and was already introduced by \citealt{Yokoyama1995} in an early 2D
  model). The phenomenon has been studied in depth by \citealt{Nobrega2017,Nobrega2018}
  using the realistic material properties and radiative transfer provided by
  the Bifrost code, which, in particular, facilitate the study of plasma at
  cool chromospheric temperatures. A snapshot of one of the experiments by
  those authors showing a temperature map and with indication of some major
  features is given in Figure~\ref{null} (right panel). In their model, 
  the magnetic field can accelerate the plasma with accelerations up to $100$ times the solar gravity for very brief periods of time after going through the reconnection site
   because of the high field line curvature and associated large Lorentz force. In the advanced phase of the surge, instead, the cool plasma basically falls with free-fall speed, just driven by gravity, as had been tentatively concluded in observations (\citealt{nelson2013}). The velocities obtained from the observations in the present chapter broadly agree with those obtained in the numerical models.

\chapter{Role of solar jets as a driver of large scale coronal disturbances}\label{c5}

\ifpdf
    \graphicspath{{Chapter5/Fig_ch5/}{Chapter5/Figs_ch5/}}
\fi
\section{Introduction}
Solar jets are occasionally associated with the large scale solar filament eruptions (\citealt{Janvier2014a}; \citealt{Chandra2017}). Solar filaments are dense and cool material 
suspended in the hot solar corona along PILs. They are found to be in magnetic dip regions. 
There are two main magnetic configurations for filaments namely sheared arcade and   FR. In the sheared arcade configuration, the arcade connects the opposite polarities on 
either sides of a PIL, whereas in the case of the FR configuration the magnetic field has helical magnetic 
structure. 
Several models have been proposed for the solar eruptions (\citealt{Aulanier2014}; \citealt{Vrsnak2014}; \citealt{Filippov2015}; \citealt{Schmieder2015}). 
Catastrophic loss of equilibrium or torus instability is the 
important mechanism for the solar eruptions (\citealt{Forbes1991}; \citealt{Kliem2006}; \citealt{Demoulin2010}).
In these models, it is assumed that the overlaying magnetic field (B$_{ex}$) decreases with the increase of the 
height (z) 
from the photosphere i.e. B$_{ex}$ $\propto$  z$^{-n}$. The FR becomes unstable when the 
decay index `n'  at its location becomes less than 
a critical value. According to simulations this value ranges from 1.3 to 1.75 
(\citealt{Torok2005}; \citealt{Isenberg2007}; \citealt{Aulanier2010}; \citealt{Zuccarello2017}). Using different observations including high resolution SDO data and 
multi--view STEREO observations it was found that  this value lies between 1 and 1.5 in observed eruptions
(\citealt{Filippov2001}; \citealt{Filippov2013}; \citealt{Zuccarello2014}; \citealt{McCauley2015}).  
In addition to failed, partial and full filament eruptions, recently two filament eruptions were observed with the
quasi-equilibrium state in the middle part of a two-step process (\citealt{Byrne2014}; \citealt{Gosain2016}). In these cases the filament 
starts  to erupt and after attaining
some height it  decelerates, stops and seems to be stable for some time. In case of \citealt{Byrne2014} this time was 
rather short i.e. $\sim$ one hour and for the case of 
\citealt{Gosain2016} the time was 15 hrs.  Such cases are very crucial and play an important role in 
understanding the Sun--Earth connections.

CMEs have attracted the solar physicists greatly as they are playing
a significant role in affecting the Earth's space environment. Usually CMEs are associated with large scale solar eruptions, {\it i.e.}, two--ribbon flares (\citealt{NJoshi2016}; \citealt{Zuccarello2017}), filament eruptions (\citealt{Schmieder2013}; \citealt{RChandra2017}), and occasionally with small scale solar eruptions, {\it i.e.}, solar jets (\citealt{Shen2012}; \citealt{Jiajia2015}; \citealt{Zheng2016}; \citealt{Sterling2018}). \citealt{Shen2012} reported two simultaneous CMEs associated with a blowout jet. One of the two CMEs was bubble-like and the other was jet--like. The authors suggested that the external magnetic reconnection produced the jet--like CME and also led to the rise of a small filament underneath the jet base. Further, they explained that the bubble--like CME is due to the internal reconnection of the
magnetic field lines. \citealt{Jiajia2015} observed a coronal jet event which led to a  high-speed CME (1000 km s$^{-1}$), suggesting that large--scale eruptions could be triggered by a small--scale jet. \citealt{Zheng2016} reported another similar event as a case study of solar jet activity which developed into a CME eruption.
However, the number of such jet--CME associated cases are very less reported in the literature to understand the mechanism and kinematic processes behind the phenomenon.

Solar jets are small scale plasma eruptions due to magnetic reconnection and act as a driver of large scale eruptions {\it i.e.} solar filament eruptions and CMEs. In this chapter two different cases are analysed for the large scale eruptions triggered by a solar jet. In the first case study a two step filament eruption from the AR NOAA 12297 on March 14-15, 2015 is studied. This filament eruption starts to erupt after triggered from the jet which initiates from the same AR and afterwards the filament remains in a meta-stable stage for more than 10 hours. Again a jet hits the meta stable filament and the filament finally erupts from the solar surface followed by the largest geomagnetic storm of solar cycle 24. In the second case study a jet event is analysed followed by a CME on April 28, 2013 which provides evidence of clear association of the jet and the CME. The jet erupted with an initial speed of $\approx$ 200 km s$^{-1}$ and developed into a CME of speed $\approx$ 450 km s$^{-1}$ together with the ambient coronal structures.


\section{Two-step filament eruption triggered by jets}

     \label{sol_obs}
The AR NOAA 12297 produced the largest geomagnetic storm (Dst index $\sim$ -223 nT) of solar cycle 24
on March 17, 2015. This geomagnetic storm was associated with a GOES C9.1 class flare of March 15, 2015 and a filament eruption, which was triggered and derived by solar jets from the same AR.
The AR NOAA 12297 was located at S22W25 on  March 15, 2015. 
Further the filament eruption was associated with a halo CME.
This filament was disturbed on March 14, 2015 by a small solar jet and finally totally erupted  with a further push by an another solar jet on March 15, 2015. 
The event was observed by AIA onboard SDO in different UV and EUV wavelengths. The event was also observed by the Global Oscillation 
Network Group (GONG) in H$_\alpha$ line center. For the magnetic field, we have used the
data from  the HMI onboard SDO.
The observational description of the filament activation and eruption on 2015 March 14-15 observed by different 
instruments is presented in following Section \ref{sdo}.

\begin{figure}
\centering
\includegraphics[width=0.9\textwidth, clip=]{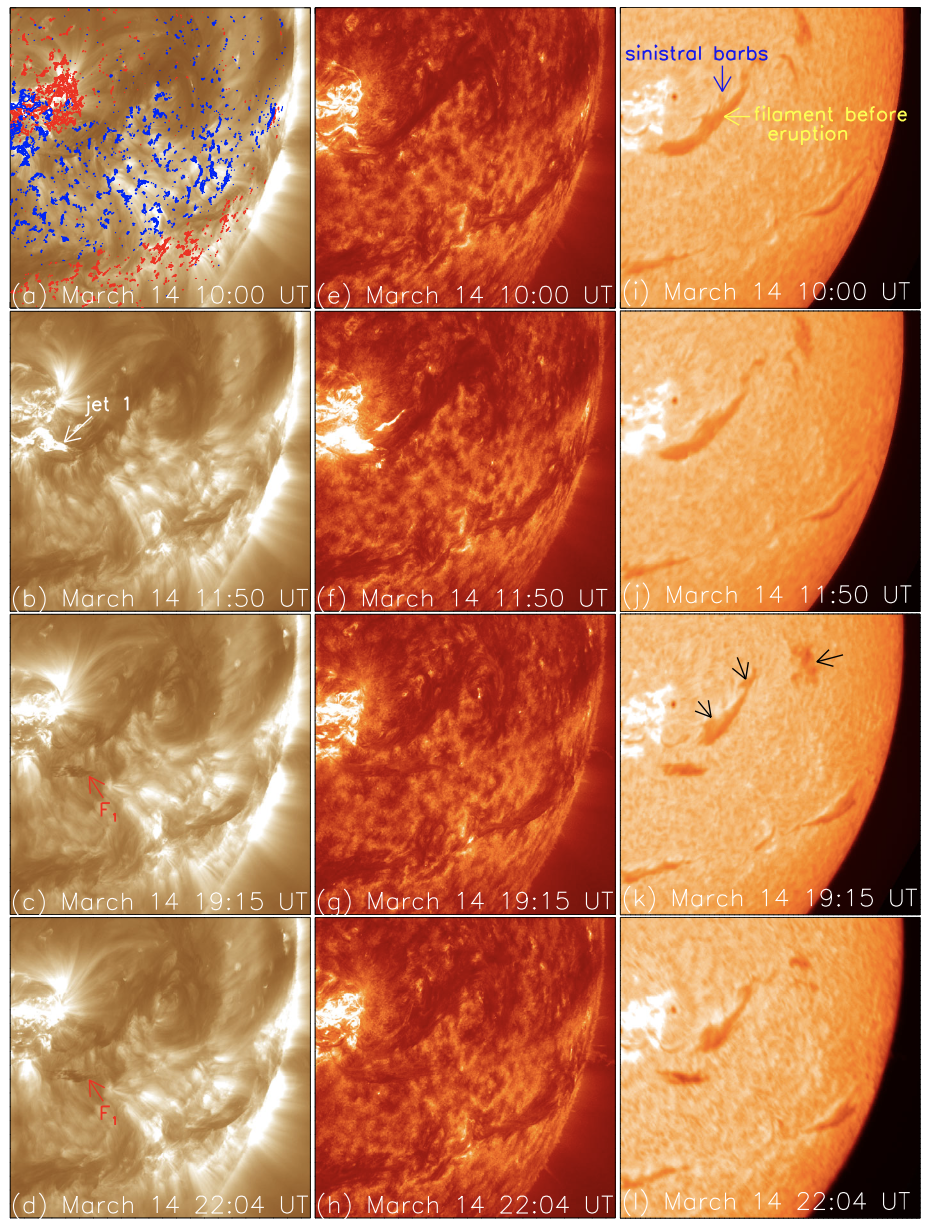}
\caption[Evolution of the filament eruption on March 14, 2015 from AR NOAA 12297.]
{Evolution of the filament on March 14, 2015. Left column : AIA 193 \AA, middle  column: AIA 304 \AA, and right column: H$_\alpha$ observations. The position of jet 1 is shown by the 
white arrow.}
\label{fig:morpho1}
\end{figure}

\subsection{Morphology of the two step filament eruption}
     \label{sdo}

Figure \ref{fig:morpho1} shows the evolution of the filament on March 14, 2015 in AIA 193 \AA\ (first column), 304 \AA\ 
(second column), and GONG H$_\alpha$ (third column). Before any activity in the 
filament it has a long sigmoidal shape (Figure \ref{fig:morpho1} a, e, f, and also yellow arrow). 
We can see the sinistral barbs in GONG H$_\alpha$ images (Figure \ref{fig:morpho1} (i)). 
This suggests the filament have a  positive twist and  hence the positive helicity. 
Since the filament is located in the southern hemisphere, this is in conformity with
the hemispheric rule of helicity. According to the helicity
hemispheric rule majority of positive/negative helicity solar features are located in the southern /northern hemisphere 
respectively (\citealt{Pevtsov1995}). 
Around 11:50 UT a jet activity started in the active 
region, we name it as jet 1 and mark by the white arrow in Figure \ref{fig:morpho1}(b). This jet activity was visible almost in all 
EUV/UV and H$_\alpha$ wavelengths. Due to this jet activity the left southern part of the sigmoidal filament was disturbed. 
Afterwards a strand of the filament separated from the main body, rose upwards and became stable after a  displacement  of $\approx$ 125 Mm.     
The broken part  of the sigmoidal filament is labeled by F$_1$ and indicated by the red arrow in Figure \ref{fig:morpho1} (c). At the 
same time, we have observed flare brightening in the AR as a GOES C2.6 class flare. The upper--right part of the filament channel, which did not disturb in 
this period is shown by blue arrows in Figure \ref{fig:morpho1}(c).
\begin{figure}
\centering
\includegraphics[width=0.9\textwidth, clip=]{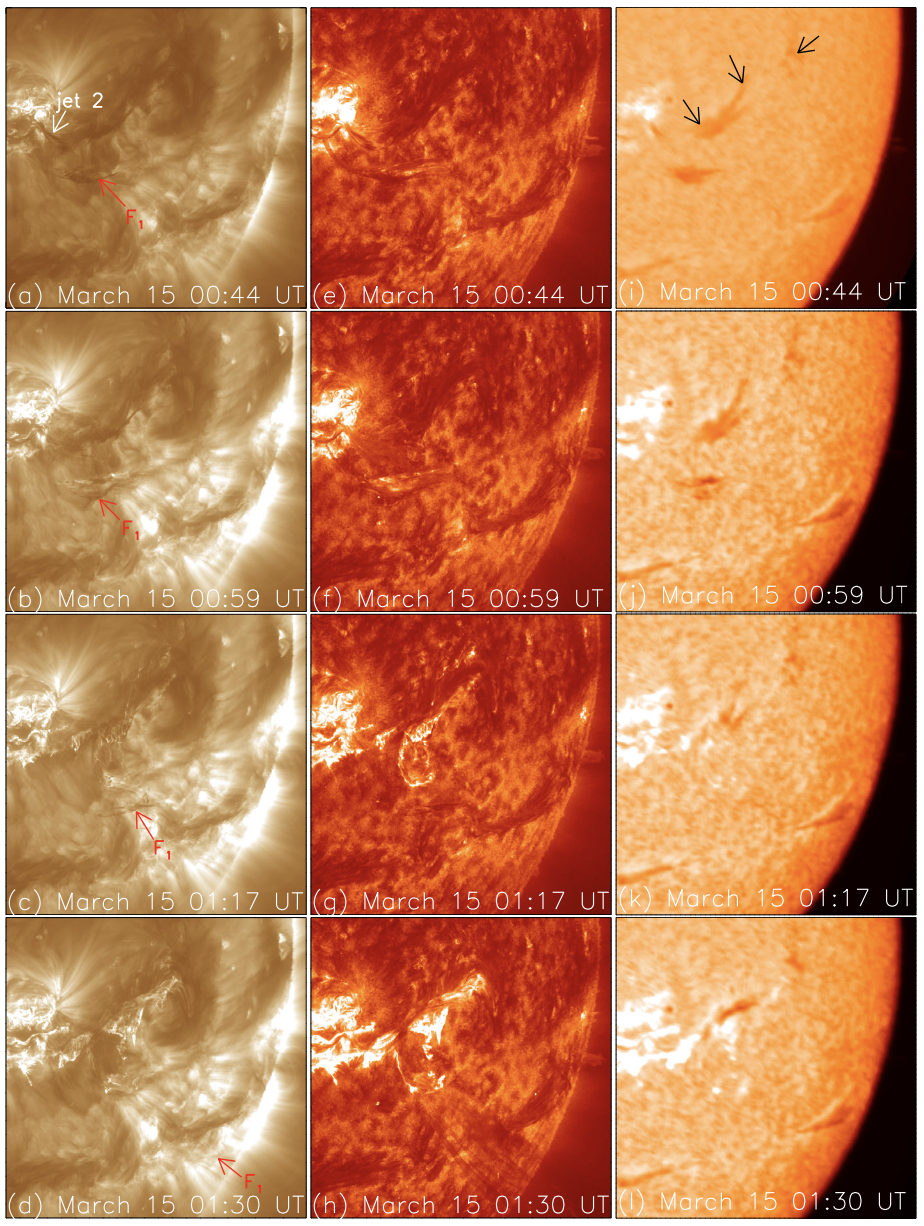}
\caption[Evolution of the filament eruption on March 15, 2015 from AR NOAA 12297.]
{Evolution of the filament on March 15, 2015. Left column : AIA 193 \AA, middle  column: AIA 304 \AA, and right column: H$_\alpha$ observations. The position of jet 2 is shown by the  white arrow.}
\label{fig:morpho2}
\end{figure}

The uplifted broken part F$_1$ from the main sigmoidal filament was stable in that 
particular location for a period of $\approx$ 12 hrs i.e upto 00:45 UT on March 15, 2015. Around 00:45 UT on March 15, 2015 we again 
observed the jet/surge activity in the AR towards the west-south direction and it interacted with the filament 
F$_1$ as well as with the big northern filament (indicated by three blue arrows). 
We label this jet as jet 2. After 00:45 UT the filament F$_1$ started to erupt 
and finally it went away from the solar surface. 
The evolution of the erupting filament is shown in Figure \ref{fig:morpho2}. The left, middle and right columns of 
Figure \ref{fig:morpho2} show the development of the erupting filament F$_1$ in AIA 193 \AA, 304 \AA, and H$_\alpha$ 
wavelengths respectively. The eruption was associated with the C9.1 class GOES flare. 
The flare has two ribbon structure visible in different EUV and H$_\alpha$ images. We can also see the ribbon separation 
as proposed by CSHKP model.
The flare was also observed in hard X-rays by the RHESSI satellite and studied by \citealt{Wang2016}.

The major upper part of the sigmoidal filament channel (indicated by blue arrows in Figure \ref{fig:morpho2}) 
was also perturbed during the ejection of jet 2. The material of the
filament rose--up and later--on it came back to the foot--points of the filaments. 
In AIA channels it becomes brighter and we can also see its twisted structure. From Figure \ref{fig:morpho2}(d), one can infer that 
the twist is right--handed, which is consistent with the sinistral filament 
chirality seen in H$_\alpha$ data. Therefore, we see here the same sign of twist in the chromosphere, the upper
solar atmosphere and in the associated MC. It seems that Jet 2 injects impulse and heat into the filament at its eastern end.
Previously static filament material comes in motion along helical field-lines of the FR. 
Thus the upper parts of the helices become visible as dark and bright (due to heating) threads. 
We observe intense field-aligned motions within the activated FR but it does not change 
significantly its position, and after the energetic phase of the event the filament restores to its 
approximately initial state. 
During the activation the filament becomes less visible in the H$_\alpha$ wavelength (H$_\alpha$ 
images in the third column of Figure \ref{fig:morpho2}) most likely for two reasons. 
The Doppler shift in the moving material can remove its H$_\alpha$ line out of the filter passband 
(dynamic disappearance), and the heating of the filament can suppress the absorption of H$_\alpha$ radiation (thermal disappearance).

\subsection{Evolution of the filament eruption}
     \label{td}
\begin{figure}[ht!]
\centering
\includegraphics[width=1.0\textwidth, clip=]{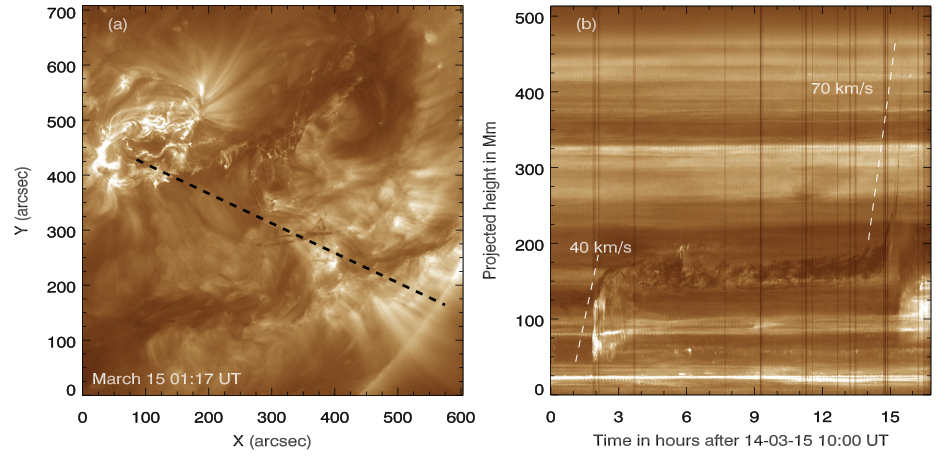}
\caption[AIA 193 \AA\ image showing the location of a slit select for the time--slice analysis.]{(a) AIA 193 \AA\ image showing the location of a slit select for the time--slice analysis. 
(b)Time-slice image of the eruption during March 14-15, 2015.}
\label{fig:slice}
\end{figure}

To investigate the height-time evolution of the filament eruption during March 14-15, 2015 we 
have selected a slit along the
direction of the filament eruption in AIA 193 \AA\ data. The position of the slit is shown by the 
dashed black line in Figure \ref{fig:slice} (a).
The time-distance diagram is given in Figure \ref{fig:slice} (b). In the time-distance diagram, we can clearly
see the two step eruption
of filament. 
In first step, the filament starts to at 
rise at $\sim$ 12:00 UT on March 14, 2015 and attains a projected height of 125 
Mm, as also discussed in Section \ref{sdo}). Due to the projection effect, this is the minimum value of height.
The real height must be of larger value. Unfortunately we do not have STEREO observations, which
could tell about the true height of the eruption. We computed the speed of this eruption
and it was found $\sim$ 40 kms$^{-1}$. After 13:00 UT the filament stops to rise and stays at the same 
height upto 00:45 UT on March 15, 2015. In the second step after March 15, 2015 00:45 UT, the filament starts 
to rise and finally fully erupts. The calculated speed of the second step eruption was 70 kms$^{-1}$. 

\begin{figure}[ht!]
\includegraphics[width=\textwidth, clip=]{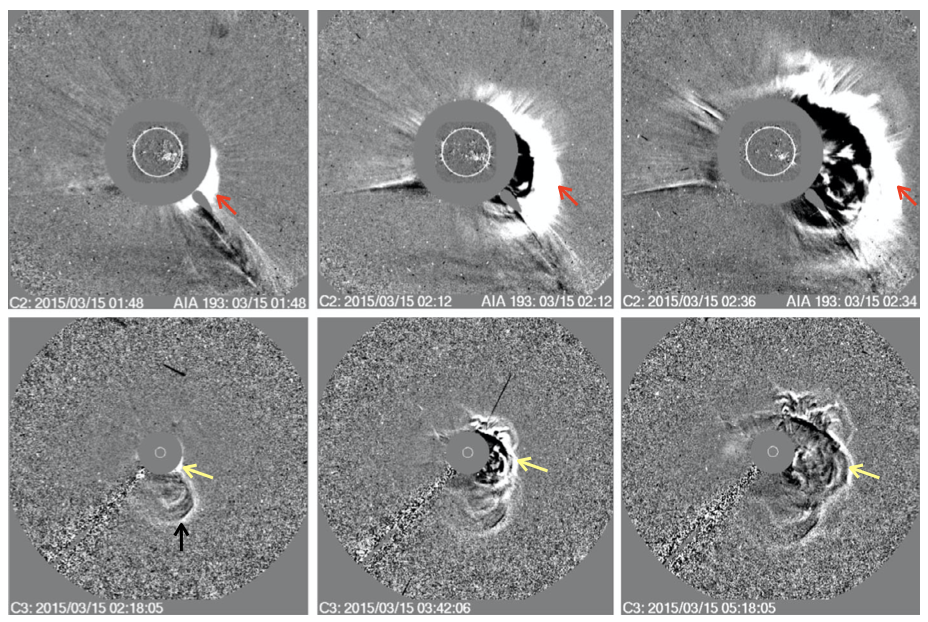}
\caption[Development of CME on March 15, 2015 observed by LASCO C2 and C3.]{Development of CME on March 15, 2015 observed by LASCO C2 (top panel, red arrows) and C3 (bottom panel, yellow arrows). 
The black arrow indicates the CME of March 14, 2015.}   
\label{fig:cme}
\end{figure}

\subsection{CME observation associated with the filament eruption}
     \label{lasco}
The CME associated with the filament eruption on March 15, 2015 was observed with the LASCO instrument.
As reported in \citealt{Wang2016}, during the C2.6 class flare of March 14, 2015 a
small filament erupted (their Figure 1 (a), F4 filament) and there was a slow CME observed in
LASCO C2 field--of--view (FOV) at $\sim$ 13:30 UT. The speed and the angular width of this CME was 208  kms$^{-1}$ and 160$^{\circ}$
respectively.
By the push from jet 1, the F$_1$ filament eruption on March 15, 2015 produced a halo CME, visible in the LASCO C2 FOV at 01:48 UT. 
The CME was visible up to 27 R$_{\odot}$ in the LASCO C3 FOV.
The running difference of the C2 and C3 coronagraph images are shown in Figure \ref{fig:cme}.
The white circle represents the solar disk occulted by the coronagraph. In the C2 images, 
the running difference of SDO AIA 193 \AA\ images are displayed inside the white circles for the same time. 
The red and yellow arrows indicate the leading 
edge of the CME in the LASCO C2 and C3 FOV respectively. The black arrow in the first image of the bottom panel points to the CME 
from the same AR on March 14, 2015. According to the LASCO CDAW Catalog (\citealt{Gopalswamy2009}) the average CME speed 
was 719 kms$^{-1}$ and the acceleration was -9.0 ms$^{-2}$. 
The mass and kinetic energy of the CME were 3.0$\times 10^{16}$ gram and 7.7 $\times 10^{31}$ ergs respectively.	
Moreover, \citealt{Liu2015} reported the maximum CME speed was 1100 kms$^{-1}$.
In C3 the FOV the CME of March 14-15, 2015 interacted at 02:18 UT. \citealt{Liu2015} considered the interaction
of these CMEs as the cause of the largest geomagnetic storm of solar cycle 24 on March 17, 2015.
\subsection{Decay index distribution for the jet driven filament eruption}
     \label{decay}
In order to find the reasons why the part F$_1$ of the initial filament erupted after 12 hrs delay in an intermediate 
state, while the remainder of the filament does not leave its position despite the strong activation, we 
should analyze the structure of the magnetic field surrounding these parts of the filament. Since the 
stability of the flux-rope equilibrium depends on the value of the decay index, we need to know the 
distribution of this parameter in the AR. In principle, the decay index should be calculated 
for the coronal field external to the FR. It is reasonable to assume that major coronal currents in the
volume of interest are contained within the FR. For the magnetic field of currents below the 
photosphere, the coronal potential field is a rather good approximation. 

We need the potential magnetic field distribution in the corona at heights of prominences, which are 
much less than a solar radius. Therefore, we can use a restricted area of a photospheric magnetogram as the 
boundary of the calculation domain and neglect its sphericity considering as a part of a flat surface 
and use the well-known solution for half-space with a plane boundary in terms of Green's 
functions (\citealt{Filippov2001}; \citealt{Filippov2013}). When we cut out a rectangular area around the 
filaments under study from the full disk magnetogram, we ignore the contribution of the magnetic sources 
outside of it. Such simplification is reasonable if the main sources of the field lie within the cut-out area. 
In our case the AR NOAA 12297 is the strongest magnetic source on the disk and is rather distant 
from other ARs. On March 14 and especially on March 15 the AR is at a considerable 
distance from the central meridian, so we need to take into account the projection effect. We construct for the 
boundary condition the data array with the equal angular size of pixels and assume the projection of the 
line-of-sight-field on the normal as the radial component. Thus we obtain the rectangular area of the magnetogram 
that looks as if the region were located at the center of the disk.

\begin{figure}[ht!]
\includegraphics[width=1.0\textwidth,clip=]{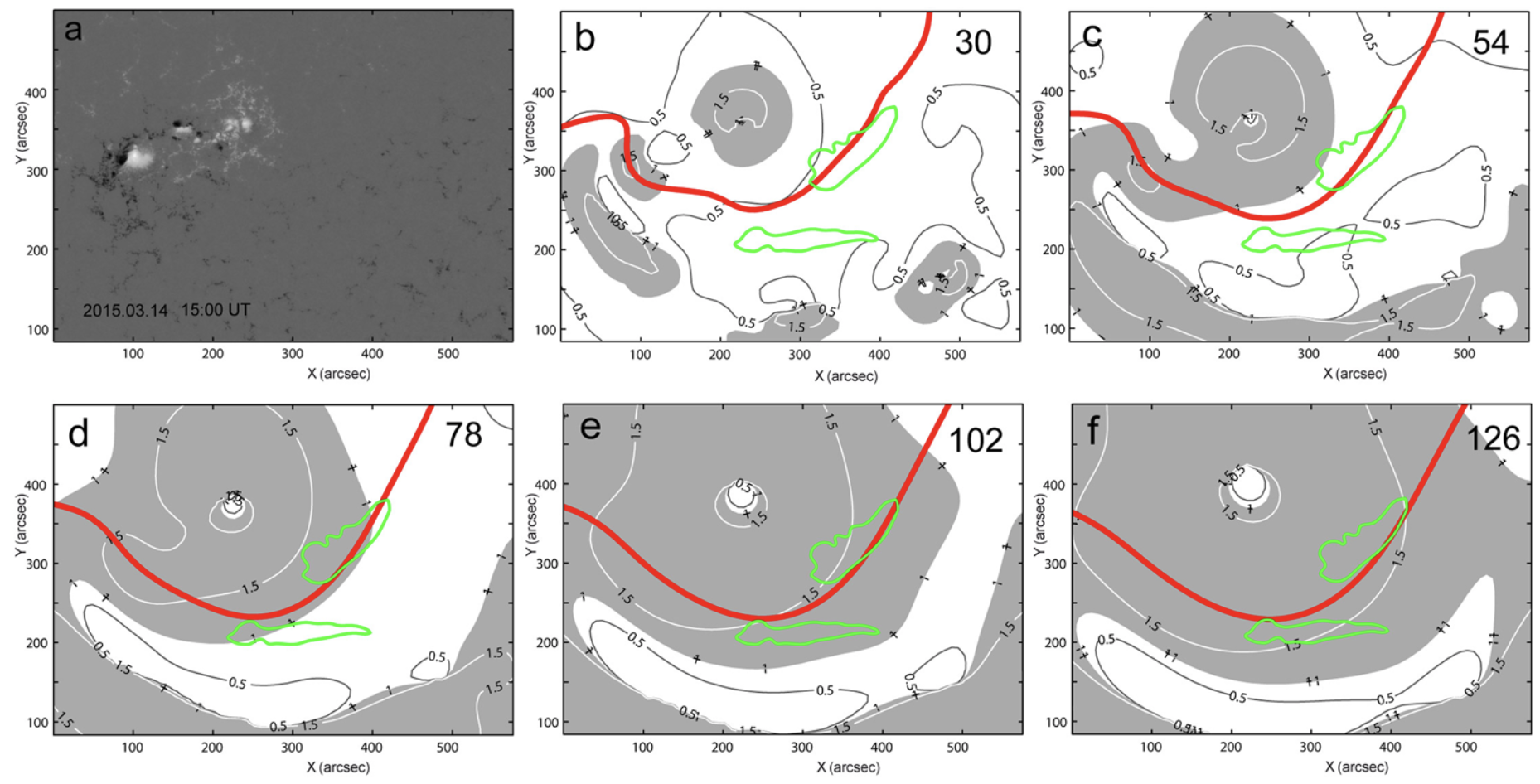}
\caption[Distribution of 
the decay index n at different heights with HMI magnetogram.]{(a) HMI magnetogram of the selected area OF THE AR. Distribution of 
the decay index `n' at different heights above the area shown in 
panel (a). Red line indicate the position of the PILs.} 
\label{fig:decay}
\end{figure}

\begin{figure}[ht!]
\includegraphics[width=1.0\textwidth,clip=]{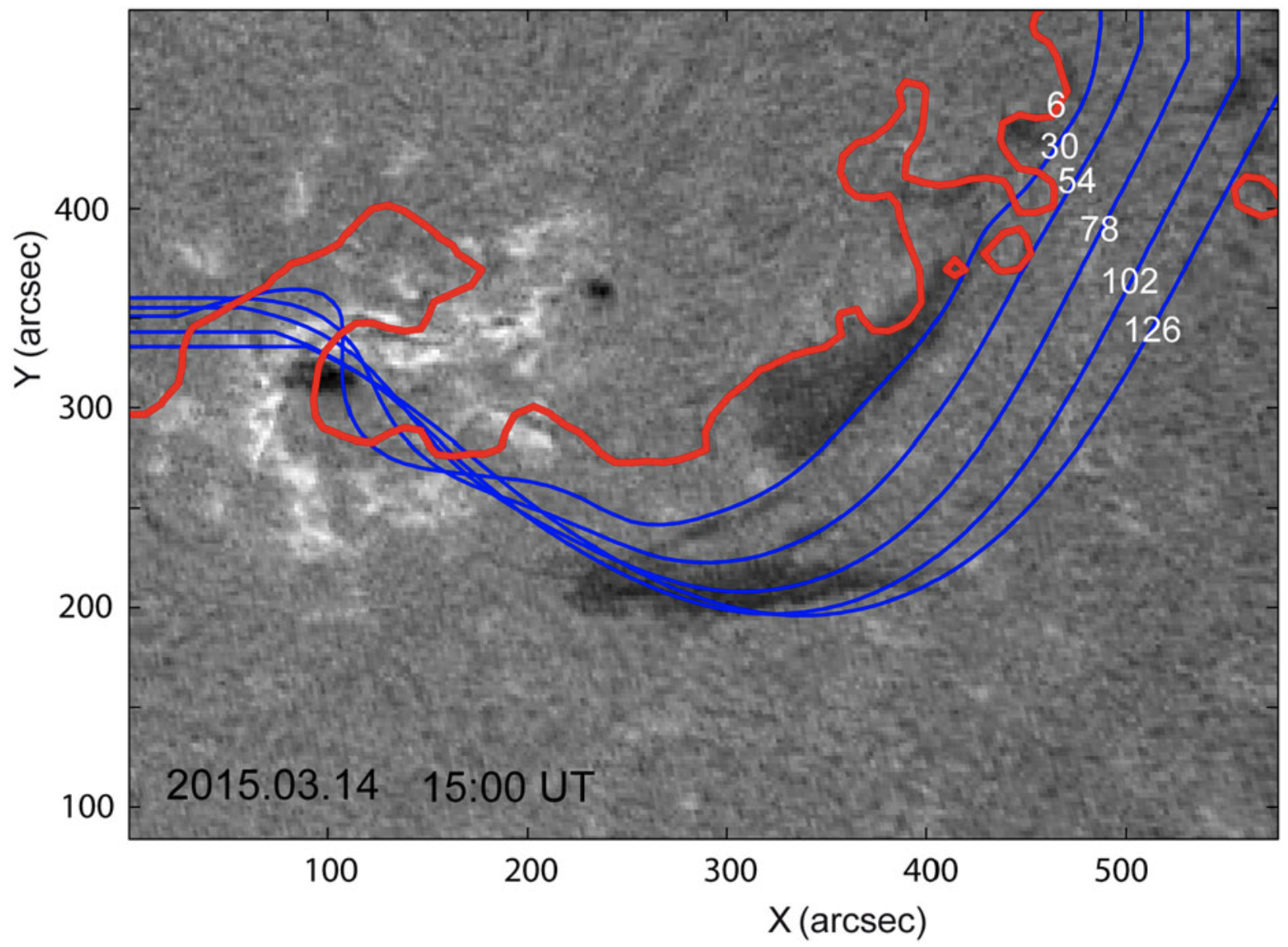}
\caption{Kanzelhoehe H$_\alpha$ filtergram of the same region as in 
Figure \ref{fig:decay} (a) taken at the same time with superposed PILs at different heights.} 
\label{fig:decay_hal1}
\end{figure}

Figure~\ref{fig:decay} (a) represents the modified fragment of the magnetogram taken by the HMI on March 14, 2015 at 15:00 UT, which was used as a boundary condition for the potential magnetic field calculations. The pixel-size in HMI magnetograms is $\sim$ 0.5$\arcsec$ or $\sim$ 0.36 Mm, which is very small compared with the expected height of a 
filament > 10 Mm. To save computational time we applied binning several times and increased pixel-size up 
to $\sim$ 10 Mm. Figure~\ref{fig:decay}(b)--(f) show the distribution of the decay index  from equation \ref{Eq1}, where B$_{ex}$ is the horizontal magnetic-field component and z (or h) is the height above the photosphere, at 
different heights above the area shown in the panel (a). The thin lines show isocontours of n = 0.5, 
1, 1.5, while the thick red lines indicate the positions of PILs at respective heights. Areas where n > 1
 are shadowed. Green contours show the position of the filament taken from the co-aligned 
Kanzelhoehe H$_\alpha$ filtergram transformed in the same way as the magnetogram (Figure~\ref{fig:decay_hal1}). Since the filaments are located at some unknown heights (we will consider this problem below) above this surface and the surface is inclined to the line-of-sight, the position of the 
filament contours does not correspond exactly to the position of magnetic features at any 
height (the filaments should be somewhat shifted to the north-west in this projection).

A PIL is a favorable place for horizontal equilibrium of a FR, because the vertical 
component of the coronal field vanishes. Any PIL at any height can be considered as a 
potential location of the FR, but it is only a necessary condition. Another necessary 
(but again not sufficient) condition for the stable equilibrium is the quantity of the decay 
index below the critical value. In fact, FRs may be found only in few places where both 
conditions are fulfilled. In Figure~\ref{fig:decay}, the segments of PILs within white areas (or at least 
outside of isocontours 1.5) are the places favorable for the flux-rope stable occurrence. 
Below 60 Mm (Figure~\ref{fig:decay} (c)) the segments of the PIL near both green contours are suitable for 
stable FRs. The contour n = 1 touches the PIL near both filaments at the height of 75 Mm 
(Figure~\ref{fig:decay} (d)), while the contour n = 1.5 touches the PIL near the southern filament at the 
height of 100 Mm (Figure~\ref{fig:decay} (e)) and near the western filament at the height of 120 Mm (Figure~\ref{fig:decay} (f)).

Unfortunately, we cannot measure directly the height of the filaments because STEREO was in an 
unfavorable position. However, we can estimate the heights using the method proposed by 
 \citealt{Filippov2016}. It is based on the confirmed by observations assumption that the material of filaments 
is accumulated near coronal magnetic neutral surfaces B$_r$ = 0 (\citealt{Filippov2016b}).
It was found also that the potential 
approximation for coronal magnetic fields is sufficient for the filament height estimations. 
Comparison of the 3D shape of the neutral surface, represented in the projection on the 
plane of the sky as a set of PILs, with the filament shape and position allows us to obtain 
information about heights of different parts of the filament including its top, or spine, which 
is the most reliable indicator of the flux-rope axis.    

Figure~\ref{fig:decay_hal1} shows the fragment of the Kanzelhoehe H$_\alpha$ filtergram of the same region as 
in Figure~\ref{fig:decay} co-aligned with the magnetogram and transformed in the same way. 
The same PILs as in Figure~\ref{fig:decay} (b)-(f) are shown but every PIL is shifted in x 
and y coordinates by values (\citealt{Filippov1999}; \citealt{Filippov2009}; \citealt{Chandra2017}):
\begin{equation} 
\Delta x=htg\lambda_0 ~~~~~~~~~~and ~~~~~~~~~~~~\Delta y=htg\varphi_0
\end{equation} 
where, h is the height of the PIL, $\lambda_0$ and $\varphi_0$ are longitude and latitude of the selected area center. 
Thus they are projected on the plane of the sky in the same way as the filament in the on-disk 
filtergram. The lowest PIL at the height of 6 Mm is red, while the other are blue. The spine 
of the western filament follows exactly the PIL at the height of 30 Mm. All filament body is 
located between this line and the red line at the height of 6 Mm. The southern filament does 
not so strictly follow any PIL, however, its spine is most likely a little bit above the PIL 
at the height of 78 Mm. For comparison figure \ref{fig:decay_hal1} shows the filament and the neutral surface on March 14 at 11 UT before the separation into two parts.  The spine of the western section 
of the filament also follows exactly the PIL at the height of 30 Mm, while the eastern section 
seems to be higher. The top of the wide part of the eastern section touches the PIL at the height 
of 54 Mm and the thin thread-like continuation of the spine crosses all PILs.

We found that a big filament located at the periphery of a strong AR undergo  a  complicated partial and two-step eruption. The idea of two-step energy release processes 
came from analyses of two-peak EUV light-curves of some flares (\citealt{Woods2011}; \citealt{Su2012}) suggested that 
two peaks in light-curves appear due to two 
stages of a single event associated with the delayed eruption of a CME. They presented AIA 
EUV observations of the limb March 8, 2011 event in which a FR accelerates in the first 
stage up to 120 km s$^{-1}$, then the speed decreases to 14 km s$^{-1}$, and in the second stage, started 
after 2 hrs after the beginning of the event, it accelerates again and becomes the CME with 
a speed of $\sim$ 500 km s$^{-1}$. \citealt{Byrne2014} also analysed this event and suggested that either 
the kink-instability or torus-instability of the FR may be the likeliest scenario. 
Since the event was at the limb and photospheric magnetic-field data were not available for 
this time, the authors did not make strong conclusions about magnetic configuration and were 
not very certain with the supposed torus instability without calculations of the decay index. 
\citealt{Gosain2016} studied two-step eruption of a quiescent filament on October 22, 2011. 
It was observed from different viewpoint by SDO, SOHO, and STEREO. The CME associated with 
the filament eruption and two bright ribbons in the chromosphere both appear 15 hrs after 
the start of the event. Computation of the decay index showed that there were zones of 
stability and instability that alternate in the corona. Below 100 Mm the equilibrium was 
stable, then the zone of instability follows from 100 to 500 Mm that gave place to the 
zone of stability again. Above a height of 600 Mm the PIL disappeared, which hinted on the 
possibility for the FR to lose the horizontal equilibrium and erupt. These results 
showed the possible scenario of the two-step eruption confirmed by observations and 
calculations. However, the FR was not clearly observed in the intermediate 
position and magnetic field calculations at great heights, above 400 Mm, were not too reliable. 

\section{Cause and kinematics of a jet–like CME} \label{sec:apj_intro}
 For the jet-CME relationn, this case study presents a jet event followed by a CME on April 28, 2013 which provides evidence of clear association of the jet and the CME. The jet erupted with an initial speed of $\approx$ 200 km s$^{-1}$ and developed into a CME together with the ambient coronal structures. 

\subsection{Observational analysis of jet and narrow CME}\label{data}
The observational data for the jet eruption and the CME is taken from SDO, STEREO, and SOHO/ LASCO.
 For the multi-thermal jet structure, we analysed the AIA data in 131 \AA, 171 \AA, 
193 \AA, 211 \AA, and in 304 \AA. For a better contrast of the hot and cool counterparts of 
the jet, we create the base and running difference images of the AIA data.
To probe the jet and CME from multiple perspectives the EUV images taken by SECCHI are analysed. For our current analysis of the jet, 
we use the EUV images of \emph{STEREO}--B in 304 \AA\ with a cadence of 10 minutes and pixel size of
1 arcsec. \emph{STEREO}--A and B were separated by 83$^{\circ}$ on April 28, 2013. The CME is well observed with {\it SOHO}/LASCO and \emph{STEREO}/COR coronagraphs. With the multi-point observations from LASCO and COR, we employ the Graduated Cylindrical Shell (GCS) model to obtain the three-dimensional height and direction of the CME (section \ref{cme_jet}). We further analyse the photospheric magnetic field using  the line-of-sight magnetograms from  HMI instrument. 
For a closer and clear view of the jet source region, we use HMI SHARP data set with a cadence of 12 minutes.
\begin{figure*}[ht!] 
\centering
\includegraphics[width=\textwidth]{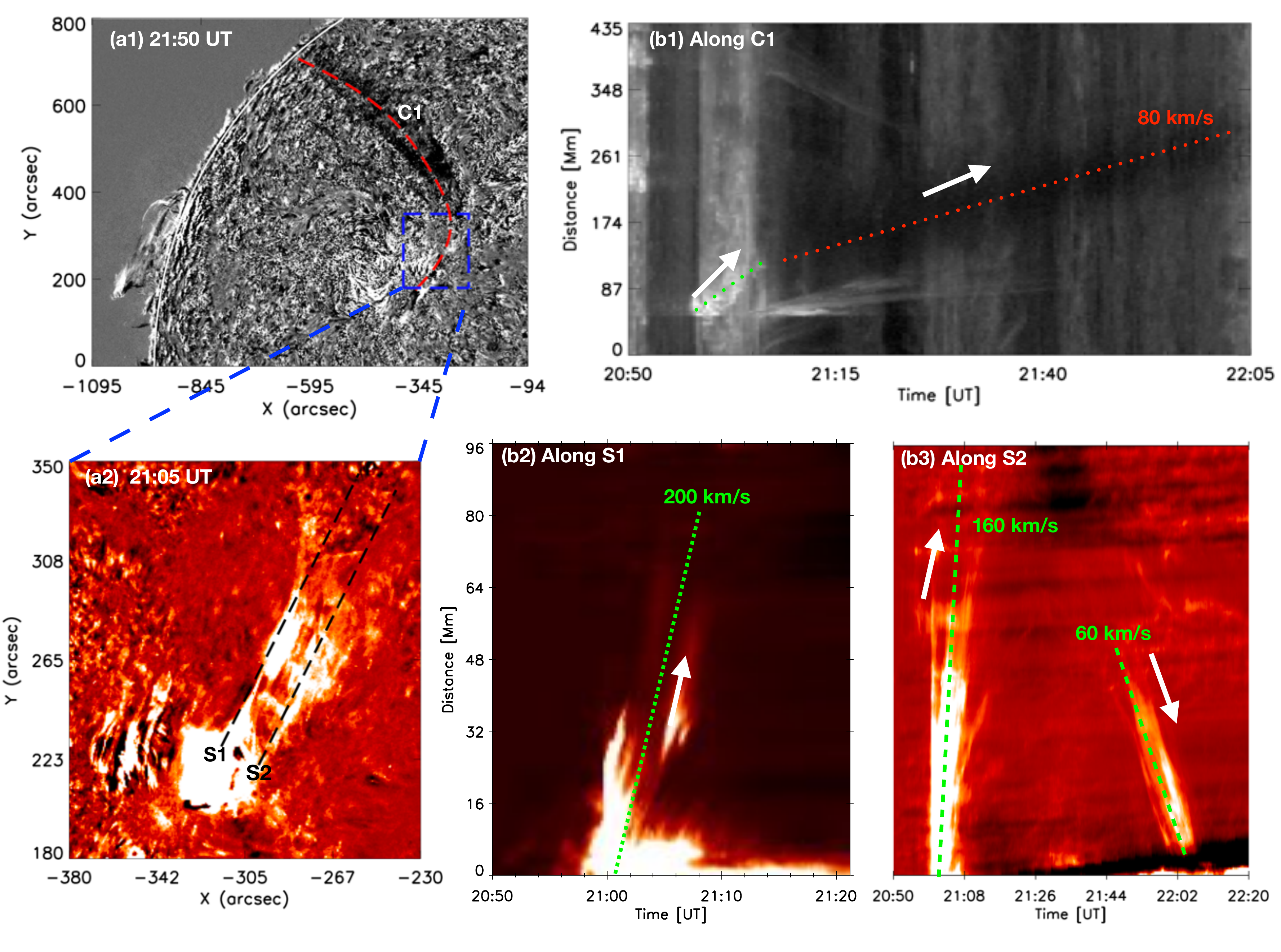}
\caption[The observed jet (a1-a2) and height--time analysis (b1-b3) with AIA 304 \AA\ on April 28, 2013 from AR NOAA 11731.]
{The observed jet (a1-a2) and height--time analysis (b1-b3) with AIA 304 \AA\ on April 28, 2013. Jet material falls back in the direction of S2 into the source region 
with a speed of $\approx$ 60 km s$^{-1}$.}
\label{timeslice_jetcme}
\end{figure*}
\subsection{Kinematics of the jet}\label{kinematics_jet}
The jet started to erupt $\approx$ 20:53 UT with a circular base, towards the northern direction from the AR NOAA 11731 (N09E23) and observed in all six AIA channels (94 \AA, 131 \AA, 171 \AA, 193 \AA, 211 \AA, and 304 \AA). After reaching to some height at about 80 Mm, the jet material was deflected from its original direction of propagation and revolved around the north--east direction. The jet was initially bright (Figure \ref{timeslice_jetcme} (a2), and afterwards followed with dark material (Figure \ref{timeslice_jetcme} (a1)), suggesting impulsively strong heating at the initial phase. The following dark material was only visible in AIA 304 \AA\ and not observed in hot channels, {\it i.e.} 171 \AA. The propagation of the whole jet in AIA 304 \AA\ is shown in Figure \ref{timeslice_jetcme} (a1) along with the red curve C1, which indicates the deflection of the jet from north direction to north--east direction towards the solar limb. The initiation of the jet from the source region is shown in panel (a2). We also observed a small jet ejection at about 21:24 UT in the eastern neighbourhood of the source region, and this jet material merged with the big jet. Panel (b1) is the height--time plot of the jet along the slit C1. The jet speed shows a two--stage profile. The speed in the later stage is about 80 km s$^{-1}$ towards the north--east direction (red dotted line). For the velocity in the initial stage, we set two slits S1 and S2 (panel (a2) of 10 pixel width in two different directions, and found that the speed in the S1 direction is 200 km s$^{-1}$ and that the other direction S2 is about 160 km s$^{-1}$
(as presented in panel (b2) and (b3)). In addition to this, we found
that a portion of the jet material falls back to the source region around 21:51 UT with a speed of $\approx$ 60 km s$^{-1}$, clearly appeared in height--time plot along S2 direction in panel (b3).
\begin{figure*}[t!] 
\centering
\includegraphics[width=\textwidth]{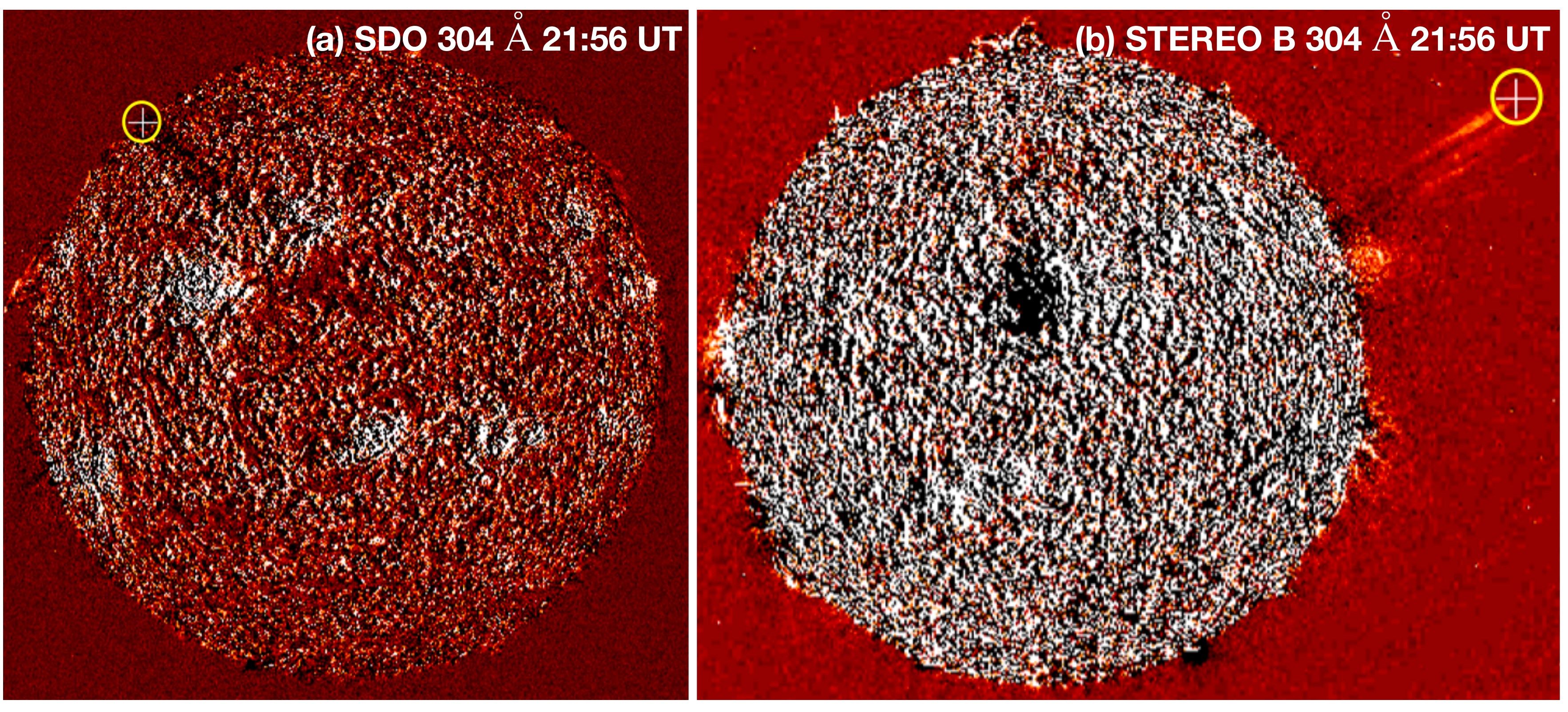}
\caption[The leading edge of the jet material in \emph{SDO}/AIA 304 \AA\ (a) and in \emph{STEREO}--B EUV 304 \AA\ (b).]
{The leading edge of the jet material in AIA 304 \AA\ (a) and in \emph{STEREO}--B EUV 304 \AA\ (b).
 The $+$ sign shows the 
location of the leading edge of the jet obtained from the SCC$_{-}$MEASURE technique.}
\label{scc}
\end{figure*}

\begin{figure*}[t!] 
\centering
\includegraphics[width=0.8\textwidth]{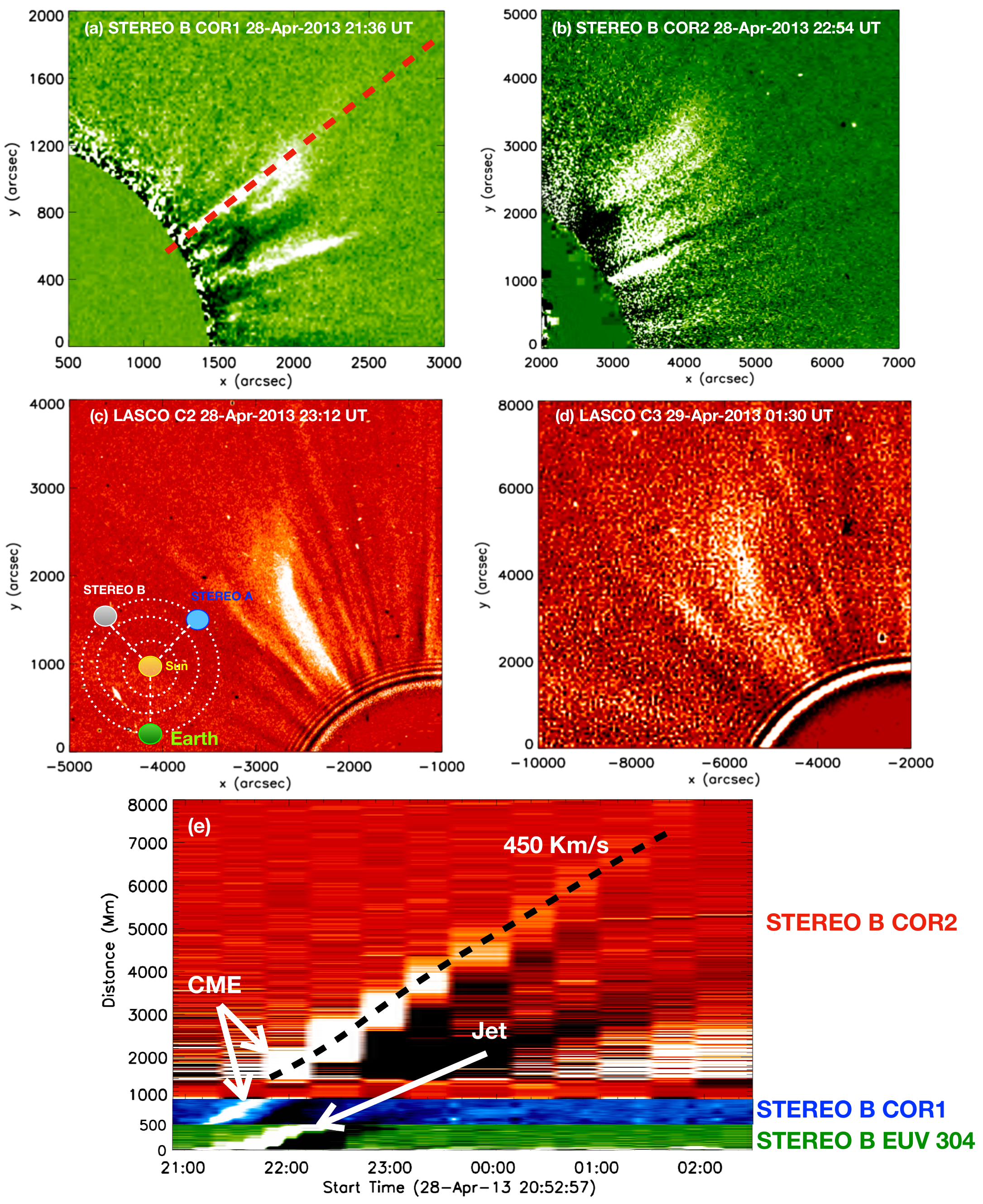}
\caption[CME associated with the jet is observed by LASCO and \emph{STEREO} coronographs.]
{CME associated with the jet is observed by LASCO and \emph{STEREO} coronographs. The direction of the slit for this height--time analysis is shown in panel (a) with red dashed line.}
\label{allcme}
\end{figure*}

From  $\approx $ 21:16 UT, \emph{STEREO}--B observed the cool counterpart of a jet in 304 \AA\ above the western limb.
The full-disk image of AIA 304 \AA\ and \emph{STEREO}--B EUV 304 \AA\ is presented in Figure \ref{scc}. The highest visible peak of the jet is indicated with a circle at the solar limb which is used to get the read jet speed. Figure \ref{allcme} (panel (c)) showed the locations of \emph{STEREO} satellite, the Earth and the Sun. With the aid of SCC$_{-}$MEASURE procedure, we get the real speed and propagation direction of the jet by clicking on the same feature in AIA 304 \AA\ and in \emph{STEREO}--B 304 \AA\ image. The real jet-speed was 200 km s$^{-1}$ towards the north--east (longitude = -18$^{\circ}$, latitude = 19$^{\circ}$) direction. However, this correction can be only applied to the second stage of the jet when it was propagating towards the north--east direction, because we do not have the stereoscopic observations for the early stage of the jet.

\begin{figure*}[ht!] 
\centering
\includegraphics[width=\textwidth]{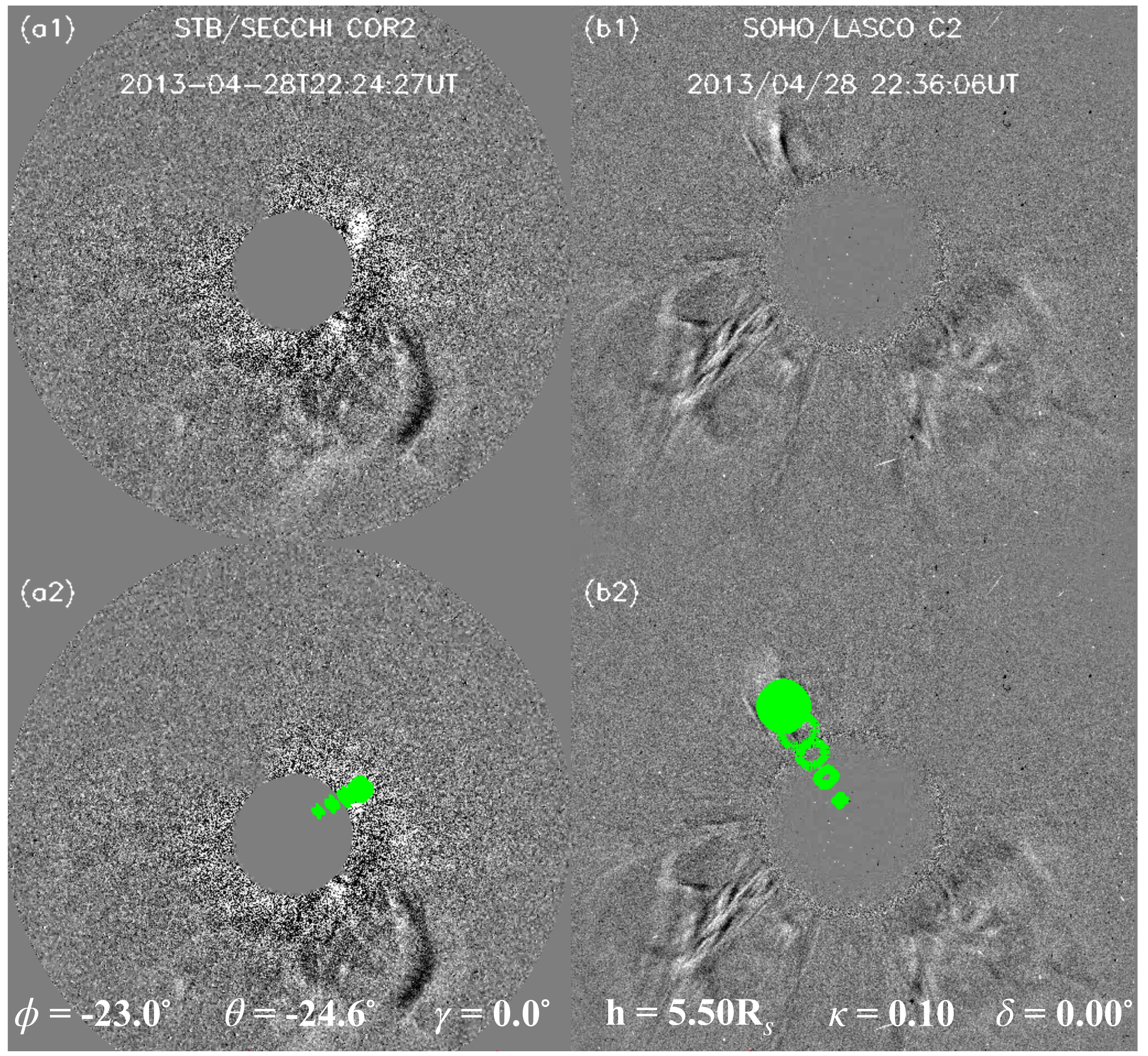}
\caption[CME associated with the jet eruption analysed with 
 GCS model.]{CME associated with the jet eruption analysed with 
 GCS model, which indicates the direction of propagation of the CME in \emph{STEREO} COR2, and LASCO C2 with a speed of 450 km s$^{-1}$.}
\label{gcs}
\end{figure*}
\subsection{Kinematics of the  CME}
\label{cme_jet}
The associated CME was well observed with {\it SOHO}/LASCO and \emph{STEREO}--B COR1 and COR2 coronographs,  as shown in Figure \ref{allcme}. The CME is not a typical one. It is narrow (width $\approx$ 25$^{\circ}$), and likes a giant jet in the corona, no matter from which perspective the CME was viewed. The jet--CME association is very much evident in \emph{STEREO}--B observations (Figure \ref{allcme}  (e)). For the continuous tracking of the solar jet in EUV channel (304 \AA) and the CME in coronagraphs, we put a slit in the jet--CME direction in \emph{STEREO}--B EUV 304 \AA, COR1 and COR2. The direction of the slit is shown in Figure \ref{allcme} (a). The continuous spatial and temporal correlation between the jet and the CME is presented in Figure \ref{allcme} (e).  The front of the CME is much higher than the jet front and the separation
between them is due to the expansion of the CME, causing the speeds of their fronts are different. If extrapolating them back to the solar surface, they almost originated from the same time.

To reduce the projection effect, we use the GCS model to get the real kinematic properties of the CME. The GCS model is developed to represent the FR structure of CMEs (\citealt{Thernisien2006}; \citealt{Thernisien2011}). It involves three geometric parameters: `h', the height of the leading edge, `$\kappa$', the aspect ratio, and `$\delta$', the half edge-on angular width, and three positioning parameters: `$\theta$', `$\phi$', and `$\gamma$', the Stonyhurst latitude and longitude of the source region, and the tilt angle of the source region neutral line respectively. The GCS model is usually used to study morphology, position, and kinematics of a CME based on the best fitting result of a CME transient recorded in white-light images. The ice-cream cone model is another model of CMEs, which composed of a ball that we call the ice-cream ball and circular cone tangent to the ball with a conic node on the solar surface (\citealt{Fisher1984}). The GCS model becomes equivalent to the ice-cream cone model when its parameter $\delta$, equals 0 (\citealt{Thernisien2011}). For our case study, we use the ice-cream cone model which is a simplified form of the GCS model and estimated the three-dimensional height and direction of the CME with LASCO C2, C3 and \emph{STEREO}--B COR2 images. The best-fitted GCS model is  displayed in Figure \ref{gcs}. The corrected CME speed from the GCS model comes out to be 450 km s$^{-1}$.

\begin{figure*}[ht!] 
\centering
\includegraphics[width=1.0\textwidth]{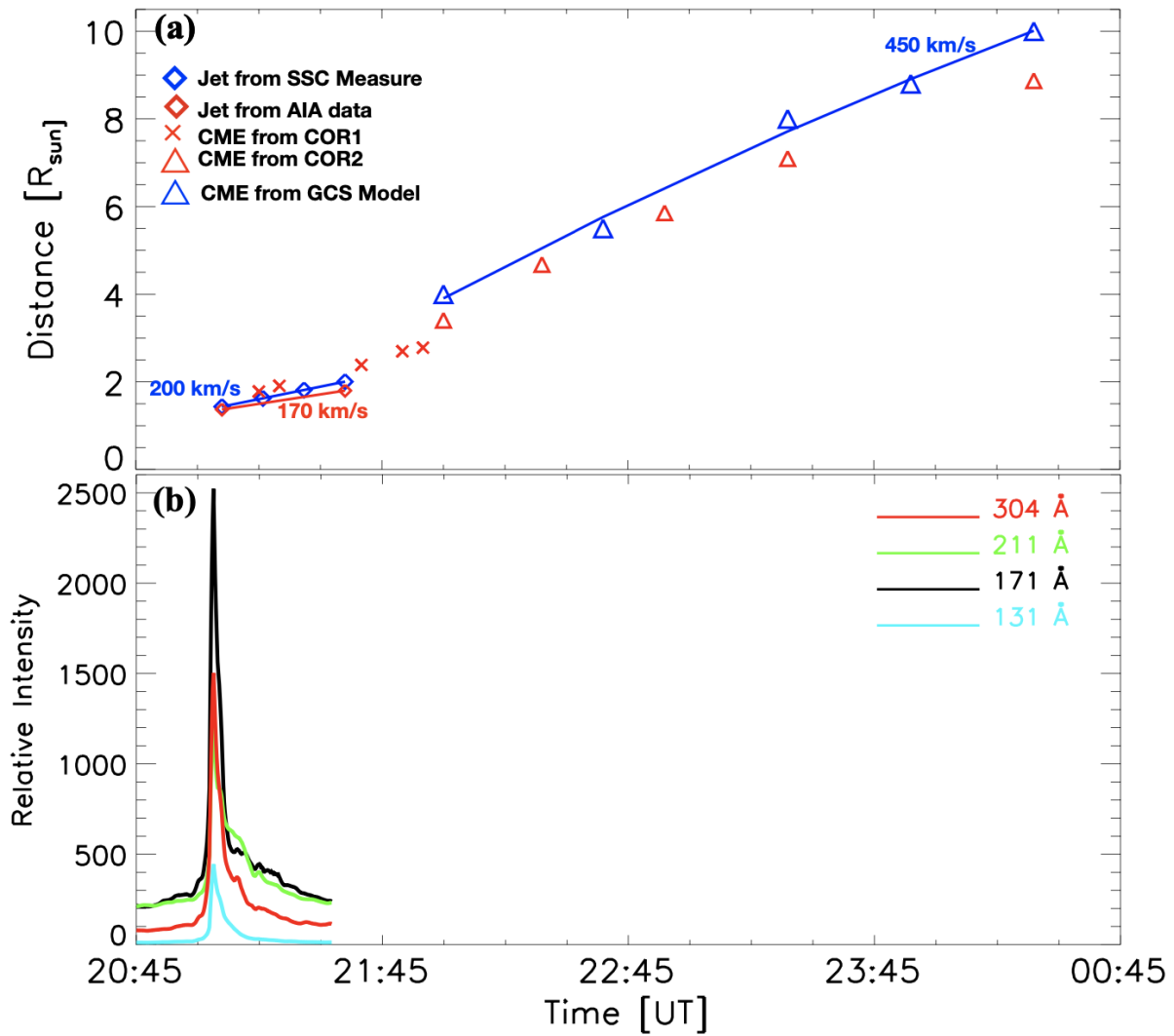}
\caption[Kinematics of the jet and the narrow CME.]
{Panel (a): The complete kinematics of the jet and the CME. The projection corrected speeds are plotted 
with blue color, while the red points are used for uncorrected data. Panel (b): Light curves for different wavelengths.}
\label{lightcurve}
\end{figure*}
Figure \ref{lightcurve} (a) depicts the complete kinematics of the jet and the CME with the different data points of various instruments.
We have corrected the projection effect for the jet and CME with SCC$_{-}$MEASURE and GCS model fitting, respectively. The corrected jet speed comes out to be $\approx$ 200 km s$^{-1}$ from SCC$_{-}$MEASURE associated with a CME of speed $\approx$ 450 km s$^{-1}$. The blue and red colors are used for corrected and uncorrected data points. This plot of temporal evolution shows the clear link between the jet and the narrow CME. Figure \ref{lightcurve} (b) shows the intensity
variation at the jet base. The impulsive peaks at the jet base show the jet peak time in various AIA wavebands. The enhancement in the light curve of EUV emission suggests that the energy injection was at the very beginning only, and not responsible for the continuous acceleration of the jet to escape from the Sun.

The speed of the CME (450 km s$^{-1}$) is much larger than that of a jet (200 km s$^{-1}$). This is because the speed of different parts of erupting structures are measured.
The speed of CME obtained from the \emph{STEREO} and LASCO observations 
is at its leading edge (v$_{front}$). It consists of the propagation speed of the CME  center (v$_{center}$) and the expansion speed (v$_{exp}$) of the CME, so v$_{front}$ = v$_{center}$ + v$_{exp}$. A cartoon illustrating the CME speed at the leading edge, which includes the CME propagation speed and expansion speed is given in \citealt{Yuming2015}.
\citealt{Gopalswamy2009} derived a relation between CME propagation speed and expansion which is confirmed in many studies till now (\citealt{Michalek2009}; \citealt{Makela2016}). With an approximation of the CME shape by a shallow ice cream cone, the relationship is defined as v$_{exp}$ = 2 v$_{front}$ sin$(w/2)$, where `$w$' is the CME width (25$^{\circ}$ in present case). Therefore v$_{exp}$ comes out to be 230 km s$^{-1}$ and v$_{center}$ should be 220 km s$^{-1}$. The jet triggered and developed into the CME and its trajectory should be followed by the CME center and not by the leading edge of the CME. Thus, the jet velocity (200 km s$^{-1}$) is comparable with v$_{center}$ (v$_{center}$ $<$ v$_{front}$). That explains the difference between the jet and CME speeds.

\subsection{Magnetic configuration of the jet source region}\label{magnetic}

\begin{figure*}[t!] 
\centering
\includegraphics[width=1.00\textwidth]{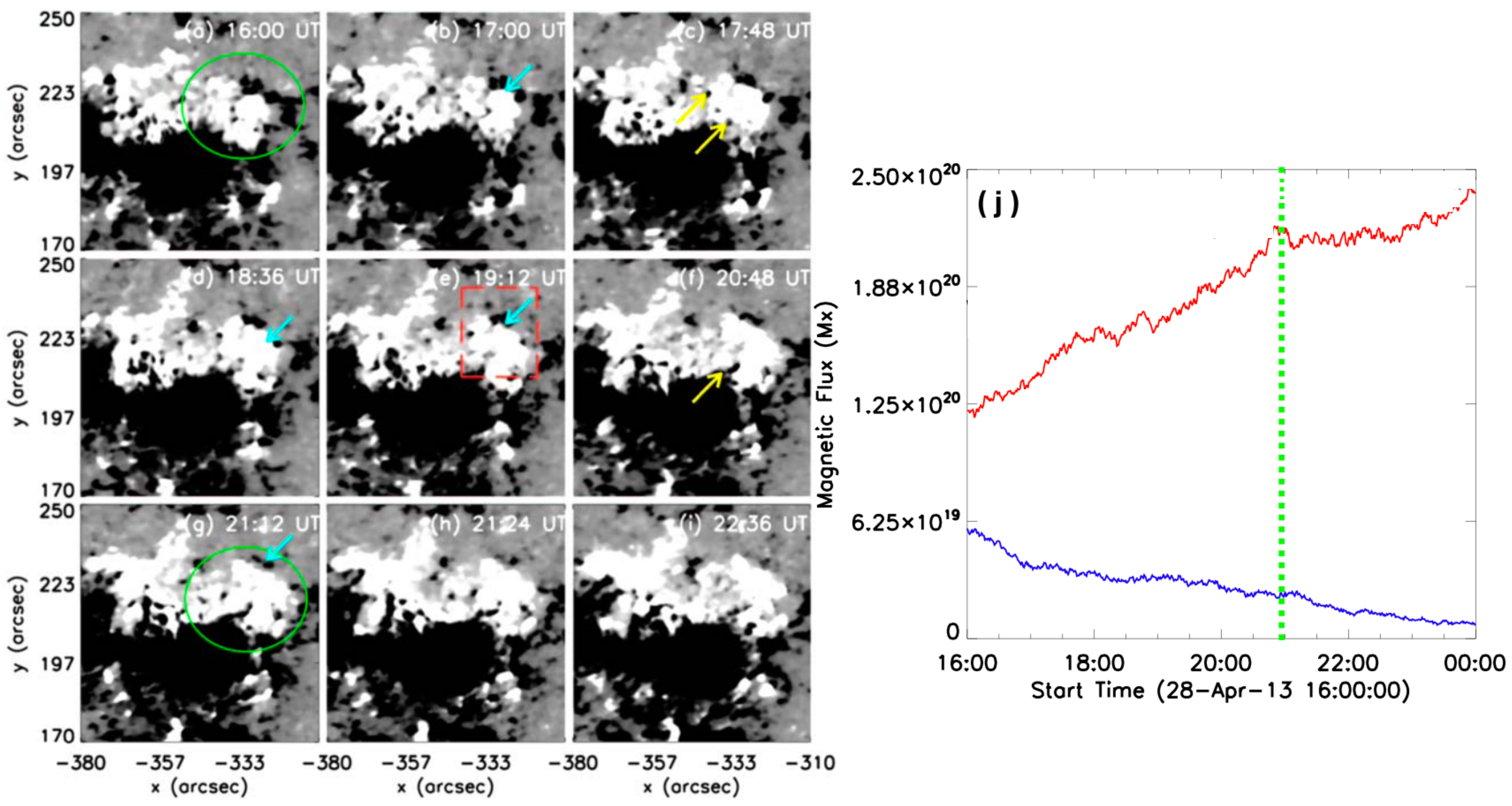}
\caption[The magnetic field configuration at the jet site]
{The left panels (a)--(i) show the magnetic field configuration at the jet site. Cyan and yellow arrows show the cancellation and emergence of negative magnetic polarity. Right panel (j) is the magnetic flux variation with time calculated at the jet source region.}
\label{hmi}
\end{figure*}

For a better understanding of the trigger mechanism of the solar jet, we did the magnetic field analysis of the source region using the HMI SHARP data of the AR 11731 on April 28, 2013. The continuous cancellation of the negative magnetic polarity by the emerging
positive magnetic spot is observed (Figure \ref{hmi}). The positive magnetic polarity ate the negative magnetic polarity which was already distributed in the jet source region (plotted inside the green circle in the panel (a) and (g)). Afterwards, small negative polarities emerges from the large negative ball and get cancelled with the big positive polarity area. The emergence of small negative polarities is shown with yellow arrows and the cancellation is indicated with cyan arrows. To look at the variation of the magnetic flux with time, we calculated the positive, and negative unsigned magnetic flux at the jet source region, which is indicated as the red rectangular box in panel (e). This is the same dimensional area we used to calculate the light curve in Figure \ref{lightcurve}(b). The flux variation with time in panel (j) shows that, there is a continuous cancellation and emergence of the negative magnetic flux (blue line) while the positive magnetic flux emerges throughout (red curve). The positive and negative magnetic flux show the simultaneous cancellation and emergence of magnetic polarities at the jet source region. The emergence of the positive magnetic flux dominated over the cancellation throughout. The initiation of jet time is shown with a green vertical line.
\begin{figure*} 
\centering
\includegraphics[width=1.0\textwidth]{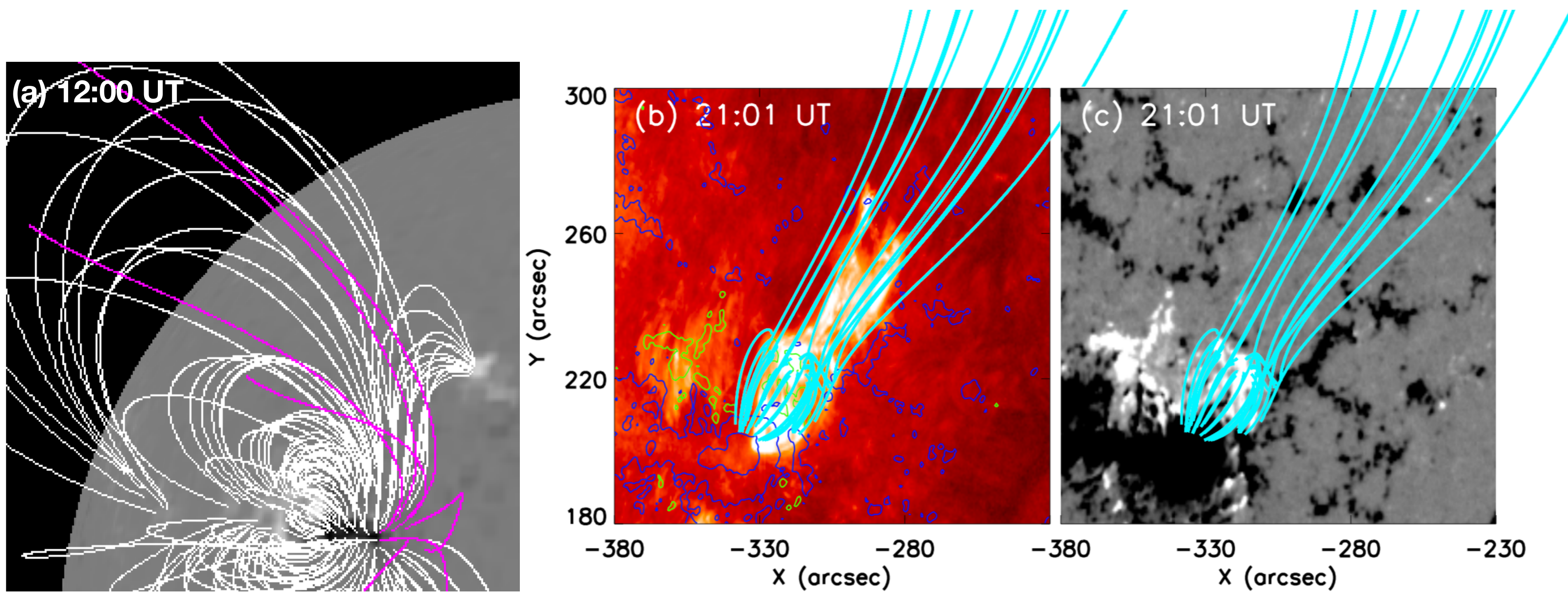}
\caption[The magnetic field extrapolation at the jet site.]
{The PFSS extrapolation of the large FOV is shown in panel (a). The white and pink lines are the closed and open magnetic field lines and the open lines are resembling of the jet propagation. Panel (b)  and (c) are showing the source region in AIA 304 \AA\ and HMI magnetogram.}
\label{pfss}
\end{figure*}
We analysed the magnetic topology at the jet location and applied two different methods of  potential extrapolation, one for the global overview of the jet eruption, and the other for the local view at the jet base region. We apply the Potential Field Source Surface (PFSS) model (\citealt{Schrijver2003}), to investigate the global magnetic topology  near the jet source region. This PFSS technique uses the HMI synoptic magnetic maps processed with a  software package available in SSWIDL.  PFSS technique is used to see the reconnection between close loops and open field lines because at large scale the corona is in potential state (\citealt{Schmieder1996}). The PFSS model for this case study is presented  in Figure \ref{pfss} (a), with open (pink) and close (white) magnetic field lines. These open field lines resemble exactly the path exactly the same as the jet, which was along the north direction in the beginning and deflected  towards the north--east afterwards.

To describe the magnetic topology of the jet base region, we extrapolate the coronal potential field  using the photospheric  LOS magnetogram as a boundary condition. The method based on the Fourier transformation (FT) method proposed by (\citealt{Alissandrakis1981}). The FT method requires the vertical component of the photospheric vector field as the input parameter. However, due to HMI vector magnetic field  limited field of view, the extrapolation hard to meet divergence--free condition. Hence, we cut a larger patch of the LOS magnetogram instead. As the AR is close to the central meridian, the LOS magnetic field could  represent the vertical field to a large extent. In the extrapolated magnetic field, we find open field lines coincide well with the extension direction of the jet shown in Figures \ref{pfss} (b) and (c).
We have tried for the NLFFF extrapolation, but it failed to reproduce the magnetic field topology of the AR. The field lines of the NLFFF did not resemble with the loops observed in the EUV passbands. It might because that the FOV of the photospheric vector magnetic field provided by HMI SHARP data is too small as it is available for significant AR patch of solar magnetic field. Hence the divergence-free condition is not completely satisfied in the extrapolation, which makes the NLFFF results unreliable. On the other hand, we mainly focus on the propagation of the jet, which is more likely to be relevant to the nearly potential, large scale magnetic field connectivity. Therefore we believe that the potential field extrapolation might be sufficient and the direction of the jet ejection is the same as of the
open magnetic field lines we have obtained from the potential field extrapolation.

  \section{Results and conclusion}
  \label{result}
     \label{dicussion}
In this chapter, the role of solar jets for triggering the large scale solar eruption is presented. Two different case studies are analysed, where in the first study the solar jet triggered a filament eruption with the largest geomagnetic solar storm of solar cycle 24 (\citealt{Chandra2017}) and in the second study the jet was directly associated with a CME (\citealt{Joshi2020ApJ}).
The main results of the study are as follows:

The initiation of filament eruption on March 14, 2015 and it's full eruption on March 15, 2015 was associated with  jet activity in the AR 12297. The decay index distribution suggests that on March 14, 2015 the filament first enters into the instability zone with a  push from a jet and after reaching some height it finds itself in the stability zone. Again on March 15, 2015 the filament enters into the instability zone and finally it erupts, when an another jet activity again hits it. The major part of filament which had not been destroyed on March 14, 2015 was activated on March 15 but could not erupt. Therefore it was a failed eruption. The coronal magnetic field calculation shows evidence that the decay index at the filament location is below the threshold of the torus instability and hence the filament fails to erupt. The observation of the same sign of the twist/helicity in the chromosphere, higher solar atmosphere and in the magnetic cloud evidence the conservation property of the helicity.

From the decay index distribution, we establish that the western section of the filament before the separation and the western filament after the separation are relatively low (30 Mm) and were
located in the zone of stability within the coronal magnetic field. The eastern section of 
the filament was less stable because on the one hand it is higher and on the other hand the 
decay index in this area was also higher. That is why, that the disturbance (solar jet) coming 
from inner parts of the AR led to the partial and failed eruption of the eastern 
section of the filament, which resulted in the separation of the filament into two parts. 
The erupted eastern section of the filament found a new equilibrium position at a 
greater height. In the new position of the eastern section, this height (about 80 Mm), considered as the southern filament F$_1$, was within the zone of stability for the decay index 
threshold n$_c$ = 1.5 and on the edge of stability for the decay index threshold n$_c$ = 1. The 
next the disturbance from the AR easily causes the start of the eruption of the 
southern filament and this eruption is full because there is no a zone of stability at heights 
above 100 Mm. So, the eastern section of the filament showed the two step eruption with the 
metastable state at the height of 80 Mm for 12 hrs. Possibly it could stay there longer if the 
strong disturbance (second jet) did not come from inner parts of the AR. The western filament was deep within the zone of stability therefore it did not erupt despite the strong activation by 
the energetic disturbance.

 The observed jet speed on April 28, 2013 computed using the multi-view point observations is about 200 km s$^{-1}$ at the height of 2 R$_\odot$. The escape velocity computed at the height of 2 R$_\odot$ comes $\approx$ 430 km s$^{-1}$. Therefore, we conclude that the complete jet cannot be escaped from the solar surface. This could be the reason we have observed the backward motion of the jet material from the propagation direction towards the source region. Even the jet speed is lower than the escape speed, we observed the clear CME associated with the jet by all the space-borne coronagraphs. The possible mechanism for the jet continuously accelerating to reach the escape speed and form the narrow CME is that the falling back material makes the upward material of the jet moving faster to keep the momentum of the whole jet conserved. We concluded that the observed speed of the CME is containing the speed of the CME center and the expansion speed, and is much larger than the jet speed, because the different parts of the erupting structures are being measured. The speed of CME center (the trajectory followed by the jet) is 220 km s$^{-1}$ and  is equivalent to the speed of the jet (200 km s$^{-1}$). This provides a clear evidence of the jet-CME association.

For the magnetic configuration at the jet origin site, two views are popular. One is the magnetic flux emergence observed in many observations and also proposed in the MHD simulations (\citealt{Shibata1992}; \citealt{Moreno2013}; \citealt{Ruan2019}; \citealt{Joshi2020MHD}). Another is the magnetic flux cancellation, which is also reported in the observations MHD simulations (\citealt{Pariat2009}; \citealt{Chandra2017}; \citealt{McGlasson2019}).
We have observed that there is a continuous emergence and cancellation of the negative magnetic flux and the positive flux is emerging throughout. Therefore, we believe that both the flux emergence and the cancellation are responsible in this case. We also observed the rotation in the jet material on April 28, 2013 when it is propagating towards the north direction 
from the source region. The untwisting of the jet suggests the injection of helicity to the upper atmosphere.

For the future study, to look forward for finding the clear {\it in situ} measurements for such association of filament eruptions with solar jets and jet--like CMEs from the newly launched Parker Solar Probe will be a major field of interest which will help to contribute for unwinding the mystery of coronal heating problem.

\chapter{Transfer of twist to a solar jet from a remote stable magnetic flux rope}\label{c6}

\ifpdf
    \graphicspath{{Chapter6/Fig_chapter6/}{Chapter6/Fig_chapter6/}}
\fi
\section{Introduction}
An overall common property for solar jet is to exhibit a twist or rotation
 (\citealt{Raouafi2016}). The twist of the jet may be due to  helical motions
(\citealt{Patsourakos2008}; \citealt{Nistico2009}). Twisting motions have been found  in  a large velocity range of jets or surges (\citealt{Chen2012}; \citealt{Hong2013}; \citealt{Zhang2014}).
In the study done by  
\citealt{Schmieder2013}, a jet analysis
revealed a striped pattern of dark and bright strands propagating along the jet, as well as apparent damped oscillations across the jet. They concluded that this
is suggestive of a (un)twisting motion in the jet, possibly an Alfv\'en wave. Spectroscopic  data also provide  signatures for detecting the twist in  jets.  
For example, blue and red shifts observed along the axis of a jet in H$_\alpha$ as well as in Mg II lines  were interpreted as confirmation of the existence of twist along the  jet (\citealt{Ruan2019}).

Spectroscopic and imaging  observations of small-scale events reveal bidirectional flows in transition  region lines at the jet base  which could correspond to an explosive reconnection (\citealt{Li2018}; \citealt{Ruan2019}).
There are different conditions for the  magnetic configuration of an AR to  trigger magnetic reconnection.  We may quote three types of conditions: magnetic flux emergence (\citealt{Archontis2004, Archontis2005}; \citealt{Moreno2008}; \citealt{Torok2009}; \citealt{Moreno2013}), magnetic flux cancellation (\citealt{Priest2018}; \citealt{Syntelis2019}), and magnetic instability (\citealt{Pariat2010,Pariat2015,Pariat2016}). The first two mechanisms predict hot and cool jets simultaneously. 
However, the presence of surges and jets is
 not frequently reported. 
 Some papers report on the X-ray jets observed by Yohkoh and associated with a surge  (\citealt{Schmieder1995}; \citealt{Canfield1996}; \citealt{Ruan2019}).
Radiative MHD simulations based on flux emergence  (\citealt{Nobrega2017,Nobrega2018}) as well as the flux cancellation model (\citealt{Syntelis2019}) show that surges can exist at the same time with hot jets. The cool plasma is advected  over the emergence domain without passing near the reconnection site and then flows  along the reconnected magnetic field lines. These models fit with the observations of X-ray jets { observed with} Hinode and with H$_\alpha$ jets from  the Swedish 1-m Solar Telescope (SST) (\citealt{Nobrega2017}).  

Recently, \citealt{Joshi2020MHD} presented a case-study of collimated hot jets and associated cool surges which fit in perfectly with the simulation of jets formed by flux emergence. The double-chambered structure found in the observations corresponds to the cool and hot loop regions found under the reconnection site in the models of \citealt{Moreno2008}.  In the model of \citealt{Wyper2019}, the overlying magnetic field is, in fact, expelled by a gentle reconnection above the closed AR via   a breakout mechanism before the instability occurs. \citealt{Pariat2015}, and \citealt{Wyper2019} show the importance of the inclination of  jets favouring the jet onset for $\theta$ = 0 - 20 degrees. 
These models are based on the instability of the system; a FR formed by shear  under the reconnection point is the trigger of the helical jet. However, based on several observations, it becomes clear that the twist is not present before the reconnection but the twist of the jet is transferred during the reconnection. For example, in  \citealt{Ruan2019}, the twist was transferred from twisted overlying  magnetic field lines remnant of the eruption of a filament two hours before the onset of the jet.
The null-point is the favourable location for the occurrence of magnetic reconnection.
 
 \citealt{Wyper2019} recently showed that reconnection can be in a region where the magnetic  field lines are tangent to the photosphere. 
This kind of region is called BP region. It  favours reconnection as a mechanism for initiating jets and surges (\citealt{Mandrini2002}; \citealt{Chandra2017}; \citealt{Zhao2017}). 
In these studies, the magnetic topology was derived by LFFF or NLFFF magnetic field  extrapolations in the corona (\citealt{Mandrini2002}; \citealt{Chandra2017}) or by directly analysing    the   observed magnetic field vector maps (\citealt{Zhao2017}).  
The  occurrence of the reconnection was clearly  taking place in the  BP regions.
It was also recently  proposed that the trigger of jets can be due to the eruption of mini-filament at the jet base (\citealt{Sterling2016}). That model fits well with the blowout jets where the  entire region below the dome of reconnection is expelled during the eruption (\citealt{Moore2010}).

In this chapter, the observations of a twisted jet, surge, and a mini-flare observed in multi-wavelengths, and with the {New Vacuum Solar Telescope} \citep[NVST,] [] {NVST2014} ground based telescope are analysed.

\section{Observations of the twisted jet}
\label{part1_ch6}
\subsection{EUV and chromospheric observations}
\label{chap6:AIA} 
In the AR NOAA 12736 a  jet along with a surge is well observed in the multi-wavelength  filters of AIA aboard SDO. AIA data consists of a sample of filters with passbands  centered at different EUV lines.
The IRIS FOV was focused on AR NOAA 12736 and the pointing of
the telescope was at 709$^{\prime}$$^{\prime}$, 228$^{\prime}$$^{\prime}$
with a FOV of
60$^{\prime}$$^{\prime}$ $\times$68$^{\prime}$$^{\prime}$ 
for the slit-jaw images (SJIs).  
The observational characteristics
are presented in Table \ref{tab:table1}.  We used the   1330 \AA\ and 2796 \AA\  SJIs  for this
study. 
There was no data for the   IRIS SJI  Si IV 1400 filter.
The SJI 1330 \AA\ includes the C \footnotesize{II}
\normalsize line formed at T= 30000 K, and the SJI 2796\,\AA{}
emission mainly comes from the Mg \footnotesize{II} \normalsize k
line. The Mg \footnotesize{II} \normalsize h and k lines
are formed at chromospheric temperatures, that is, between 8000 K and
15000 K (\citealt{Pontieu2014}; \citealt{Alissandrakis2018}).  The co-alignment
between the different optical channels of IRIS was achieved
by using the {\it drot\_map} in solar software to correct the differential
rotation.  The SJIs in the broadband filters (1330\,\AA{}, and
2796\,\AA{})
 were taken at a cadence of 14 s. IRIS performed medium  coarse rasters  of 4 steps from 01:43:27 UT to
02:42:30 UT on March 22, 2019. 
The raster
step size is 
2$^{\prime}$$^{\prime}$ so each spectral raster spans a field of view
of 6 $^{\prime}$$^{\prime}\times$62
$^{\prime}$$^{\prime}$.  The nominal spatial resolution is 0.$^{\prime}$$^{\prime}$33. 
IRIS  provides line profiles in  Mg II k 2796.4\,\AA{} and  Mg II h 2803.5\,\AA{},  Si IV (1393.76 \AA, 1402.77 \AA) and C II  (1334.54 \AA, 1335.72 \AA) lines  along four slit positions. 
Calibrated level 2 data are used in this study, with corrected dark current subtraction (\citealt{Pontieu2014}). 

\subsection{Magnetic field observations}
\label{obs_HMI}
 The longitudinal magnetic field is provided by the HMI team with a cadence of 45 s and a pixel size of 0.5$\arcsec$. To obtain the magnetic field vectors in full, we inverted the HMI level-1p IQUV data, averaged on a  12 minute cadence, by applying the Milne-Eddington inversion code UNNOFIT (\citealt{Bommier2007}). We selected  a large  area   covering the AR 12736 and applied a solar rotation compensation to select the same region over more than six hours of observation. We thus treated 22 maps from March 21, 2019 at 23:00 UT to March 22, 2019 at 03:12 UT and three later maps of the same region from 05:00 UT to 05:24 UT. The specificity of UNNOFIT is that a magnetic filling factor is introduced to take into account the unresolved magnetic structures as a free parameter of the Levenberg-Marquardt algorithm that fits the observed set of profiles with a theoretical one. However, for further application, we used only the averaged field, that is, the product of the field with the magnetic filling factor, as recommended by \citealt{Bommier2007}. The interest of the method lies in a better determination of the field inclination. After the inversion, the 180$^{\circ}$ remaining azimuth ambiguity was resolved by applying the ME0 code developed by Metcalf, Leka, Barnes, and Crouch (\citealt{Leka2009}) and available at \url{http://www.cora.nwra.com/AMBIG/}. After resolving the ambiguity, the magnetic field vectors were rotated into the local reference frame, where the local vertical axis is the $o_z$ axis.

\subsection{H$_\alpha$ observations}
The H$_\alpha$ observations were taken with the NVST telescope in China, pointed at the AR  12736 at N09 W60 on March 22, 2019 from 00:57:00 UT to 04:37:00 UT. We used the line--center H$_\alpha$ observations at 6562.8 \AA \ that were obtained in a field of view 
(FOV) of 126$\arcsec$ $\times$ 126$\arcsec$ with a cadence of 29 seconds. 
It displays  the H$_\alpha$ fine structures and  shows their evolution very well.
For the current analysis, we used the level 1+ data.
To focus on the jet region, we cut the data cube after rotating it  with north upwards, as in the space data (AIA  and HMI observations) and used the data of FOV of  65$\arcsec$ $\times$ 65$\arcsec$.

\begin{table} 
\caption{
IRIS observation of AR NOAA 12736 on March 22, 2019.}
\bigskip
\centering
\setlength{\tabcolsep}{16pt}
\begin{tabular}{llll}
\hline
Location& Time & Raster& SJI\\
 & (UT) &  &
\\
\hline
x=709$\arcsec$& 01:43 - &  FOV: 6$\arcsec$ $\times$ 62$\arcsec$& FOV: 60$\arcsec$ $\times$ 68$\arcsec$\\
 y=228$\arcsec$& 02:42 &  Steps: 4 $\times$ 2$\arcsec$& C II 1330 \AA \\
 & &  Spatial & Mg II 2796 \AA \\
 & &  Resolution: 0.${\arcsec}$33 & Time \\
 & &  Cadence: 3.6 s & Resolution: 14 s \\
 
\hline
 \end{tabular} 
\label{tab:table1}
\end{table} 
\begin{figure*}[ht!]
\centering
\includegraphics[width=1.0
\textwidth]{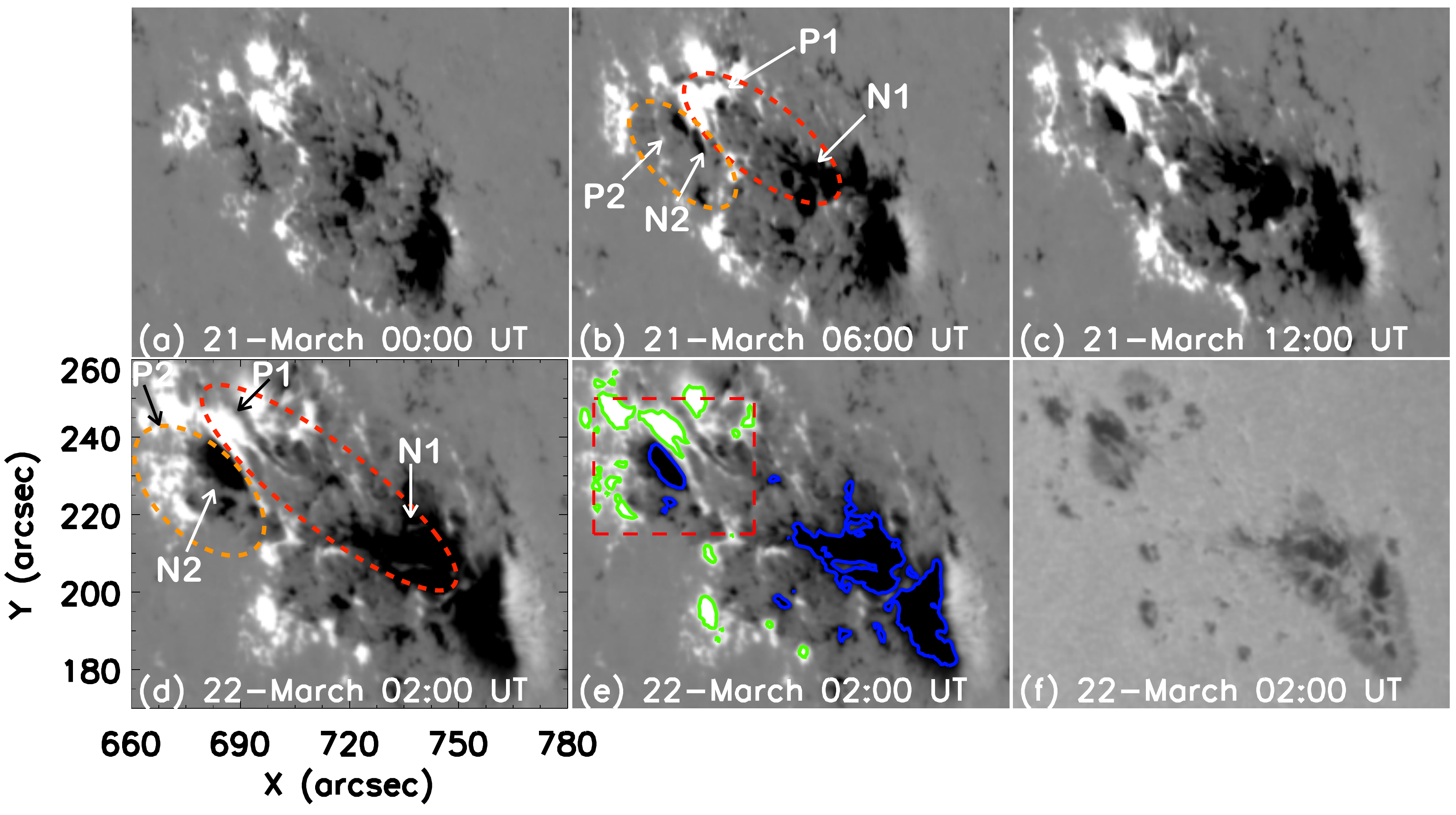}
\caption[HMI longitudinal  magnetograms  of AR  NOAA 12736 showing the evolution of the magnetic polarities.]
{Panels (a--e): HMI longitudinal  magnetograms  of AR  NOAA 12736.  Reconnection is occurring between the two large emerging flux areas EMF1 (P1, N1) and EMF2 (P2, N2) encompassed in the  two ovals drawn in panels (b) and  (d).}

\label{chap6_fig1}
\end{figure*}

\begin{figure*}[ht!]
\centering
\includegraphics[width=1.0
\textwidth]{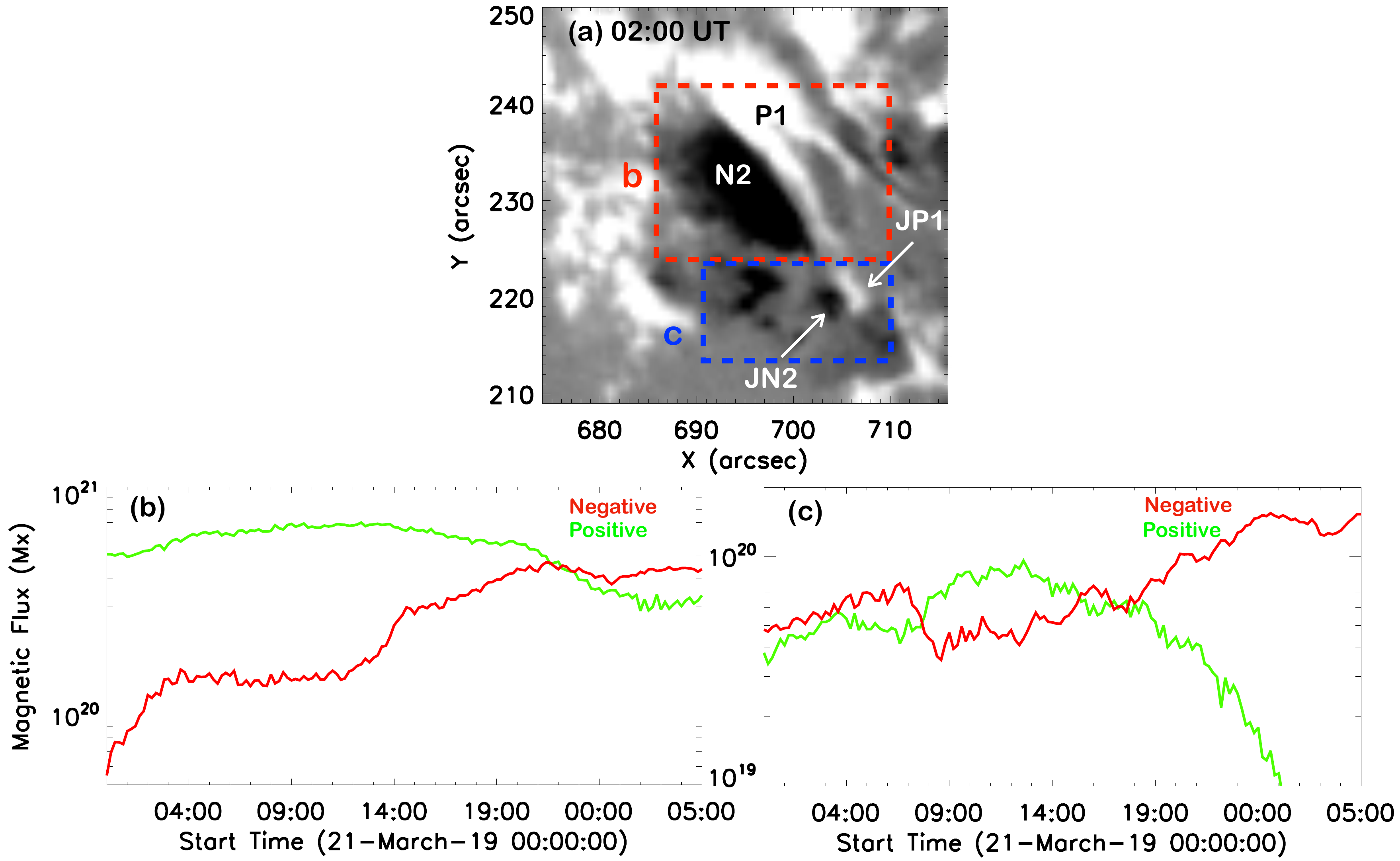}
\caption[Magnetic flux cancellation in two areas including  the major  bipole and the small jet bipole.]
{Magnetic flux cancellation in two areas including  the major  bipole (P1, N2) (red box (b))  and the small jet bipole  (JP1, JP2) (blue box (c)) respectively. Panels (b and c): variation of the magnetic flux in the red  and blue boxes.}
\label{flux}
\end{figure*}
\section{Birth of  the AR}\label{obs}
\label{sec:mag} 
A  mini-flare (B6.7 X-ray class) and its associated jet was initiated in AR NOAA 12736 located  at N09 W60 on March 22, 2019 around  02:02 UT.
The AR 12736 was emerging progressively since March 19, 2019. On March 21, we note two emerging flux regions elongated along the  north-east to south-west  direction (the ovals in Figure \ref{chap6_fig1} (b and d)). The first emerging flux region (EMF1) is the main component of the AR, with negative leading polarity and positive following polarity. The second emerging flux region (EMF2) consists of  many fragmented negative polarities which are travelling very fast as the emerging flux  is  expanding  towards  south  and  squeeze the positive polarity of EMF1.  Consequently, a very high magnetic field  gradient is observed perpendicularly to the polarity inversion line (PIL) between the squeezed polarities: `P1' positive polarity belonging to EMF1 and `N2' negative polarity belonging to EMF2 (the red box in Figure \ref{chap6_fig1} (e)). The negative polarity N2 is surrounded by positive polarities P1 on the right side and P2 on the left side and top. This topology is classical with the aim of getting a null point, as we see discuss further on in this chapter.  In the HMI observations, we note the fast sliding motion of  negative polarity N2 towards  the south and the motion of  positive polarity P1 in the opposite direction, which  creates a strong shear between them. Along this PIL, we distinguish that at the time of the flare observations, we principally observe two bipoles emanating from the two EMFs in the diagonal of the box (NE-SW):  a large north bipole (P1, N2) and a very tiny bipole  (JP1, JN2) in the south, which was detached  progressively from the northern N2 polarity a few hours before (explained in Section \ref{reconnection}).  We computed the flux budget for these two bipoles and we found a significant decrease of the positive flux in the two  boxes, each of them including a bipole (Figure \ref{flux}). We interpret these decreases by magnetic cancelling flux. The tiny bipole is labeled  with `J' like ``jet" because this is the
 the location where the jet took place.

\begin{figure*}[!ht]
\includegraphics[width=1.0
\textwidth]{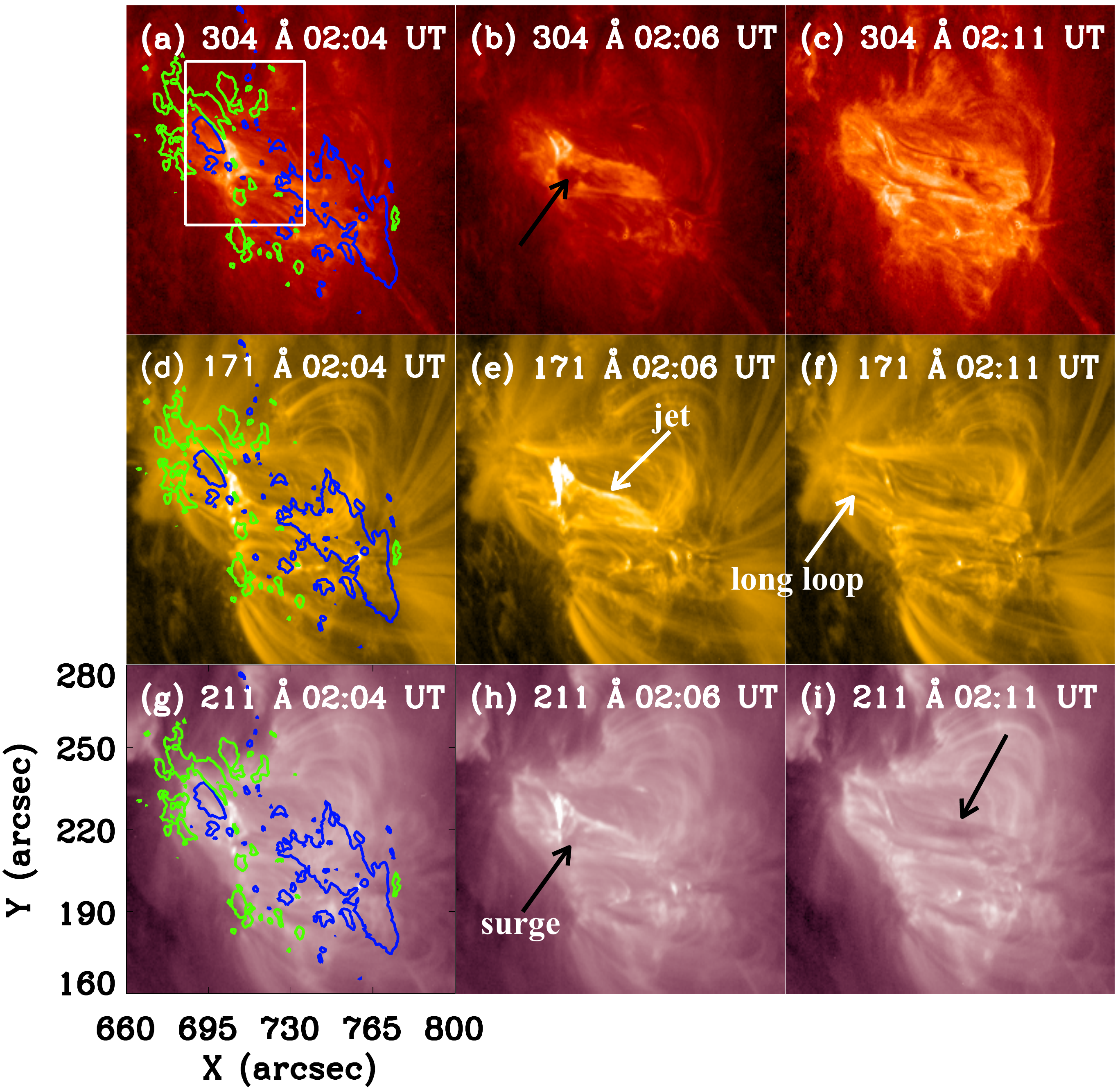}
\caption[Solar jet  and surge observed in different AIA/EUV channels (304 \AA, 171 \AA, and 211 \AA) on March 22, 2019.]
{Solar jet  and surge observed in different AIA/EUV channels (304 \AA, 171 \AA, and 211 \AA) on March 22, 2019. The black arrows in (b) and (h) indicate a dark area corresponding to a surge, the white arrow points the jet in (e).}
\label{AIA1}
\end{figure*}

\begin{figure*}[!ht]
\includegraphics[width=1.0
\textwidth]{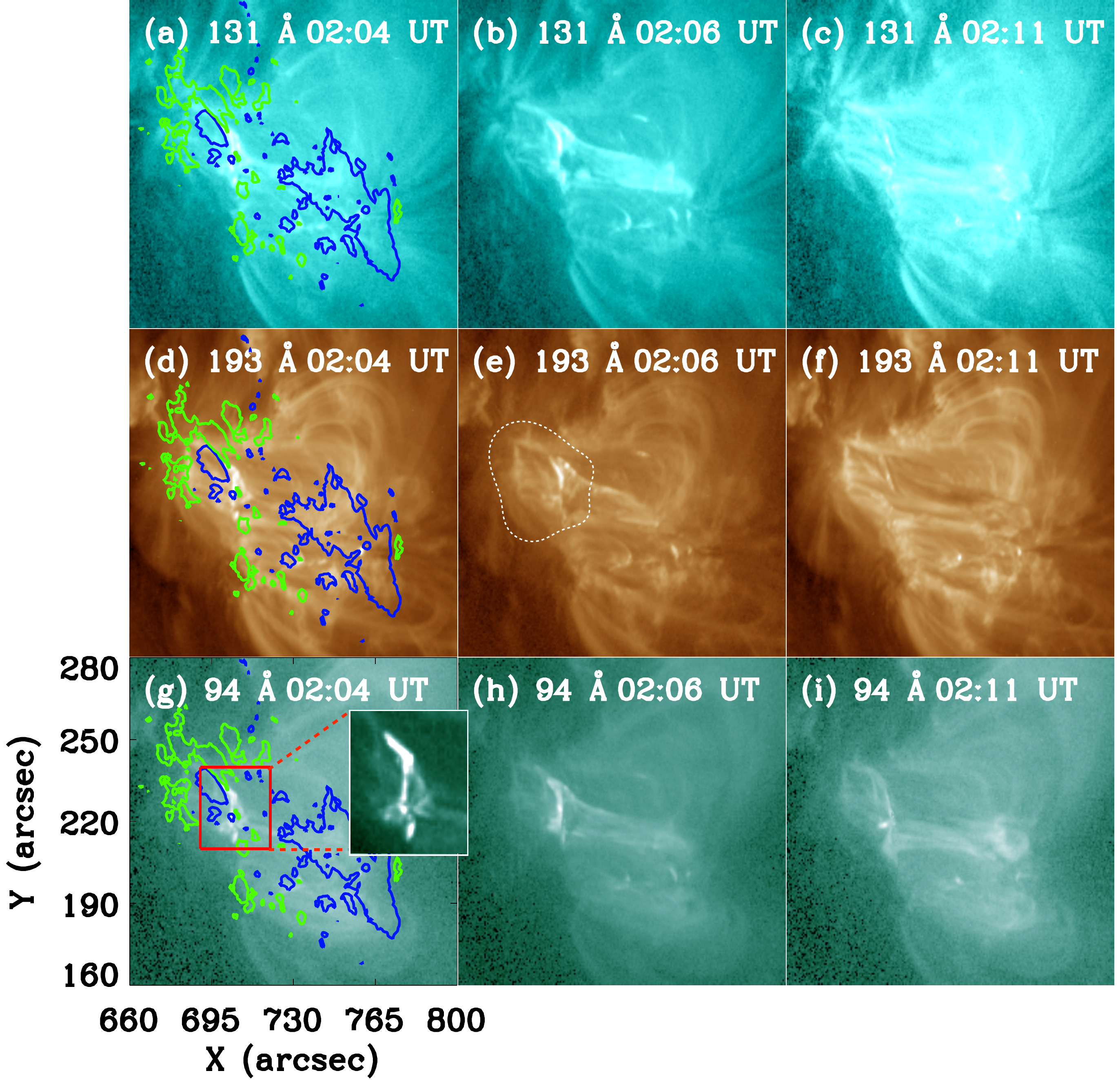}
\caption[Solar jet  and surge observed in the other AIA/EUV channels (131 \AA, 193 \AA and 94 \AA).]{Solar jet  and surge observed in the other AIA/EUV channels (131 \AA, 193 \AA and 94 \AA). Panel (e) shows a dome structure at the jet base highlighted by a white dashed contour.}
\label{AIA2}
\end{figure*}

 \label{sec:AFS}
  The AIA observations cover the  AR and the full development of the jet in  multi-temperatures provided by all the sample of  AIA filters from 304 \AA\ to 94 \AA, all along the range of temperatures from 10$^5$ K to 10$^7$ K (Figure \ref{AIA1} and \ref{AIA2}).  Contours  of  longitudinal magnetic fields ($\pm$ 300 Gauss) are overlaid on the AIA images to specify the location of the small bipole JP1-JN2 at the jet base.
 Arch filament system (AFS) are well visible over the  two emerging flux EMF1 and EMF2 with cool and hot low lying loops joining the positive  and negative polarities for each of them, P1 and N1 on the  west side and  P2 and N2 on the east side.  
 Filaments  belonging to these AFS are 
 particularly  visible as dark structures due to absorption mechanism in 171 \AA,   131 \AA,  and 193 \AA\ at 02:06 UT (Figure \ref{AIA1} panel e and Figure \ref{AIA2} panels b, e). These filaments are parallel to each other, oriented more or less NE to SW
 from P1 to N1 and from P2 to N2 in the direction of the extension of each  EMF.   
  They do not lie along any PIL and, therefore, they do not correspond to the usual definition of filaments; rather, they  are  more or less  perpendicular to the PIL in each EMF.  Filters AIA 171 \AA\, and 193 \AA\, are good proxies for detecting cool structures visible in H$_\alpha$. 
  At these wavelengths, the EUV emission is absorbed by the  hydrogen and helium continua  (\citealt{Anzer2005}). The opacity of the hydrogen and helium resonance continua at 171 \AA\, is almost two orders of magnitude lower than the  Lyman continuum  opacity  at 912 \AA\, 
  and thus similar to the H$_\alpha$ line opacity (\citealt{Schmieder2004}). 
  We confirm the presence of H$_\alpha$ filaments/AFS by looking at the  H$_\alpha$ images (Figure \ref{NVST}). The two AFS over the two EMFs are well identified.  On the west side the AFS are
  dense and long, with some  narrow arch filaments overlying the bright corridor of the PIL (N2-P1)  and the dome of EMF2 before the burst  (panel b). The AFS over EMF2 on the east side have a fan structure with an anchorage all around the negative polarities N2  and the other in the positive polarities P1 and P2. It gives the impression of an `anemone' structure which is frequently observed for jets  triggered  by emerging flux  (\citealt{Shibata1982}; \citealt{Schmieder2013}; \citealt{Joshi2020MHD}).
In order to follow the jet development with AIA,  we focus on the FOV covering 
  the two bipoles identified in the previous section.
 
 \subsection{Morphology of the twisted solar jet} 
\label{jet_AIA} 
\begin{figure*}[!t]
\includegraphics[width=1.0
\textwidth]{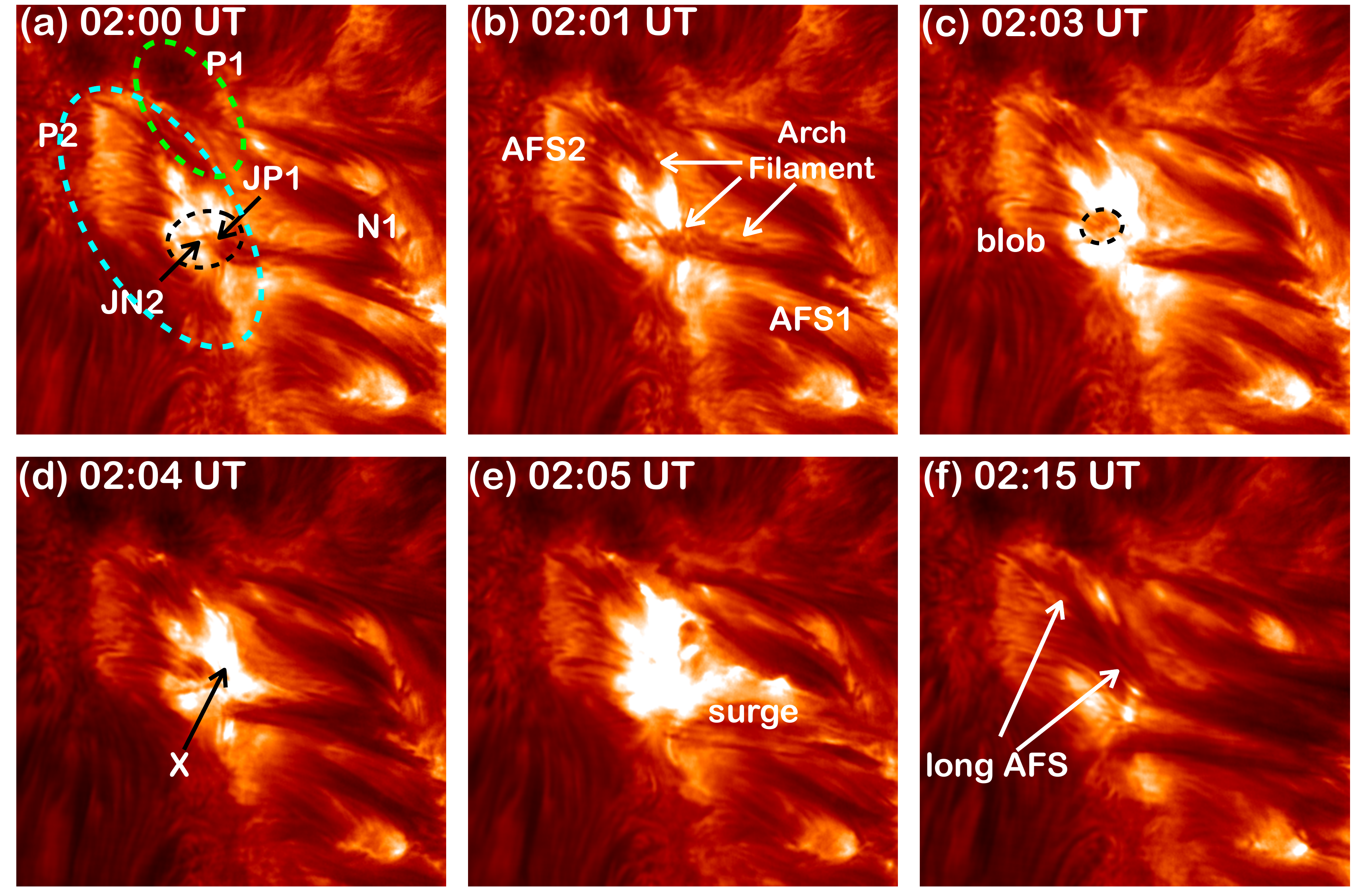}
\caption[H$_\alpha$ line center observations of the AR NOAA 12736 with the NVST telescope.]
{H$_\alpha$ line center observations of the AR NOAA 12736 with the NVST telescope for four times before (a-d), during (e), and one after (f) the surge extension. }
\label{NVST}
\end{figure*}

 It is interesting to see that activity had started before the onset of the jet, with very bright north-south tiny   threads observed above the PIL between the two EMFs and, more precisely, between the part of the  PIL in the northern bipole (P1-N2)  and continuing into the south tiny bipole (JP1-JN2)  around 02:01 UT until 02:04 UT. The bright signature around  a  dome structure overlying the EMF2 at the jet base is highlighted by a white dashed contour around it, as seen in Figure \ref{AIA2} (e). This dome is highlighted by the small fibrils with an  asymmetrical anemone shape visible in the NVST images (Figure \ref{NVST} b). Along the west side of the dome, the brightening with an north-south arch-shape is visible  in all AIA channels
 before the  burst  indicates already the presence of  hot plasma (T between 10$^4$ to 10$^6$ K). This region between the two EMFs corresponds to  {\it QSLs} (\citealt{Demoulin1996}), where  a strong high electric current  develops and heats the plasma, as shown here  by the arch-shape brightening. This is subsequently confirmed with the analysis of the photospheric vector magnetic field  maps in Section \ref{reconnection}. These {\it QSLs} have been calculated in \citealt{Yang2020} and are well-identified in this region. QSLs are robust structures but their  localisation is not commonly defined with any substantial accuracy (\citealt{Dalmasse2015}; \citealt{Joshi2019}). In our case, the moving polarities is a problem for the exact localisation  of {\it QSLs}.
 
 \begin{figure*}[t!]
\centering
\includegraphics[width=1.0
\textwidth]{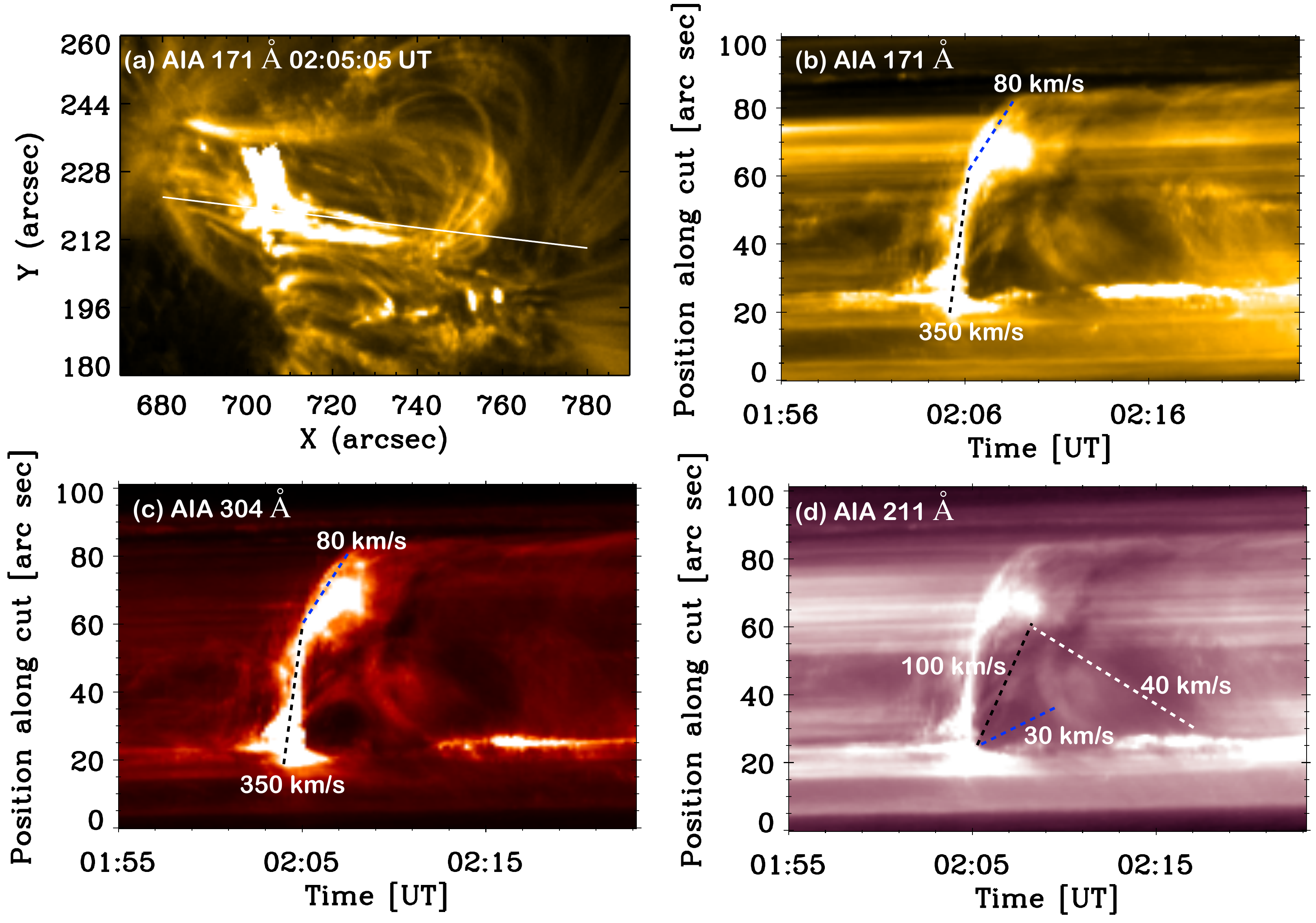}
\caption[Height-time profile for the jet in different AIA wavelengths.]
{Height-time plots for the jet in different AIA wavelengths. The surge is erupted with a maximum speed of $\approx$ 100 km s$^{-1}$ and the dark material came back with a speed of $\approx$ 40 km s$^{-1}$.}
\label{timeslice1}
\end{figure*}
 
At the same time (02:04 UT), a jet with two branches inserting a surge is also observed.
  Dark absorbing material is visible 
   at $\approx$ 02:02 UT, resembling a blob with no really  defined shape in the southern part of the arch-shaped brightening (Figure \ref{NVST} panel c). It then extends to the north and, finally, goes along the jet direction. The surge appears as a dark area in the images because of  the absorption of the UV emission. Therefore, the dark part observed in 171 \AA\, and in the other AIA filters, that is, (211, 193, 94 \AA) should  correspond to cool plasma as seen in H$_\alpha$ (\citealt{Schmieder2004}). It is why 
   we call it a `surge' (Figure \ref{AIA1} and \ref{AIA2}). The surge appears as a bright structure in the NVST images (Figure \ref{NVST} e). However,   the surge   consists of  cool plasma because H$_\alpha$ formation  temperature is lower than 1.5 $\times$ 10$^4$ K.  The jet base  on the east  side  of the EMF1  is extended  along a more or less  north-south  direction, along
    the PIL between EMF1 and EMF2 (P1 and N2; Figure \ref{AIA2} g) and the  jet top  on the west  side  of the EMF1 is limited at the location of  
    N1. At 02:11 UT  and  over a few minutes up until 02:18 UT, we can observe long  bright and dark AFS striding over the two EMFs (right columns of Figure \ref{AIA1} and Figure \ref{AIA2}). 
  The characteristics of the jet are the following: length around 50 Mm, the base width between 15-20 Mm. The jet lifetime is 
  between 02:02 UT to 02:11 UT.

  With  AIA 304 \AA\ data, we made several observations of a mini-flare, approximately at the same location, nearly one every hour and generally not accompanied by such a wide jet. The detail of the recurrent mini-flares is as follows: at the beginning of the movie,  a  mini-flare is visible at 20:00 UT, then  at  20:27 UT, 21:28 UT, 22:51 on March 21 and  at 00:39 UT, 01:25 UT,  01:39 UT, 02:03 UT on March 22. Regularly, before each  burst, we can clearly see  two AFSs:  one over EMF1 in the west and one  over EMF2 in the east. After the burst, long-arch filaments connect the extreme eastern polarity P2 to the extreme western polarity N1. Prior to our mini-flare and jet (around 01:59 UT), the two AFSs were separated by an area with mixed bright and dark patches. Then at 02:09 UT, there is a long system of arch filaments. At 02:28 UT,  
   when the phase of the activity is over, the initial  configuration with the two distinguished AFSs, just as before the jet, is restored. This chain of mini-flare and ejection is recurrent.

 
  To analyse the evolution of the observed jet, we created the projected height-time plots of the jet in different AIA wavebands (171 \AA, 211 \AA, and 304 \AA), which are presented in Figure \ref{timeslice1}. To obtain this height-time plot, we chose a broad slit (width = 8 pixels) to cover the plasma outflow. 
 The average jet speed along the slit direction shows two different slopes: a steep slope in the starting phase (till 02:05 UT) of the jet eruption with an average speed of 350 km s$^{-1}$ and a slow phase of 80 km s$^{-1}$ in the later stage from 02:05 UT to 02:08 UT.
 The slow phase may be due to the presence of loop system in the path of the jet. It seems that when the jet material is passed through this loop system, it decelerated and,
 finally, it stopped. The cool material visible as an absorbing feature in AIA channels is detected about one minute after the hot jet with no well defined speed. The cool material  appears to be present along the LOS in small patches but it is not moving towards the west as the jet is doing. Later, the cool material (or surge) is escaping in  slightly different directions than the jet. Among it, we could identify some blobs with different projected speeds
 ($\approx$ 100 and 30 km s$^{-1}$). Then the cool material  came back at a speed of $\approx$ - 40 km s$^{-1}$ (Fig \ref{timeslice1} (d)). According to the location of the AR (W60), these velocities are underestimated by a factor of cos(60). The positive velocities correspond to the material that is heading away from the observer's view.

\begin{figure*}[ht!]
\centering
\includegraphics[width=1.0
\textwidth]{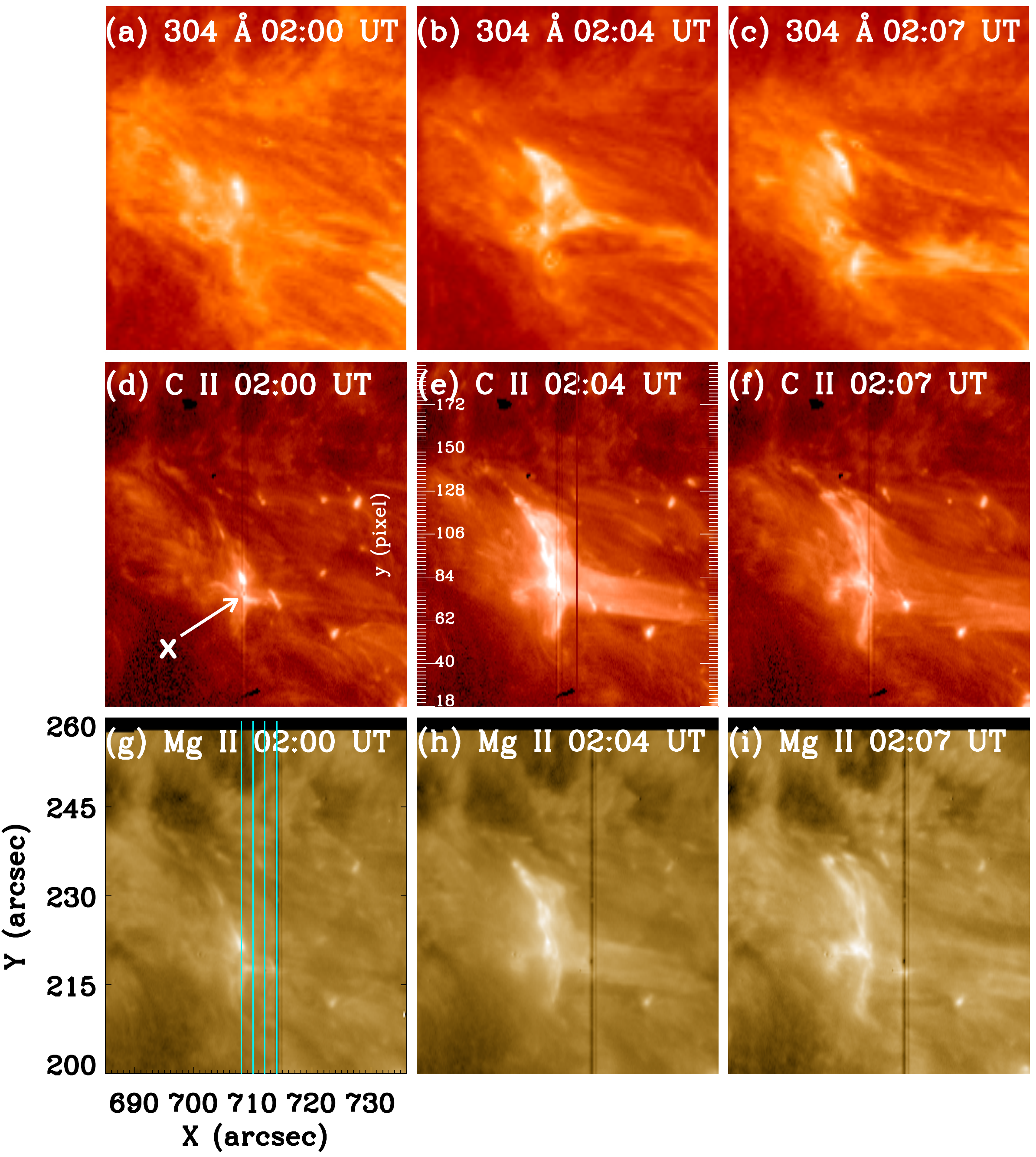}
\caption[Solar jet observed in  co-aligned  images of EUV AIA 304 \AA\, of IRIS C II  SJI, and of IRIS Mg II SJI.]
{Solar jet observed in  co-aligned  images of EUV AIA 304 \AA\ (top), of IRIS C II  SJI (middle), and of IRIS Mg II SJI  (bottom). The  four positions of the slit in the raster mode  is shown with vertical cyan lines in panel (g). In IRIS CII  SJI images  (middle row) the reconnection point is  indicated.}
\label{AIA_IRIS1}
\end{figure*}
 \subsection{Comparison between IRIS SJIs and AIA 304 \AA\ observations}
  \label{comparison}
  The  analysis of IRIS data,  all along the evolution of the jet, shows a good correspondence between the structures visible in AIA 304 \AA\  and in IRIS C\footnotesize{II} \normalsize SJIs. This correspondence is  summarised in Figure \ref{AIA_IRIS1}. We note that the nominal coordinates of IRIS  in the file headers do not correspond to the nominal coordinates of AIA. Therefore, we had to shift  the FOV of AIA by 4$\arcsec$ in x-axis and 3$\arcsec$ in y-axis to obtain a good co-alignment.The IRIS slit,  with its four positions, crosses the bright zone corresponding to the jet base, namely, the dome top, which is supposed to be the reconnection site  along a few pixels between around pixels 60 to 120  in the left slit position (Figure \ref{AIA_IRIS1} (e)). Around 02:00 UT, in the 304 \AA\, image  as well as in C\footnotesize{II}  \normalsize and Mg \footnotesize{II} \normalsize IRIS  SJIs,  small bright threads along two vertical paths that are mixed with 
 tiny round-shape  darker areas are visible in the middle of the FOV  where the reconnection  occurred  (Figure \ref{AIA_IRIS1} a, d, and g). It is clear that in this small zone, there is no 
north-south  filament  along the PIL (N2-P1)   which would be visible by absorption in AIA 193 \AA.  The very  light-dark  filament-type structure with a vague sigmoidal shape in the NVST images that is localised at this place is, in fact, part of the AFS  (Figure \ref{NVST} b) because there is no sigmoid  visible in the hot channels of AIA, where plasma should be heated   due to high electric currents along a sigmoid (\citealt{Barczynski2020}). In the north of this zone,  long-lying,   more or less east-west  AFS, as well as, on both sides of the zone (pixels 60-120), the short AFS-overlying EMF1 and EMF2 are visible. The short AFS-overlying EMF2  have  a dome  shape like the asymmetrical  anemone formed by the fibrils visible
in the NVST images (Figure \ref{NVST}). The location of the onset of the  mini-flare
 is indicated by  the point `X'  at the crossing location between the  arch-shape QSL and an east-west bright  line in Figure \ref{AIA_IRIS1} (d). The location of the `X' point   in IRIS observation   is at 709$\arcsec$, 218$\arcsec$ and  in AIA it  is at 705$\arcsec$, 215$\arcsec$. In  Figure   \ref{AIA_IRIS1} (top panels), we translated the AIA images to obtain a good co-alignment with IRIS images.
 Around 02:03 - 02:04 UT, the arch-shape  QSL  was brightening  and the flare started 
 with  the onset of the jet  ejection visible  in  the C\footnotesize{II} \normalsize and Mg\footnotesize{II} \normalsize SJIs (Figure \ref{AIA_IRIS1} b,e). The bright jet was obscured by dark material in front of it, which is the surge; both the jet and the surge were extending at the same time around 02:04 UT.
 From 02:05 UT to 02:07 UT, the jet  extended along  two bright branches with  a dark area between them (Figure \ref{AIA_IRIS1} (c),(f)). 
 At 02:07 UT, AIA 304 \AA\ image shows the extension of the  surge covering all the bright jet. The dark area is  due to the absorption of the 304 \AA\,  emission by He continua (\citealt{Anzer2005}) (Figure \ref{AIA_IRIS1} (c)). In  the C\footnotesize{II} \normalsize and Mg\footnotesize{II} \normalsize filters, it is not so pronounced because of the large band-pass of IRIS filters relative to the width of the lines  and the low emission of the lines in the jets (Figure \ref{AIA_IRIS1} (f),(i)). Nevertheless, we  can still distinguish a bright EW-elongated jet in the south  and  some dark area above it that might correspond to the surge. This is confirmed in the NVST images (Figure \ref{NVST}).

\section{Magnetic environment of the AR} \label{reconnection}
In Section \ref{sec:mag}, we  follow the birth of the AR using HMI longitudinal magnetic field. Here, we analyse 
the magnetic topology of the AR  using the full vector magnetic field to understand the orientation of the magnetic field  lines inside the two bipoles (P1-N2 and JP1-JN2) 
involved in the mini-flare and the jet to confirm the existence of a BP.
\subsection{HMI Magnetic field vector maps}
\label{sec_vector}
\begin{figure*}[t!]
\centering
\includegraphics[width=1.00
\textwidth]{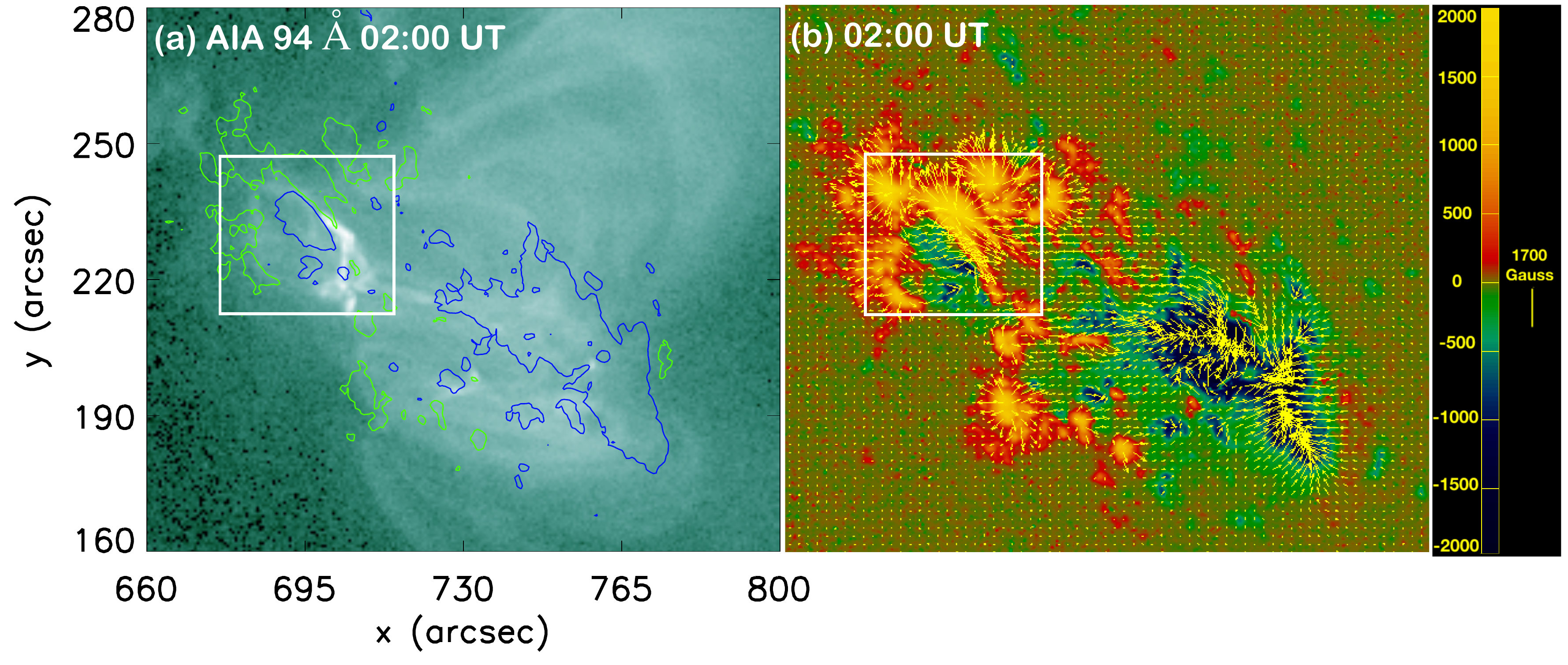}
\caption[The jet base appeared as brightening with an arch-shape in between the positive (green contours) and negative (blue contours) in AIA 94 \AA.]
{Panel (a): The jet base appeared as  brightening with an arch-shape in between the positive (green contours) and negative (blue contours) in AIA 94 \AA. Panel (b): HMI Vector magnetic field map with the same FOV as of the (a). The white square is the FOV for the Figure \ref{vector} (a-b).}
\label{full_HMI}
\end{figure*}
The HMI SHARP longitudinal magnetic field movie, with its high cadence, shows
the fast evolution of the EMF2. The negative polarities N2-JN2 were continuously sliding along the positive polarity P1 and  initiating bright points from time to time. We used  the  HMI vector magnetic field map at  the closest time of the reconnection at 02:00 UT. Figure \ref{full_HMI} presents in the right panel the  magnetic field  vector map  computed  with the  UNNOFIT inversion code 
at 02:00 UT  and the corresponding full AR as a contextual image meant to  show  the brightening at the base of the jet in AIA 94 \AA.  The  vector magnetic field maps represent the  full magnetic field vectors with their three components in the solar local reference frame, 
  generally referred to as the heliospheric reference frame. The vertical component in this reference frame is represented via a colour table. The two horizontal components are associated to form an horizontal vector, which is represented by arrows. However, the pixel dimension is viewed along the LOS in order to be able to co-align the FOV  with the AIA images. A box indicates  the small FOV  encircling the region which contains the brightening at the jet base  corresponding to  the  QSL at the reconnection site. 
We carried out a zoom analysis to probe the nature of magnetic field vectors in this jet region (Figure \ref{vector} a).
The length of the arrows represent the strength of the  horizontal magnetic field.

\subsection{FR vector pattern and formation of small bipole}
\begin{figure*}[ht!]
\centering
\includegraphics[width=\textwidth]{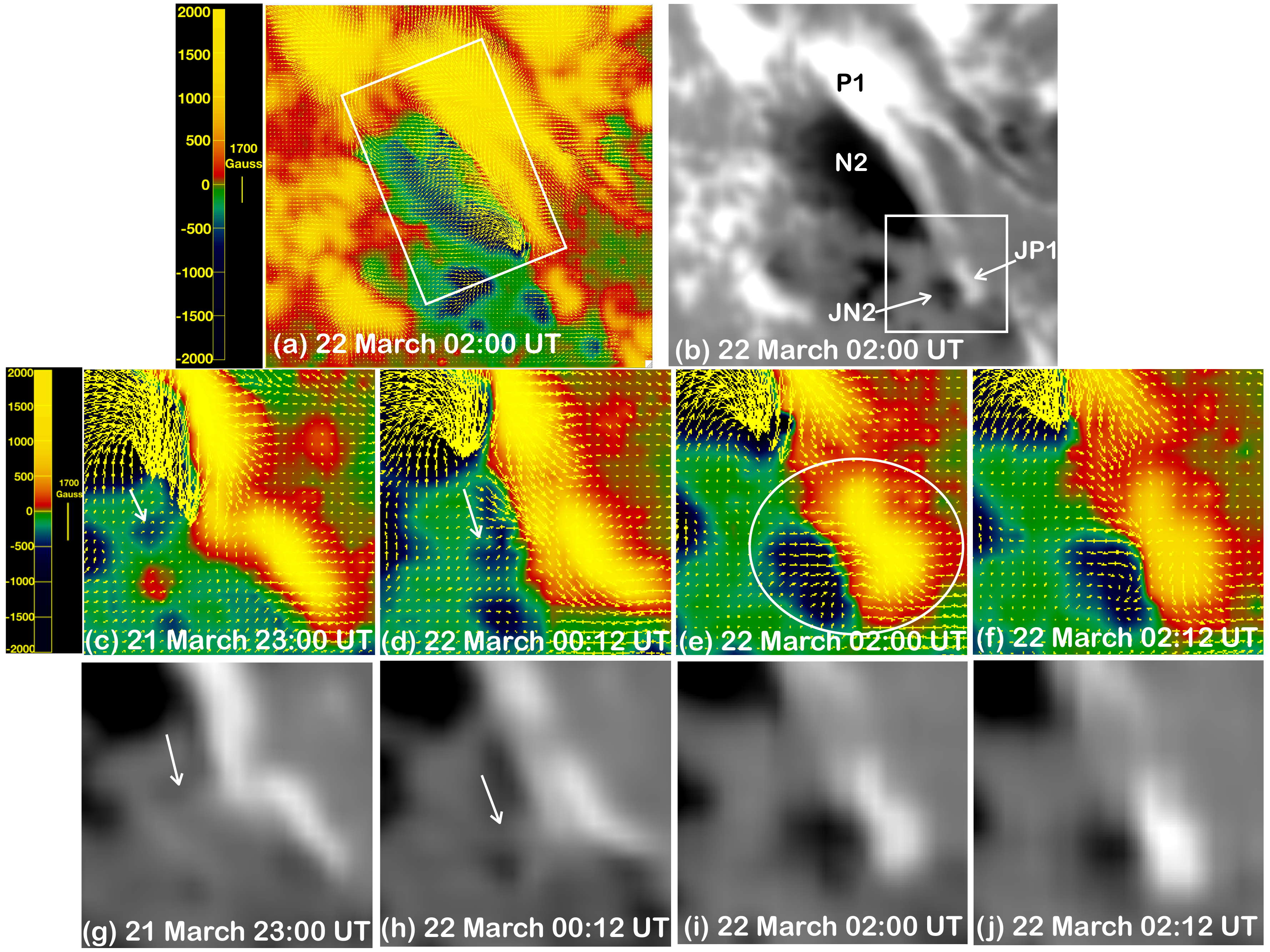}
\caption[Vector magnetic field configuration of a  part of AR NOAA 12736 where the jet was initiated.]
{Panel (a): Vector magnetic field configuration of a  part of AR NOAA 12736. Panel (b): LOS magnetic configuration at the AR including the two bipoles P1-N2, JP1-JP2. Color bar indicates the vertical magnetic field strength and the arrow shows the strength of the horizontal magnetic field.}
\label{vector}
\end{figure*}

In the long region between P1 and N2,  we  make note of  a  characteristic  pattern of the magnetic field  vectors that  suggests the presence of  
a twisted FR with  vectors converging together in the PIL in the middle part   (between P1 and N2)  and vectors
  turning  at both ends, in the top and bottom parts of the FR, resembling the hooks of a FR (Figure \ref{vector} (a)).  In the vicinity of the FR,  there is 
  an interface that separates the regions of turning and returning of the vectors,
  which represent the boundary between  the FR  and the arcades over the FR and  its  surrounding area.
This pattern is  relatively stable according with the 22 maps computed around the jet time. On March 21  at 20:00 UT, the  FR was already created and  was continuously observed until March 22 at 05:00 UT.  At first glance, the FR does  not seem to participate to the  formation of the jet. A very detailed study shows that, in fact, this was a very particular case involving a transfer of  twist from the FR to the jet  during  the FR  extension towards the south  before the   reconnection.

The relationship of the FR and the jet  is detected in the HMI movie of the longitudinal magnetic field where the formation of the small bipole (where the jet was initiated) is observed.
The longitudinal HMI movie shows a stress  created by the sliding of  the two opposite polarities (P1 and N2). These two polarities come from the two opposite  magnetic emerging regions (EMF1 and EMF2). On March 21  at 23:00 UT,  a few hours before the jet, a negative polarity  part of N2  detaches 
and moves towards the south,
sliding along the  positive polarity P1 to form the small bipole JN2-JP1  (Figure \ref{vector} g-j). The  new bipole  is formed 
with the  small positive (JP1) and  the negative (JN2) polarity
encircled in Figure \ref{vector} (e). This small bipole  is formed by  collision  of two opposite sign polarities  coming from two  different magnetic systems and not by  direct magnetic flux emergence.

\subsection{BP magnetic configuration and twist transfer}
Looking at the direction of the magnetic field
vectors  between JN2 and JP1,  we find that they are oriented  from the negative polarity to  the positive polarity, which is evidence that it is a BP region, more generally, that it is a region with magnetic field lines that exhibit  a dip
grazing  the surface at the PIL
(Figure \ref{vector} e). We note that the BP is observed only at this precise time (02:00 UT) -- and not before  and not after (Figure \ref{vector}  c, d, f).

We arrive at the conclusion that with the extension of the FR {towards  the south}, it is possible that the arcades of FR interact with the overlying magnetic field.
Some part of the twist of  the FR could be  transferred to the jet, however, there is still a remnant component in the small bipole, as we see in Figure \ref{vector} (f).  The rotation of the structure at the base could  explain this transfer of twist.
To make certain of the existence of  the FR, we compare this finding with  MHD  simulations.

\subsection{Comparison between MHD models and observations}
\label{OHM}
    We used  the MHD simulations  of  \citealt{Zuccarello2015},  where the physical conditions are used to create a FR in an AR. Starting from an asymmetric, bipolar AR, as in \citealt{Aulanier2010},  they investigated different  classes of photospheric motions that are capable of forming  a FR.  Here, we consider the results of the simulations with regard to  converging  motions  towards the PIL of the AR with  magnetic flux cancellation. Progressively twisted magnetic field lines  were  globally  wrapping  around an axis and, eventually, formed a FR. The dynamics of the FR is modelled by using a version of the {\it Observationally driven High-order scheme Magnetohydrodynamic} (OHM) code (\citealt{Aulanier2005OHM,Aulanier2005,Aulanier2010}). It is a $\beta$ = 0 simulation, so the plasma conditions are not studied. 
    The OHM code solves the standard zero-$\beta$ MHD equations in the Cartesian coordinates  with line-tied and open boundary conditions. The line-tied reflective boundary conditions ensure that the foot-point of magnetic field line can only perform the horizontal motion on to the boundary. The pressureless ($\beta$ = 0) time-dependent MHD equations for an ionized and collisional plasma are given as (\citealt{Aulanier2005OHM}):

    \begin{equation}
        \frac{\partial \rho}{\partial t} = - \vec{\nabla} \cdot (\rho ~\vec{u})
    \end{equation}
    
    \begin{equation}
        \rho \frac{\partial \vec{u}}{\partial t} = - \rho (\vec{u} \cdot \vec{\nabla})~\vec{u} + \vec{J} \times \vec{B} + \rho~ \mathcal{D}~\vec{u}
    \end{equation}
    
    \begin{equation}
        \frac{\partial \vec{B}}{\partial t} = \vec{\nabla} \times (\vec{u}\times \vec{B}) + \mathcal{R}~\vec{B}
    \end{equation}
    
    \begin{equation}
      \vec{\nabla} \times \vec{B} = \mu \vec{J}
    \end{equation}
    
    \begin{equation}
        \vec{\nabla}\cdot \vec{B} = 0
    \end{equation}

    where, $\rho$ is the mass density, $\vec{u}$ is the plasma velocity, $\vec{J}$ is the electric current density, $\vec{B}$ is the magnetic field, $\mu$ is the magnetic permeability, $\mathcal{D}$ is the diffusion operator for the velocity, and $\mathcal{R}$ is the diffusion operator for the magnetic field.
    OHM code solves these equations using Einstein's notation for special derivatives 
    (\citealt{Aulanier2005OHM, Aulanier2005}). The detail on the numerical method for OHM code is well explained in \citealt{Aulanier2005OHM}.
    The results from these MHD simulations have  already been validated by testing different flare activities, such as sigmoid currents  of FR (\citealt{Aulanier2010}), electric current density increase in flare ribbons (\citealt{Janvier2014}), and electric current density decrease at CME footpoints
(\citealt{Barczynski2020}). In this study, we want to test if the footprints of the FR in the HMI magnetic vector (vec B) maps have a similar pattern as  
    the footprints of the  theoretical FR in these MHD simulations.
    
    
    
    
    
    

\begin{figure}[ht!] 
\centering
\includegraphics[width=0.9\textwidth]{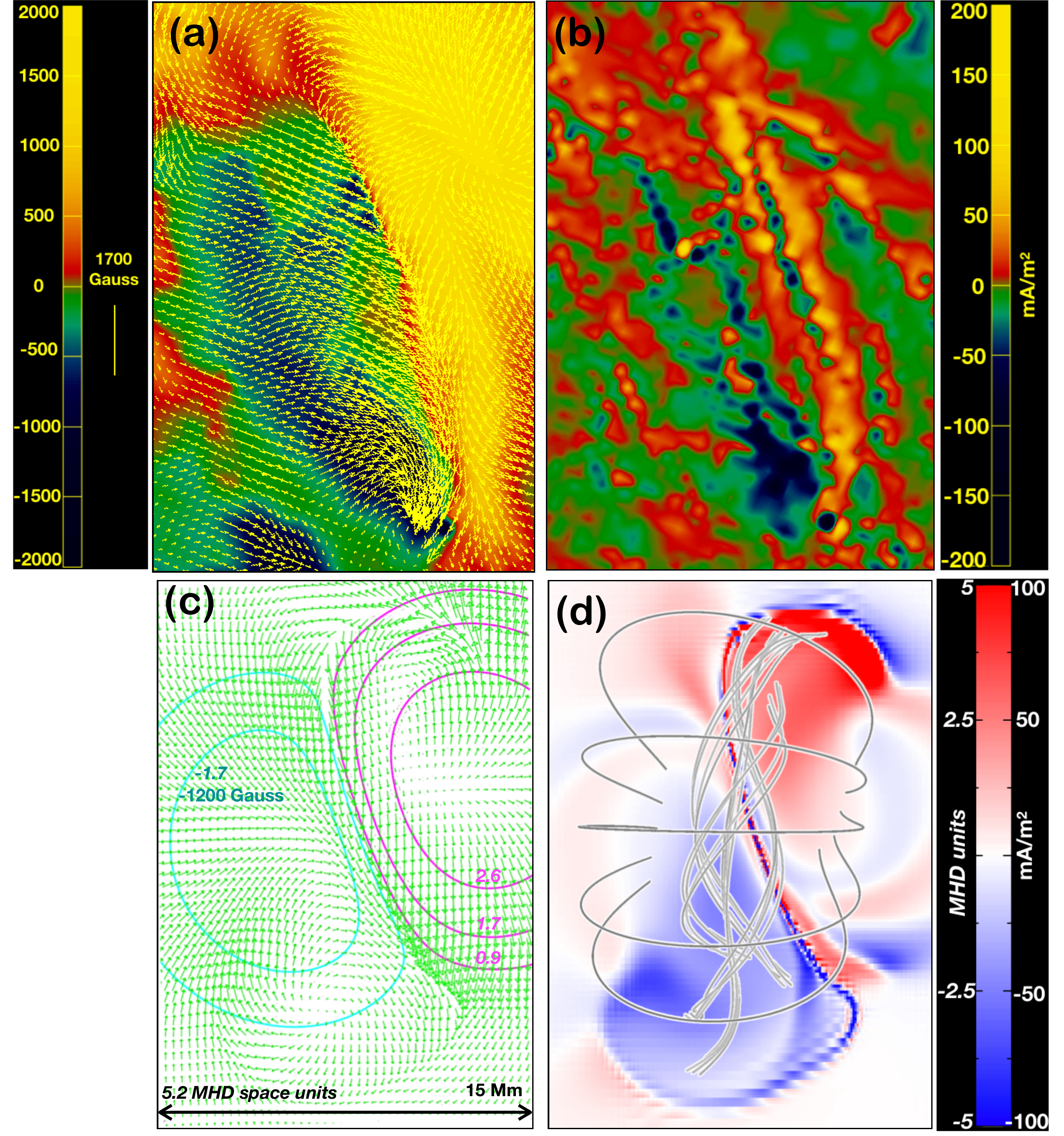}
\caption[FR evidenced  in the HMI  observations and comparison with the images from MHD simulations.]
{FR evidenced  in the HMI  observations (a,b) and comparison with  MHD simulations (c,d). Panel (a-b): Vector magnetic field and current density maps computed with the UNNOFIT code. Panel (c): From MHD simulations the  iso-contours of vertical magnetic field with vectors. Panel (d): The magnetic field lines are plotted with grey color and red/blue contours are electric currents.}
\label{model}
\end{figure}

The comparison between our observations (panels a-b) and MHD simulations (panels c-d) is presented in Figure \ref{model}. We rotated our observation in panel (a-b) by 30$^{\circ}$ in the clockwise direction for an improved comparison with the MHD simulations. It is very clear that a  sheared magnetic field is generated along the PIL and the vectors are strongly inclined along with PIL. We have also evidence for swirling of the magnetic field in the top and bottom part of the FR.
The sheared vec B that converges towards the PIL   is a characteristic motion to create a BP.  This pattern can also be seen in \citealt{Barczynski2019}. It is due to the summed effects of: (i) the shear that creates a BP with vec B in the negative  polarity pointing towards the positive polarity; and (ii) the asymmetry of the photospheric flux concentration with a stronger positive polarity (in  the model and observations) which is due to the  magnetic pressure pushing all the fields towards the (weaker) negative  polarity. Hence it is  leading to some sheared vector within the positive  polarity to point  towards the negative polarity. These two effects lead to the convergence. 
polarity. These two effects lead to the convergence. The swirling motions visible at both ends of FR  are well-represented by vec B, which display  similar angles and  similar spatial gradients  at the edges of the swirlings, which 
separate the swirling vec B from the surrounding magnetic field that has more potential, that is, exhibiting more radial from the center of the magnetic polarity. For example, this separation is visible at the top right in Figure \ref{model} (a and c) in the positive polarity, where radial vectors are close to turning vectors to the left, and at the  bottom  of the negative polarity, there is a similar separation between radial vectors and vectors turning to the left. This kind of separation is  reminiscent of a QSL, just as in the MHD models (\citealt{Janvier2013}; \citealt{Aulanier2019}).  Moreover, 
those swirls correspond  exactly to the footpoints of sigmoidal field lines, even though they are not visible in the extreme ultraviolet (EUV).
Finally, the similarity  of all these characteristics structures 
(e.g. BP, QSL, sigmoidal field line) between the MHD models and our observations leads us to infer the existence of a FR in the immediate vicinity of the jet.

 In addition to the pattern of the photospheric horizontal fields, a relatively good match is also found for the vertical current densities.
Both the HMI observation and the MHD simulation display a dominance of the J$_z$ and B$_z$ of the same signs in each polarity of the bipole, with an elongated double-peaked J$_z$ pattern all along the PIL, as well as more extended patches at the ends of the sheared PIL. In the model, those extended patches correspond to the footpoints of the FR field lines (Figure \ref{model} d). One difference, however, between the observation and the model is that with HMI, the extended patches in the negative polarity is more clearly visible than in the positive polarity (Figure \ref{model} b). We argue that this difference is minor since it may be due to that fact that in the positive polarity, the swirling patterns of the vector fields are located in relatively weaker vertical fields than in the negative one (i.e. 500G in the former vs. 1500G in the later, Figure \ref{model} a). With the same twist in both polarities, the weaker fields result in weaker current densities in the positive polarity. Another difference is that in the MHD simulation, some strong QSL-related current sheets surround the FR footpoint related extended patches (\citealt{Janvier2013}; \citealt{Aulanier2019}). These are not visible with HMI and we argue that this is due to the limitations of the HMI data, from which current sheets can only be extracted during flares and with some processing of the data, as in \citealt{Janvier2014};  \citealt{Barczynski2020}.
%

Comparing the magnitudes of  current densities between  models and  observations requires us to scale the model to physics (solar) units. The reason behind this is that the model was calculated with dimensionless units, resulting in maximum current densities on the order of five units. Such a scaling has already been done for the estimation of flare energies (\citealt{Aulanier2013}). Here, they need to be adjusted to this specific observed bipole. Still, we should bear in mind that this can only be done approximately given the differences in shape between the observed and modelled flux concentrations. 
To convert the non-denationalized j MHD-units to real j solar-units, we calculate as (\citealt{Aulanier2005OHM}):
\begin{equation}\label{eq:comparison}
    J_{solar} (A m^{-2}) = j_{0} \times \frac{1}{\mu} \times \frac{B_{solar}}{B_{0}} \times \frac{L_{0}}{L_{solar}} 
\end{equation}

Thus, using HMI as a reference (Figure \ref{model} a), we attributed a magnetic field amplitude of $\pm 1200 G$ to the B$_z$ isocontour $B_z = \pm1.7$ and a bipole size of 15 Mm to the width of 5.2 space units as displayed in Figure \ref{model} c. Then we scaled the OHM model using a magnetic unit $B_0$ = 700G and a spatial unit $L_0 = 2.9 \times 10^{6}$ m. We reset the magnetic permeability from unity in the simulation to its real value $\mu$ = 4 $\pi \times 10^{-7}$ N $A^{-2}$. As a result, the  dimensionless current-densities that we model here have to be multiplied by $B_0/(\mu L_0)$ to be expressed in A/m$^{2}$ as in equation \ref{eq:comparison}. With these settings, the currents reached up to 100 mA/m$^2$ at the FR footpoints. This value is only half of what is measured with HMI, so the modelled currents are in qualitative agreement with the observed ones. 
The difference in magnitude may be attributed  to the existence of a stronger twist in the observed bipole than  the twist in the model.
Yet it is arguably more likely due the different aspect ratios of the observed and modelled bipoles, the latter being less elongated than the former (Figure \ref{model} a and Figure \ref{model} c).

\subsection{Magnetic shear: consequence of the jet}
When the twisted FR fieldline (the one initially rooted in JN2 and stretched by the footpoint motion) eventually reconnects with the large western untwisted loop (rooted in JP1 and far west in the negative polarity) then two new field lines are formed (Figure \ref{cartoon}):
\begin{enumerate}
    \item The first one is the untwisting jet field line. Composed of a long truncated-FR part at the east and a long truncated-loop at the west.
    \item The second one is like a flare-loop forming below the `X'-point. It is rooted between JN2 and JP1. It is composed at the east of the truncated leg of the FR field line, and at the west it is the former leg of the loop. That's the JN2-JP1 structure, shown by a small loop in Figure \ref{cartoon} (d). 
\end{enumerate}

 Just after these two lines have reconnected, they are composed of a eastern part where field aligned currents are present (which are associated to twist in the FR), and of a western part where no (or weak) electric currents exist ({\it i.e.} a potential field). After reconnection these lines are not in force-free condition. A force-free field satisfies (\citealt{Demoulin1997}; \citealt{Aulanier2010}; \citealt{Aulanier2014}):
 \begin{equation}
     (\vec{B} \cdot \vec{\nabla})\alpha=0
 \end{equation}
 where, $\vec{j}$ is the electric current density and $\vec{B}$ is the magnetic field. $\alpha$ is a constant related with the intensity of the coronal electric current density (\citealt{Demoulin1997}; \citealt{Chandra2011}).
These equations justify that in force free fields currents are co-linear with magnetic fields ({\it i.e.} field-aligned currents) and the ratio between currents and magnetic field is constant along a field line.
In the present case the products of reconnection have a finite $\alpha$ at the east, and zero $\alpha$ at the west. So they are not force free. Therefore, some (torsional) Alfv\'en waves must be launched to reach a new balance, in which twist is eventually redistributed all along the previously-reconnected field-lines. After the relaxation, $\alpha$ must has been redistributed along the loop (shown in panel d), so the whole reconnected JN2-JP1 loop must be current-carrying. These currents are either associated to shear or twist. The shear was not present before the jet onset, but develops during and after the jet onset. So the shear is not the source of the jet, but it is a consequence of the jet.

The clues 
of our interpretation are the identification of a non-eruptive FR, from which some twist is carried away and eventually reconnects into the jet at 
`X'-point current-sheet.
The transport of twist away from the FR towards a BP is supported by the HMI observations of a moving negative flux-concentration whose transverse fields point towards a positive one. The twist is transported at a long distance of the FR which remains non-eruptive. The tilt observed in the IRIS spectra in the four positions of the slit which by chance are exactly at the site reconnection  confirmed the transfer of twist at the jet base.

\section{Results and conclusion}
\label{discussion_ch6}

In this chapter, the observations of a twisted jet, a surge, and a mini-flare which occurred  in the AR 12736 on March 22, 2019 at 02:05 UT are studied. The event was observed in multi-wavelengths with AIA and IRIS instruments and detailed in the magnetic  field vector maps obtained by HMI and  computed with the UNNOFIT code. The MHD simulations were used to validate the vec B observations (\citealt{Aulanier2010}; \citealt{Zuccarello2015}).
The main results and conclusions are following :

The AR consisted of the collapse of two EMF regions, each of them overlaid by an AFS. The jet and surge  reconnection site is along the PIL between these two AFS. The AFS over the east side  evolved rapidly due to photospheric surface motions. Prior to the reconnection, the AFS exhibit a dome shape. After the reconnection, long AFS overlying both EMFs are observed. This  is confirmed in  the NVST H$_\alpha$ images. A large FR in the vicinity of the jet region is detected. The patterns of transverse fields and vertical current densities, as observed by HMI and appearing without being constrained  a priori in an MHD simulation of non-eruptive 
FR  formation with flux-cancellation of sheared loops, show a  good accordance.
The location of the FR  is  fully supported by HMI vec B  and electric currents J$_z$ maps.
The magnetic topology of the AR demonstrates  a BP region  due to the particular formation of the bipole by collision of opposite polarities, which is  dynamically  transformed to an `X'-point  current sheet. The fast extension of the FR towards the site of reconnection  due to photospheric surface motions offers the possibility for the FR arcades to reconnect with magnetic pre-existing field lines at the `X'-point current sheet without the eruption of the  FR. The extension of the FR may transmit twist to the jet.

\begin{figure}[ht!]
\centering
\includegraphics[width=\textwidth]{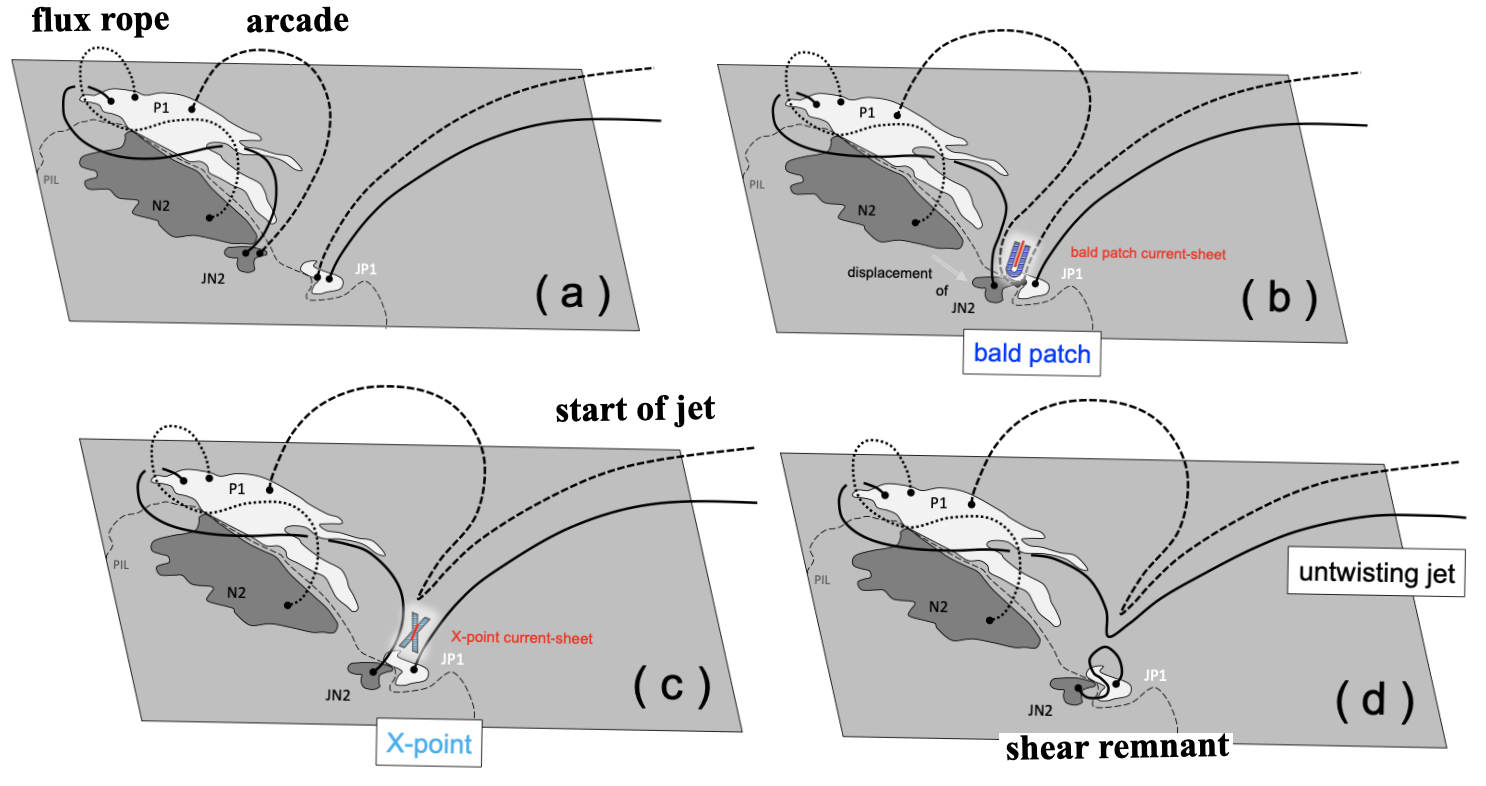}
\caption[Sketch  of the formation of the jet and  transfer of the twist from the FR to the jet during reconnection.]
{Sketch  of the formation of the jet and  transfer of the twist from the FR to the jet during reconnection.
Panel (a) magnetic configuration before  the reconnection, panel (b)  formation of the BP current sheet, panel (c) X-point current sheet, panel (d) the untwisting jet after the reconnection and the remnant twist in the bipole JP1 and JN2.}
\label{cartoon}
\end{figure}

A cartoon is proposed where the FR between P1 and N2 is represented by the solid  twisted line  (Figure \ref{cartoon}, a). 
It is extended to the  south, creating the bipole JN2-JP1. The BP  current sheet is generated  between the overlying arcade of FR and the magnetic field line of the west  emerging flux P1-N1 (panel b).  At this time, a first reconnection occurs at a localised point that is very deep in the atmosphere. 
The Mg II  profiles resemble  those found in IRIS bombs (IB)  with extended wings (\citealt{Peter2014}), which are proposed to have been formed during   BP current sheet reconnection (\citealt{Zhao2017}). 
Such chromospheric wide profiles   have been  modelled in MHD simulations (\citealt{Hansteen2019}).
It has also been shown that a BP could be  transformed immediately at a  null point.
We propose in panel (c) that  the reconnection occurs in the null point (`X'-point) that is formed dynamically  along a  current sheet  or a flat spine-surface  above a dome that is not depicted in the cartoon  panel (c). Cool material trapped in the BP during its formation is  expelled with a large blueshift,  
as revealed in IRIS Mg II line profiles with extended blue  wings. The spectra shows an evident tilt, which indicates the presence of helical motions.  The reconnection  site  is heated at all the temperatures and the  hot jet is expelled towards the west  side in twisted field lines (panel d). The cool material follows different paths than the hot and acts as a wall in front of the hot jet. It resembles the  surges that accompany  jets in the MHD simulations of \citealt{Moreno2008}; \citealt{Nobrega2016, Nobrega2018}.

Our magnetic analysis benefit from the treatment of the HMI vector magnetic field by the UNNOFIT code which uses a filling factor which takes into account the non resolved structures. In each pixel there is an equilibrium between magnetized regions and non-magnetized regions which implies a better determination of the magnetic field inclination \citep{Bommier2016}. This is an important aspect  for  regions with weak magnetic field. It is the case in the small bipole where our jet  reconnection takes place  and where we have detected the BP.
A more important aspect for this case is the chance to have the IRIS spectra exactly at the reconnection site. IRIS spectra shows directly the transfer of twist between two stable systems at the reconnection point by unveiling a helical structure. The spectral analysis of IRIS C II, Si IV, and Mg II lines will probe the multi-thermal atmosphere of solar flares and will provide a precise calibration of magnetic reconnection height.

 \chapter{Spectroscopic analysis of the multi thermal atmosphere of a flare and jet}
\ifpdf
    \graphicspath{{Chapter7/Figures/}{Chapter7/Figures/}}
    \ifpdf
    \graphicspath{{Chapter6/Fig_chapter6/}{Chapter6/Fig_chapter6/}}
\label{c7}
\section{Introduction}
Solar jets are commonly observed  with IRIS and the  multi wavelength AIA instrument. IRIS spectroscopic and imaging  observations of jets reveal bidirectional outflows in transition region lines at the base of the jets implying  explosive  magnetic  reconnection processes  (\citealt{Li2018}; \citealt{Ruan2019}).
Bidirectional outflows in the LOS are detected  by the extended wings in chromospheric and transition line profiles (\citealt{Innes1997}; \citealt{Tian2018}; \citealt{Ruan2019}). With IRIS  the chromospheric C II and Mg II lines are frequently observed not only in the UV bursts but also in the quiet chromosphere as well as in solar flares and jets (\citealt{Leenaarts2013a}; \citealt{Rathore2015}). They  are optically--thick chromospheric 
  lines, which need  a radiative transfer approach to determine the physical quantities of  plasma. The Mg II  lines are formed in multi layer atmosphere.
Simulations in the  quiet chromosphere has been carried out by several authors (\citealt{Athay1968}; \citealt{Milkey1974}; \citealt{Ayres1976}; \citealt{Uitenbroek1997}; \citealt{Lemaire2004}; \citealt{Leenaarts2013a,Leenaarts2013b}; \citealt{Pereira2013}; \citealt{Grubecka2016}). The core of the line is formed just under the transition region (T$<$ 20,000 K), the wings at the minimum of temperature (T = 5000 K). 
IRIS  spectral data allow to make many  progresses on the plasma diagnostics in flares and shows the blue and redshifts. The blue asymmetry can be explained by down-flowing plasma absorbing the red peak emission and not by strong blueshift emission
 (\citealt{Berlicki2005}).
Chromospheric response to intense heating, even in the 1D model, is complicated. The shape of the emission line profiles depends sensitively on the physical conditions of the plasma and its dynamics, in particular the plasma flows that arise at the line core formation heights. 
They may have symmetrical profiles. Moreover the highest Near Ultraviolet (NUV) continuum enhancements  observed in strong flares are most
likely because of the Balmer continuum  formed by Hydrogen recombination (\citealt{Kleint2017}) and consequently  flares can be assimilated to  white light flares, commonly observed in optical continuum  where the energy  deposit is localized at  the minimum temperature  region. 
The ratio of  IRIS transition  region lines is also a good diagnostics for the determination of  the plasma density in flares (\citealt{Polito2016}). The electron density ({\it N}{$_e$}) in flare ribbons can be enhanced by two orders of magnitude more than in plage region ({\it N}{$_e$} $>$ 10 $^{13}$ cm$^{-3}$). 
For multiple flaring kernels, chromospheric  lines show a rapidly evolving
double-component structure: an enhanced emission component at rest, and a broad, highly red-shifted
component of comparable intensity. \citealt{Graham2020} interpreted such observations by beams penetrating very deep in the atmosphere. The red-shifted components migrate from redshifts towards
the rest wavelength. The electron beams would  dissipate their
energy higher, driving an explosive evaporation, and a counterpart condensation is created as a very dense layer.

In this present chapter, we analysed a twisted jet and a flare of B6.7 GOES class (that we call mini flare) from the spectroscopic point of view. 
Mg II, Si IV, and C II spectra and line profiles at the reconnection site of the jet, 
are analysed  leading to   
a sketch of  dynamical reconnection.
We proposed on a possible multi thermal reconnection  model with multi layers from   very deep layers
in the atmosphere, e.g. at  the minimum temperature region, to the corona.
We propose a sandwich model with stratification of multi layers to explain the observations during the reconnection.
\begin{figure*}[!ht]
\centering
\includegraphics[width=1.0
\textwidth]{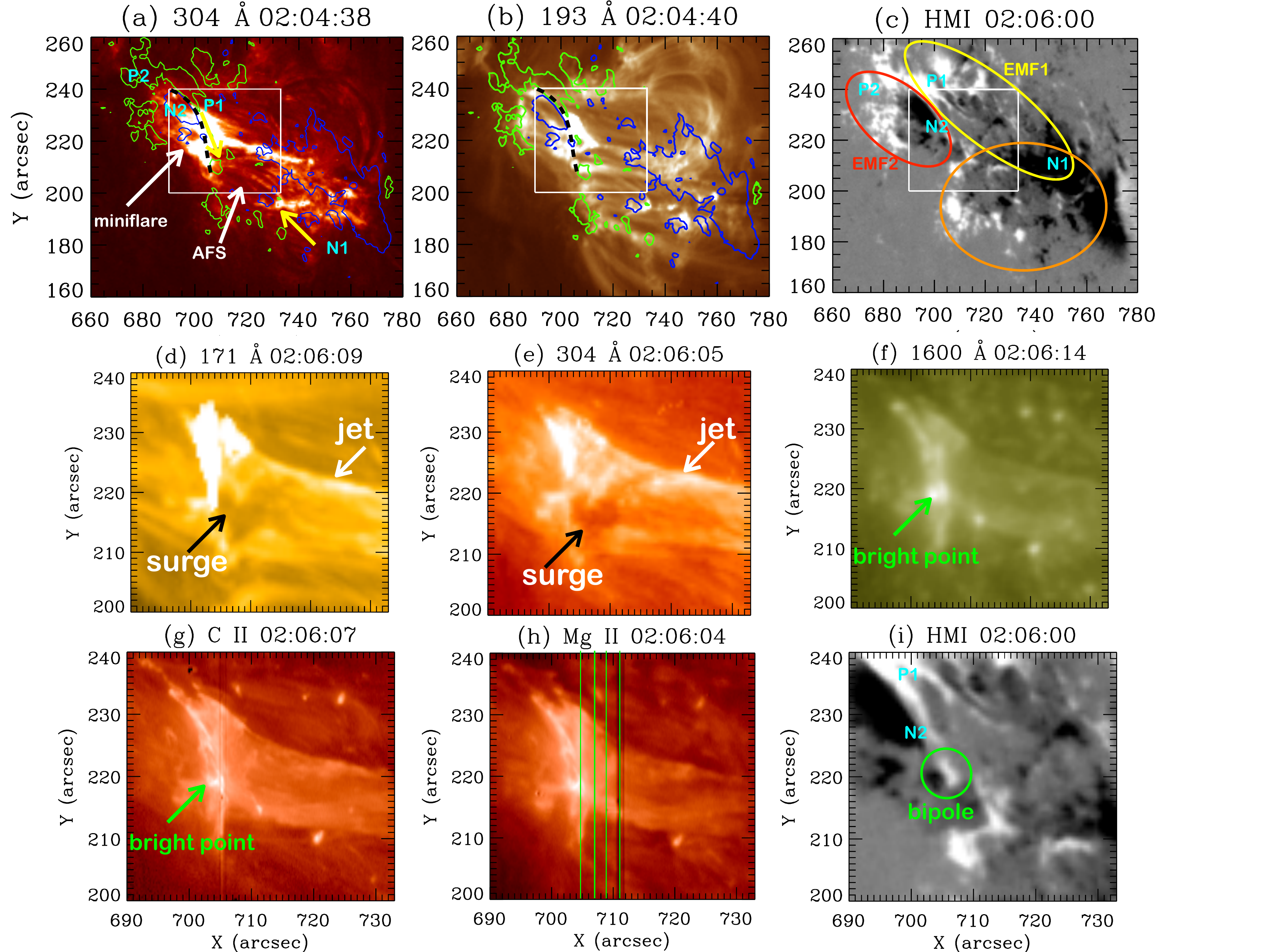}
\caption[Multi-wavelength observations of the  solar jet and mini-flare in different AIA and IRIS wavebands on March 22, 2019.]
{Multi-wavelength observations of the  solar jet and mini-flare in different AIA and IRIS wavebands. Panel (c): longitudinal magnetic field configuration consisting of EMF1 and EMF2 and an earlier EMF. 
In panel (i), the bipole  where the reconnection takes place
is presented.}
\label{AIA_IRIS}
\end{figure*}

   \begin{figure*}[ht!]
\includegraphics[width=1.0
\textwidth]{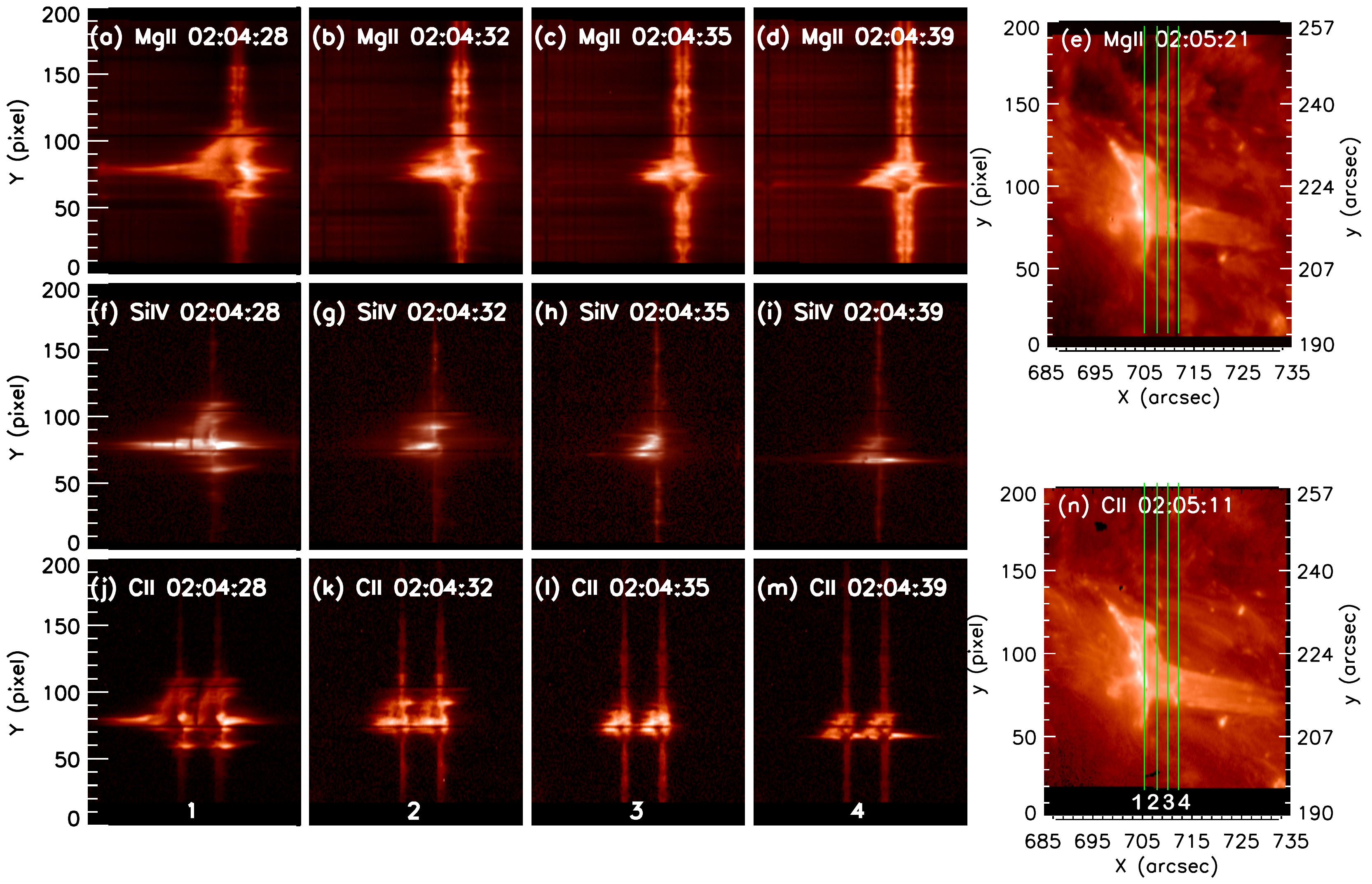}
\caption[IRIS spectra of the Mg II k line at 2796.35 \AA\ (a-d), Si IV 1402.77 \AA\ line (f-i), and   C II 1330 \AA\ line (j-m) at the four slit positions.]{({\it Left columns 1- 4}) IRIS spectra of the Mg II k line at 2796.35 \AA\ (a-d), Si IV 1402.77 \AA\ line (f-i), and  C II 1330 \AA\ line (j-m) at the four slit positions.} 
\label{spectra}
\end{figure*}
  
\section{Spectroscopic observations}
\label{obs2}
 The AR  has been  formed by successive emerging fluxes during 24 hours before the jet observations. The AR  magnetic configuration at the time of the mini flare  consists of three EMFs: an earlier one (orange oval) and  two very active EMFs:  EMF1 (P1-N1) and EMF2 (P2-N2) 
  highlighted  by  the yellow and the red  ovals (Figure \ref{AIA_IRIS} panel (c)). 
  The contours of the longitudinal magnetic field ($\pm$ 300 Gauss) is overlaid  on AIA 304  \AA\ and
  193 \AA\ images (Figure \ref{AIA_IRIS} a-b). The  polarity inversion line  (PIL)
 between these two  EMFs  (between P1 and N2 more precisely)
 is shown by a dashed dark line in panel (a-b).
  The images of second and third rows in Figure \ref{AIA_IRIS} present a zoom view of the mini flare at the jet base at 02:06:05 UT observed with AIA.  
  In panels (d-f) the jet is seen to develop westwards while the mini flare corresponds to a  North-South arch-shape  brightening along the PIL and 
  a bright point
  in its middle (Figure \ref{AIA_IRIS} panel f-g).

IRIS  provides line profiles in  Mg II k  and h lines (2796.4\,\AA{} and 2803.5\,\AA{} respectively),  Si IV (1393.76 \AA, 1402.77 \AA) and C II  (1334.54 \AA, 1335.72 \AA) lines  along the four slit positions (slit length of 202 pixels equivalent to 62$\arcsec$). The Mg II h and k lines are formed at chromospheric temperatures, {\it e.g.} between 8000 K and 20000 K (\citealt{Pontieu2014}; \citealt{Heinzel2014b}; \citealt{Alissandrakis2018}). C II is formed around T = 30,000 K and Si IV around 80,000 K.
Many other  chromospheric and photospheric lines have been identified in the spectra of the mini flare (Table \ref{table2}).
 \begin{table*}[ht]
\caption{Identification of  the lines in IRIS wavelength ranges of  C II, Si IV, and Mg II lines observed in the mini flare at the jet base; (bl) means blended.}
\bigskip
\centering
\setlength{\tabcolsep}{10.0pt}
\begin{tabular}{llllllll}
\hline
Ion   & $\lambda$ (\AA)    && Ion &$\lambda$ (\AA)&  & Ion &$\lambda$ (\AA)\\
\hline
C II &1334.54       &&  O IV &1399.776&& Mg II triplet   & 2791.6 \\
C II  &1335.72           &&  O IV&1401.163 && Mg II k  &  2796.4  \\
Fe II & 1392.817   &&  Si IV &1402.77  && Mg II triplet & 2797.9\\
Ni II  &1393.33              && O IV &1404.806 (bl) &&   &2798.0 \\
  Si IV &    1393.589                               && Si IV&1404.85 (bl) & &Mg II h &2803.5 \\
 Si IV&     1393.76                            &&S IV &1406.06 &&  &   \\
 \hline
 \label{table2}
 \end{tabular}
\end{table*}
\begin{figure*}[ht!]
\includegraphics[width=1.0
\textwidth]{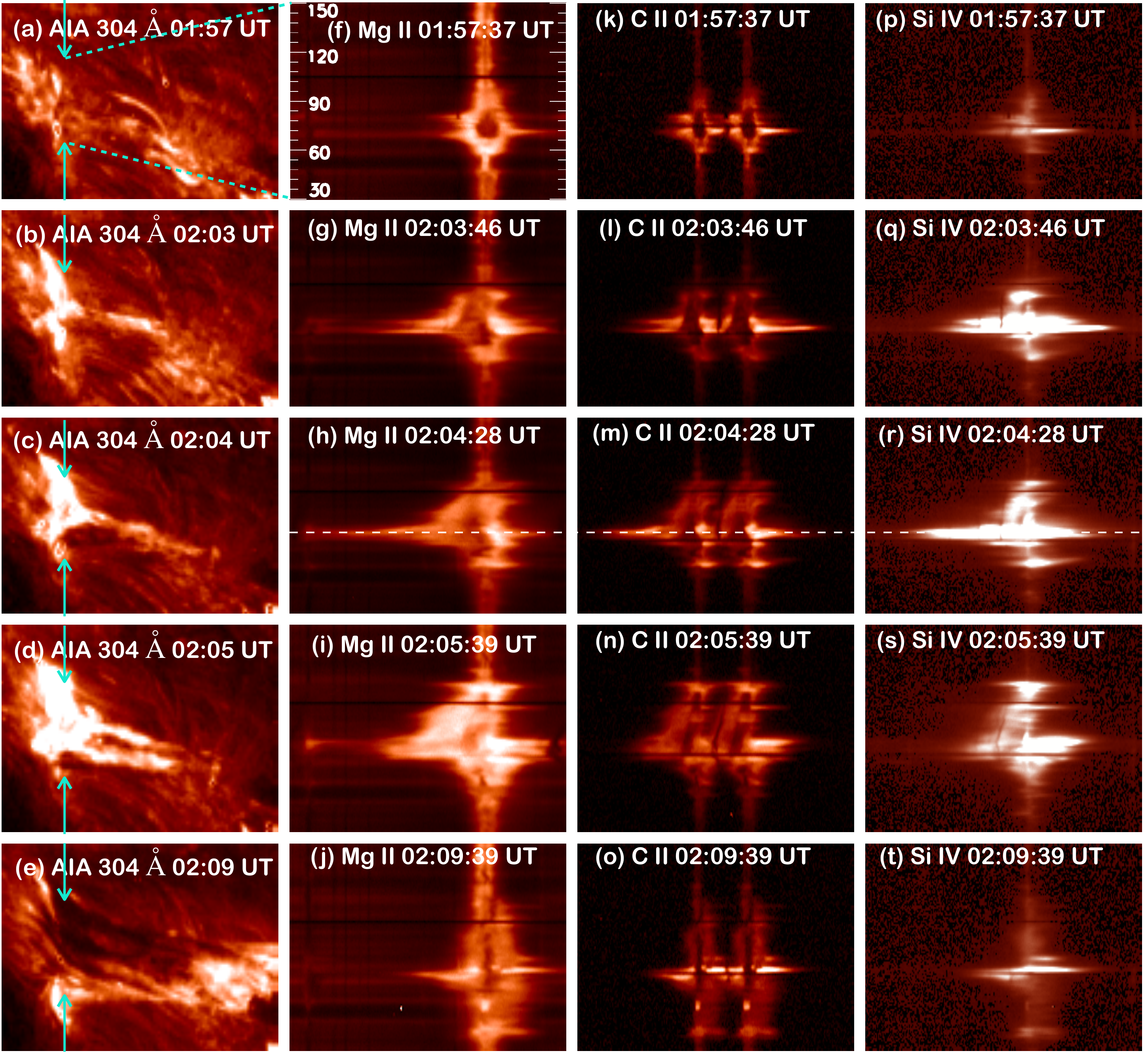}
\caption[Jet reconnection base and jet evolution between 01:57 UT to 02:09 UT.]
{Jet reconnection base and jet evolution  between (from top to bottom). First column presents the images in AIA 304 \AA. Second, third, and last columns show IRIS spectra  of the jet reconnection  site at slit 1 in the Mg II k 2796.35 \AA\  line , C II doublet, and in Si IV  1393.76 \AA\ respectively. The vertical cyan arrows  and inclined dashed lines  in the first column indicate the position of the slit.}

\label{AIA_IRISspectra}
\end{figure*}
\subsection{Mini flare and jet  observed with AIA and IRIS} 
 \label{miniflare_AIAIRIS}
 An example of 
 IRIS SJIs in 1330 \AA\ and 2796 \AA\, is presented in 
 Figure \ref{AIA_IRIS} (g-h). 
 The FOV of IRIS SJIs includes
  the mini flare (bright point in panels (f-g)) and a part of the wide jet base.  The bright point is considered as 
 as  the reconnection site (or `X' point) at the jet base. The four  positions of the slit scanned the mini flare site around x= 705$\arcsec$ and y = 220$\arcsec$ and the arch-shape  brightening  at the base of the jet (panel (h)).
  Globally the structures visible in IRIS SJIs  are similar to those in AIA 304 \AA\ (50,000 K).  The FOV of IRIS has been shifted by 4$\arcsec$ in x axis and 3$\arcsec$ in y axis to be co-aligned with AIA coordinates.  
  
 In AIA 304 \AA\, between  02:04:09 UT to 02:06:09 UT we see that the 
 jet  base has a triangular shape.
 Between the two external sides of the jet's triangular base there are two slightly bright patches in an East-West direction.
In  C II observations we can follow the formation of small kernels at 02:04:28 UT, 02:05:25 UT,  02:05:39 UT, and 02:06:07 UT, travelling from one side to the other side of the triangle following these bright patches (from east to west). 
  In  AIA 304 \AA\ images,  the development of the surge is well visible in between  02:04 and 02:07 UT (Figure \ref{AIA_IRIS} panel (e) at 02:06:05 UT). However, the surge is not so well visible in 
  IRIS SJIs 1330 \AA\ and 2796 \AA\  taken at the corresponding times (Figure \ref{AIA_IRIS} g-h).  This can be explained because the  absorption of the UV emission is only  efficient   for lines with  wavelengths below the hydrogen Lyman continuum limit ($\lambda$ $<$ 912 \AA)  (\citealt{Schmieder2004}).
  Moreover, the non-visibility of the surge can be due to the large wavelength ranges of the IRIS SJIs filters where  the full line profiles  are integrated and the line emission in the jet was not strong enough. 

    \subsection{IRIS spectra of mini flare}
  \label{Dopplershift}
  \label{IRIS_miniflare}  
  \label{time_IRIS}  
To process the IRIS Mg II h and k data, we used the
spatial and wavelength information in the header of the
IRIS level-2 data and derived the rest wavelengths of
the Mg II k  2796.35 (4) \AA, and  Mg II h 2803.52 (6) \AA\, from the reversal positions of the averaged spectra at the disk. 
For C II and Si IV lines the zero velocity is defined in a similar way  (Table \ref{table2} for the rest wavelengths used in the present work).
  We show one example of spectroscopic data
  obtained between 02:04:28 UT  and 02:04:39 UT with  the four slit positions 1, 2, 3, 4 for the three different elements Mg II, C II, and Si IV (Figure \ref{spectra}). 
 The correspondence between pixels  along the slit and arcsecs in SJIs is shown as y coordinates of 
  the SJIs (Figure  \ref{spectra} panels e and n).
  The slit  position 1 shown in panel (n) crosses the bright zone between 60 to 105 pixels  (around 210-230 arcsec) corresponding to the jet base.
  In the middle of the zone, the brightest point
  along the slit   is
  the reconnection site (y $\approx$ pixel 79- 80 corresponding to the position `X' (705$\arcsec$, 220$\arcsec$) in Figure \ref{AIA_IRIS} (g) and \ref{spectra} (e,n)).
  At the reconnection site the spectra shows very complex structures that we will analyse in the next sections.  We note that in all the slit positions 1-4, similar features are shown, but they are more pronounced in the slit position 1, which seems to be exactly at the reconnection site  for this time. We will mainly restrict our study to the slit position 1.
  It is not really possible to reconstruct  an adequate spectroheliogram image with  only four positions  distant in x  of 2$\arcsec$ each. 
  
\begin{figure*}[ht!]
\includegraphics[width=1.00
\textwidth]{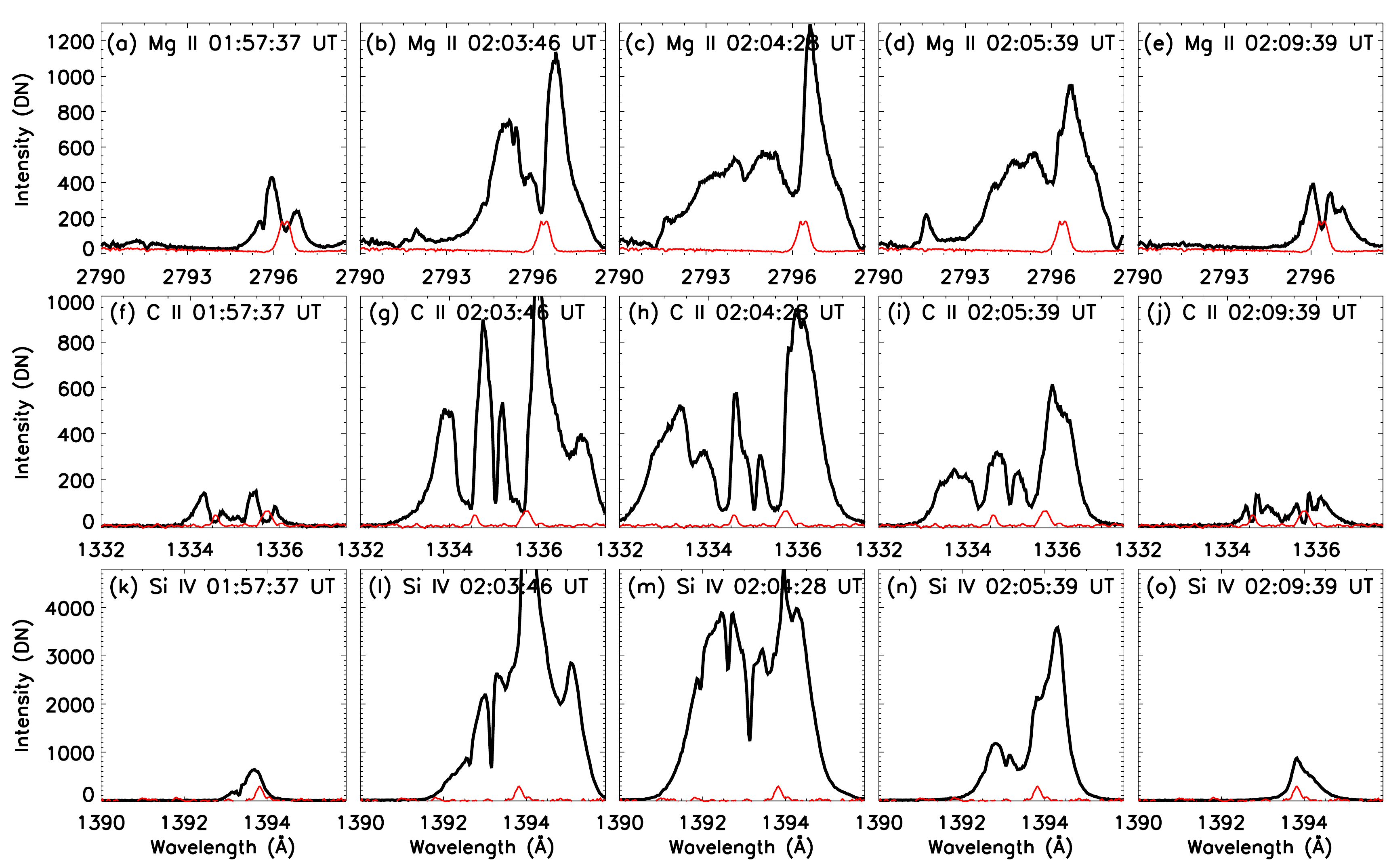}
\caption[Evolution of jet reconnection site  (UV burst) between 01:57 UT and 02:09 UT through the profiles of the Mg II k, C II, and Si IV  1394 lines.]
{Evolution of jet reconnection site (UV burst) through the profiles of the Mg II  k (a-e), C II (f-j), and Si IV  1394  (k-o) lines observed in slit 1. The location of the y point in the different spectras is shown in Figure \ref{AIA_IRISspectra} with a white dashed line.}
\label{threeProfiles}
\end{figure*}  
  We observed that AIA 304 \AA\ images have a better contrast than the IRIS SJIs to show the cool structures visible by absorption.
  Therefore  we co-align carefully the 
  images in AIA 304 \AA\ with the IRIS SJIs in order to indicate
  exactly the position of each pixel of the slit in the AIA 304 \AA\, images  to be able to 
  discuss the evolution of the structures visible in the 304 \AA\, images jointly with the spectra shape of IRIS lines using both coordinates the pixels along 
  the slit  and the AIA  coordinates.
  The evolution of the structures 
 visible  in AIA 304 \AA\ images: mini flare, jet and surge
 are  summarized 
 in five different times in Figure \ref{AIA_IRISspectra}, corresponding to: pre reconnection time (first row), reconnection times (second and third rows), jet base extension (forth row), after reconnection time (fifth row). Between the two vertical blue arrows in the AIA images (left column),  a  section of slit at position 1 is located. The right columns present the spectra in this section for the three  elements Mg II, C II and Si IV.
   Table \ref{table4} gives a  detail about the characteristics of these  typical profiles in the four slit positions during the different phases of the jet time observations.  They are changing very fast and it is rather complicated to analyse all of them.
   
  \subsection{Characteristics of the IRIS spectra}
  \label{comparison2}
  For each  IRIS spectra shown in Figure \ref{AIA_IRISspectra} we select the pixel 79  corresponding to the reconnection site and  draw the   line profile  of the three elements  for the five times (Figure \ref{threeProfiles}). The line profiles help  to  interpret  the nature and evolution of the structures during the different phases of the reconnection. We focus our analysis of Mg II and C II chromospheric lines in the following subsections during the three phases of the jet reconnection. Then we analyse in details the line profiles of the three elements  during the reconnection times.

 Around 01:57 UT in AIA 304 \AA\ image, tiny vertical bright 
 areas along the inversion line are visible (panel (a1) in Figure \ref{AIA_IRISspectra}). The corresponding  C II and Mg II  spectra show  very large central dip which could represent the presence of cool material at rest which  absorbs the incident radiation (Figure \ref{AIA_IRISspectra} panels (b1 and c1)) and the corresponding line profiles in  
Figure \ref{threeProfiles}  panels (a and f).
The Mg II k and C II line profiles at this  position  and around (pixels= 70, 76, 79) are presented  again  but with a zoom  and  
with the x-axis in Dopplershift units in km s$^{-1}$  (Figure \ref{Mg_CII_lines}). They are  very broad  with a central dip (FWHM more than 1 \AA\  which corresponds to $\pm$50 km s$^{-1}$) while the peaks of  the Mg II and C II lines are equally distant (100 km s$^{-1}$). The central dip would imply that cool material absorbed the incident radiation
more or less at the rest. Such cool material could be due to parts of  arch filaments 
trapped in  the magnetic field lines  between the two EMFs (EMF1 and EMF2) in the vicinity of the bright point region before the reconnection.

 Around 02:03 - 02:05 UT the  mini flare (UV burst at the `X' point) starts in the middle of this bright area  with  the onset of the jet  ejection (Figure \ref{AIA_IRISspectra} with  304 \AA\,images in panels (b,c,d), Mg II spectra in panels (g,h,i) and C II spectra  in panels (l,m,n)) and their corresponding line profiles in 
Figure \ref{threeProfiles}  panels (b,c,d and g,h,i). The bright jet is obscured by a surge, a  set of dark (cool) materials in front of it; both the jet and the surge are extending toward the West at the same time. In AIA 304 \AA\  image at 02:05 UT the jet is extended along two bright  branches with  a dark area in between. During this time the spectra show very broad blue  wings along the slit in the same zone $\pm$ 10 $\arcsec$ around y=220$\arcsec$. 
 At y= 79 pixel (approximately at 220$\arcsec$), the profiles in all the lines are the  most extended. We notice that the spectra along the slit show a tilt at the northern bright branch (Figure \ref{AIA_IRISspectra} panels i,n), which could indicate some rotational motion there (Section \ref{sec:tilt} for more details).
 The wavelength positions of the dark absorption core of the Mg II and C II line profiles 
 along the slit show a clear zigzag pattern of the blue and red shifts  which could correspond to cool plasma motion with different  velocities  along the slit.
  In the next sections (Sects. \ref{UV_burst} and \ref{sec:cloud}) we analyse the profiles of the three lines
  to obtain  quantitative values of the Dopplershifts of the plasma in  the  reconnection zone of the jet.

 At 02:09 UT, long dark East-West filament structures in the North of the reconnection site are observed
 in the 304 \AA\ images (panel (e) in Figure \ref{AIA_IRISspectra}).
 Their corresponding spectra show a dark core and  weak emission in the   red  wings  all along the slit (y(pixel)=70-105) which could correspond to  cool plasma absorbing the red peak emission of the jet. This cool plasma may be plasma of the surge or to the AFS  going  away of the observer with Dopplershifts of less than 30 km s$^{-1}$ (Figure \ref{AIA_IRISspectra}).

  \begin{figure*}[!t]
\centering
\includegraphics[width=1.0 \textwidth]{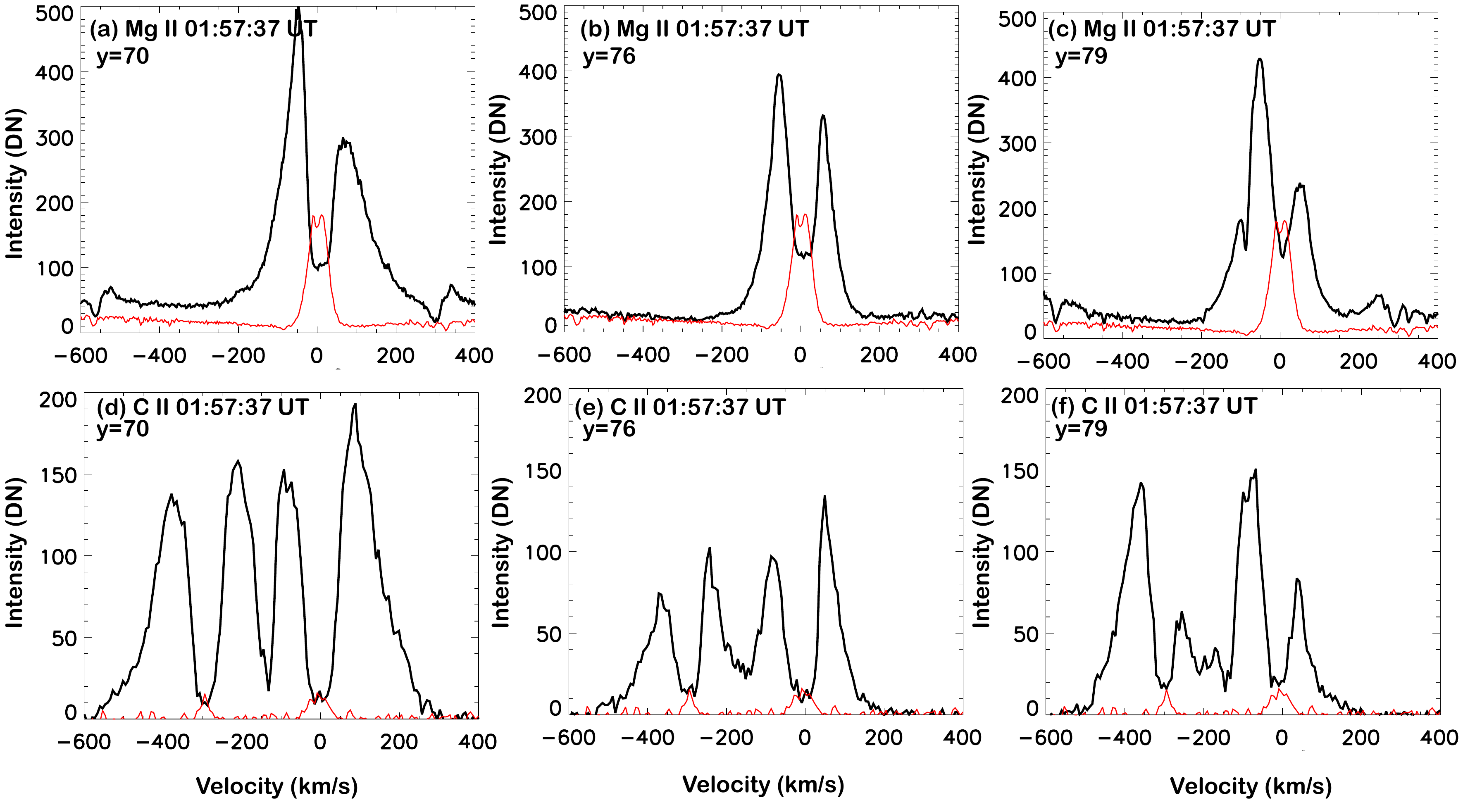}
\caption[Comparison of the profiles of Mg II k and the two C II lines in
three pixels along the slit at the position of the UV burst.]
{Comparison of the profiles of Mg II k (top row) and  the two C II (bottom row) lines in
three pixels along the slit at the position of the UV burst 
at 01:57:37 UT before the burst. The red profiles are reference profiles, which were used to determine the rest wavelengths.}
\label{Mg_CII_lines}
\end{figure*}
\begin{figure*}[ht!]
\centering
\includegraphics[width=1.0 \textwidth]{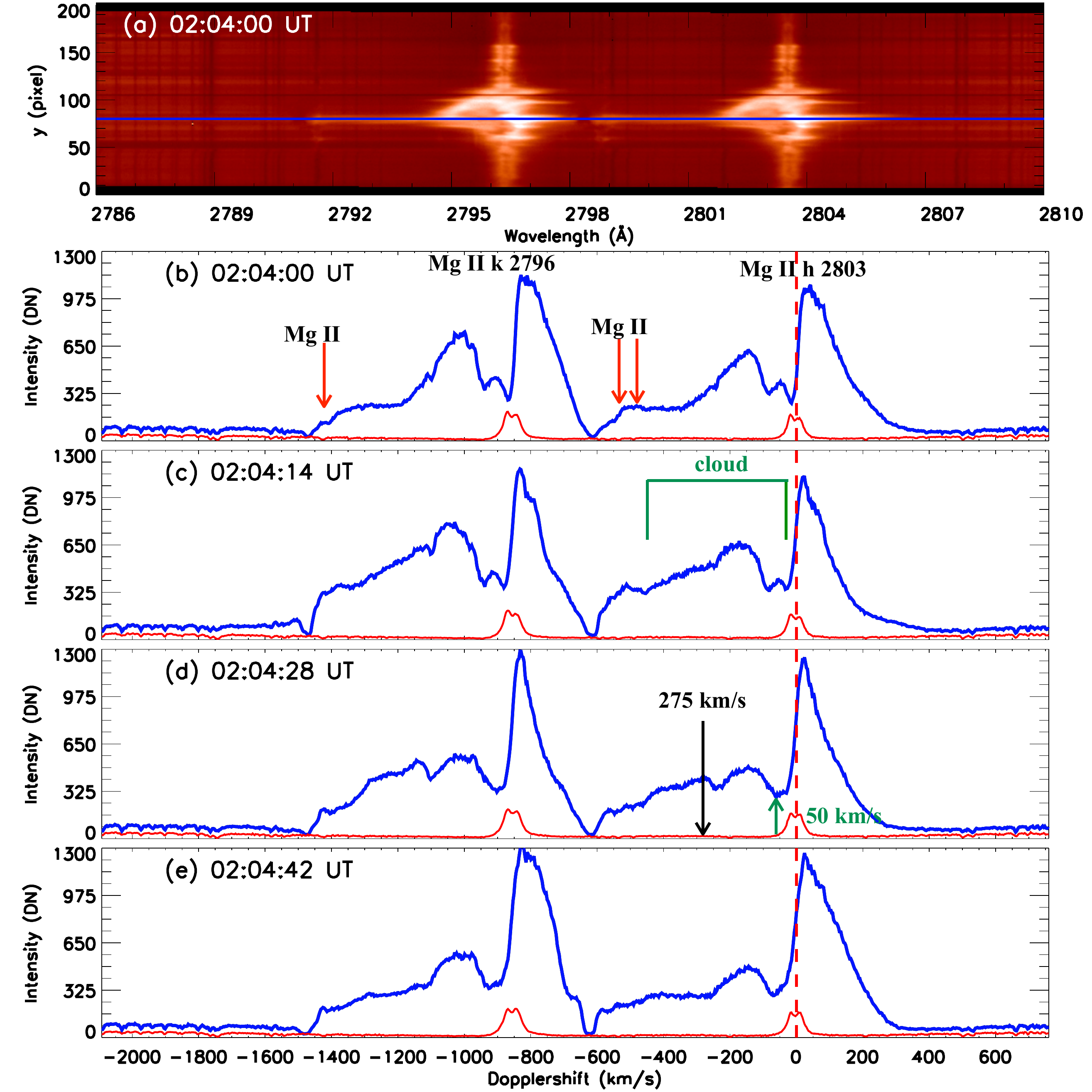}
\caption[Mg II spectra at 02:04:00 UT along the slit position 1 before the UV burst.]
{(a) Mg II spectra at 02:04:00 UT along the slit position 1 before the UV burst  (Figure \ref{AIA_IRISspectra} g). Panels (b-e) from top to bottom:  evolution of the Mg II k and h line profiles showed by a profile every 14 sec during less than one minute time.}
\label{MgII_evolution}
\end{figure*}
  \subsection{UV burst in `X' point }
  \label{UV_burst}

The IRIS slit at position 1
crosses the mini flare (UV burst at `X' point) so that the
spectra along the slit bring many information about the dynamics of the UV burst as explained in the previous section (Section \ref{comparison2}) and in 
Figure \ref{AIA_IRISspectra}. Figure \ref{threeProfiles}
shows the evolution of the UV burst 
  using the three lines (Mg II, C II, and Si IV) profiles for  y(pixel)=79 (220$\arcsec$) with a time scale of  one  minute. 
 The  profiles change very fast on this time scale.  We  analyse  these profiles at each time in order to derive  the characteristics   (velocity and temperature) of  structures which are  integrated along the LOS.
  
\begin{figure}[ht!]
\centering
\includegraphics[width=0.80 \textwidth]{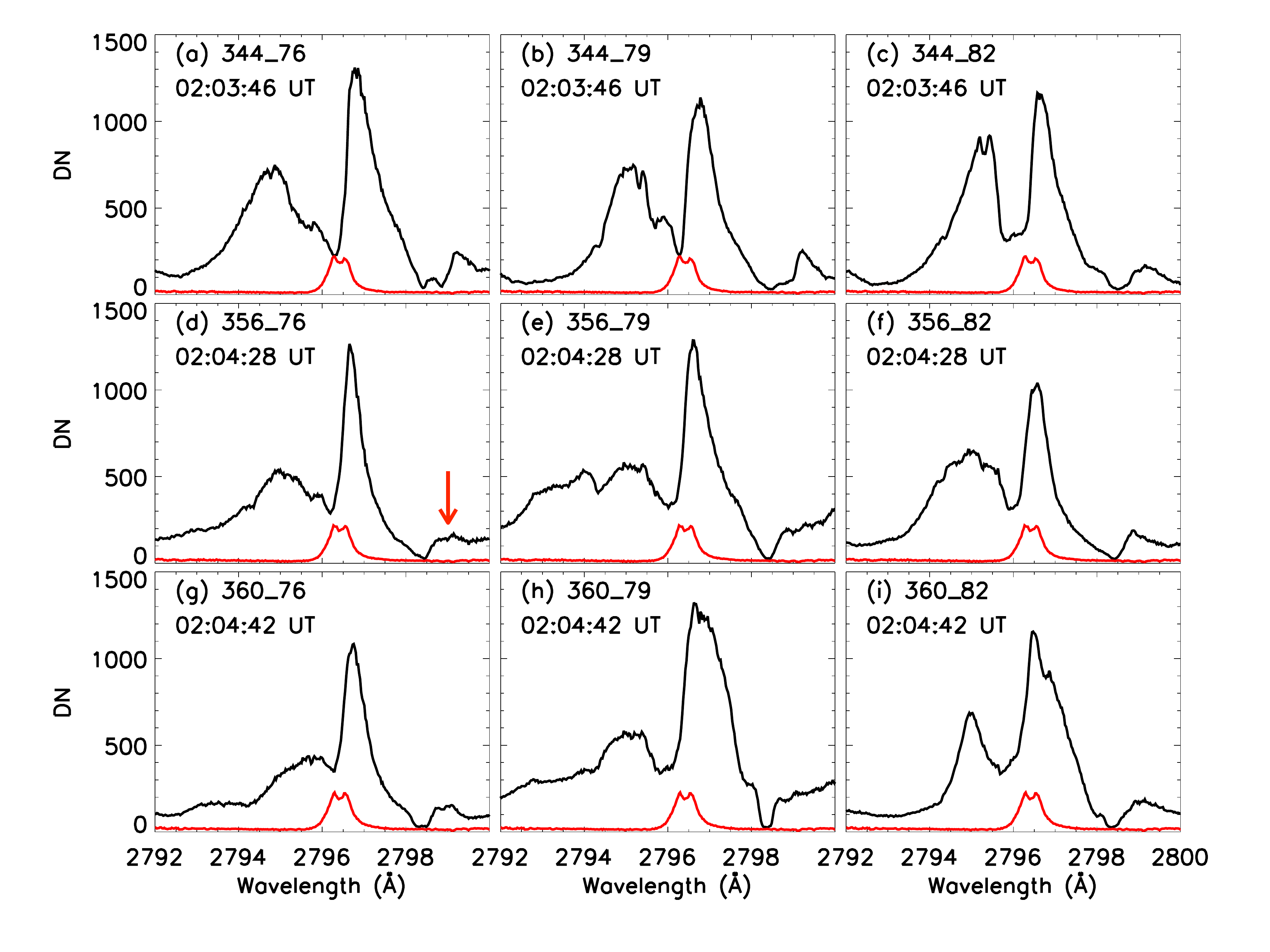}
\caption[Evolution of Mg II k line profiles  inside the UV burst.]{Evolution of Mg II k line profiles inside the UV burst.  The red arrows indicate the emission of the Mg II triplet lines. \label{burst_MgII}}
\end{figure}
\begin{figure}[ht!]
\centering
\includegraphics[width=0.80 \textwidth]{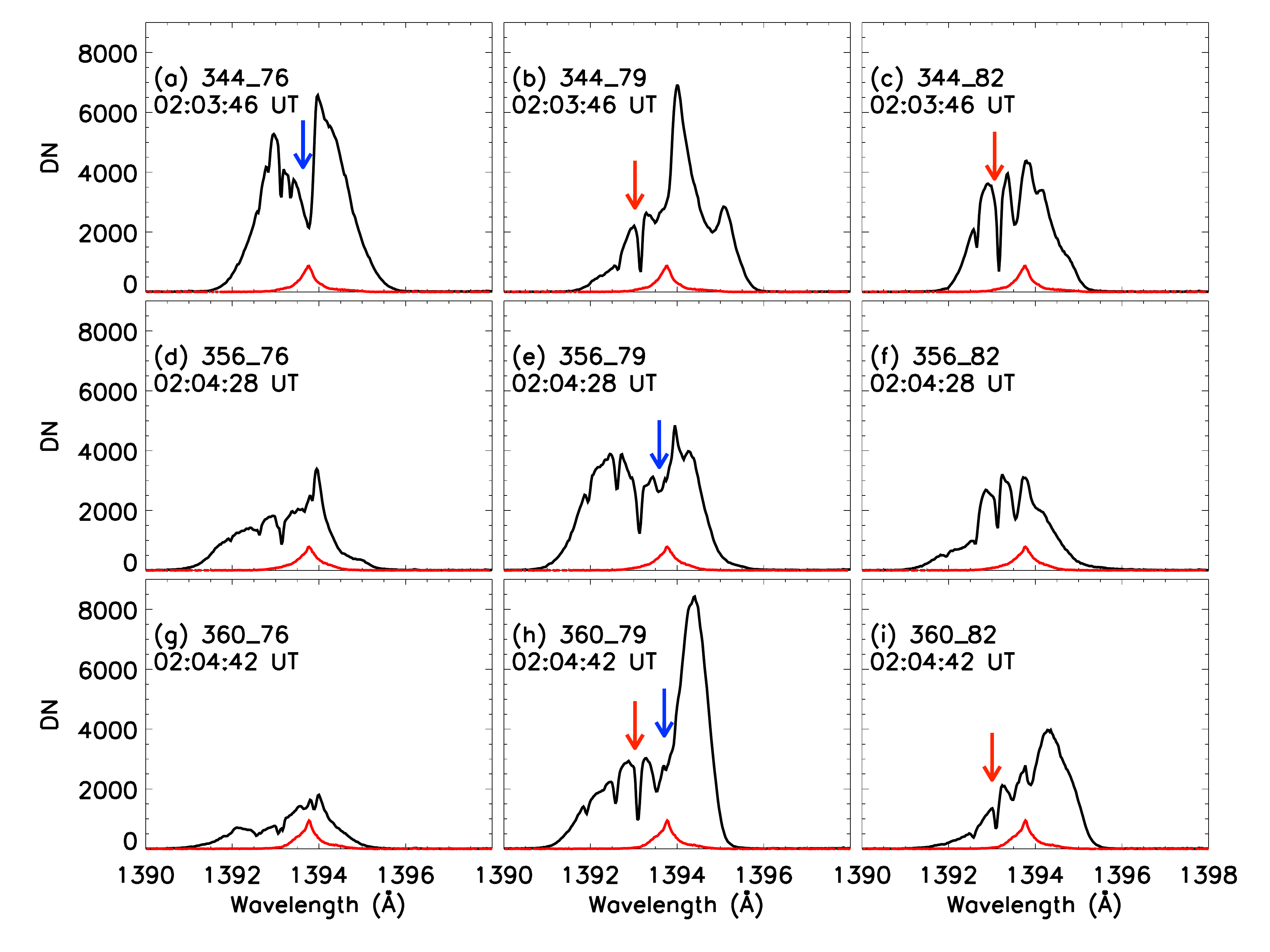}
\caption[Evolution of  Si IV 1393.76 \AA\ line profiles  inside the UV burst.]{Evolution of  Si IV 1393.76 \AA\ line profiles inside the UV burst. The UV burst in located in slit 1. The red arrows indicate the absorption by the  Ni II 1393.33 \AA\, and Fe II 1393.589 \AA.}
\label{burst_SiIV}
\end{figure}
At 02:03:46 UT  in very localized pixels  inside the burst   Mg II, C II and Si IV  profiles have   more or less symmetrical profiles with high peaks  with extended blue and red wings ($\pm$ 200 km s$^{-1}$, Figure \ref{threeProfiles} panels 
b, g, i and Figure \ref{burst_MgII}, \ref{burst_SiIV}). 
With  such  extended wing  profiles in a few pixels we may think of bilateral outflows  of reconnection 
(\citealt{Ruan2019}). Such  outflows 
with   super Alfv\'enic speeds were  observed in a    direction perpendicular to the jet initiated by the reconnection like in our observations. 
\begin{figure*}[!ht]
\centering
\includegraphics[width=1.0 \textwidth]{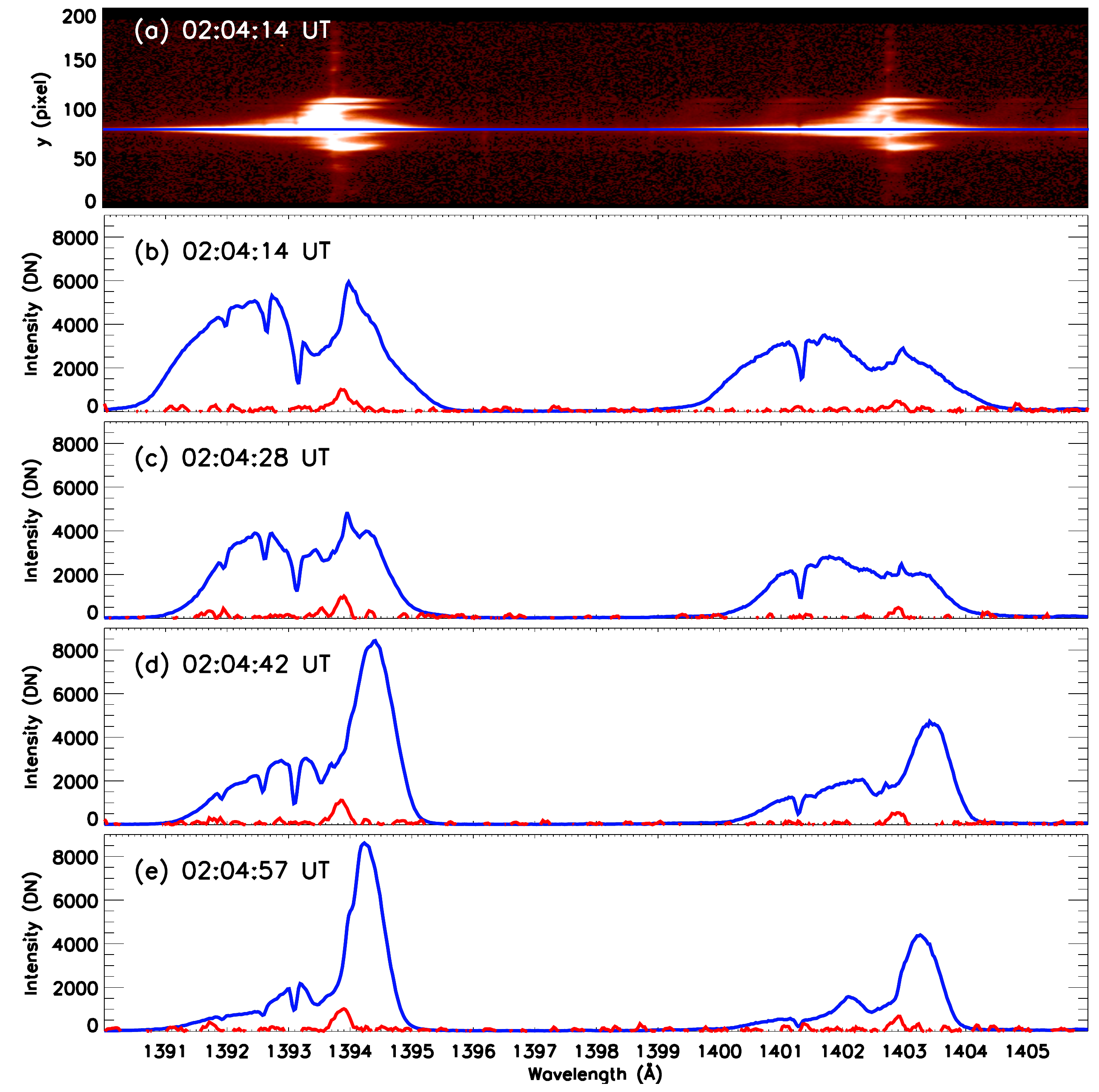}
\caption[Si IV spectra at 02:04:14 UT at the start of the UV burst.]
{Panel (a): Si IV  spectra at 02:04:14 UT at the start of  the UV burst. Panels (b-e): from top to bottom: fast evolution  during less than one minute, one  Si IV  profile every 14 s at slit position 1 between  02:04:14 UT and 02:04:57 UT. The reference profiles are shown in red.}

\label{SiIV_evolution}
\end{figure*}
For the time of reconnection  around  02:04:28 UT, Mg II, Si IV and C II line spectra are presented for the four slit positions (Figure \ref{threeProfiles}).
 The profiles at this time at y=79 pixel exhibit a  high  peak of   
  emission with strong blue shift extended wing (third column (panels c,h,m) in Figure \ref{threeProfiles}) although 
  the evolution of the profiles are shown with a low cadence. In Figures \ref{MgII_evolution} and \ref{SiIV_evolution} we show  the details of the fast evolution of the UV burst between 02:04:14 UT and 02:04:28 UT taking advantage of the high cadence of IRIS. 
  
  The Mg II profiles of the UV burst during  this time scale did not evolve drastically, contrary to the Si IV profiles in the same time interval. The Si IV profiles are very broad during the UV burst maximum with a FWHM of the order of 4 \AA.  A few seconds later at  02:04:57 UT the Si IV profiles consist only  of  one peak  with a FWHM of 1 \AA\, and  an  intensity increasing about a factor of 100. 
  We analyse  first the Mg II profiles of the  mini flare  (UV burst)  to understand the composition  and the dynamics of the plasma along the LOS.
  The zero velocity is  defined, as we explained earlier, by  the dip in the reverse profile of Mg II line profile observed in the chromosphere.
  The Mg II profiles are  also very broad with a FWHM  of the order of 5 \AA\, and asymmetric with a high red peak and a very extended blue wing (Figure \ref{MgII_evolution}).  
 The  blue peak  is much lowered compared to the red one. This characteristic of the profiles can be produced by the  absorption of the blue peak  emission by a cloud  centered around 50 km s$^{-1}$. In the far blue wing an emission  is detected until -5 \AA, which might  come from a  second  cloud centered around a higher value (intuitively determined around - 275 km s$^{-1}$) and  the emission of
 the Mg II triplet at 
 2797.9 and 2798 \AA\ which are  effectively at -5 \AA\ from Mg II h. All the  Mg II triplet lines have been identified in the spectra. The profiles of   Mg II, C II, Si IV lines at the UV burst  plotted in Dopplershift units relative to the rest
wavelength show similar velocities, 
  which indicates that they correspond to real plasma moving with high flow speed (Figure \ref{MgII_SiIV}).
  Although the extended  blue wings could also  be  interpreted as due to a velocity gradient inside the cloud along the LOS instead of a moving cloud. In the next section we apply a cloud model technique to have quantitative results.
  \begin{figure*}[!ht]
\centering
\includegraphics[width=1.0 \textwidth]{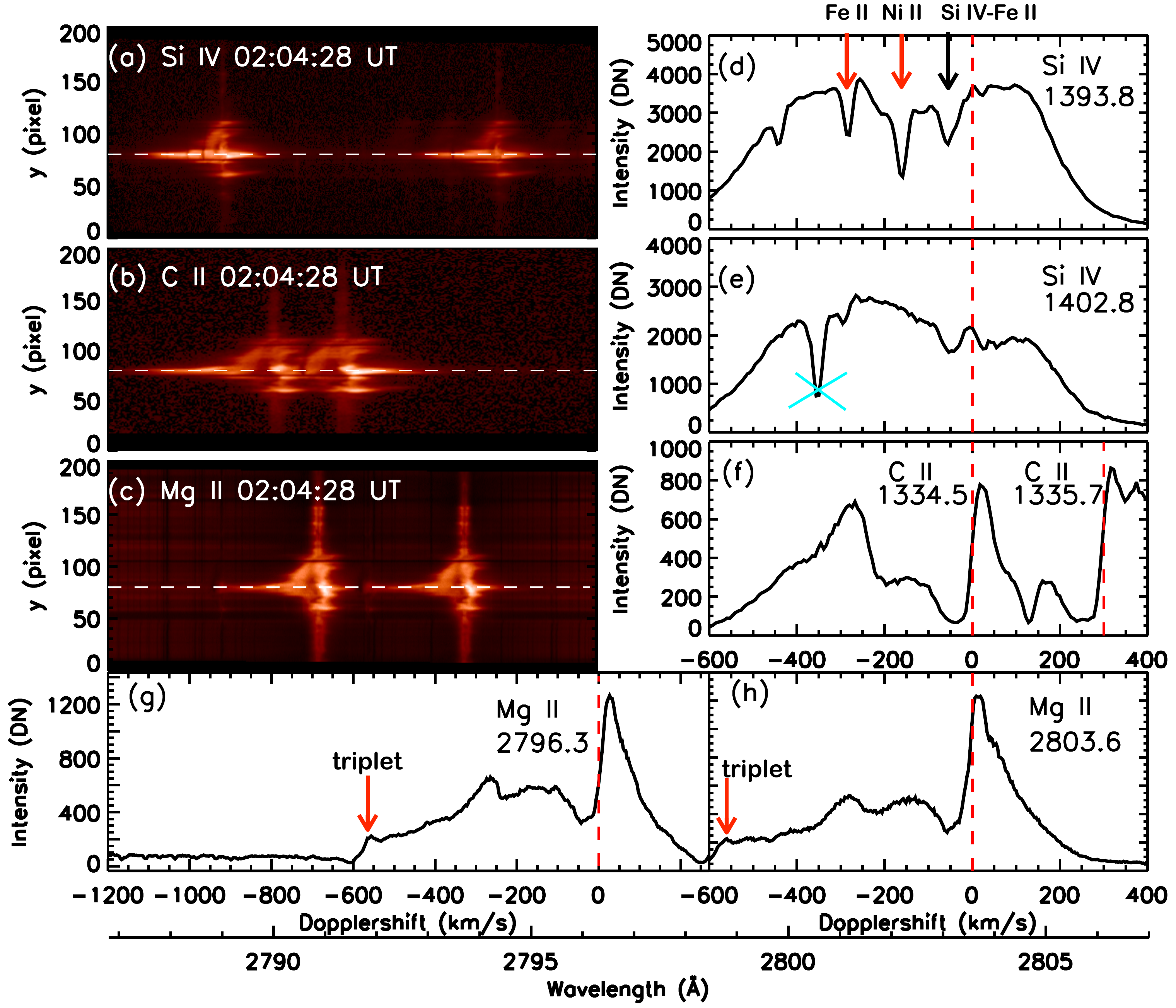}
\caption[Spectra of the jet base (UV burst) showing  the extended blue wing of 
 Si IV  line, C II line and Mg II line]
 {Spectra of the jet base (UV burst) showing  the extended blue wing of  Si IV  line (panel (a)), C II line  (panel (b)) and Mg II  line (panel (c)) at 02:04:28 UT, the white horizontal dashed lines in these three panels indicate the position where the profiles are drawn in panels (d-h).}

\label{MgII_SiIV}
\end{figure*}
  \section{Cloud model method for Mg II lines}
  \label{sec:cloud}
 Cloud model method was first introduced by \citealt{Beckers1964} for understanding asymmetric line profiles in the chromosphere. The structure overlying the chromosphere is defined by four constant parameters: optical thickness, source function, Doppler width  and radial velocity.  Moreover, \citealt{Mein1988} developed the  cloud method by  considering non constant  source function and velocity gradients. Therefore this technique was  applied for different structures  with large velocities, mainly observed in the H$_\alpha$ line, {\it e.g.} post flare loops (\citealt{Gu1992}; \citealt{Heinzel1992}),  spicules on the disk (\citealt{Heinzel1994}), and  atmospheric structures in the quiet-Sun (\citealt{Mein1996}; \citealt{Chae2020}), even using   
  multi clouds (\citealt{Tziotziou2007}). 
  This new development  allows to derive dynamical models in the chromosphere (\citealt{Heinzel1999}). This technique is valid for a  chromospheric structure with a large discontinuity (high radial velocity), overlying the chromosphere along the LOS.
  
Recently \citealt{Tei2018} 
 applied the cloud model  technique with constant source functions to  the Mg II lines  which present complex profiles because of their  central reversal. Considering  multi clouds, Mg II complex profiles of off-limb spicules were  successfully fitted (\citealt{Tei2020}). Cloud model technique applied to Mg II lines allows us to  unveil the existence of moving clouds over the chromosphere. For the present analysis, this is how during the peak phase of the reconnection two clouds
  overlying the region of reconnection are considered to fit the asymmetric Mg II profiles observed in the UV burst region.
  Mg II  asymmetric line profiles are assumed to be the result of the presence of two overlapping clouds 
{\it c}1 and {\it c}2 located above a background atmosphere along the LOS. We suppose the background atmosphere is symmetric with high peaks in the Mg II lines.
We consider a situation where the cloud {\it c}2 is located above the cloud {\it c}1 along the LOS. Assumptions for the two clouds are as follow;
 \begin{enumerate} 
 \item {The absorption profile of a cloud  has a Gaussian shape.}
 \item {The  two clouds have generally different physical properties.} 
 \item {The source function, 
 the LOS velocity, the temperature, the turbulent velocity in each cloud are independent of depth (constant in the cloud).}
  \end{enumerate} 

The total observed intensity $I_m(\Delta\lambda)$ emitted, when there is one cloud (m=1) or there are two clouds (m=2) on the background atmosphere of intensity $I_0(\Delta\lambda)$ along the LOS, is given by the relation (\citealt{Mein1988}; \citealt{Heinzel1999}; \citealt{Tei2018}):
\begin{equation}
I_m(\Delta\lambda)=I_{m-1}(\Delta\lambda){\rm e}^{-\tau_m(\Delta\lambda)}+S_m[1-{\rm e}^{-\tau_m(\Delta\lambda)}],
\label{eq:m-cloud}
\end{equation}
where
$S_m$ is constant the source function  and
\begin{equation}
\tau_m(\Delta\lambda)\equiv \tau_{0, m}\exp\left[-\left(\frac{\Delta\lambda-\Delta\lambda_{{\rm LOS}, m}}{\Delta\lambda_{{\rm D}, m}}\right)^2\right]
\end{equation}
 is the optical thickness of the cloud {\it c}1 (m=1 case) or {\it c}2 (m=2 case) with the Doppler width (\citealt{Tziotziou2007}; \citealt{Chae2020}):
\begin{equation}
\Delta\lambda_{D, m}\equiv \frac{\lambda_0}{c}\sqrt{\frac{2k_{\rm B}T_m}{m_{\rm Mg}}+V_{{\rm turb}, m}^2}.
\end{equation}
Here,
$\Delta\lambda =\lambda-\lambda_0$ is the difference between the wavelength, $\lambda$, and the rest wavelength of the Mg II 
line considered, 
$\Delta\lambda_{{\rm LOS}, m}\equiv\lambda_0 V_{{\rm LOS}, m}/c$ is the shift of wavelength corresponding to the LOS velocity of the cloud of number {\it m}, $V_{{\rm LOS}, m}$ ($c$ is the light speed);
$T_m$ and $V_{{\rm turb}, m}$ are the temperature and the turbulent velocity of the cloud of number {\it m}, respectively;
$k_{\rm B}$ is the Boltzmann constant;
$m_{\rm Mg}$ is the atomic mass of magnesium (\citealt{Tei2020}; \citealt{Joshi2020IRIS}).
Combining the equation (\ref{eq:m-cloud}) of $m=1$ and the one of $m=2$, the total observed 
intensity $I_2(\Delta\lambda)$ emitted by two clouds is given by the relation:
\begin{equation}
\begin{aligned}
I_2(\Delta\lambda)=I_0(\Delta\lambda){\rm e}^{-\tau_1(\Delta\lambda)}{\rm e}^{-\tau_2(\Delta\lambda)}+S_1[1-{\rm e}^{-\tau_1(\Delta\lambda)}]{\rm e}^{-\tau_2(\Delta\lambda)}+ S_2[1-{\rm e}^{-\tau_2(\Delta\lambda)}].
 \end{aligned}
\end{equation}
\begin{figure}[ht!]
\centering
\includegraphics[width=0.75 \textwidth]{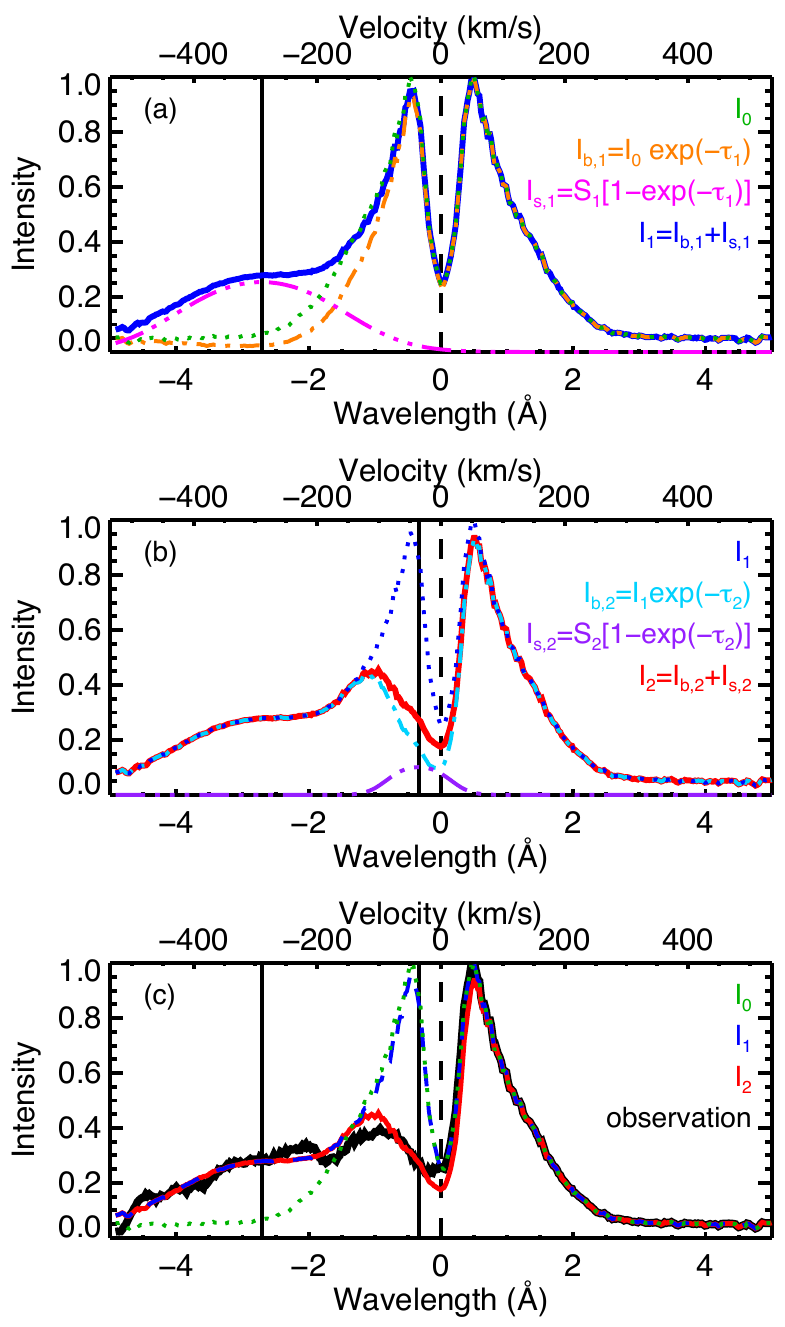}
\caption[Two-cloud model calculation of the Mg II k  
line profile at 02:04:28 UT.]
{Two-cloud model of Mg II k  
line profile at 02:04:28 UT. Panel (a): Detail of the $I_1$ profile. (b) Detail of the $I_2$ profile. (c) Comparison of observed (solid, black) and modeled (solid, red) profiles.}
\label{cloud1}
\end{figure}

In the present work, we adopt $T_1 = T_2 = 10^4$ K since the cloud temperatures do not affect the result as long as we use a temperature lower than 20000 K  at which Mg II is ionized.
This is because Mg atom is relatively heavy and the thermal width is small compared to the non-thermal velocity in this situation.
For the background intensity, $I_0(\Delta\lambda)$, we use a symmetric line profile constructed from the red-side of the observed  profile, as done by  \citealt{Tei2018}.
In addition, $\alpha_m$ is defined as the ratio of the source function of the cloud of number {\it m} to the background intensity at the line center [$\alpha_m\equiv S_m/I_0(\Delta\lambda=0)$].
We consider a situation where a low velocity component is in the foreground ({\it c}2) along the LOS in order to lower the  peak intensity as it is observed.
Figure \ref{cloud1} shows the result of a two--cloud model fitting. The values of the free parameters are summarized in Table \ref{tab:cloud}.
Two clouds have been detected, one with strong blueshifts (-290 km s$^{-1}$) and the other with a large optical thickness but lower blueshift (-36 km s$^{-1}$); these values are not far from our approximate estimation (Section \ref{UV_burst}). The turbulent velocity derived for the cloud {\it c}1 is large (150 km s$^{-1}$). This could correspond to the existence of a large velocity gradient inside the cloud, which has not be considered in the assumptions where on the contrary all the parameters are constant. On the other hand, the assumption of a  symmetrical  Mg II profile background for  the flare  does not influence  the fast cloud existence, since the wavelength range of this component is very far in the wing (Figure \ref{cloud1}). 

\begin{table}
\centering
  \caption{Results of two-cloud modeling (c1 and c2).}
\label{tab:cloud}
\setlength{\tabcolsep}{18pt}
\begin{tabular}{ccccc}
     \hline
      Cloud & $\alpha$ & $\tau_0$ & V$_{\rm LOS}$ & V$_{\rm turb}$\\
      & & & (km s$^{-1}$)& (km s$^{-1}$)\\
      \hline
      c1 & 1.6& 0.99& -290& 150 \\
      c2& 0.5& 1.6& -36& 50 \\
   \hline
    \end{tabular}
\end{table}

  \section{Spectral tilt profiles}
  \label{sec:tilt}
  Spectra 
  of the Mg II, C II, and Si IV lines show a spectral
tilt at 02:05:39 UT (Figure \ref{AIA_IRISspectra} panels (i,n,s)). Figure \ref{gradient_spectra_Mg} details the spectra of Mg II and Si IV lines for different times 02:05:25 UT and 02:05:39 UT  in two slit positions distant of 6$\arcsec$. The tilt is visible in these two positions, the profiles  have dominant
red wings in the southern part of the brightening
(y(pixel) = 50 to 79), they become roughly symmetric
in the middle of the brightening (y = 79) with large extended blue wing nevertheless
and show dominant blue wings in the northern
part  with decreasing blueshifts  until being symmetrical profiles (y = 79 to 120).
 The tilt is well visible in  O IV lines, lines in emission  identified in the vicinity of Si IV 1402.77 \AA\, (Table \ref{table2}).
We could quantify the displacement of the line according to the position along the slit.
These types of spectra
are well known and are typically associated
with twist (\citealt{Pontieu2014}) or rotation (\citealt{Rompolt1975}; \citealt{Curdt2012}) or the presence of 
 plasma in helical structures (\citealt{Li2014}). 
The tilt  observed in our  spectra can be explained by  the presence of an helical structure at the base during the reconnection process
due to  transfer of twist from a FR in the vicinity of the jet.
 \begin{figure*}[ht!]
\centering
\includegraphics[width=1.0 \textwidth]{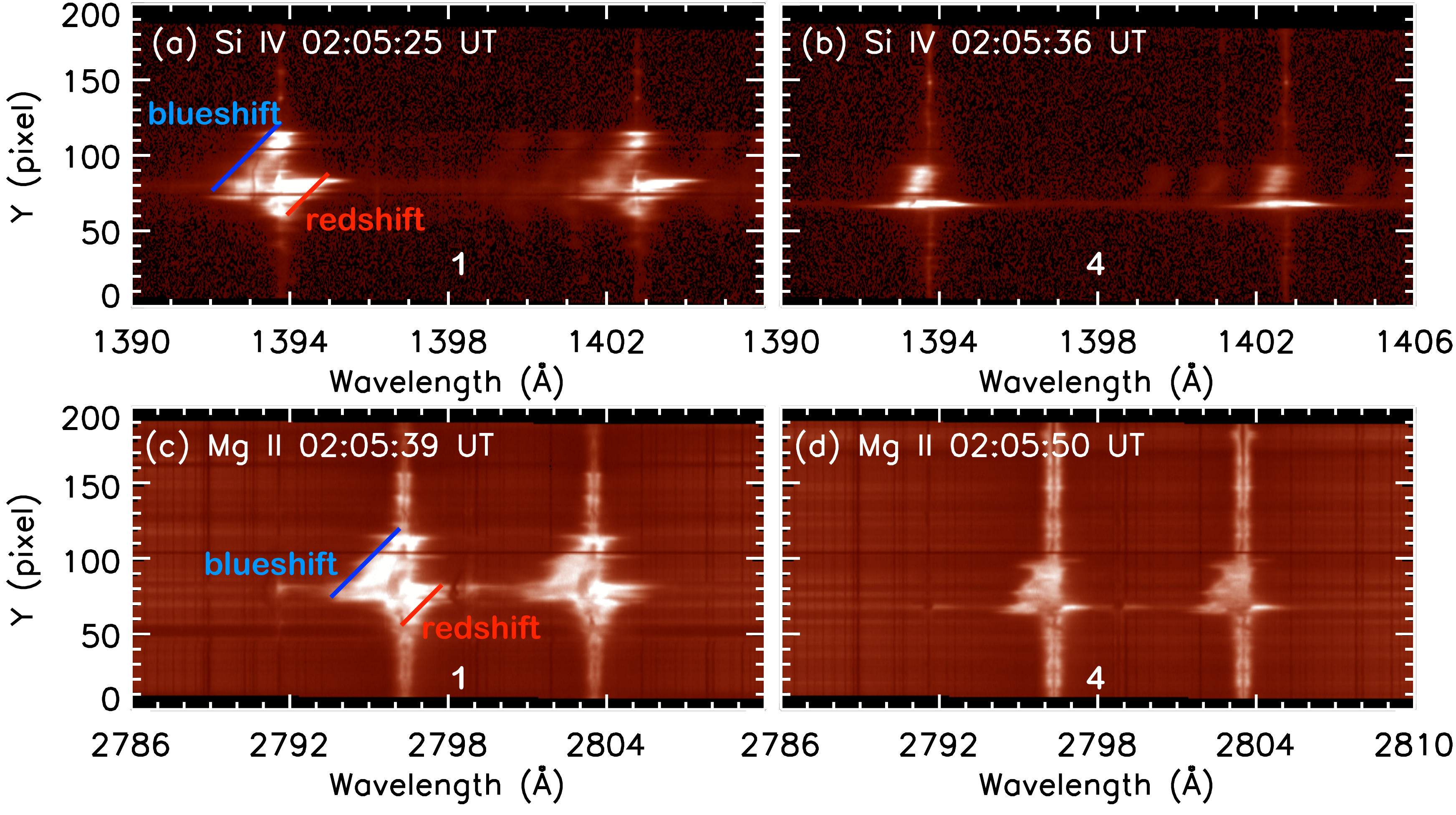}
\caption[Tilt observed in the Si IV and Mg II spectra during the GOES flare time]{Tilt observed in Si IV and Mg II spectra during the GOES flare time at slit positions 1 and 4,  distant of 6$\arcsec$ (panels a-b for Si IV lines, panels c-d for Mg II lines) (Figure \ref{AIA_IRISspectra} panels d, i, s).
The blue and redshift are shown with the solid lines in the spectra of Si and Mg at slit position 1.
}\label{gradient_spectra_Mg}
\end{figure*}
   \begin{figure}[ht!]
\centering
\includegraphics[width=\textwidth]{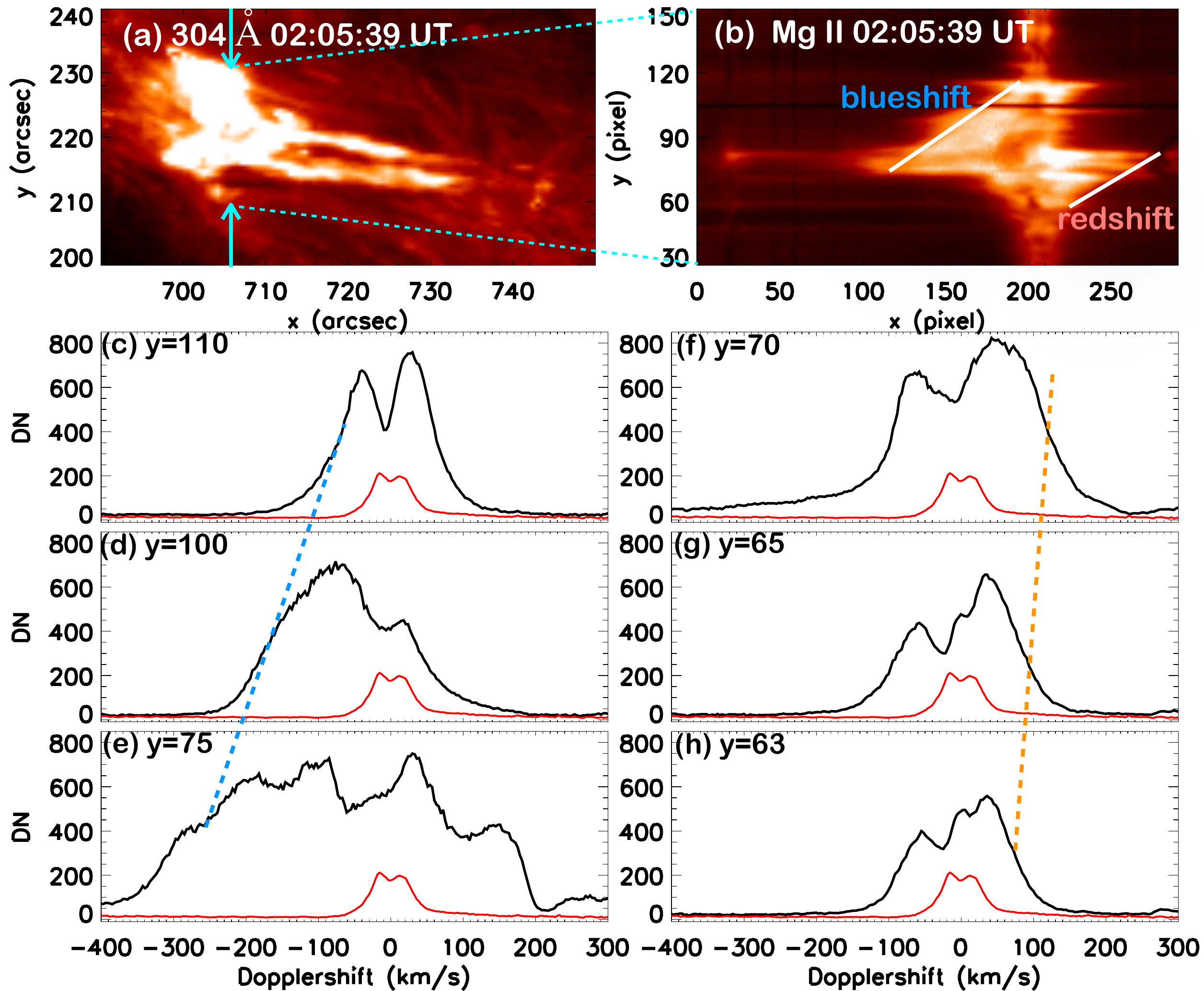}
\caption[Mini flare and the  bright jet  with two branches  inserting a cool dense surge.]
{Mini flare and the  bright jet  with two branches inserting a cool dense surge in AIA 304 \AA\, (panel a).
The red and blue shift  wings
are shown by the white tilted lines on the left (blueshift) and right (redshift) in the spectra.}

\label{tilt}
\end{figure}
\begin{figure*}[ht!]
\centering
\includegraphics[width=\textwidth]{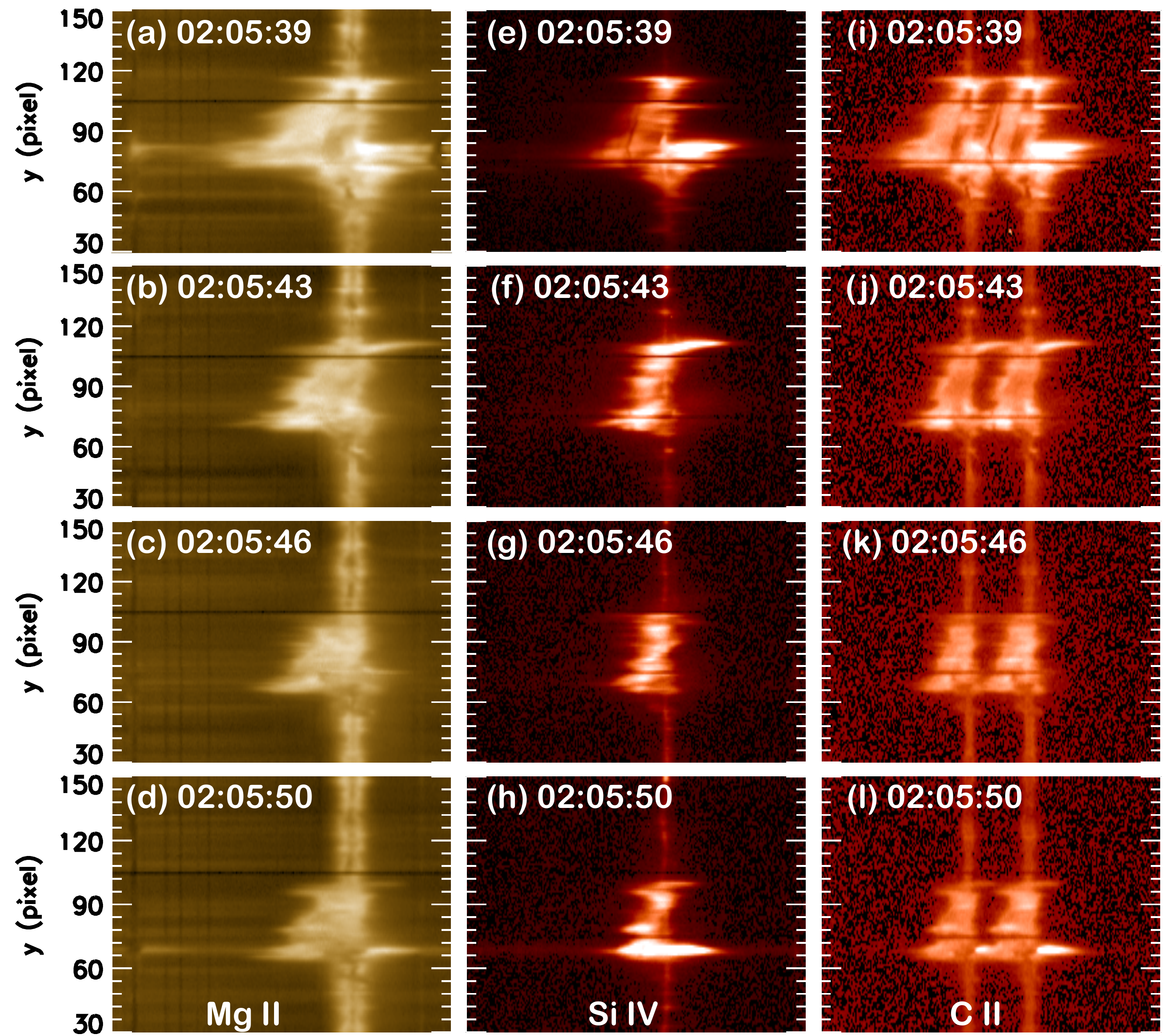}
\caption[Tilt observed in the three lines Mg II, C II, and Si IV observed with IRIS instrument.]{Tilt observed in the three lines Mg II (left column), C II (middle column), and Si IV (right column) observed with IRIS instrument.}
\label{3_tilt}
\end{figure*}

 Over the course of the reconnection phase (starting at 02:04 UT) the Mg \footnotesize{II} \normalsize , C \footnotesize{II,} \normalsize and Si \footnotesize{IV} \normalsize spectra show extended blue and red wings around the pixel value of 80. As an example, we show  Mg \footnotesize{II k} \normalsize line spectra (Figure \ref{tilt} (b)).  
 At the pixel value 80, the  Mg\ footnotesize{II} \normalsize line profile is the most extended one on the blue and red sides like in bidirectional outflows. This kind of bidirectional flow has been interpreted as being the site of reconnection in some events (\citealt{Ruan2019}). Therefore,  we consider  this zone (around the pixel value 80) to be  the reconnection site. Rapidly (in less than one minute), we see an extension of the brightening  of the wings of Mg \footnotesize{II k} \normalsize spectra in pixels along the slit in the central zone (pixel value 60 to pixel value 120) (Figure \ref{tilt} (a)).
 In the north and south parts of the reconnection site, the spectra shows a tilt. The Mg II profiles are not symmetrical all along the slits;  they present some bilateral flows only in a few pixels around y = 79. Otherwise, they have extended blue wings for example for pixel y $>$ 85  and not corresponding red wings. It is the reason that we do not consider that the bilateral flows exist all along the slit (10 Mm long). Therefore,  the  existence of this gradient and tilt is obvious.  The tilt is characterised by the gradient of the Dopplershifts that exist for profiles along the slit at a given  time.
The line profiles of  Mg \footnotesize{II k} \normalsize line
  show  important  extensions of the wings   at 02:05:39 UT (Figure \ref{tilt} (c-h)). 
The predominantly show an  extended blue wing in the northern
part (pixel value 75 to pixel value 110 (c-e)) with decreasing blueshifts 
at y = 110; they are roughly symmetric
in the middle of the brightening (y = 80) with  more extended blue wings nevertheless (until -300 km s$^{-1}$).
In the southern part of the brightening, the profiles have a dominant
 enhancement in the  red wing 
(pixel value 63 to pixel value 70 (f-h)).
The x-axis of the Figure \ref{tilt} shows  Dopplershifts in km s$^{-1}$.  
These Dopplershifts do not really correspond to up and down flows because  the region is located at 60${^\circ}$ in the west.  Therefore, the blue-shifted material is going, in fact, to the left of the reconnection site over the EMF2 and not in the direction of the jet. This means that all  cool material visible at -300 km s$^{-1}$ for which the  emission is relatively high in the Mg \footnotesize{II} \normalsize wings is going to the east and the redshifted material is expelled toward the west side as is the jet, with a maximum velocity of 80 km s$^{-1}$. The transverse velocity of the cool material along the west side  has dispersed values of between 30 km s$^{-1}$ and 100 km s$^{-1}$.  This means that one part of the cool red-shifted material could be  nearly normal to the solar surface while the other part would be inclined like  the jet.
A similar behaviour is observed in the four positions of the slit for Mg II, C II, and Si IV lines (Figure \ref{3_tilt}). The tilt in  the four  Si IV spectra is even more readily visible because Si IV is a transition  region line   with only one  emission peak when compared to chromospheric  lines with two peaks.

This  type of tilt spectra along a slit was first observed for prominences (\citealt{Rompolt1975}) and interpreted as rotating prominences before eruption. Thanks to the Solar Ultraviolet Measurements of Emitted Radiation (SUMER) spectrograph onboard SOHO, and now also with the IRIS spectrograph, such a tilt behaviour in the spectra is frequently observed. They are well-known and typically associated
with twist (\citealt{Pontieu2014})  or rotation (\citealt{Curdt2012}),
or flows of  plasma in helical
structures (\citealt{Li2014}).  In \citealt{Li2014},  the 
long filament crossed by the  IRIS slits changed the direction of its rotation in the middle of the filament. In our observations,  the jet  is rotating   in the same direction in all four positions. The  slit  scanned only 6 arc sec of the jet, mainly capturing the jet base with the dome shape. The tilt in our spectra finally reach a length around 60 pixels, which represents around 15 Mm. We interpret this tilt by the rotation of a structure crossed by the slit, the structure being the base of the jet or, possibly, cool plasma that follows helical structures.
The profiles of the Mg \footnotesize{II k} \normalsize line with extended wings resemble the profiles of the IRIS bombs (IBs) that were discovered by \citealt{Peter2014}  and analysed by \citealt{Grubecka2016} and \citealt{Zhao2017}. \citealt{Grubecka2016} found that the IBs were formed in the very low atmosphere between 50 to 900 km in the chromosphere. The magnetic configuration of the reconnection site is similar to that of the Ellerman bombs (EBs) in BP regions, where there is no vertical magnetic field (\citealt{Georgoulis2002}; \citealt{Zhao2017}). We conjecture that between the two EMFs in the QSL region, there is a BP reconnection region as in IBs (\citealt{Zhao2017}).  The  BP topology in the  region  of the present jet is confirmed in the topological analysis.
The cool material which is expelled towards the east could correspond to the dark blobs that previously existed in this area, which were trapped in the BP region while the BP was forming between   two mini-flare events.

\section{Optical thickness and the electron density} 	

{ Using  IRIS transition region lines (O IV, S IV and Si IV) electron density may  be computed (\citealt{Polito2016}; \citealt{Dudik2017}; \citealt{Young2018b}). We note that in the spectra corresponding to the reconnection site at the reconnection time, O IV lines are 
detected, even the emission is relatively weak (Figs. \ref{SiIV_evolution}, \ref{gradient_spectra_Mg} top panels and  \ref{SiIV_profiles_slit4}). Si IV line profiles vary drastically according to  time or location as we have already mentioned in Sect. \ref{UV_burst}.  
The great variety of shapes of profiles of the  Si IV lines 
rises a question about the variations of the optical thickness throughout the observed mini-flare area. For this investigation, we employ the  method involving
the intensity ratio of the {Si} {IV} 1393.75\,\AA~and 1402.77\,\AA~resonance lines (\citealt{Delzanna2002}; \citealt{Kerr2019}).
Since a long time this 
technique  exists  for stars  to determine the amount of opacity in the Si IV lines and is  very powerful 
 for providing the physical dimensions of the scattering  layer 
\citep{Mathioudakis1999}. For computing the intensity Si IV ratio   we select two observing times
: one time  during reconnection (02:04:28 UT) with slit 1 at the reconnection location (Fig. \ref{SiIV_evolution}), and the other  time at one minute later (02:05:36 UT) in slit 4 at the jet base (6$\arcsec$ away from the reconnection point).}

\begin{figure}[ht!]
\centering
\includegraphics[width=\textwidth]{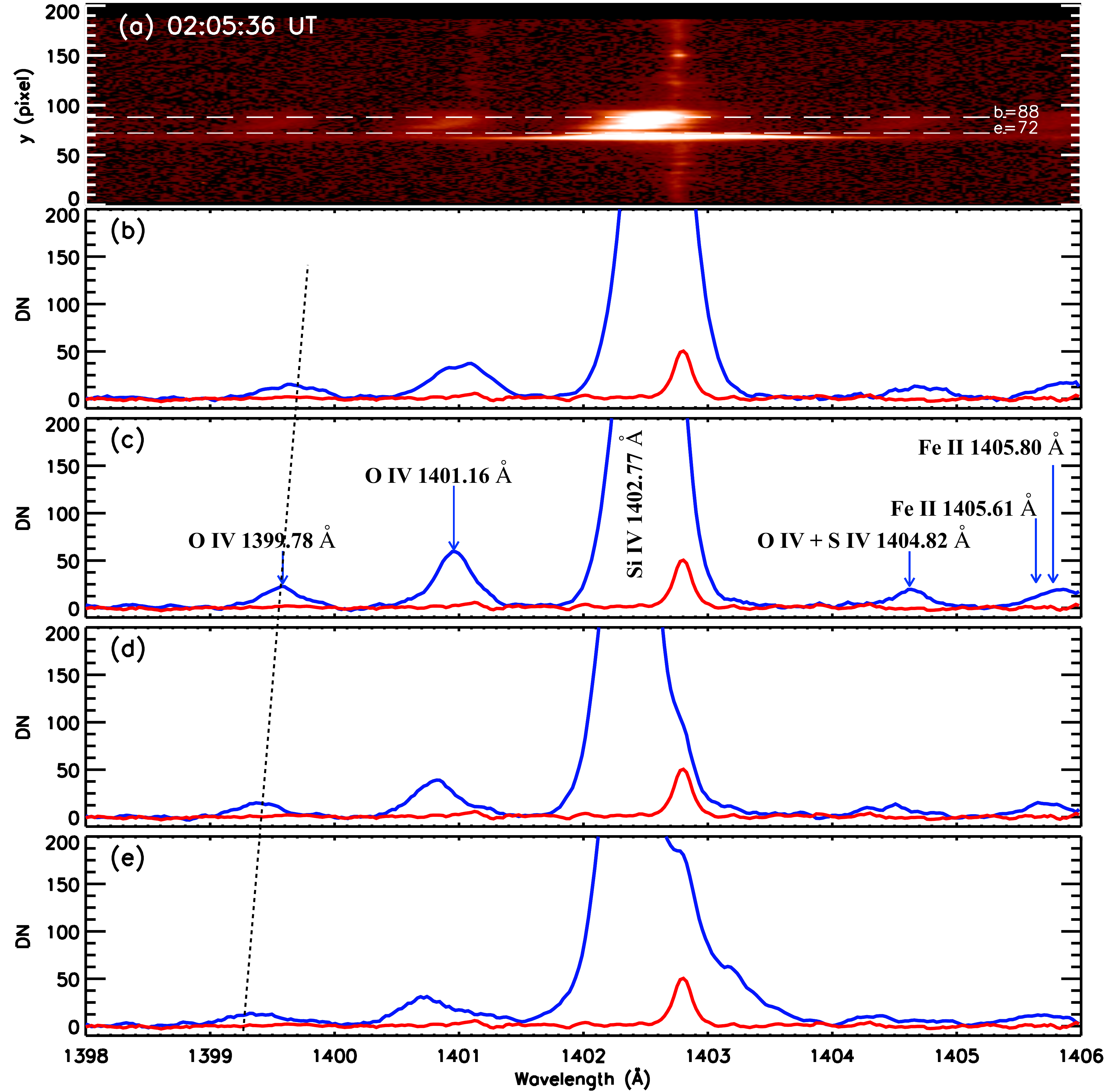}
\caption[Si IV profiles in slit  position 4 at 02:05:36 UT for pixels y = 72.]{Si IV profiles in slit  position 4 at 02:05:36 UT. The positions of these pixels lie between the two dashed lines in panel (a). The vertical dashed line in panels (b-e) represents the tilt of the spectra. } 
\label{SiIV_profiles_slit4}
\end{figure}
\subsection{
Si IV integrated intensity}	
During the reconnection phase in slit position  1  at 02:04:28 UT, the extended profiles of the {Si} {IV} resonance lines were obtained by summing the intensities observed at the different wavelengths throughout the profiles. 
We note that the absorption features were excluded from the analysis.
However the Si IV  1402 \AA\  are so wide (4 \AA),  overlying other transition region lines (O IV lines). We are not able to remove the contribution of these lines from  the integrated Si IV line intensity values. Based on the relatively-low intensities of these blending lines, we do not expect the uncertainty of this method to exceed 10\%.

Away from the reconnection region, in slit position 4 at 02:05:36 UT, we  compute the line integrated intensities  by fitting the observed profiles with Gaussian functions 
using the \texttt{xcfit.pro} fitting routine.
Typically, two Gaussian functions were needed to fit each line profile  of Si IV, in the mini-flare area 
between y = 76 and 88.  The two   Gaussian functions   
are separated by up to $\approx$0.4\,\AA, one function has  a narrow Full Width half maximum (FWHM)  and the other one  an extended  FWHM but with a lower intensity \citep{Dudik2017}. The  two Si IV line profiles showed
similar asymmetries. 
Particularly exceptional profiles of both lines were observed at y $\approx$80, where the `bumps' in line red wings were considerably weaker. Based on this, we suggest that the bumps present in the profiles do not originate in blends, but in the motion of the emitting plasma. To check  this assumption, we calculated synthetic spectra using CHIANTI v7.1 (\citealt{Dere1997}; \citealt{Landi2013}) for log($N_{\text{e}}$ [cm$^{-3}$]) = 11, flare DEM. Even though we found several lines blending both lines of {Si} {IV}, their contributions were found to be negligible.

\begin{figure}[ht!]
  \centering    
    \includegraphics[width=\textwidth]{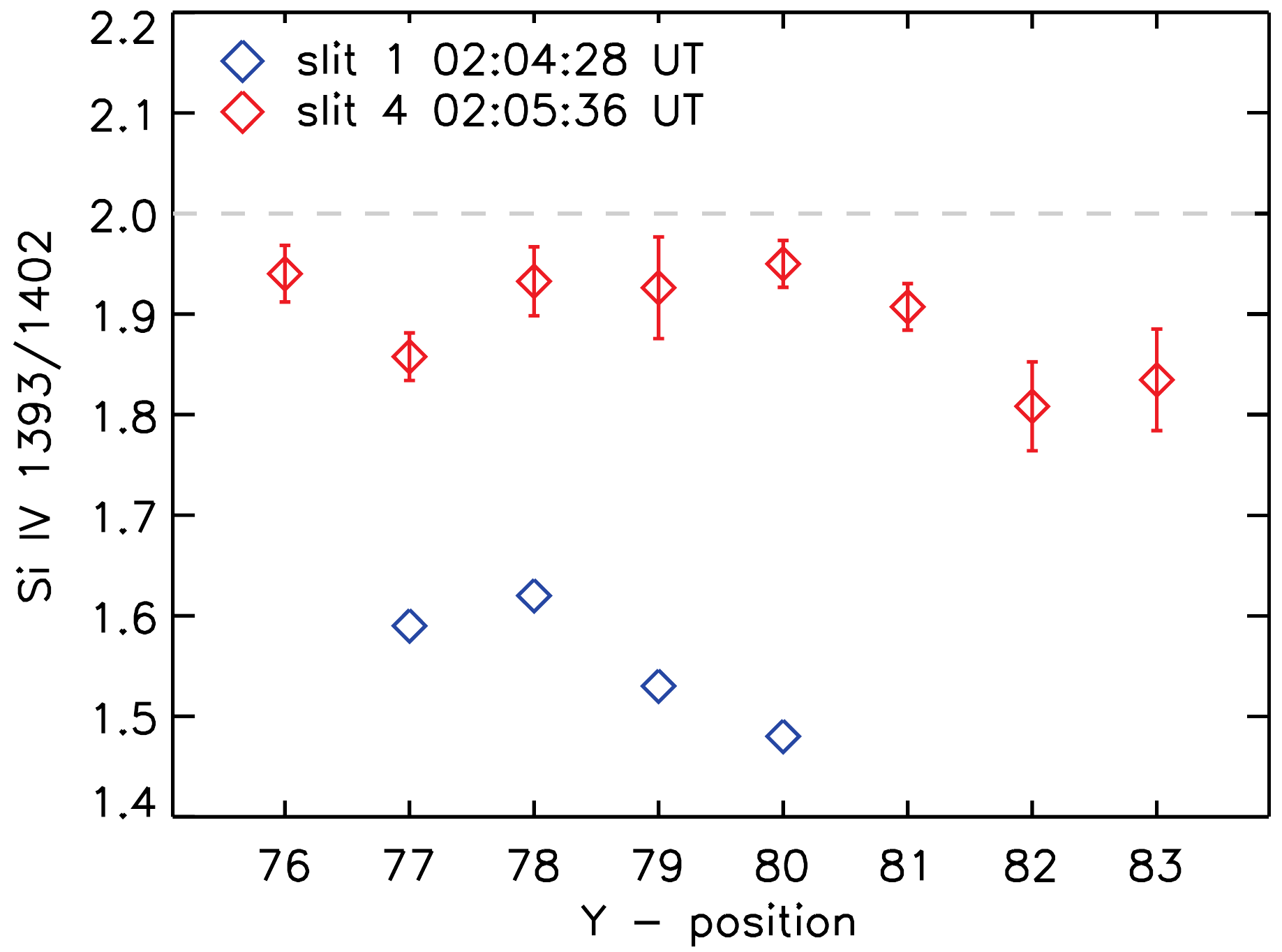}
  \caption{Variations in the Si IV 1393.75\,\AA~and 1402.77\,\AA\ line ratio used as a probe for measuring the optical thickness. Blue diamonds indicate the ratios measured in the reconnection region, while red ones away from it and later on. \label{fig_si_ratios}}
\end{figure}
\subsection{Si IV line ratios} 
The ratios of the {Si} {IV} 1393.75\,\AA~and 1402.77\,\AA~resonance lines are shown in Fig. \ref{fig_si_ratios}. There, the diamonds indicate the measured ratios at different positions along the slit. They were color-coded in order to distinguish between the ratios measured in (blue) and away (red) from the reconnection region. For optically thin plasma, this ratio should be equal  to 2 \citep{Delzanna2002, Kerr2019}, which we indicated using grey dashed line in Fig. \ref{fig_si_ratios}. Even though all of the observed ratios are below this value, the ratios measured farther from the reconnection region (red diamonds) are consistently closer to 2 than those measured in the reconnection region (blue diamonds). The computed  ratio range for the red points,
  away of the reconnection site  is  between $\approx$1.72 and 1.95. The   maximum  values  are  around 1.9  for points 79 and 80 which infers to us  the  ability to derive the electron density.
For the blue points  in the mini-flare area at the reconnection time,   the low ratio  value of Si IV lines (1.48 to 1.62) with a high uncertainty  does not allow us to compute  the electron density
  \citep{Kerr2019}. 
 More-less the Si IV line profile shapes are similar to the Mg II line shapes (Fig. \ref{MgII_SiIV})  with   similar extended blue wings which means that such large profiles has several components like the Mg II lines containing  certainly the emission of the flare plus the emission of a cloud with high velocities as we have concluded by analysing the Mg II profiles with the cloud model. It is nearly impossible to distinguish the two components in each Si IV profile and therefore to compute  the optical thickness of the cloud and the flare region.  
  The behaviour of Si IV line in flares  is exactly in a similar way as predicted in the theoretical models \citep{Kerr2019}, that there is a stratification in the heights at which the various lines  (Si IV, C II, Mg II) form, which varies with time in the flare. Initially the core of the Si IV 1393.75 line forms highest in altitude. Toward the end of the heating phase the compression of the chromosphere results as the lines formation in a very narrow region, which persists into the cooling phase. We believe to this stratification of heights of formation of Si IV, and C II lines during the flare. As it is
  suggested 
  by \citealt{Kerr2019} the Si IV intensity ratio represents the ratio of the source function of the lines, the magnitude of the lines depending on the temperature of the layer where they are formed.  The thermalization of   both Si IV  lines  would occur higher in the atmosphere where the temperature is larger.  Tests using the RADYN  code should be used to understand such low ratio values in term of opacity effect. 
  One minute later after the flare this effect is negligible.

\subsection{Computation of the electron density}
\begin{figure}[t!]
\centering
\includegraphics[width=0.9\textwidth]{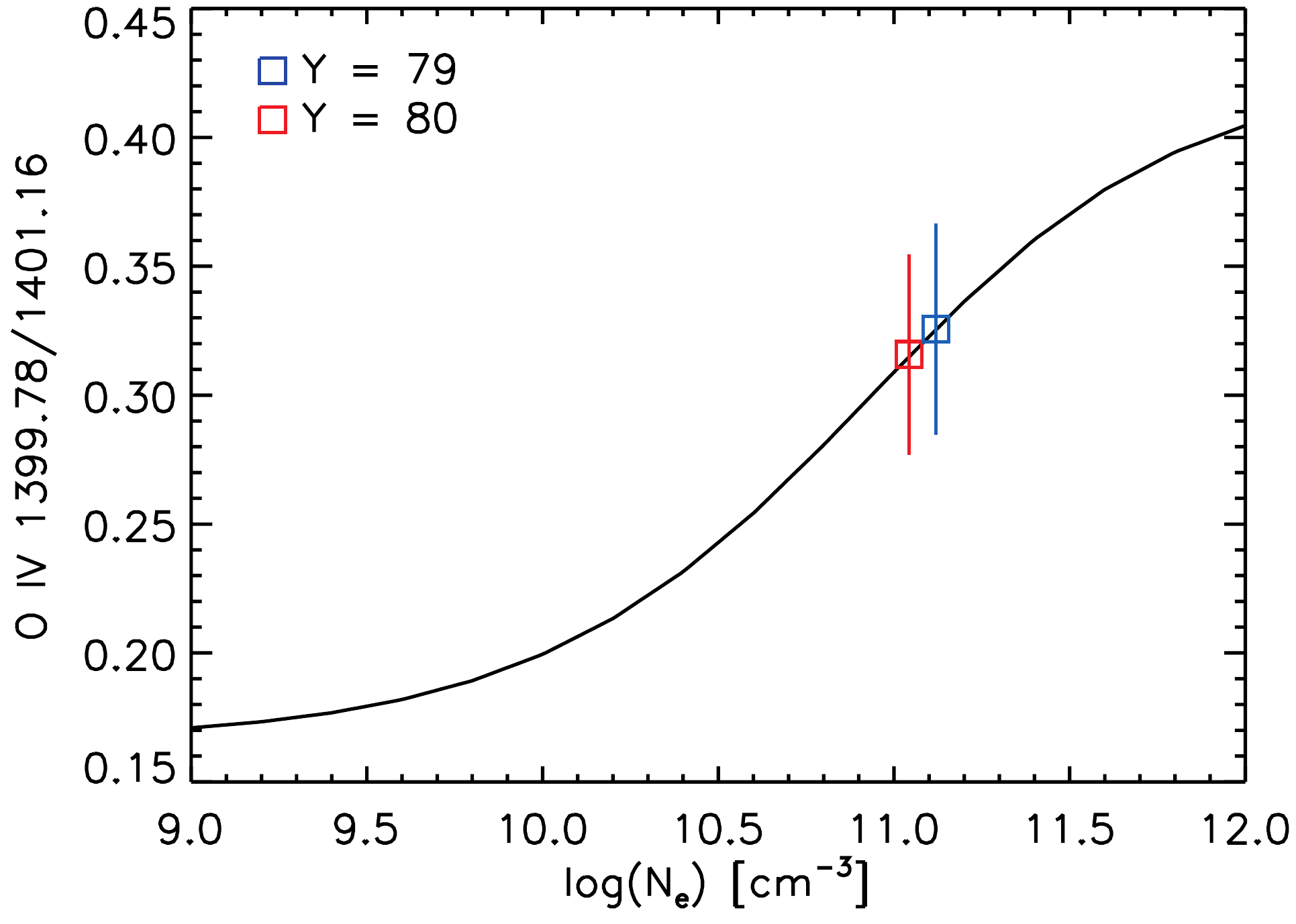}
  \caption{Estimation of the electron density using the intensities of the intercombination lines of O IV measured at y = 79 (blue), 80 (red). \label{fig_density}}
\end{figure}
The electron density could be computed in the mini flare area, away from the reconnection region (in slit 4) and observed a minute after the reconnection itself.
In order to obtain the electron densities 
we used density-sensitive ratios of line intensities. IRIS routinely observes multiple inter-combination lines which provide ratios useful for this purpose as they are not affected by opacity issues \citep{Polito2016,Young2018b}. Here, we are able to obtain reliable fits of lines composing only  one ratio, being the {O} {IV} 1399.78\,\AA~and 1401.16\,\AA~lines. The former line was however still weak, with reliable fits only at slit 4 observed at 02:05:36 UT and  for y = 79 and 80. The measured ratios are presented in Fig. \ref{fig_density}. The emissivities are calculated using the \texttt{dens\_plotter.pro} routine contained within the CHIANTI package for the peak temperature of formation of the ion (log($T$ [K]) = 5.15). At both locations, the resulting densities are roughly log($N_{\text{e}}$ [cm$^{-3}$]) = 11 $\pm$ 0.3. Note that these values correspond to those obtained by \citealt{Polito2016} in a plage and a bright point. The mixed ratio proposed by \citealt{Young2018b} with Si IV and O IV lines could not be computed due to the low counts.
\subsection{Path length}
Measured electron densities can then be used to calculate $\tau_{\text{0}}$ at the center of the line using e.g. Equation (21) of \citealt{Dudik2017}:
\begin{equation}
	\tau_{\text{0}} \approx 0.26f \frac{\left\langle N_{\text{e}} \right\rangle}{10^{10} \text{ cm}^{-3}}, 
\end{equation}
where $f$ is the path length ($\triangle s$) filling factor in an $\textit{IRIS}$ pixel defined as $f = \triangle s/0.33 \arcsec$. The numerical factor of $0.26$  was by \citealt{Dudik2017} calculated using thermal line widths of the resonance lines of {Si} {IV} for the Maxwellian distribution. However, the observed profiles of the {Si} {IV} lines are much broader. The widths resulting from fits are typically $>$0.5\,\AA, which reduces the numerical factor down to 0.012. Still, $\tau_{\text{0}}$ cannot be calculated unless the value of $f$ is known. If we for simplicity assume $f=1$ and utilize the measured density of log($N_{\text{e}}$ [cm$^{-3}$]) = 11, the modified formula leads to $\tau_{\text{0}} \approx 10^{-1}$. Note that at the same time and positions, ratios of the resonance lines were consistently closer to 2 compared to the values measured in the reconnection region, indicating relatively optically-thinner plasma. The formula for $\tau_{\text{0}}$ can easily be rewritten in terms of the path length $\triangle s$. Using the relation for $f$ and measurements of the line widths and electron densities, we obtain (in kilometres):
\begin{equation}
	\triangle s \approx 2000 \tau_{\text{0}}.
\end{equation}
Since we relaxed the assumption of $f=1$, the estimate for $\tau_{\text{0}}$ does not hold any further. However, as indicated by the resonance line ratios, outside of the reconnection region it is most likely $<1$. The possible path lengths are thus determined using the derived linear relation, while their upper boundary is $\approx$2000 km. As it has been demonstrated for flares in stars \citep{Mathioudakis1999}
if the electron density in the atmosphere is
known, opacity can provide important information on the linear
dimensions of the scattering layer.

 \section{Multi temperature layers for magnetic reconnection}
 \label{multi-temps}
 \subsection{Sketch of the dynamical plasma motion}
  \label{sec:sketch}
  We propose  a sketch to explain the dynamics of the plasma during the reconnection (Figure \ref{sketch}).  We draw the field lines in solid black lines, the flow directions are indicated with  blue or red  arrows, with solid thick arrows for blueshifts or redshifts. Before the  jet onset (panel a) the magnetic topology of the region consists of  two emerging flux:  EMF1 and EMF2  overlaid  by AFS. Between EMF1 and EMF2 the bipole (P1-N2) is located, where the reconnection takes place.  Just before the  reconnection (panel b), there is the formation of a suspected BP region in the middle of the bipole with cool plasma trapped inside (Section \ref{time_IRIS}).
 At the time of reconnection (panel c) cool plasma clouds are ejected  with strong blueshifts (Section \ref{comparison2}). As the region  is located  at W 60, the clouds with blueshifts  are in fact  ejected over the emerging flux EMF2.  At the same time the BP is transformed in an `X' null-point current-sheet with bilateral outflows (panel c). Simultaneously there is the  ejections of the jet and  surge  on the right side of the reconnection site over EMF1 
   (panel d). After the reconnection long AFS and hot loops are formed  overlying the region (panel e).
  This sketch is in the line of the cartoon proposed in \citealt{Joshi2020FR} for explaining the reconnection in a X-current sheet.
\begin{figure}[t!]
\centering
\includegraphics[width=0.70 \textwidth]{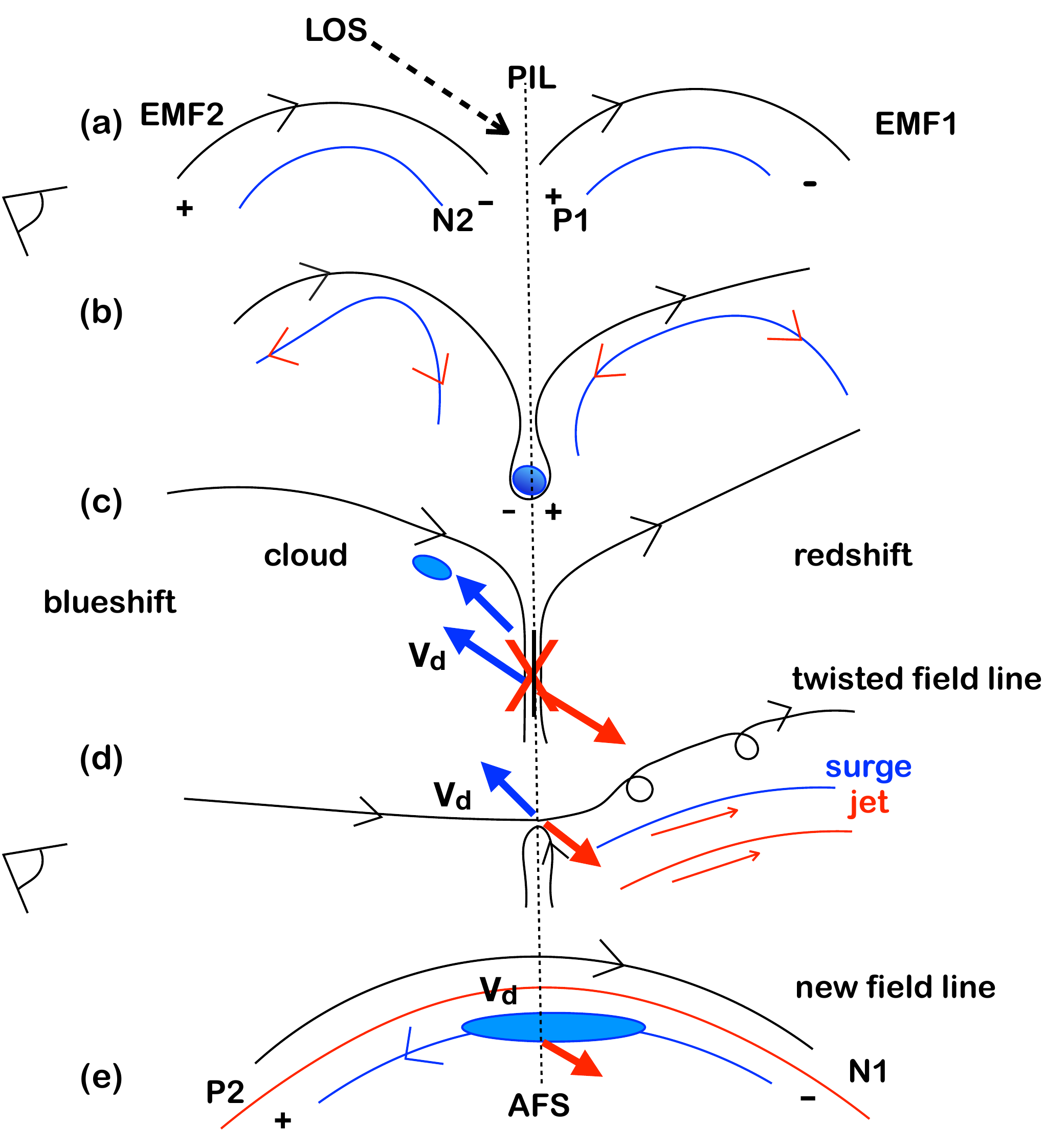}
\caption{Sketch of the dynamics of the plasma  in 2D (x,z) plane at the reconnection site before, during, and after the magnetic reconnection.}
\label{sketch}
\end{figure}

  \subsection{Presence of cool material over hot atmosphere}
   \label{cool_over}
During the reconnection time (around 02:04 UT) the IRIS spectra in the mini flare show the presence of cool material over hot plasma like in IBs (Section \ref{comparison2}).
 This is how the Mg II  large extended blueshift profiles have been  interpreted by the existence of two cool clouds over the reconnection site at the time of the reconnection.  One part  of the trapped cool material could be  ejected with   a low velocity  while the other part   is ejected with  a fast  upward velocity during approximately one minute.
  Cool material is  propelled  to a distance  of 20,000 km along the LOS in one minute (300  km s$^{-1}$ $\times$ 60 sec).  Moreover the large dark dip in the Mg II line centers at this time   could be considered as the cool plasma of the  surge which exhibits low Dopplershifts but high transverse velocity.

  The presence  of such cool plasma over the heated atmosphere at the reconnection site is also  confirmed by the presence of chromospheric lines  striping the Si IV profiles. 
   Si IV  1393.8 \AA\, profiles in the UV burst  reveal 
  that 
  the presence of absorption
lines from singly ionized species (Fe II and Ni II)
(Figure \ref{MgII_SiIV} (d), and Table \ref{table2}).  
The
presence of  such lines superimposed on emission lines implies that cool chromospheric  material
is stacked on top of hot material.
The cool material would  come from the BP region when the  magnetic field lines were tangent to the solar surface at the photosphere (Figure \ref{sketch} panel b). During the BP reconnection cool material is ejected with more   or less  fast  flows as we have shown in the explanation of the Mg II profiles.
The wavelengths of these Fe II and Ni II lines are located  at around  0.5  to 1 \AA\ far from the Si IV line center. Therefore  only when the Si IV line profiles are broad enough with extended wings,  
these chromospheric lines  are shown
themselves in absorption. It has been already observed in the  IBs (\citealt{Peter2014}) and IRIS UV bursts (\citealt{Yan2015}).

 Moreover at the time of the mini flare (02:04:28 UT) a narrow dip in the profile of Si IV at its
rest wavelength is  observed, 
 blended by Fe II line (Figure \ref{MgII_SiIV}).  It could be due  multi-component flows like in the IBs (\citealt{Peter2014}). If that would be the
case, then one would expect the line profile to
be composed of two or more Gaussians at different
Dopplershifts representing the different
flow components. However  it looks not to be  the case and  the dip is deeper for  the strongest Si IV line 1394 \AA, and the dip is always at the rest wavelength.  Therefore this dip in Si IV could  be the signature of 
 opacity effects as in the UV burst presented in \citealt{Yan2015}.
 The former authors show similar profiles of Si IV  with self absorption and with chromospheric lines  visible as  absorption lines.  They explain  these profiles by  the  superposition of different structures, in the deep atmosphere the reconnection leads to a significantly enhanced brightness and width of Si IV,  the  light passes through the overlaying cool structures where Ni II leads to absorption. Higher up there is emission of Si IV in overlying cool loops which lead to narrow self-absorption of Si IV. 
 This scenario is possible in our observations.  

In the UV burst bilateral outflows 
and expelled clouds with super Alfv\'enic flows of the order 200 km s$^{-1}$ have been observed  in Mg II and Si IV (Section \ref{UV_burst}). These profiles are similar to those of IBs found by \citealt{Peter2014}; \citealt{Grubecka2016}. 
However,  we detected also O IV lines in the vicinity of  Si IV 1393.76 \AA\ (Table \ref{table2} and Figure \ref{SiIV_profiles_slit4}). O IV are forbidden lines formed just below 0.2 MK.  Our UV burst is definitively not exactly an IB  where O IV lines were not detected (\citealt{Peter2014}) and  did not support the long  debate about the  temperature of the formation on Si IV line.
 Si IV could be  out of ionization equilibrium  in high velocity flow plasma and the nominal formation temperature of Si IV could  be in fact lower  than 80,000 K \citealt{Dudik2014}; \citealt{Nobrega2018}). However when we observe simultaneously   O IV line emission  as well as Si IV  which is also  formed at transition region temperatures it confirms that the plasma  is heated and Si IV is not at  chromospheric temperature. In fact AIA observations showed the  mini flare  in its hotter filters until 10$^7$ K with 211 \AA\, filter.
 \begin{figure}[ht!]
\centering
\includegraphics[width= \textwidth]{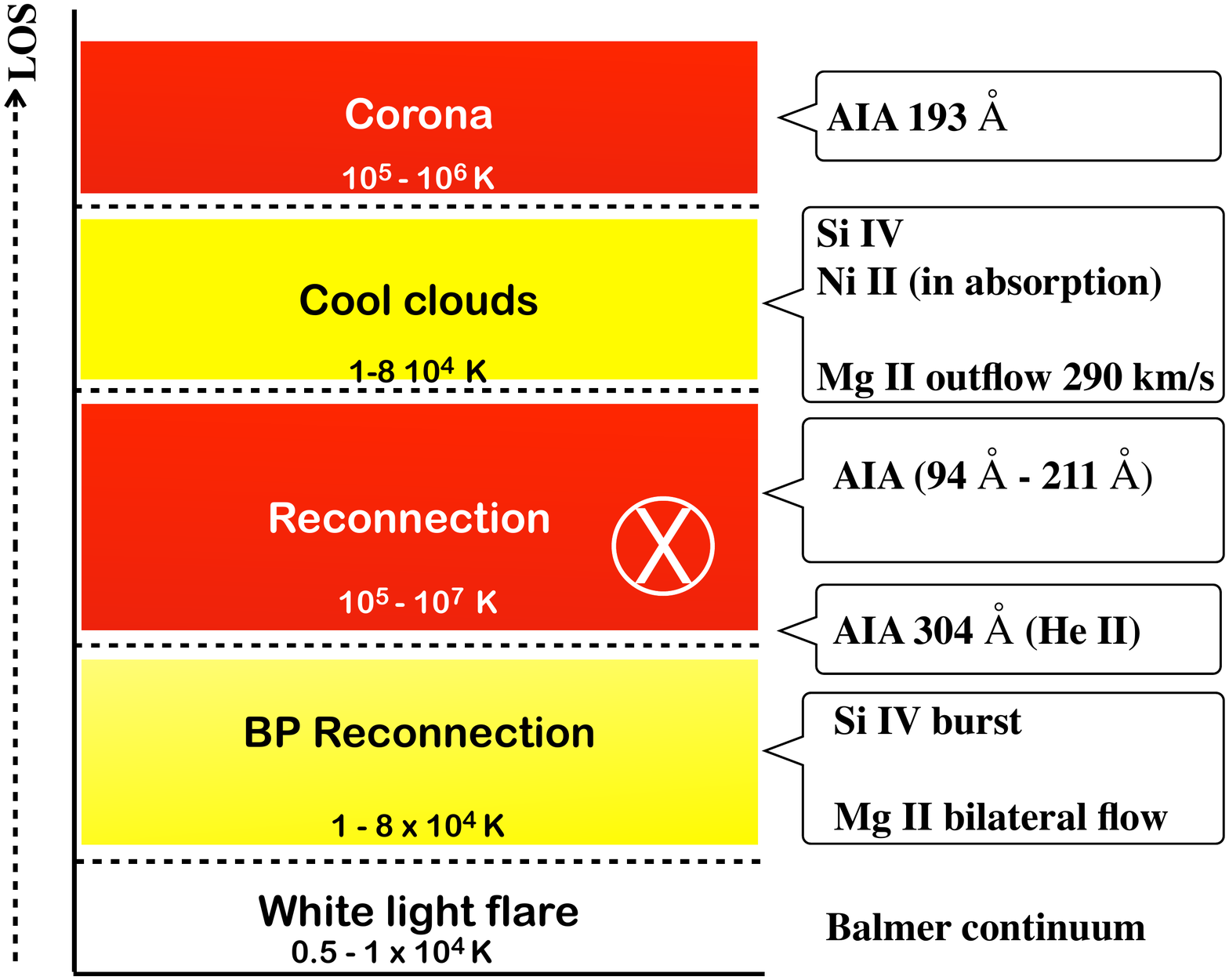}
\caption[Model of multi-layers of the mini flare atmosphere during the jet  reconnection in a BP region.]
{Model of multi-layers of the mini flare atmosphere during the jet  reconnection in a BP region. The model is  based on the observations of emission or absorption of the IRIS lines and continua, and the images of AIA. 
}
\label{sandwich}
\end{figure}

   \begin{figure}[ht!]
\centering
\includegraphics[width=1.00 \textwidth]{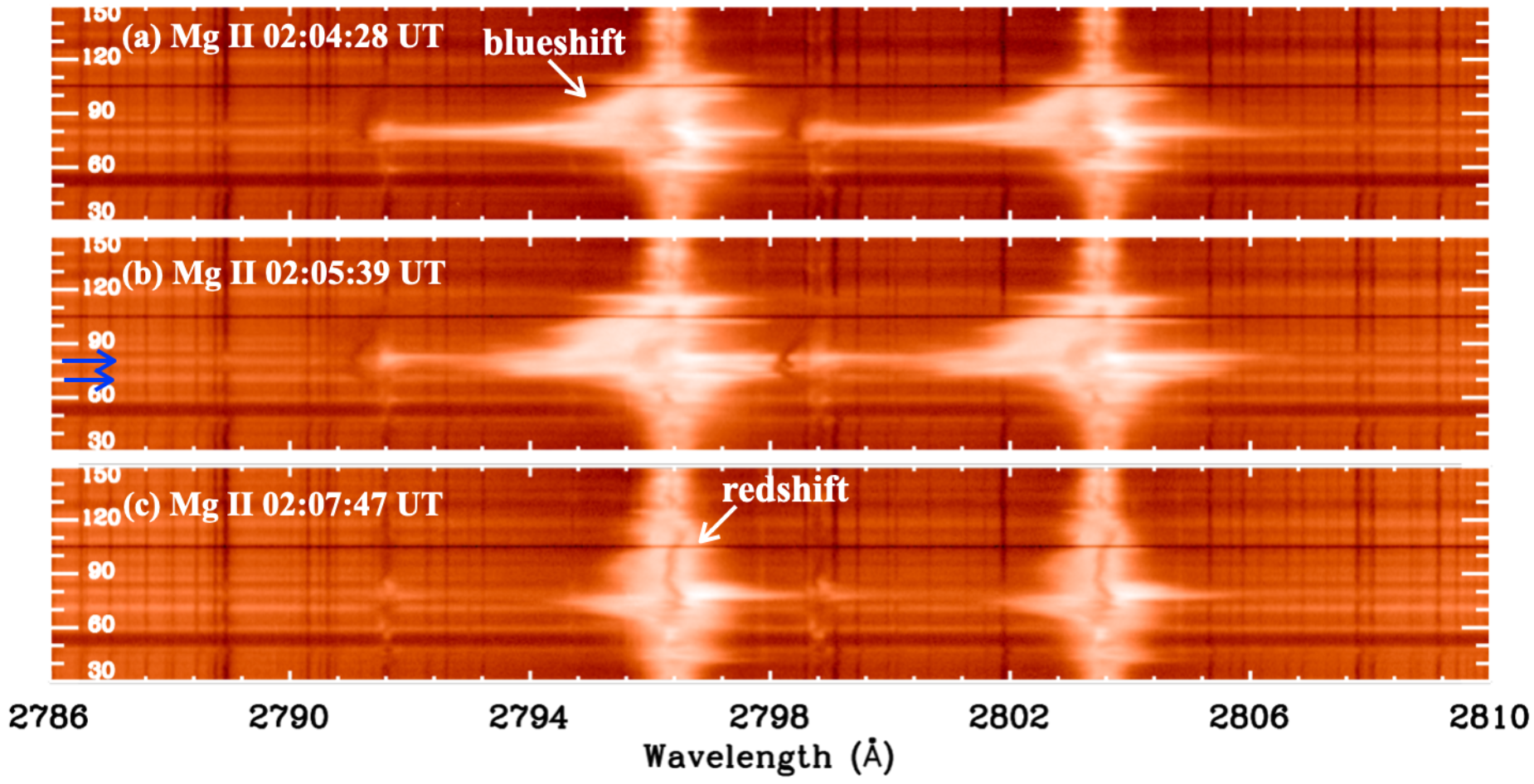}
\caption[Mg II k and h and continuum visible during and after the reconnection.]{Mg II k and h and continuum visible during (panel a) and after 
(panel b-c) the reconnection. The blue and redshift observed at different time are shown with white arrows in panels (a and c).}
\label{Balmer}
\end{figure}
\subsection{Heating  at the minimum temperature and white light flare}
\label{white_light}
At the reconnection site we observed  an  enhancement of the Balmer continuum implying that there are
non-thermal 
electron bombardment during reconnection. There was no RHESSI data to confirm such acceleration of particles of high energy.
White light flares are commonly observed  in  optical  continuum emission which dominates the energetics of flares. Balmer continuum enhancement as signature of white light flare was discovered by
\citealt{Heinzel2014} using IRIS spectra around Mg II lines. \citealt{Kleint2017} have shown that it could be detected also in some cases in the IRIS SJI at 2832 \AA.
The Balmer continuum emission is optically thin in the chromosphere. The enhancement of the Balmer continuum  is marginal in the line core and inner wings where the line intensity is high (\citealt{Liu2014}). However it is significant in the extended wings where the line emission is less visible. Balmer continuum is constant over the whole NUV Band. Therefore during our UV burst multi temperature flare the Balmer continuum was not detectable due to  the high contrast between the emission in the wings and  peaks of Mg II line.
When  our  flare burst emission was  decreasing, then it is possible to detect the emission of the continuum in the Mg II  far wings   at the reconnection site (Figure \ref{Balmer}). Therefore it is not possible to give the precise  time when the non thermal particle bombardment occurs. As in \citealt{Tei2018} an other  scenario is possible  in which the  upflow of cool plasma is lifted up by expanding hot plasma owing to the deep penetration of non-thermal electrons into the chromosphere. Heating  the minimum temperature region is consistent with the brightening found in AIA 1700 \AA\, and 1600 \AA\, at  the reconnection site, even transition region lines like {\it e.g.} C IV are very important,
(\citealt{Simoes2019}), and  may be responsible for the enhancement of intensity. The spatio-temporal variations of the Mg II k and Si IV profiles are displayed
for three times and for three different pixel values 
in Figures \ref{burst_MgII}. The profiles vary fast on these time and spatial scales (in 30 sec. and 1$\arcsec$ respectively).

\begin{table*}[ht!]
\caption[Characteristics of the  evolution of the pattern and  Mg II spectra of the mini flare at the jet  base.]{
Characteristics of the  evolution of the pattern and  Mg II spectra of the mini flare at the jet  base (reconnection site at `X' point).
}
\centering
\resizebox{\textwidth}{!}{
\begin{tabular} {ccccc}
\specialrule{.1em}{1em}{0.5em}
Time (UT) & AIA 304 \AA\ & \multicolumn{3}{c}{IRIS Mg II spectra} \\
& & (slit 1)& (slit 2-3)& (slit 4)\\
\specialrule{.1em}{1em}{0.5em}
01 :51 :15 & Arch Filament Systems (AFS) & redshift at pixel 80 & high peaks & blue shift at pixel 80\\
       & over EMF1 and EMF2 & strong  central absorption & central absorption & central absorption\\
& & & &  \\
01 :54 :33& bright threads in X & blue/red shift at pixel 80 & broad lines& \\
& between the 2 AFS & central absorption & &\\
& & & &  \\
01 :56 :26& AFS with bright ends &long wings red and blue& large central absorption& thin blue\\
& & & &  \\
01 :59 :34& bright  threads in X  &ten y pixels with broad& \\
 & mixed with dark kernels &central absorption & & \\
 & & & &  \\
02 :01:10 & preflare in X & bright blue along 20 pixels & broad central absorption& thin \\
& & & &  \\
02 :02 :21 & bright NS arch &extended blue  wing  profile &blue  and broad & thin\\
& & & &  \\
02: 02: 59 & onset of surge&strong  central absorption& shift towards blue& \\
& & & &  \\
02 :03 :32 & kernels & very bright peaks symetrical & blue& blue \\
& & & &  \\
02 :04 :28& bright triangle jet base & bright  tilt blue to red  & large bright blueshift  & \\
& & & &  \\
02 :05 :39 & mini flare  & very bright blue tilt  & bright tilt &bright tilt\\
&double jet and surge&extended blueshift & & \\
& & & &  \\
02 :06 :22& expansion &zigzag central absorption & zigzag & elongated thin wing \\
& & & &  \\
02 :07 :04&  long surge over the jet&weak blue peak,  redshift& weak broad profiles & \\
& & & &  \\
02 :10 :21 & long loops& thin blue pixel& broad weak& continuum\\
 & bright and dark& weak extended redshift& & \\
 & & & &  \\
02 :14 :24 & long loops & one pixel extended  peaks & weak broad profiles& extended  thin blue-red  \\
   & long AFS & long red wing extension in y & red &\\
  & & & &  \\
02 :18 :24 & end  &  symmetric profiles& weak broad profiles& continuum \\
\specialrule{.1em}{1em}{0em}

\label{table4}
\end{tabular} 
}
\end{table*}

\section{Results and conclusion}
 In the present study, we have not only the signatures of IBs with IRIS but also the enhancement of Balmer continuum detected in  IRIS spectra like in a white light mini flare. The mini flare is  also visible as  brightenings  
in all  AIA filters with multi temperatures from 10$^5$ K to 10$^7$ K.
This mini flare, called mini flare due to its small area and GOES strength flux (B6.7) is in fact a  very energetic  flare belonging to the category of white light flares. The chance is that  the slit of IRIS with its high spatial and temporal resolution was exactly at the site of the reconnection. Therefore it is the first time that we have such important information on a white light  mini flare. Our main results and conclusions are as follows:

The magnetic reconnection height of the mini flare at the jet base is very important parameter to probe about the reconnection mechanism. When jets are observed over the limb    the reconnection point  is clearly visible in the corona (e.g 10 Mm in the case of \citealt{Joshi2020MHD}). For events occurring on the disk it is difficult to derive the altitude of reconnection.  \citealt{Grubecka2016,Reid2017} and  \citealt{Vissers2019} used a NLTE radiative transfer code in a 1D atmosphere model to derive the altitude of formation of the  reconnection in UV burst visible in Mg II lines. The altitude range spans the high photosphere and chromosphere (50 to 900 km).  With IRIS it was also found  in UV bursts or IRIS bombs (IBs) that cool plasma  emitting in  chromospheric lines could overlay  hotter plasma at  Si IV  line temperature \citep{Peter2014}. After a long debate about the temperature of the Si IV lines in possibly non equilibrium state,  3D simulation based on Bifrost code \citep{Gudiksen2011} coupled with the MULTI3D code \citep{Leenaarts2009} succeeded to mimic the large observed Si IV profiles (very similar to our UV  Si IV burst) and the  extended wing of the Ca K line with synthetic profiles both formed at different altitudes simultaneously due to an extended vertical current sheet  in  a strong magnetized atmosphere \citep{Hansteen2019}.  They proposed that  ``the current sheet is located in a large bubble of emerging
magnetic field, carrying with it cool gas from the photosphere''. 
This scenario is certainly valid for our observations. 
We attempted  to compute the optical thickness and the electron density in the mini flare area during the reconnection process  \citep{Kerr2019,Judge2015}. The transition lines  (Si IV and O IV) usually used for such estimations had so perturbed profiles that no reliable numbers could be put forward at the time of the reconnection. One minute later the Si IV lines were estimated  to  be  nearly  optical thin  and the computed electron density arises to log($N_{\text{e}}$ [cm$^{-3}$]) = 11 $\pm$ 0.3, value which corresponds to a plage or bright point, similarly in the paper of \citealt{Polito2016}.

We have discussed about the signatures of the different elements and  proposed a dynamical model for the magnetic reconnection. The analysis of HMI magnetogram of the AR shows that the region consists of several EMFs. The jet occurs between two of them when the negative polarity of the following EMF was collapsing with the  positive polarity of the leading EMF. Such magnetic topology  leads to 
a bald patch magnetic configuration where the magnetic field lines are tangent to the photosphere. This topology was confirmed by the first reconnection  signature in  Mg II lines with symmetric extended wings at 02:03:46 UT similar to  those in IBs 
commonly occurring in bald patches \citep{Georgoulis2002,Zhao2017}.  At the reconnection site  these bilateral outflows ($\pm$ 200 km s$^{-1}$) observed in a few pixels were  interpreted as reconnection jet. Less than one minute later (02:04:28 UT) extended Mg II line blue wing suggests super Alfv\'enic flows  (Figure \ref{threeProfiles} panel c).  With the cloud model technique, which represents a formal solution of the transfer equation under some assumptions, two plasma clouds, one with high speeds  (blueshifts of 290 km s$^{-1}$) and one with medium speed (blueshifts of 36 km s$^{-1}$) were detected. The identification of ``explosive'' i.e. 290 km s$^{-1}$  flow  is unique in such circumstance of reconnection.  We conjecture that cool plasma was trapped between the two EMFs  which could correspond to arch filament plasma and was expelled during the reconnection. The second cloud with lower velocity is certainly due to the surge plasma accompanying the jet.

With the cloud model technique, which represents a formal solution of the transfer equation under some assumptions, two plasma clouds, one with high speeds  (blueshifts of 290 km s$^{-1}$) and one with medium speed (blueshifts of 36 km s$^{-1}$) were detected. The identification of ``explosive'' {\it i.e.} 290 km s$^{-1}$  flow  is unique in such circumstance of reconnection.  We conjecture that cool plasma was trapped between the two EMFs  which could correspond to arch filament plasma and was expelled during the reconnection. The second cloud with lower velocity is certainly due to the surge plasma accompanying the jet. Such models are used to derive  true velocities which usually differ from
those obtained from Doppler shifts (this is the basic idea behind the cloud
model). Then the question is what is the nature of  the obtained velocities,
 expelled  plasma blobs like in surges \citep{Moreno2008,Nobrega2018} as we suggested, upflows (evaporation), downflows (chromospheric condensations) as it is proposed for flares \citep{Berlicki2005,DelZanna2006}. However using the cloud-model technique cannot help us for the proper understanding of the nature of detected flows, it is just a diagnostics method. Numerical simulations using RADYN and RH \citep{Kerr2019} or  Flarix RHD codes \citep{Kasparova2019}  would certainly give more insights  on the physical process involved.  We  found similarities between such simulation models and our  observations like the high variability of the Si IV lines depending strongly on opacity effects. These observations could be the boundary conditions of future simulations.

More analysis of solar jets/surges, association with the bright points, with spectroscopic techniques to probe the Dopplershifts, bidirectional outflows, and twisting in solar jets are needed to explain their triggering and driving mechanisms. In future, we plan to include the observations from the newly launched Solar orbiter (\citealt{Anderson2020}; \citealt{Rochus2020}) and from Swedish 1 m Solar Telescope \citep[SST,][]{Scharmer2003} to explore the nature of cool and hot jets and their comparison with MHD flux emergence/cancellation models. Such observations can serve to validate the numerical experiments of the theoretical scientists. 



\begin{spacing}{0.9}



\bibliographystyle{apalike}
\cleardoublepage

\bibliography{References/references} 


\end{spacing}


\begin{appendices} 

\end{appendices}

\printthesisindex 

\end{document}